\documentclass[11pt, twoside, reqno]{amsart}
\usepackage[thm_section]{macros}
\definecolor{cellfill}{HTML}{C7E3E2}
\definecolor{cellline}{HTML}{21918C}
\definecolor{copyone}{HTML}{5EC962}
\definecolor{copytwo}{HTML}{3B528B}
\definecolor{samplegray}{HTML}{888780}
\definecolor{gridgray}{HTML}{D3D1C7}

\begin{document}
\begin{titlepage}
\begin{spacing}{1}
\title{\textbf{\Large Order-Explicit Linearization of High-Dimensional $U$-Statistics$^*$}}
\author{
\begin{tabular}[t]{c@{\extracolsep{4em}}c} 
\large{David M. Ritzwoller} &  \large{Vasilis Syrgkanis}\vspace{-0.7em}\\ \vspace{-1em}
\small{Stanford University} & \small{Stanford University} \\ \vspace{-0.7em}
\end{tabular}%
\\
}
\date{%
\today\\ $^*$Email: ritzwoll@stanford.edu, vsyrgk@stanford.edu. We thank Jiafeng Chen, Victor Chernozhukov, Harold Chiang, John Duchi, Matthew Gentzkow, Han Hong, Guido Imbens, Lihua Lei, Joseph Romano, Brad Ross, Jann Spiess and the audience at the Paris Econometrics Seminar for helpful comments and conversations. Ritzwoller gratefully acknowledges support from the National Science Foundation Graduate Research Fellowship and the Sloan Foundation Graduate Fellowship on the Fiscal and Economic Effects of Innovation and Productivity Policies, awarded through the National Bureau of Economic Research. Syrgkanis gratefully acknowledges support from the National Science Foundation through grant IS-2337916. Computational support was provided by the Data, Analytics, and Research Computing (DARC) group at the Stanford Graduate School of Business (RRI:SCR\_022938).}
                      
\begin{abstract}
\smalltonormalsize{We give an order-explicit large deviation bound for the difference between a high-dimensional $U$-statistic and its H\'{a}jek projection. In particular, we show that any $U$-statistic of order $b$ on $n$ observations, with a $d$-dimensional kernel whose coordinates have $\psi_1$-Orlicz norm at most $\phi$, has a maximum deviation from its H\'{a}jek projection of order $O_p(\phi b n^{-1}\log^2(dn))$. The proof relies on the development of novel order-explicit moment inequalities for higher-order Hoeffding components. We show that this rate is unimprovable, up to the polynomial factor on the logarithmic term. As corollaries, we obtain new Bernstein-type concentration and Gaussian approximation results for high-dimensional $U$-statistics. We apply these results to establish the consistency of a set of resampling-based simultaneous confidence intervals built around a class of nonparametric regression estimators constructed with subsampled kernels. This class encompasses several forms of random forest regression, including Generalized Random Forests.} 
\end{abstract}
\end{spacing}
\end{titlepage}
\maketitle
\thispagestyle{empty}
\setcounter{page}{1}
\begin{spacing}{1.3}

\section{Introduction}

Let $\mathbf{D}_n = (D_i)_{i=1}^n$ collect independent observations from a distribution $P$ on a set $\mathcal{D}$. For each symmetric, $d$-dimensional kernel function $u :\mathcal{D}^b  \to \mathbb{R}^d$, define the $b$-order $U$-statistic
\begin{align}
    U_{n,b}(u) = {n \choose b}^{-1} \sum_{\mathsf{s} \in \mathcal{S}_{n,b}} u(D_{\mathsf{s}})~,\label{eq: U general intro}
\end{align}
where the set $\mathcal{S}_{n,b}$ collects all of the subsets of $[n]=\{1,\ldots,n\}$ of size $b$ and the quantity $D_{\mathsf{s}}$ collects the subset of the observed data $\mathbf{D}_n$ with indices in the set $\mathsf{s}$. 

The canonical approach for studying the large-sample behavior of statistics with this structure, due to \cite{hoeffding1948class}, is based on the observation that the projection of the average \eqref{eq: U general intro} onto the space of statistics that decompose additively into the contributions of individual observations takes the form
\begin{align}
    \frac{b}{n} \sum_{i=1}^n u^{(1)}(D_i)~,
    \quad\text{where}\quad
    u^{(1)}(D) = \mathbb{E}[u(D_{[b]})\mid D_1=D]
    - \mathbb{E}[u(D_{[b]})]~.\label{eq: Hajek intro}
\end{align}
The average \eqref{eq: Hajek intro} is known as the H\'{a}jek projection of the $U$-statistic \eqref{eq: U general intro}, following \cite{hajek1968asymptotic}. The essential idea is that, if the residual from this projection can be shown to be of a lower stochastic order, then the problem reduces to the analysis of an average of i.i.d.\ random variables. In particular, under classical regularity conditions, \cite{hoeffding1948class} establishes that if the dimension $d = 1$ and the order $b$ is fixed, then
\begin{align}
    \sqrt{\frac{n}{\sigma^2_{b} b^2}} (U_{n,b}(u) - \mathbb{E}[u(D_{[b]})]) 
    \overset{d}{\to} \mathsf{N}(0,1)~,\quad\text{where}\quad \sigma^2_{b} = \Var(u^{(1)}(D_i))~.\label{eq: hoef clt intro}
\end{align}
See \cite{serfling1980approximation}, \cite{lee1990u}, and \cite{van2000asymptotic} for textbook treatments.

In this paper, we extend this approach to settings where the order $b$ and the dimension $d$ are non-negligible relative to the sample size $n$. Formally, we obtain the following order-explicit large deviation bound on the residual from the H\'{a}jek projection. Throughout, the quantities $c$ and $C$ denote universal positive constants, whose values are allowed to change in each appearance. We write $A \lesssim B$ if $A \leq C B$.

\begin{theorem}\label{lem: U stat linearization}
Consider a symmetric kernel function $u :\mathcal{D}^b  \to \mathbb{R}^d$. If the order $b$ satisfies $b \log(dn) \leq cn$ and each component $u_j(\cdot)$ of $u(\cdot)$ satisfies
\begin{align}
    \|u_j(D_{[b]}) \|_{\psi_1} \leq \phi~,\label{eq: orlicz norm bound state}
\end{align}
then the large deviation bound
\begin{align}
    \bigg\| U_{n,b}(u) 
    - \mathbb{E}[u(D_{[b]})] - \frac{b}{n}\sum_{i=1}^n u^{(1)}(D_i) \bigg\|_{\infty} \lesssim 
     \frac{b}{n} \phi\log^2 (dn)\label{eq: main Hajek bound}
\end{align}
holds with probability greater than $1-C/nd$.
\end{theorem}

\noindent We give the proof in \cref{sec: prf of U stat linearization}. Moreover, we show that the bound \eqref{eq: main Hajek bound} is unimprovable up to the polynomial factor on the logarithmic term. 

\cref{lem: U stat linearization} generates two corollaries through the application of the idea developed in \cite{hoeffding1948class}, each closing an open problem in the large-sample approximation of high-dimensional $U$-statistics. The first corollary is the following order-explicit Bernstein-type large deviation bound. The proof follows by combining \cref{lem: U stat linearization} with a standard Bernstein inequality for i.i.d.\ sums, and is deferred to \cref{app: proofs for u-stat}.

\begin{cor} \label{cor: general u stat bernstein}
Define the H\'{a}jek projection variance $\sigma^2_{b,j} = \Var(u^{(1)}_j(D_i))$ and set $\bar{\sigma}_b = \max_{j\in[d]} \sigma_{b,j}$. If the conditions of  \cref{lem: U stat linearization} hold, then the inequality
\begin{equation}\label{eq: general u stat moment}
\Big\| U_{n,b}(u) - \mathbb{E}[u(D_{[b]})] \Big\|_{\infty} 
\lesssim \sqrt{\frac{b^2 \bar{\sigma}^2_{b}\log(nd)}{n}}
+ \phi \frac{b}{n} \log^2(nd)
\end{equation}
holds with probability greater than $1-C/nd$.
\end{cor}

\noindent  Large deviation bounds for $U$-statistics were first given in \cite{hoeffding1963probability} (see Lemma A.5 of \cite{song2019approximating} for a modern statement). However, the leading term in the bound given there takes the form
\begin{align}
    \sqrt{\frac{b \bar{\nu} \log(dn)}{n}}~, \quad \text{where}\quad \bar{\nu} = \max_{j\in[d]} \nu_j
    \quad\text{and}\quad \nu_j = \Var(u_j(D_{[b]}))~.\label{eq: Hoeffding orig norm}
\end{align}
As the bound $ b\sigma^2_{b,j} \leq \nu_j$ follows from Hoeffding orthogonality, the quantity \eqref{eq: Hoeffding orig norm} is larger than the leading term in \eqref{eq: general u stat moment}.\footnote{In part motivated by the deficiency of the \cite{hoeffding1963probability} bound, \cite{arcones1993limit}, \cite{arcones1995bernstein}, \cite{gine2000exponential}, \cite{1adamczak2008tail}, and \cite{peel2010empirical} establish a series of refined large deviation bounds for high-dimensional $U$-statistics that use appropriate normalizing factors (among many other related results). However, the constants used to express these bounds depend implicitly on the order $b$. \cite{maurer2019bernstein} and \cite{minsker2023u} give large deviation bounds for $U$-statistics that can converge when the order is increasing, but place much stronger assumptions on the regularity and smoothness of the kernel under consideration.} Seen differently, through comparison to the asymptotic approximation \eqref{eq: hoef clt intro}, the leading term of the non-asymptotic bound \eqref{eq: general u stat moment} uses the correct normalizing quantity. 

The difference between the leading term in the bound \eqref{eq: general u stat moment} and the quantity \eqref{eq: Hoeffding orig norm} can be large. For example, \cite{song2019approximating} consider a non-parametric regression estimator based on a random forest construction that can be represented as a $b$-order $U$-statistic with $\sigma^2_{b,j} \asymp b^{-2}$ and $\nu_j \asymp 1$. In this case, the ratio $b\bar{\sigma}^2_{b} / \bar{\nu}$ is proportional to $b^{-1}$ and so the leading term in the bound \eqref{eq: general u stat moment} is negligible relative to the quantity \eqref{eq: Hoeffding orig norm} in regimes where $b$ is increasing. In more extreme, although not uncommon, cases the H\'{a}jek projection variance can be exactly zero, giving rise to a degenerate $U$-statistic. See, for instance, the test statistics considered in \cite{gretton2012kernel} and \cite{shekhar2023permutation}. In these settings, the bound \eqref{eq: general u stat moment} improves on the \cite{hoeffding1963probability}  bound by a factor of $(b/n)^{1/2}$, up to logarithmic terms.

The second corollary is the following order-explicit central limit theorem for high-dimensional $U$-statistics. The proof follows by combining \cref{lem: U stat linearization} with the high-dimensional central limit theorem for i.i.d.\ sums stated in \cite{chernozhuokov2022improved}, and is deferred to \cref{app: proofs for u-stat}.\footnote{The boundedness condition \eqref{eq: boundedness intro} can be relaxed to the bound $\mathbb{E}[ (u_j^{(1)}(D_i))^4] \leq \phi^2 \sigma_{b,j}^2$ for each $j$.}

\begin{cor}\label{cor: general u stat clt} Continue the notation introduced in \cref{lem: U stat linearization}. Define $\underline{\sigma}_b = \min_{j\in[d]} \sigma_{b,j}$. If the order $b$ satisfies $b \log(dn) \leq cn$ and the kernel $u(\cdot)$ satisfies the bound
\begin{align}
    \| u(D_{[b]}) - \mathbb{E}[u(D_{[b]})]\|_{\infty} \leq \phi\label{eq: boundedness intro}
\end{align}
almost surely, then the inequality
\begin{align}\label{eq: general u stat clt}
\sup_{\mathsf{R} \in \mathcal{R}} \bigg\vert 
P\left\{\sqrt{\frac{n}{b^2}}\Sigma_b^{-1/2} (U_{n,b}- \mathbb{E}[u(D_{[b]})]) \in \mathsf{R}\right\} - P\left\{ \Sigma_b^{-1/2}Z \in \mathsf{R}\right\}\bigg\vert \lesssim \left(\frac{ \phi^2 \log^5(dn)}{\underline{\sigma}^2_b n}\right)^{1/4}
\end{align}
holds, where $\Sigma_b = \mathsf{Diag}(\sigma^2_{b,1},\ldots,\sigma^2_{b,d})$, $Z$ denotes a centered Gaussian vector in $\mathbb{R}^d$ with the same covariance matrix as $u^{(1)}(D_i)$, and $\mathcal{R}$ denotes the set of hyper-rectangles in $\mathbb{R}^d$.
\end{cor}

\noindent The \cite{hoeffding1948class} univariate central limit \eqref{eq: hoef clt intro} was extended only recently to the regime where $b$ is non-negligible relative to  $n$. Under the condition that $b^{-1} \lesssim \sigma_b^2$, \cite{diciccio2022clt} give a result that applies to the regime $b = o(n^{1/2})$. \cite{wager2018estimation} and \cite{peng2022rates} strengthen this result to the full regime $b=o(n)$. The results developed in each of these papers apply only to the case that $d = 1$. In parallel, \cite{chen2018gaussian}, \cite{chen2019randomized}, and \cite{song2019approximating} give results that apply to the case where $d$ can be large relative to $n$. Of these papers, only \cite{song2019approximating} gives results with explicit dependence on the order $b$. However, their results are only applicable to the regime $b^2 \underline{\sigma}_b^{-2}\log^7(dn) = o(n)$. Thus, in settings where $b^{-1} \lesssim \underline{\sigma}^2_b$, this result applies to the regime $b^3\log^7(dn) = o(n)$. By contrast, in this case, the central limit theorem \eqref{eq: general u stat clt} applies to the full regime $b\log^5(dn) = o(n)$. 

This refinement can be essential for the analysis of modern statistical methods that aggregate algorithms computed on subsamples of a larger data set. To illustrate this point, in \cref{sec: random forest}, we apply \cref{cor: general u stat clt} to establish the consistency of an approach for constructing resampling-based simultaneous confidence intervals around nonparametric regression estimators based on subsampled kernels, including subsampled nearest-neighbor regression
\citep{fix1989discriminatory,khosravi2019non,demirkaya2024optimal} and random forest regression
\citep{breiman2001random,mentch2016quantifying,wager2018estimation}.

Our analysis again follows the agenda set out by \cite{hoeffding1948class}, and makes essential use of \cref{cor: general u stat clt}. We first show that, under regularity conditions, the estimators under consideration are well-approximated by $U$-statistics of order $b$ with $d$-dimensional kernels that satisfy the bound $b^{-1} \lesssim \sigma_b^2$. However, these estimators are only consistent if the subsample size $b$ satisfies $n^{p/(p+c)}=o(b)$, where $p$ is the dimension of the available covariates. This places the relevant regime outside of the range $b=o(n^{1/3})$ covered by available high-dimensional central limit theorems, while scalar results that allow for larger values of $b$ cannot account for growth in the kernel dimension $d$. \cref{cor: general u stat clt} closes this gap by showing that high-dimensional Gaussian approximation holds at the subsample sizes required for consistency.

Our treatment of this application builds on \cite{oprescu2019orthogonal}, who augment the Generalized Random Forests framework proposed by \cite{athey2019generalized} to study pointwise inference for subsampled random forest estimators with nuisance parameters. Our results are comparable to the bounds on the accuracy of Gaussian multiplier bootstrap confidence regions for nonparametric regression and $Z$-estimation given in \cite{chernozhukov2014anti} and \cite{belloni2018uniformly}. We generalize these results, in the sense that we treat inference for conditional $Z$-estimators whose score function is potentially unknown to the researcher.

The improvement of \cref{lem: U stat linearization} over existing results is the consequence of a set of new order-explicit moment inequalities for higher-order components of Hoeffding decompositions. In the scalar case, existing arguments rely on pairing the total orthogonality of the Hoeffding decomposition with Chebyshev-type bounds. This approach is insufficient for the high-dimensional case, which requires exponential tail bounds. Existing high-dimensional results instead build on the decoupling and symmetrization arguments surveyed in \cite{de2012decoupling}. However, available decoupling bounds incur poor dependence on the order $b$, which becomes prohibitive when $b$ is non-negligible. 

Our main technical innovation is a new order-explicit one-sided decoupling bound for $U$-statistics that is sufficient to resolve this tension. We show that the quantity that results from this bound can be controlled by combining a symmetrization argument due to \cite{sherman1994maximal} with a specialization of a moment inequality from \cite{gine2000exponential}. This approach yields exponential tail bounds for higher-order Hoeffding terms that have the correct scaling in the regime where both $b$ and $d$ are large. 

Additional applications of large order $U$-statistics to split-sample testing, robust estimation, and kernel density estimation are reviewed in \cite{diciccio2022clt}, \cite{minsker2023u}, and \cite{song2019approximating}. Omitted proofs are given in \cref{app: proofs for u-stat,sec: generic,app: kernel verify pf}. \cref{app: additional,app: bin boot proofs,app: supplemental figures} give additional results and discussion that will be introduced at appropriate points throughout the paper.

\section{Proof of \cref{lem: U stat linearization}\label{sec: prf of U stat linearization}}

In this section, we outline the proof of \cref{lem: U stat linearization}. Throughout the proof and supporting Lemmas, we maintain $\mathbb{E}[u(D_{[b]})] = 0$ without loss. The result is obtained in two steps. First, we decompose the difference between the $U$-statistic under consideration and its H\'{a}jek projection through a Hoeffding expansion \citep{hoeffding1948class,efron1981jackknife}. We adopt the following formulation of Hoeffding's expansion based on the use of stochastic differences, stated as Part (i) of the following Lemma (see e.g., \cite{lachieze2017new}). Part (ii) gives the well-known  specialization of this result to $U$-statistics (see e.g., Lemma 5.1.5 A of \cite{serfling1980approximation}). 

\begin{lemma}[Hoeffding's Expansion]\label{lem: Diff H decomposition}
Fix a deterministic function $f : \mathcal{D}^m \to \mathbb{R}$. 

\noindent{\textbf{(i)}} Let $D^\prime_{[m]}$ denote an independent copy of $D_{[m]}$. Construct $D^{(i)}_{[m]}$ by replacing $D_i$ with $D_i^\prime$ in $D_{[m]}$, leaving all other entries unchanged. For each index $i$, define the difference operator
\begin{equation}
    \nabla_i f(D_{[m]}) = f(D^{(i)}_{[m]}) - f(D_{[m]})~.
\end{equation}
If the moment $\mathbb{E}[f(D_{[m]})^2]$ is finite, then the representation
\begin{align}
    f(D_{[m]})
    & = \mathbb{E}[f(D_{[m]})]
    + \sum_{l=1}^m \sum_{\mathsf{s} \in \mathcal{S}_{m,l}}f^{(l)}(D_{\mathsf{s}})~,\quad\text{where} 
    \label{eq: H decompotion}  \\
    f^{(l)}(D_{i_1}, ..., D_{i_l}) 
    & = (-1)^l \mathbb{E}[\nabla_{i_1}\cdots \nabla_{i_l} f(D_{[m]})\mid D_{i_1}, ..., D_{i_l}]~, \nonumber
\end{align}
holds, the non-constant summands appearing in \eqref{eq: H decompotion} are mean-zero, and all $2^m$ summands are uncorrelated.

\noindent{\textbf{(ii)}} For each integer $l\leq n$, define the operator
    \begin{align}
        U_{n,l}(h) = {n \choose l}^{-1} \sum_{\mathsf{s} \in \mathcal{S}_{n,l}} h(D_{\mathsf{s}})
    \end{align}
    on the set of measurable functions $h :\mathcal{D}^l\to \mathbb{R}$. If the statistic $f(D_{[m]})$ is mean-zero and has a finite variance and the function $f(\cdot)$ is symmetric in its arguments, then the decomposition
    \begin{align}
        U_{n,m}(f) = \sum_{l=1}^m {m \choose l} U_{n,l}(f^{(l)})\label{eq: Hoeff decomp apply to U}
    \end{align}
    holds and all $m$ summands appearing in \eqref{eq: Hoeff decomp apply to U} are uncorrelated.
\end{lemma}

\noindent In particular, specialized to our setting, we obtain the decomposition
\begin{align}
    U_{n,b}(u) - \frac{b}{n}\sum_{i=1}^n u^{(1)}(D_i)
    & = \sum_{m=2}^b {b \choose m} U_{n,m}(u^{(m)})\label{eq: break up Hajek} 
\end{align}
by applying Part (ii) of \cref{lem: Diff H decomposition}. 

Second, and more substantively, we bound each of the summands of the decomposition \eqref{eq: break up Hajek} by applying the following new higher-order moment bound.

\begin{lemma}\label{lem: individual degenerate}
    Fix an integer $1 \leq m \leq \min\{ b, n/2\}$. If the function $f:\mathcal{D}^b \to \mathbb{R}$ is symmetric in its arguments and satisfies the bound 
    \begin{align}
        \| f(D_{[b]})\|_{\psi_1} \leq \phi~,
    \end{align}
    then the higher-order moment inequality
    \begin{align}
        \mathbb{E}\left[ 
        \bigg\vert {b \choose m} U_{n,m}(f^{(m)}) \bigg\vert^q \right]^{1/q} & \leq
        \phi q \left( \frac{C qb}{n} \right)^{m/2}
    \end{align}
    holds for each $2 \vee 2\log(n) \leq q \leq c n/b$.
\end{lemma}
\noindent 
The essential novelty of the bound is that the dependence on the order $m$ is kept explicit. The proof, given in the following subsection, develops a new one-sided decoupling bound for $U$-statistics of an arbitrary order, which avoids introducing order-dependent constants present in previous work (see, for comparison, \cite{de1995decoupling} and Chapter 3.1 of \cite{de2012decoupling}). The argument then proceeds by combining symmetrization, Bonami's inequality, and a Rosenthal-type decomposition due to \cite{gine2000exponential} to control the resulting quantity. The final bound is obtained through the application of a new estimate for higher-order moments of conditional expectations of Hoeffding components, derived from a further application of Hoeffding orthogonality.

To complete the proof, observe that by setting $q = 2 \vee 2\log(nd)$ and the constant $c$ sufficiently small, the assumption $b \log(nd) \leq c n$ guarantees that $q \leq c n / b$ and $b\leq n/2$, so that \cref{lem: individual degenerate} applies. Thus, for each integer $m \in \{2,\ldots,b\}$ and each component $u_j(\cdot)$ of the kernel $u(\cdot)$, Markov's inequality yields
\begin{align}
    & P\left\{ 
    \bigg\vert 
    {b \choose m} U_{n,m}(u_j^{(m)})
    \bigg\vert 
    \geq 2 e \phi \log(nd) \left( \frac{Cb \log(nd)}{n} \right)^{m/2}
    \right\} \nonumber \\
    & \leq \exp(-2\log(nd))
    \frac{\mathbb{E}\left[ 
        \bigg\vert {b \choose m} U_{n,m}(u_j^{(m)}) \bigg\vert^{2\log(nd)} \right]}{\left( \phi\log(nd) \left( \frac{C b \log(nd)}{n} \right)^{m/2} \right)^{2\log(nd)}} \nonumber \\
      &\leq \exp(-2\log(nd)) = \frac{1}{n^2d^2}~.
\end{align}
Moreover, by taking the universal constant $c$ sufficiently small, the condition $b \log(dn) \leq c n$ again implies
\begin{align}
    \phi \log(dn) \sum_{m=2}^b \left( \frac{C  b \log(dn)}{n} \right)^{m/2}
    \lesssim \frac{\phi b \log^2(dn)}{n}~.
\end{align}
Consequently, the decomposition \eqref{eq: break up Hajek} and a union bound imply that
\begin{align}
    & P\left\{ 
    \bigg\| U_{n,b}(u) - \frac{b}{n}\sum_{i=1}^n u^{(1)}(D_i) \bigg\|_{\infty} 
    \geq C \frac{\phi b \log^2(dn)}{n}
    \right\} \nonumber \\
    & \leq \sum_{j=1}^d \sum_{m=2}^b
    P\left\{
    \bigg\vert {b \choose m} U_{n,m}(u_j^{(m)}) \bigg\vert
    \geq 2 e \phi \log(nd) \left( \frac{C b \log(nd)}{n} \right)^{m/2}
    \right\} \nonumber \\
    & \leq \frac{b-1}{n^2 d}
    \leq \frac{C}{nd}~,\label{eq: obtain final bound}
\end{align}
as required.\hfill\qed

\begin{remark}\label{rem: lower bound}
    The bound \eqref{eq: obtain final bound} is unimprovable up to the polynomial factor on the logarithmic term. In particular, in \cref{app: proofs for u-stat} we show that, for each $n \geq b \geq 2$ and $d \geq 1$, there exists a symmetric kernel function $u :\mathcal{D}^b  \to \mathbb{R}^d$ whose coordinates satisfy the bound \eqref{eq: orlicz norm bound state} and such that the lower bound
    \begin{align}
        P\left\{
        \bigg\| U_{n,b}(u) - \mathbb{E}[u(D_{[b]})]
        - \frac{b}{n}\sum_{i=1}^n u^{(1)}(D_i) \bigg\|_{\infty}
        \geq c
        \frac{\phi b \log(nd)}{n}
        \right\} \geq \frac{c}{nd}~.
    \end{align}
    holds. That is, the factor $\phi b/n$ and at least one logarithmic factor in $nd$ are unavoidable.\hfill$\blacksquare$
\end{remark}

\subsection{Proof of \cref{lem: individual degenerate}\label{sec: orf of individual degenerate}}

Classical bounds for moments of higher-order Hoeffding components are obtained from an instance of a decoupling and symmetrization argument. The motivating idea is that a $U$-statistic would be easier to analyze if each of the arguments of its kernel were computed on a separate, independent sample. To see this, let $\mathbf{D}_n^{(1)}, \ldots, \mathbf{D}_n^{(m)}$ denote $m$ independent copies of the data $\mathbf{D}_n$ and consider the decoupled statistic
\begin{align}
    \widetilde{U}_{n,m}(f^{(m)})
    =
    \binom{n}{m}^{-1} \sum_{(i_1, \ldots, i_m) \in \mathcal{S}_{n,m}}
    f^{(m)}(D_{i_1}^{(1)}, \ldots, D_{i_m}^{(m)})~.
    \label{eq: decoupled u statistic}
\end{align}
Panels A and B of \cref{fig: index sets} visualize the summands of the
statistics $U_{n,2}(g)$ and $\widetilde{U}_{n,2}(g)$, respectively. As distinct observations from the sample $\mathbf{D}_n^{(l)}$ always appear in distinct summands, by conditioning on all copies of the data other than the $l$th, the statistic \eqref{eq: decoupled u statistic} can be viewed as an average of independent random variables. Standard moment bounds for independent sums, in particular symmetrization and Khintchine-type inequalities, can therefore be applied over the randomness induced by the $l$th coordinate by conditioning in this way, while leaving the independence structure of the remaining coordinates intact. By iteratively applying a moment bound of this form, once in each coordinate, with each application contributing a factor of order $n^{-1/2}$, it is possible to obtain a moment bound for $\widetilde{U}_{n,m}(f^{(m)})$ of order $n^{-m/2}$.\footnote{Each application of the symmetrization inequality requires
that the summands be mean-zero conditional on the remaining
coordinates. This holds at every iteration because the Hoeffding
component $f^{(m)}$ is completely degenerate, in the sense that it has
mean zero conditional on any strict subset of its arguments. The
argument is therefore not applicable to $U$-statistics with general
kernels.} See, e.g., \citet{de1992decoupling}, \citet{arcones1993limit}, and \citet{sherman1994maximal}.

\begin{figure}[t]
\begin{centering}
\caption{Decoupling Construction}
\label{fig: index sets}
\medskip{}
\begin{tabular}{c}
\begin{tikzpicture}[
  every path/.append style={arrows=-},
  panel title/.style={font=\small},
  panel sub/.style={font=\footnotesize, text=black!55},
  axis label/.style={font=\footnotesize},
  panel letter/.style={font=\small},
  bar/.style={line width=2.2pt},
]
 
\begin{scope}[shift={(0cm,0cm)}, x=0.22cm, y=0.22cm]
  \fill[cellfill] (0,0) rectangle (12,12);
  \foreach \k in {0,...,11} \fill[white] (\k,{11-\k}) rectangle ++(1,1);
  \draw[gridgray, line width=0.3pt, step=1] (0,0) grid (12,12);
  \draw[cellline, line width=0.5pt] (0,0) rectangle (12,12);
  \draw[bar, samplegray] (0,-0.9) -- (12,-0.9);
  \draw[bar, samplegray] (-0.9,0) -- (-0.9,12);
  \node[axis label, below] at (6,-1.5) {$i_1$};
  \node[axis label, left]  at (-1.5,6) {$i_2$};
  \node[panel title] at (6,16.9) {$U_{n,2}(g)$};
  \node[panel sub]   at (6,14.9) {One sample};
  \node[panel sub]   at (6,13.4) {Distinct indices};
  \node[panel letter] at (6,-4.4) {(a)};
\end{scope}
 
\begin{scope}[shift={(3.95cm,0cm)}, x=0.22cm, y=0.22cm]
  \fill[cellfill] (0,0) rectangle (12,12);
  \foreach \k in {0,...,11} \fill[white] (\k,{11-\k}) rectangle ++(1,1);
  \draw[gridgray, line width=0.3pt, step=1] (0,0) grid (12,12);
  \draw[cellline, line width=0.5pt] (0,0) rectangle (12,12);
  \draw[bar, copyone] (0,-0.9) -- (12,-0.9);
  \draw[bar, copytwo] (-0.9,0) -- (-0.9,12);
  \node[axis label, below] at (6,-1.5) {$i_1$};
  \node[axis label, left]  at (-1.5,6) {$i_2$};
  \node[panel title] at (6,16.9) {$\widetilde{U}_{n,2}(g)$};
  \node[panel sub]   at (6,14.9) {Distinct samples};
  \node[panel sub]   at (6,13.4) {Distinct indices};
  \node[panel letter] at (6,-4.4) {(b)};
\end{scope}
 
\begin{scope}[shift={(7.9cm,0cm)}, x=0.22cm, y=0.22cm]
  \fill[cellfill] (0,6) rectangle (6,12);
  \draw[gridgray, line width=0.3pt, step=1] (0,0) grid (12,12);
  \draw[cellline, line width=0.5pt] (0,0) rectangle (12,12);
  \draw[bar, copyone] (0,-0.9) -- (6,-0.9);
  \draw[bar, copytwo] (-0.9,6) -- (-0.9,12);
  \node[axis label, below] at (6,-1.5) {$i_1$};
  \node[axis label, left]  at (-1.5,6) {$i_2$};
  \node[panel title] at (6,16.9) {$\bar{U}_{n,2}(g)$};
  \node[panel sub]   at (6,14.9) {Distinct samples};
  \node[panel sub]   at (6,13.2) {Repeated indices allowed};
  \node[panel letter] at (6,-4.4) {(c)};
\end{scope}
 
\begin{scope}[shift={(11.85cm,0cm)}, x=0.22cm, y=0.22cm]
  \foreach \i in {2,3,5,8,9,12}
    \foreach \j in {1,4,6,7,10,11}
      \fill[cellfill] ({\i-1},{12-\j}) rectangle ({\i},{13-\j});
  \draw[gridgray, line width=0.3pt, step=1] (0,0) grid (12,12);
  \draw[cellline, line width=0.5pt] (0,0) rectangle (12,12);
  \foreach \i in {2,3,5,8,9,12}
    \draw[bar, samplegray] ({\i-1},-0.9) -- ({\i},-0.9);
  \foreach \j in {1,4,6,7,10,11}
    \draw[bar, samplegray] (-0.9,{12-\j}) -- (-0.9,{13-\j});
  \node[axis label, below] at (6,-1.5) {$i_1$};
  \node[axis label, left]  at (-1.5,6) {$i_2$};
  \node[panel title] at (6,16.9) {$U_{n,2}(g,\mathsf{R})$};
  \node[panel sub]   at (6,14.9) {One sample};
  \node[panel sub]   at (6,13.4) {Random distinct indices};
  \node[panel letter] at (6,-4.4) {(d)};
\end{scope}
 
\begin{scope}[shift={(0.6cm,-1.8cm)}]
  \fill[cellfill] (0,0) rectangle (0.24,0.24);
  \draw[cellline, line width=0.4pt] (0,0) rectangle (0.24,0.24);
  \node[anchor=west, font=\footnotesize] at (0.32,0.12)
    {Included pair $(i_1, i_2)$};
  \draw[bar, samplegray] (4.5,0.12) -- (4.82,0.12);
  \node[anchor=west, font=\footnotesize] at (4.9,0.12)
    {Sample $\mathbf{D}_n$};
  \draw[bar, copyone] (7.8,0.12) -- (8.12,0.12);
  \node[anchor=west, font=\footnotesize] at (8.2,0.12)
    {Copy $\mathbf{D}_n^{(1)}$};
  \draw[bar, copytwo] (10.9,0.12) -- (11.22,0.12);
  \node[anchor=west, font=\footnotesize] at (11.3,0.12)
    {Copy $\mathbf{D}_n^{(2)}$};
\end{scope}
 
\end{tikzpicture}\tabularnewline
\end{tabular}
\par\end{centering}
\medskip{}
\justifying
{\footnotesize{}Notes: \cref{fig: index sets} illustrates the construction of the statistics considered in the proof of \cref{lem: individual degenerate}, in the case $m = 2$ and $n = 12$. Each panel displays a grid enumerating the pairs of indices $(i_1, i_2)$. Shaded cells indicate the pairs included in the corresponding sum. As each kernel under consideration is symmetric, ordered pairs are displayed. The bars adjacent to each axis indicate the sample indexed by the corresponding coordinate. Positions without a bar do not enter the sum. Panel A displays the $U$-statistic $U_{n,2}(g)$, which averages over all pairs of distinct indices of a single sample. Panel B displays the decoupled statistic $\widetilde{U}_{n,2}(g)$, defined in \eqref{eq: decoupled u statistic}, which retains the same index set, but evaluates each coordinate of the kernel on an independent copy of the data. Panel C displays the decoupled statistic $\bar{U}_{n,2}(g)$, defined in \eqref{eq: decoupled Us prime}, which averages over all pairs of the first $\lfloor n/2 \rfloor$ indices of two independent copies of the data. As the coordinates displayed in Panel C index distinct samples, pairs with $i_1 = i_2$ reference distinct observations. Panel D displays the statistic $U_{n,2}(g, \mathsf{R})$, defined in \eqref{eq: decoupled Us}, which averages over the product of the disjoint random cells $\mathsf{r}_1$ and $\mathsf{r}_2$ of a single sample. The conditional expectation of the statistic displayed in Panel D, given the data $\mathbf{D}_n$, is equal to the statistic displayed in Panel A. The statistics displayed in Panels C and D are identically distributed.}{\footnotesize\par}
\end{figure}

The core challenge, then, is to show that the moments of the decoupled statistic \eqref{eq: decoupled u statistic} can be used to bound the moments of the Hoeffding component $U_{n,m}(f^{(m)})$. This is the role of a decoupling inequality. Inequalities of this form are the subject of a large literature, surveyed in Chapter~3 of \citet{de2012decoupling}. The sharpest available bound is due to \citet{de1995decoupling}, who show that
\begin{align}
    \mathbb{E}\big[ \vert U_{n,m}(g) \vert^q \big]^{1/q}
    \leq
    h(m) \,
    \mathbb{E}\big[ \vert \widetilde{U}_{n,m}(g) \vert^q \big]^{1/q}
    \label{eq: dlp and ms bound}
\end{align}
holds for any symmetric kernel $g$, where the sequence $h(m) = m^{\Theta(m^2)}$ depends only on the order $m$. The deficiency of this bound is its dependence on $m$. If the order $m$ is fixed, then $h(m)$ can be absorbed into universal constants. In our setting, however, the decomposition \eqref{eq: break up Hajek} involves components of every order $2 \leq m \leq b$. If $m$ is non-negligible, the factor $m^{\Theta(m^2)}$ can grow much more quickly than the factor $n^{-m/2}$ shrinks, and so bounds obtained from \eqref{eq: dlp and ms bound} are useful only if $b$ grows very slowly with $n$.

We resolve this issue by developing the following new one-sided decoupling bound for higher-order moments of $U$-statistics. In contrast to the classical bound \eqref{eq: dlp and ms bound}, the multiplicative factor appearing in this inequality is equal to one.

\begin{lemma}\label{lem: one-sided decoupling}
Define the decoupled statistic
\begin{align}
    \bar{U}_{n,m}(g) & =
    \frac{1}{\lfloor n/m\rfloor^m}
    \sum_{i_1=1}^{\lfloor n/m\rfloor} \cdots \sum_{i_m=1}^{\lfloor n/m\rfloor}
    g(D_{i_1}^{(1)}, \ldots, D_{i_m}^{(m)})\label{eq: decoupled Us prime}
\end{align}
for each symmetric kernel $g : \mathcal{D}^m \to \mathbb{R}$. The inequality
    \begin{align}
        \mathbb{E}\left[ 
        \vert  U_{n,m}(g) \vert^q \right]^{1/q}
        \leq 
        \mathbb{E}\left[ 
        \vert  \bar{U}_{n,m}(g)\vert^q \right]^{1/q}\label{eq: apply decouple}
\end{align}
holds.
\end{lemma}
\begin{proof}
    Let $\mathsf{R} = (\mathsf{r}_1,\ldots,\mathsf{r}_m,\tilde{\mathsf{r}})$ be a random partition of $[n]$, such that each cell $\mathsf{r}_i$ has cardinality $\lfloor n/m \rfloor$ and the remainder cell $\tilde{\mathsf{r}}$ has cardinality $n - m \lfloor n/m \rfloor$. Define the statistic
    \begin{align}
    U_{n,m}(g, \mathsf{R}) & = 
    \frac{1}{\lfloor n/m\rfloor^m}
    \sum_{i_1 \in \mathsf{r}_1} \cdots \sum_{i_m\in \mathsf{r}_m}
    g(D_{i_1}, \ldots, D_{i_m})~.\label{eq: decoupled Us}
    \end{align}
    Observe that the statistics \eqref{eq: decoupled Us prime} and
    \eqref{eq: decoupled Us} are identically distributed; compare Panels C and D of \cref{fig: index sets}. In turn, observe that the elements of the fixed set $\mathsf{s}\in \mathcal{S}_{n,m}$ are assigned to distinct cells of the partition $\mathsf{r}_1,\ldots,\mathsf{r}_m$ with probability 
    \begin{equation}
    m! \frac{\lfloor n/m \rfloor}{n} \frac{\lfloor n/m \rfloor}{n-1}\cdots \frac{\lfloor n/m \rfloor}{n - m + 1} = \lfloor n/m \rfloor^m {n \choose m}^{-1}~.
    \end{equation}
    Thus, we can evaluate
    \begin{align}
    \mathbb{E}\left[ 
    U_{n,m}(g, \mathsf{R})
    \mid \mathbf{D}_n\right]
    & = {n \choose m}^{-1}
    \sum_{\mathsf{s} \in \mathcal{S}_{n,m}} g(D_{\mathsf{s}})
    = U_{n,m}(g)~.
    \end{align}
    Consequently, as $q \geq 1$, Jensen's inequality implies that
    \begin{align}
        \mathbb{E}\left[ 
        \vert U_{n,m}(g) \vert^q \right]^{1/q}
        \leq 
        \mathbb{E}\left[ 
        \vert U_{n,m}(g, \mathsf{R}) \vert^q \right]^{1/q}~.
    \end{align}
    But then, as \eqref{eq: decoupled Us prime} and \eqref{eq: decoupled Us} are identically distributed, the bound \eqref{eq: apply decouple} also holds.\hfill
\end{proof}

\begin{remark}\label{rem: hoeffding representation}
The construction applied in the proof of \cref{lem: one-sided decoupling} is reminiscent of the representation
\begin{align}
    U_{n,m}(g)
    =
    \frac{1}{n!} \sum_{\pi \in \mathcal{P}_n}
    \bigg\lfloor \frac{n}{m} \bigg\rfloor^{-1}
    \sum_{l=1}^{\lfloor n/m \rfloor}
    g(D_{\mathsf{s}_{\pi,l}})~,
    \label{eq: Hoeff perm rep}
\end{align}
given in \citet{hoeffding1948class}, where $\mathcal{P}_n$ denotes the set of permutations of $[n]$ and $\mathsf{s}_{\pi,l} = \{\pi((l-1)m +
1), \ldots, \pi(lm)\}$. In particular, Jensen's inequality and the representation \eqref{eq: Hoeff perm rep} imply that
\begin{align}
    \mathbb{E}\big[ \vert U_{n,m}(g) \vert^q \big]^{1/q}
    \leq
    \mathbb{E}\Bigg[ \bigg\vert
    \bigg\lfloor \frac{n}{m} \bigg\rfloor^{-1}
    \sum_{l=1}^{\lfloor n/m \rfloor}
    g(D_{\mathsf{s}_{\pi,l}})
    \bigg\vert^q \Bigg]^{1/q}~,
    \label{eq: Hoeffding Jensen}
\end{align}
where, again, the multiplicative factor is equal to one. Bounds of this form are used to obtain the classical \citet{hoeffding1963probability} inequality, whose leading term is displayed in \eqref{eq: Hoeffding orig norm}.\hfill$\blacksquare$
\end{remark}

The essential feature of the bound \eqref{eq: apply decouple} is that the dominating statistic \eqref{eq: decoupled Us prime} retains the structure that made the classical decoupled statistic \eqref{eq: decoupled u statistic} tractable; compare Panels B and C of
\cref{fig: index sets}. Each argument of the kernel is again computed on its own independent copy of the data, and distinct observations from the $l$th copy again appear in distinct summands. The following Lemma illustrates that the iterative symmetrization argument described at the beginning of this section can thus be applied to $\bar{U}_{n,m}(g)$, without incurring a factor analogous to $h(m)$. We defer the proof to \cref{app: proofs for u-stat}.

\begin{lemma}\label{lem: symm and khint}For each $l$ in $[m]$, let $V^{(l)}=(V^{(l)}_i)_{i=1}^n$ denote a collection of i.i.d.\ Rademacher variables.\\
    \noindent \textbf{(i)} 
    If the symmetric function $g : \mathcal{D}^m \to \mathbb{R}$ satisfies the condition
    \begin{equation}
        \mathbb{E}[g(D_1, D_2, \ldots, D_m) \mid D_2, \ldots,D_m] = 0\label{eq: completely degenerate}
    \end{equation}
    almost surely, then the moment inequality
    \begin{align}
        & \mathbb{E}\left[ 
        \vert  \bar{U}_{n,m}(g) \vert^q \right]^{1/q} \nonumber \\
        & \leq
        \frac{2^m}{\lfloor n/m\rfloor^m}
        \mathbb{E}\left[ 
        \bigg\vert  
        \sum_{i_1=1}^{\lfloor n/m\rfloor} \cdots \sum_{i_m=1}^{\lfloor n/m\rfloor}
        V_{i_1}^{(1)} \cdots V_{i_m}^{(m)}
        g(D_{i_1}^{(1)},\ldots,D_{i_m}^{(m)})
        \bigg\vert^q \right]^{1/q}
    \end{align}
    holds for each $q\geq 1$.
    
    \noindent \textbf{(ii)} Fix the deterministic real-valued array $\{g_{(i_1, ...,i_m)}: i_j \in [\lfloor n/m\rfloor]\text{ for all } j\in[m]\}$. The moment inequality
    \begin{align}
        & \mathbb{E}\left[
        \bigg\vert
        \sum_{i_1=1}^{\lfloor n/m\rfloor} \cdots \sum_{i_m=1}^{\lfloor n/m\rfloor}
        V_{i_1}^{(1)} \cdots V_{i_m}^{(m)} g_{(i_1, ...,i_m)}
        \bigg\vert^q
        \right]^{1/q} \nonumber \\
        & \leq 
        q^{m/2}
        \left(
        \sum_{i_1=1}^{\lfloor n/m\rfloor} \cdots \sum_{i_m=1}^{\lfloor n/m\rfloor}
        g_{(i_1, ...,i_m)}^2
        \right)^{1/2}
    \end{align}
    holds for each $q > 2$. 
\end{lemma}

\noindent Part (i) is obtained by an iterated application of a standard symmetrization inequality for sums of independent, centered random variables; see e.g., Section 6.4 of \citet{vershynin2018high}. Part (ii) is a direct specialization of the Bonami inequality for homogeneous Rademacher chaos, stated as Theorem 3.2.2 of \citet{de2012decoupling}; see also \citet{sherman1994maximal}.\footnote{A closely related symmetrization inequality, also based on \citet{sherman1994maximal}, is applied by \citet{song2019approximating} directly to the Hoeffding components $U_{n,m}(f^{(m)})$, without decoupling. Paired with the Bonami inequality, this argument again attains the scale $n^{-m/2}$ for each component. However, a quadratic dependence on $b$ in the resulting bounds arises due to their treatment of the analogue of the sum of squares appearing in \eqref{eq: apply symm and khint}. There, each summand is bounded by a quantity that does not depend on the order $m$, obtained from the tail condition placed on the kernel, so that the bound obtained for the sum is proportional to the number of summands. In fact, Hoeffding
orthogonality implies that the moments of the squared component $(f^{(m)})^2$ decrease at the rate $\binom{b}{m}^{-1}$ as the order $m$ increases. \cref{lem: bound parts} exploits this decay, through an argument that relies on the decoupled structure of the statistic $\bar{U}_{n,m}(f^{(m)})$. \citet{minsker2023u} applies the same symmetrization inequality in a way analogous to \citet{song2019approximating}, with bounds on exponential moments in place of bounds on first moments, under substantially stronger smoothness conditions on the kernel.}

To apply \cref{lem: symm and khint}, recall that the Hoeffding projection $f^{(m)}$ is completely degenerate, in the sense of  the condition \eqref{eq: completely degenerate}. Thus, Parts (i) and (ii) of \cref{lem: symm and khint} and the law of iterated expectation imply that
\begin{align}
    &\mathbb{E}\left[ 
        \bigg\vert {b \choose m} \bar{U}_{n,m}(f^{(m)}) \bigg\vert^q \right]^{1/q}\nonumber\\
        &\leq
        \frac{2^m}{\lfloor n/m\rfloor^m}
        \mathbb{E}\left[ 
        \bigg\vert  {b \choose m}
        \sum_{i_1=1}^{\lfloor n/m\rfloor} \cdots \sum_{i_m=1}^{\lfloor n/m\rfloor}
        V_{i_1}^{(1)} \cdots V_{i_m}^{(m)}
        f^{(m)}(D_{i_1}^{(1)},\ldots,D_{i_m}^{(m)})
        \bigg\vert^q \right]^{1/q} \nonumber \\
        & \leq
        C^m q^{m/2} \lfloor n/m\rfloor^{-m} {b \choose m}
        \mathbb{E}\left[
        \bigg\vert
        \sum_{i_1=1}^{\lfloor n/m\rfloor} \cdots \sum_{i_m=1}^{\lfloor n/m\rfloor}
        f^{(m)}(D_{i_1}^{(1)},\ldots,D_{i_m}^{(m)})^2
        \bigg\vert^{q/2}
        \right]^{1/q}~.\label{eq: apply symm and khint}
\end{align}
Hence, by combining \cref{lem: one-sided decoupling} and \eqref{eq: apply symm and khint}, we obtain the inequality
\begin{align}
    & \mathbb{E}\left[ 
        \bigg\vert {b \choose m} U_{n,m}(f^{(m)}) \bigg\vert^q \right]^{1/q} \label{eq: post decouple and symm}\\
    & \leq
    C^m q^{m/2} \lfloor n/m\rfloor^{-m/2} {b \choose m} \nonumber \\
       & \quad\quad \cdot \mathbb{E}\left[
        \bigg\vert
        \frac{1}{\lfloor n/m\rfloor^m}
        \sum_{i_1=1}^{\lfloor n/m\rfloor} \cdots \sum_{i_m=1}^{\lfloor n/m\rfloor}
        f^{(m)}(D_{i_1}^{(1)},\ldots,D_{i_m}^{(m)})^2
        \bigg\vert^{q/2}
        \right]^{1/q}~,\nonumber 
\end{align}
for each $q > 2$.

Thus, it suffices to give a suitable bound for the moment 
\begin{align}
    \mathbb{E}\left[
        \bigg\vert
        \frac{1}{\lfloor n/m\rfloor^m}
        \sum_{i_1=1}^{\lfloor n/m\rfloor} \cdots \sum_{i_m=1}^{\lfloor n/m\rfloor}
        f^{(m)}(D_{i_1}^{(1)},\ldots,D_{i_m}^{(m)})^2
        \bigg\vert^{q/2}
        \right]^{2/q}~.\label{eq: generic post symm}
\end{align}
We do this through the application of the following Lemma, whose proof is again deferred to \cref{app: proofs for u-stat}.

\begin{lemma}\label{lem: bound parts}
\textbf{(i)} 
Let $g : \mathcal{D}^m \to \mathbb{R}_{+}$ be a symmetric, non-negative kernel. For each integer $0\leq t \leq m$, define
\begin{align}
    \Gamma_{t,q}(g) = 
    \mathbb{E}\left[ \bigg\vert
    \mathbb{E}\left[ 
    g(D_{[m]})
    \mid D_1, \ldots, D_t
    \right]\bigg\vert ^q
    \right]^{1/q}~,
\end{align}
where we adopt the convention that $\Gamma_{0,q}(g) = \mathbb{E}[g(D_{[m]})]$. The inequality
\begin{align}
    \mathbb{E}\left[
        \bigg\vert
        \frac{1}{\lfloor n/m\rfloor^m}
        \sum_{i_1=1}^{\lfloor n/m\rfloor} \cdots \sum_{i_m=1}^{\lfloor n/m\rfloor}
        g(D_{i_1}^{(1)},\ldots,D_{i_m}^{(m)})
        \bigg\vert^{q}
        \right]^{1/q} \nonumber
        \leq C^m \sum_{t=0}^m \left(\frac{q}{\lfloor n/m\rfloor}\right)^t
        \Gamma_{t,q}(g)
\end{align}
holds for each $q \geq 1 \vee \log(\lfloor n /m\rfloor)$. 

\noindent\textbf{(ii)}
Let $f: \mathcal{D}^b\to \mathbb{R}$ be a symmetric kernel that satisfies the bound $\|f(D_{[b]})\|_{\psi_1}\leq \phi$. Then, for each integer $m \leq b$ and $0 \leq t \leq m$, the inequality
\begin{align}
    \Gamma_{t,q}((f^{(m)})^2) \leq \phi^2 (Cq)^2 C^t 
    {b \choose m}^{-1}
    \left( \frac{b}{m} \right)^t
\end{align}
holds for each $q \geq 1$.

\end{lemma}

\noindent Part (i) is a specialization of  Proposition 2.1 of \cite{gine2000exponential} to large higher-order moments of symmetric non-negative kernels. The bound controls moments of the form \eqref{eq: generic post symm} with a geometrically weighted series of conditional moments obtained by fixing subsets of arguments of the kernel under consideration. Part (ii) then gives a suitable bound for these conditional moments. The result follows by fixing the first $t$ arguments of the kernel $f$ and identifying the corresponding section of $f^{(m)}$ as the $(m-t)$th Hoeffding component of the induced kernel. This representation then facilitates a further application of Hoeffding orthogonality. We are not aware of a direct antecedent for this estimate, although Hoeffding orthogonality has been used in a related way to control the variance of higher-order terms in Hoeffding expansions. See, for instance, \cite{athey2019generalized} and \cite{peng2022rates}.

To complete the proof, observe that as $\log(n) \leq q/2$, Parts (i) and (ii) of \cref{lem: bound parts} imply
\begin{align}
    & \mathbb{E}\left[
        \bigg\vert
        \frac{1}{\lfloor n/m\rfloor^m}
        \sum_{i_1=1}^{\lfloor n/m\rfloor} \cdots \sum_{i_m=1}^{\lfloor n/m\rfloor}
        f^{(m)}(D_{i_1}^{(1)},\ldots,D_{i_m}^{(m)})^2
        \bigg\vert^{q/2}
        \right]^{2/q} \nonumber \\
    & 
    \leq C^m \sum_{t=0}^m \left(\frac{q/2}{\lfloor n/m\rfloor}\right)^t
        \Gamma_{t,q/2}((f^{(m)})^2) 
    \leq C^m 
    \phi^2 q^2     {b \choose m}^{-1}
    \sum_{t=0}^m \left(
    \frac{q/2}{\lfloor n/m\rfloor}
     \frac{C b}{m} 
     \right)^t~.\label{eq: apply partial bounds}
\end{align}
Consequently, as $q \leq c n/b$ and $b \leq n/2$, we obtain the bound
\begin{align}
    \mathbb{E}\left[ 
        \bigg\vert {b \choose m} U_{n,m}(f^{(m)}) \bigg\vert^q \right]^{1/q} & \leq
    C^m q^{m/2 + 1} \phi \bigg\lfloor \frac{n}{m} \bigg\rfloor^{-m/2} {b \choose m}^{1/2}
\end{align}
by plugging \eqref{eq: apply partial bounds} into \eqref{eq: post decouple and symm} and applying the geometric series formula. Hence, as 
\begin{align}
    {b \choose m} \leq \left( \frac{eb}{m}\right)^m
    \quad\text{and}\quad
    \bigg\lfloor \frac{n}{m} \bigg\rfloor^{-m/2}
    \leq \left(\frac{2m}{n} \right)^{m/2}~,
\end{align}
we obtain the bound 
\begin{align}
    \mathbb{E}\left[ 
        \bigg\vert {b \choose m} U_{n,m}(f^{(m)}) \bigg\vert^q \right]^{1/q} & \leq
        \phi q \left( \frac{C qb}{n} \right)^{m/2}
\end{align}
for all $2\log(n) \leq q \leq c n/b$, as required.\hfill\qed

\section{Simultaneous Inference for Subsampled Kernel Regression\label{sec: random forest}}

In this section, we apply \cref{lem: U stat linearization} to establish the consistency of an approach for constructing resampling-based simultaneous confidence intervals around nonparametric regression estimators based on subsampled kernels. We begin in \cref{sec: construction} by defining and illustrating the confidence region under consideration. In \cref{sec: sub kernel def,sec: nuisance and moment} we define subsampled kernel regression and state several regularity conditions. The main result is given in \cref{sec: kernel coverage}.  Throughout, for any function $f(x)$ and vector $\bm{x}^{(d)}=(x^{(j)})_{j=1}^d$, we let $f(\bm{x}^{(d)})$ denote the vector $(f(x^{(1)}),\ldots,f(x^{(d)}))$.

\subsection{Construction\label{sec: construction}}

The target for inference is the parameter vector $\theta_0(\bm{x}^{(d)})$, where $\theta_0(x)$ is defined as the unique solution in $\theta$ to the conditional moment equation
\begin{equation}\label{eq: estimating equation}
M(x; \theta, g_0) = \mathbb{E}\left[m(D_i;\theta,g_0) \mid X_i = x\right] = 0~.
\end{equation}
Here, $X_i$ is a sub-vector of the observation $D_i$, $\bm{x}^{(d)}=(x^{(j)})_{j=1}^d$ is a vector of query points, and $g_0$ is a nuisance parameter. We focus on estimators $\hat{\theta}_n(x)$ obtained as solutions to the empirical moment equation
\begin{equation}\label{eq: subsampled kernel}
M_{n}(x; \theta,\hat{g}_n, \mathbf{D}_n) = \sum_{i=1}^n K(x, X_i) m(D_i; \theta, \hat{g}_n) = 0~,
\end{equation}
where $\hat{g}_n$ is some first-stage estimator of $g_0$ and $K(\cdot,\cdot)$ is a data-dependent kernel constructed with subsampling.

\begin{figure}[t]
\begin{centering}
\caption{CATE Estimates}
\label{fig: cate}
\medskip{}
\begin{tabular}{c}
\includegraphics[scale=0.4]{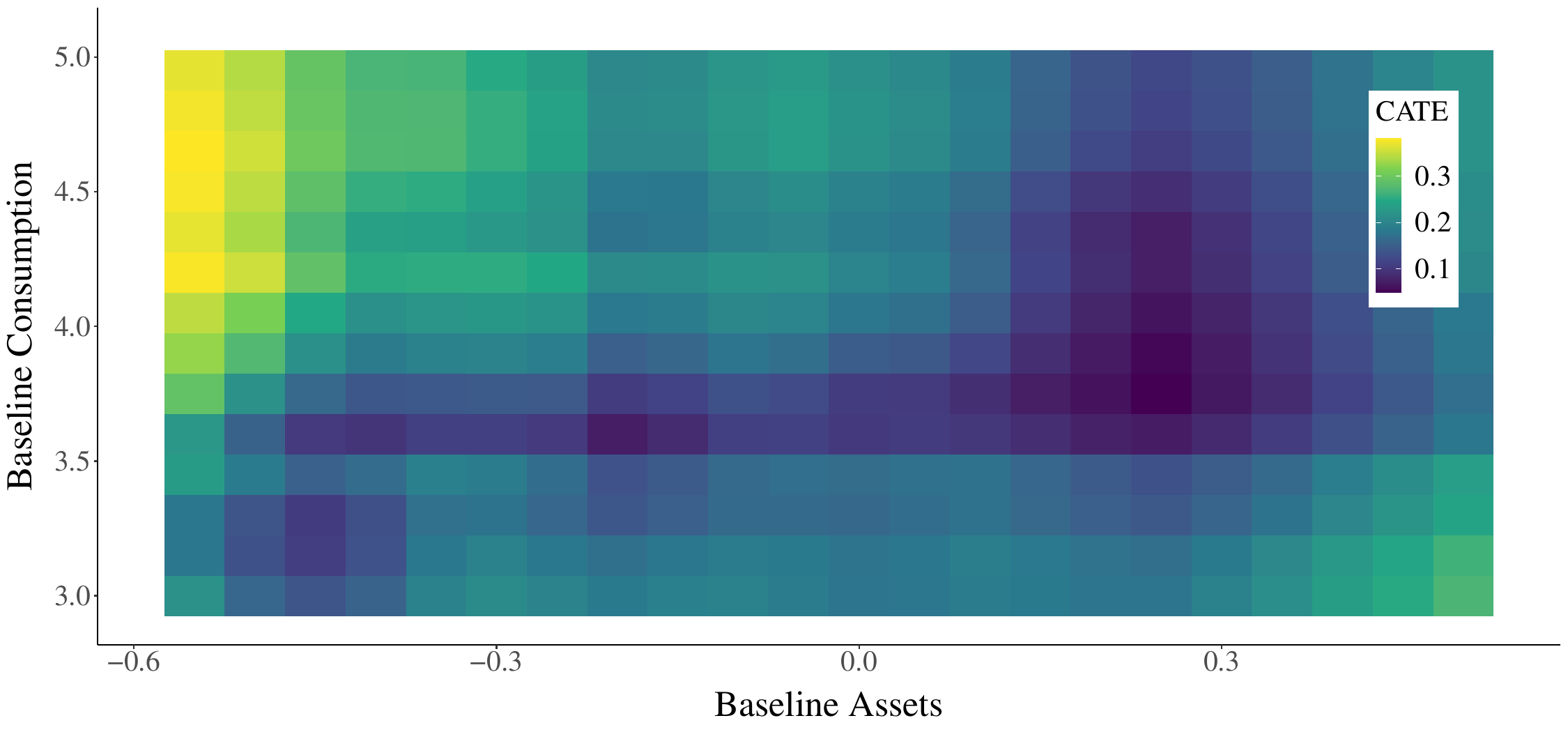}\tabularnewline
\end{tabular}
\par\end{centering}
\medskip{}
\justifying
{\footnotesize{}Notes: \cref{fig: cate} displays a heat map giving CATE estimates for the intervention studied in \cite{banerjee2015multifaceted} on post-treatment assets. The color of each rectangle indicates the estimate of the CATE queried at the rectangle's central point. The horizontal and vertical axes display the baseline monthly consumption, normalized to 2014 US dollars and measured in logs base 10, and the baseline value of an index for total assets. CATE estimates are obtained by solving the empirical moment equation \eqref{eq: subsampled kernel} for each value $x$ on an evenly spaced grid on both axes. See \cref{sec: data} for further details.}{\footnotesize\par}
\end{figure}

\begin{ex}\label{ex: banerjee}
To fix concepts, consider \cite{banerjee2015multifaceted}, who study the effects of a poverty alleviation program implemented in Ghana.\footnote{\cite{banerjee2015multifaceted} study data collected from several similar graduation programs. We focus on the data from their evaluation of the program implemented in Ghana. \cref{sec: data} gives further details.} For each individual in their sample, they observe the data $D_i = (Y_i, W_i, Z_i)$, where $Y_i$ is a measurement of total assets taken two years after the implementation of the program, $W_i$ is an indicator denoting assignment to the program, and $Z_i$ is a vector of covariates. A broad aim of the study is to determine the conditions under which recipients of aid experience lasting improvements in welfare. One quantity that can inform this determination is the conditional average treatment effect (CATE)
\begin{equation}\label{eq: CATE}
\theta_0(x) = \mathbb{E}_P \left[ Y_i(1) - Y_i(0)  \mid X_i = x\right]~,
\end{equation}
where $Y_i(1)$ and $Y_i(0)$ are the potential outcomes generated by the intervention $W_i$ and $X_i$ is some chosen subvector of $Z_i$. Many modern approaches to estimating CATEs are premised on the observation that $\theta_0(x)$ is the solution to the conditional moment equation
\begin{equation}\label{eq: CATE moment}
M(x; \theta, g_0) = \mathbb{E}\left[(\mu_1(Z_i) - \mu_0(Z_i)) + \beta(W_i, Z_i) (Y_i - \mu_{W_i}(Z_i)) - \theta \mid X_i = x\right]=0~,
\end{equation}
where the nuisance parameter $g_0$ collects the conditional outcome regression and Horvitz-Thompson weight
\begin{equation}\label{eq: CATE nuisance}
\mu_{w}(z) = \mathbb{E}_P\left[Y_i \mid W_i = w, Z_i = z\right]\quad\text{and}\quad \beta(w, z) = \frac{w}{\pi(z)} - \frac{1- w}{1-\pi(z)}~,
\end{equation}
for the propensity score $\pi(z) = P\{W_i = 1 \mid Z_i = z\}$. See \cite{semenova2021debiased}, \cite{foster2023orthogonal}, and \cite{kennedy2023towards}, for further discussion.

\cref{fig: cate} displays CATE estimates for the experiment studied in \cite{banerjee2015multifaceted}, where the chosen conditioning covariates $X_i$ are pretreatment  measurements of monthly consumption and total assets. The nuisance parameter estimate $\hat{g}_n$ and the kernel $K(x,x^\prime)$ are constructed with the implementation of random forest regression made available through the ``GRF'' R package \citep{athey2019generalized}. The graduation program appears to be most effective for individuals with a high level of baseline consumption and a low level of baseline assets.\footnote{The quartiles of baseline log consumption are 3.33, 3.76, and 4.20. The quartiles of baseline assets are -0.45, -0.71, and 0.03. Panel A of \cref{fig: density} displays a scatter plot of baseline log consumption and assets.} These results are suggestive of a poverty trap: individuals with an opportunity to increase their assets are able to do so only if they have a high level of baseline consumption \citep{balboni2022people,kraay2014poverty}.\hfill$\blacksquare$
\end{ex}

We consider a method for assessing the statistical significance of estimates typified by \cref{fig: cate}. In particular, we study a simple procedure for constructing confidence intervals $\hat{\mathcal{C}}(\bm{x}^{(d)})$ centered around the solutions to \eqref{eq: subsampled kernel} that satisfy the bound
\begin{equation} \label{eq: asymptotic validity}
\sup_{P \in \mathbf{P}}
\big\vert P\left\{ \theta_0(\bm{x}^{(d)}) \in \hat{\mathcal{C}}(\bm{x}^{(d)})  \right\} - (1-\alpha) \big\vert  \leq  r_{n,d}
\end{equation}
for a sequence $r_{n,d}$ that converges to zero in regimes where $d$ may increase more quickly than $n$ and $\mathbf{P}$ is some statistical family that contains the distribution $P$.\footnote{For the sake of simplicity, we consider only the case that the moment $m(\cdot ;\theta,g_0)$ and the parameter $\theta$ are scalar-valued. The vector-valued case is relevant, e.g., when estimating CATEs with multiple treatment variables. Our results generalize to the vector-valued case at the cost of additional notation.} 

To this end, let $\mathsf{h}$ denote a random element of $\mathcal{S}_{n,n/2}$, i.e., a random half-sample of $[n]$ and let $\hat{\theta}_\mathsf{h}(\bm{x}^{(d)})$ denote the vector of the solutions to \eqref{eq: subsampled kernel}, with the data $\mathbf{D}_n$ replaced by the half-sample $D_\mathsf{h}$. Our construction is premised on approximating the distribution of the root
\begin{equation}
  R_n(\bm{x}^{(d)}) 
   = \hat{\theta}_n(\bm{x}^{(d)}) - \theta_0(\bm{x}^{(d)})\label{eq: root}
\end{equation}
with the conditional distribution of the half-sample bootstrap root
\begin{equation}
R^*_n(\bm{x}^{(d)}) = \hat{\theta}_\mathsf{h}(\bm{x}^{(d)}) - \hat{\theta}_n(\bm{x}^{(d)})~.\label{eq: bootstrap root}
\end{equation}
The nuisance parameter estimator $\hat{g}_n$ does not need to be re-estimated when computing \eqref{eq: bootstrap root}. A version of the half-sample bootstrap is implemented by default in the GRF R package \citep{athey2019generalized}. The half-sample bootstrap is an instance of subsampling \citep{politis1994large,politis2012subsampling}. 

Let $\hat{\lambda}^2_{n,j}$ denote the variance of $\sqrt{n}R^*_{n}(x^{(j)})$, conditioned on the data $\mathbf{D}_n$, and let $\hat{\mathsf{cv}}_{n}(\alpha)$ denote the $1-\alpha$ quantile of the distribution of the studentized statistic
\begin{equation}\label{eq: critical value quantity}
\hat{S}^*_{n}(\bm{x}^{(d)}) = \sqrt{n} \left\| \hat{\Lambda}_n^{-1/2}R^*_n(\bm{x}^{(d)}) \right\|_\infty~,
\end{equation}
again conditioned on the data $\mathbf{D}_n$. Here, $\hat{\Lambda}_n$ denotes the diagonal matrix with elements $\hat{\lambda}^2_{n,j}$.\footnote{The quantities $\hat{\lambda}^2_{n,j}$ and $\widehat{\mathsf{cv}}_{n}(\alpha)$ are easily approximated by resampling the bootstrap root \eqref{eq: bootstrap root}. To simplify exposition, we omit explicit consideration of residual randomness induced by this approximation.} We consider confidence intervals with the following structure.

\begin{defn}\label{def: uniform ci} Define the intervals
\begin{equation}
\hat{\mathcal{C}}(x^{(j)}) = \hat{\theta}_n(x^{(j)}) \pm n^{-1/2}\hat{\lambda}_{n,j} \widehat{\mathsf{cv}}_{n}(\alpha)\quad\text{for each}\quad j \text{ in }[d]~.
\end{equation}
The level-$\alpha$ half-sample confidence region for $\theta_0(\bm{x}^{(d)})$ is given by $\hat{\mathcal{C}}(\bm{x}^{(d)})$.
\end{defn}

\noindent Confidence intervals with the same structure, based on different choices of bootstrap root, are studied in, e.g., \cite{chernozhukov2014anti} and \cite{belloni2018uniformly}. 

The essential feature of the bootstrap root \eqref{eq: bootstrap root} is that it can be computed without knowing anything about the structure of the estimator $\hat{\theta}_n(\bm{x}^{(d)})$. In particular, approaches based on the Rademacher or Gaussian multiplier bootstrap rely on knowledge of a linear approximation to $\hat{\theta}_n(\bm{x}^{(d)})$. To gain intuition, suppose that the estimator $\hat{\theta}_n(\bm{x}^{(d)})$ satisfies the exact linear representation
\begin{equation}\label{eq: exact linear}
\hat{\theta}_n(\bm{x}^{(d)}) = \frac{1}{n} \sum_{i=1}^n \bar{u}(\bm{x}^{(d)}, D_i)
\end{equation}
for some function $\bar{u}(\cdot,\cdot)$. Let $V_1,\ldots,V_n$ denote a collection of random variables, where $V_i$ takes the value $1$ if the index $i$ is an element of the random set $\mathsf{h}$ used to define the half-sample bootstrap root \eqref{eq: bootstrap root}, and takes the value $-1$ otherwise. Observe that
\begin{flalign}
R^*_n(\bm{x}^{(d)}) = \hat{\theta}_\mathsf{h}(\bm{x}^{(d)}) - \hat{\theta}_n(\bm{x}^{(d)}) 
&= \frac{2}{n} \sum_{i=1}^n \mathbb{I}\{ i \in \mathsf{h}\} \bar{u}(\bm{x}^{(d)}, D_i) - \frac{1}{n} \sum_{i=1}^n \bar{u}(\bm{x}^{(d)}, D_i)\nonumber\\
&= \frac{1}{n}\sum_{i=1}^n V_i \left(\bar{u}(\bm{x}^{(d)}, D_i) - \theta_0(\bm{x}^{(d)})\right)~.
\label{eq: half as weight}
\end{flalign}
The weights $V_i$ are exchangeable Rademacher random variables, i.e., they are uniformly distributed on $\{1,-1\}$. If the weights were fully independent, the representation \eqref{eq: half as weight} reduces to the more familiar Rademacher bootstrap. The representation \eqref{eq: half as weight} is due to \cite{yadlowsky2023evaluating}, who draw on a similar observation made in the context of two-sample testing in \cite{chung2013exact}.

In practice, many widely applied estimators are not perfectly linearly decomposable. Often, however, estimators do satisfy an approximate linear decomposition, in the sense that the equality \eqref{eq: exact linear} holds with a remainder term of order, say, $o_p(n^{-\gamma})$ for some positive constant $\gamma$. As we will see, estimators constructed with subsampled kernels are approximately linear around some unknown function $\bar{u}(\cdot,\cdot)$. If an estimator $\hat{\theta}_n(\bm{x}^{(d)})$ is approximately linear, then the subsampled estimate $\hat{\theta}_{\mathsf{h}}(\bm{x}^{(d)})$ is \emph{immediately} also approximately linear.\footnote{Approximate linearity does not immediately imply a representation analogous to \eqref{eq: half as weight} for a root constructed with an empirical bootstrap, as approximate linearity would not necessarily hold under the empirical distribution.} Thus, for approximately linear statistics, the representation \eqref{eq: half as weight} continues to hold, now with a remainder term of order $o_p(n^{-\gamma})$.

\begin{figure}[t]
\begin{centering}
\caption{Half-Sample Confidence Region}
\label{fig: half}
\medskip{}
\begin{tabular}{c}
\textit{Panel A: Upper Bound}\tabularnewline
\includegraphics[scale=0.40]{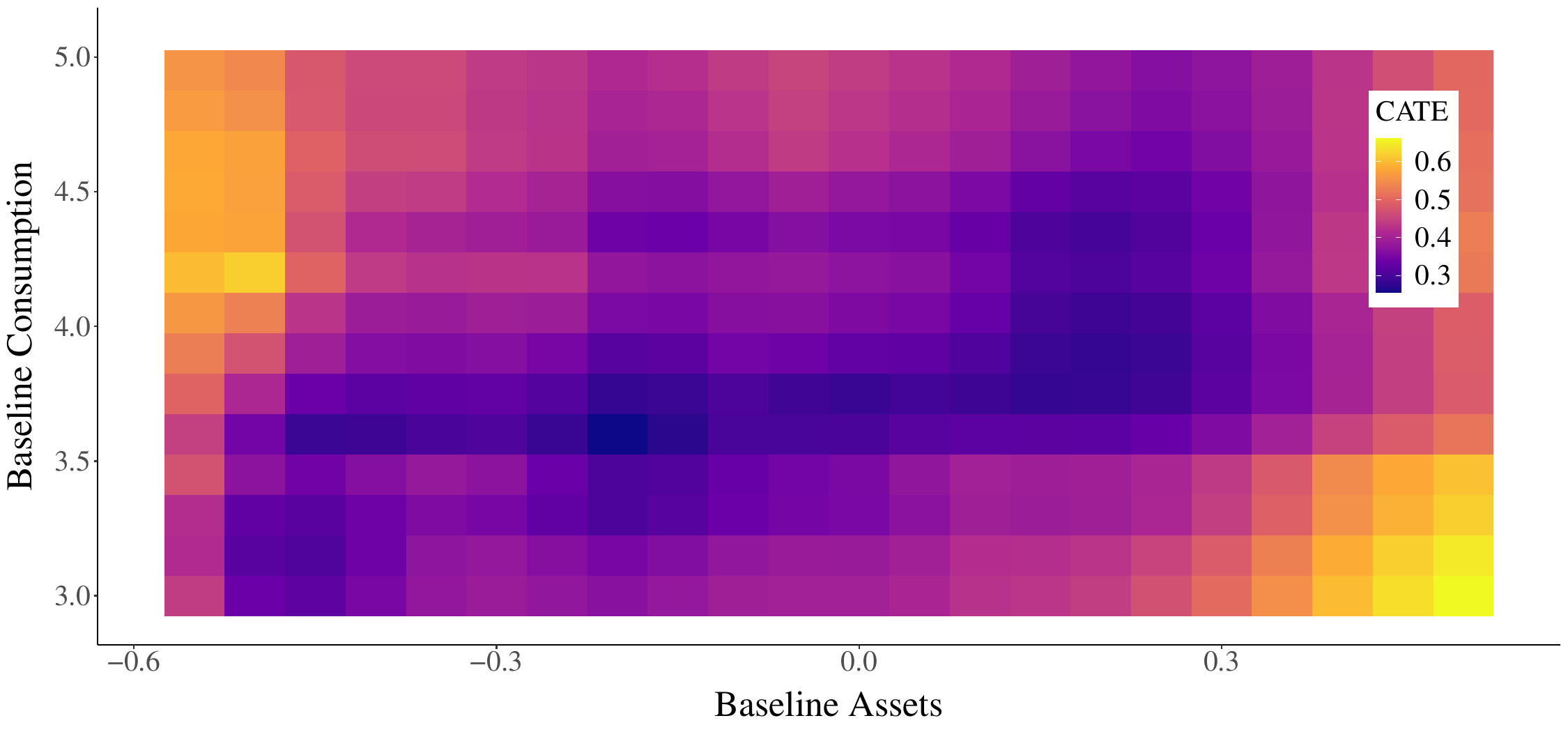}\tabularnewline
\textit{Panel B: Lower Bound}\tabularnewline
\includegraphics[scale=0.40]{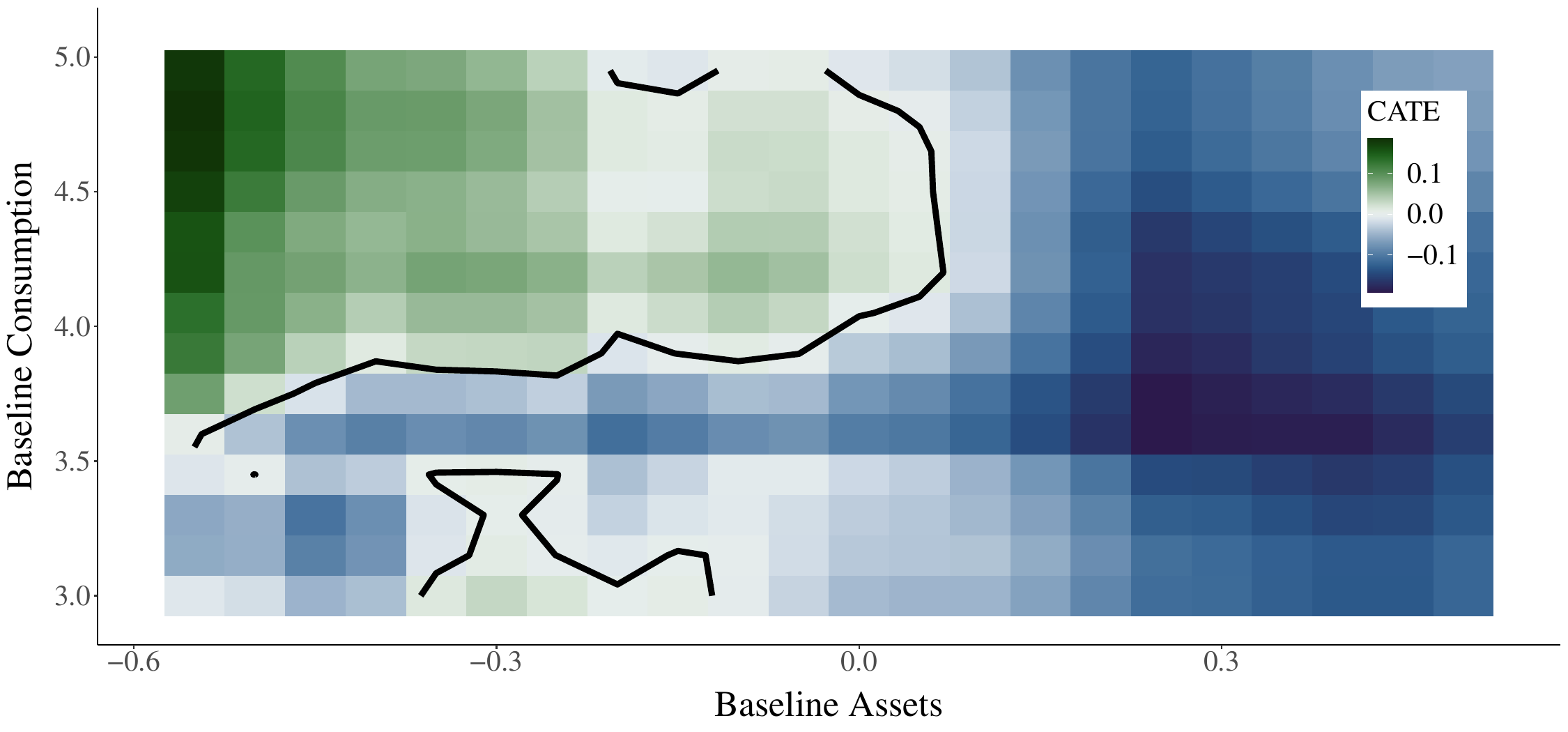}\tabularnewline
\end{tabular}
\par\end{centering}
\medskip{}
\justifying
{\footnotesize{}Notes: \cref{fig: half} displays heat maps giving half-sample upper and lower confidence bounds for the CATE of the intervention studied in \cite{banerjee2015multifaceted} on post-treatment total assets. The confidence bounds are constructed at level $\alpha = 0.1$. The upper and lower bounds are displayed with different color palettes to emphasize the use of different scales. A contour line has been superimposed over the lower bound to demarcate where the bound crosses zero. The axes and estimator are the same as in \cref{fig: cate}.}{\footnotesize\par}
\end{figure}

\begin{ex}[continues=ex: banerjee]
We now return to the data studied in \cite{banerjee2015multifaceted}. \cref{fig: half} displays upper and lower half-sample confidence bounds for the CATE \eqref{eq: CATE} on post-treatment assets. The null hypothesis that the CATE is equal to zero is only rejected for individuals with low baseline assets. The lower bounds are meaningfully larger than zero only for individuals with low baseline assets and high baseline consumption. That is, the graduation program has a positive impact on individuals who do not have many assets to begin with, but who do have access to a stable source of consumption. On the other hand, the confidence regions contain zero for individuals who have low baseline consumption or high baseline assets. In \cref{sec: simulation}, we measure the coverage and average length of the confidence region $\hat{\mathcal{C}}(\bm{x}^{(d)})$ in a simulation calibrated to this setting.\hfill$\blacksquare$
\end{ex}

\subsection{Subsampled Kernel Regression\label{sec: sub kernel def}} 

Subsampled kernel regression is a broad class of algorithms for solving regression problems of the form \eqref{eq: subsampled kernel}, based on constructing a data-driven kernel function $K(x,x^\prime)$ with subsampling. Formally, fix some positive integer $r$ and let $(\mathsf{s}_q)_{q=1}^r$ collect a sequence of subsets of $[n]$ drawn independently and uniformly from $\mathcal{S}_{n,b}$. Let $\xi$ denote some auxiliary source of randomness and let $(\xi_{\mathsf{s}_q})_{q=1}^r$ collect a set of independent random variables with the same distribution as $\xi$. We study conditional empirical moment estimators of the form \eqref{eq: subsampled kernel}, where the kernel function $K(x,x^\prime)$ admits the decomposition
\begin{equation}\label{eq: subsampled kernel def}
K(x, X_i) = \frac{1}{r}\sum_{q=1}^r \mathbb{I}\{i \in \mathsf{s}_q\} \kappa(x, X_i, D_{\mathsf{s}_q}, \xi_{\mathsf{s}_q})
\end{equation}
for some known kernel $\kappa(\cdot,\cdot,D_{\mathsf{s}}, \xi_{\mathsf{s}})$. Several widely applied instances of subsampled kernels are as follows.\footnote{The half-sample bootstrap is particularly computationally efficient for estimators constructed with subsampling, as the estimator and the half-sampled estimator can be constructed using the same collection of subsamples. See Section 4.1 of \cite{athey2019generalized}.}

\begin{ex}\label{ex: knn}
Subsampled nearest-neighbors regression is a simple example of a kernel with the structure \eqref{eq: subsampled kernel def} \citep{fix1989discriminatory}. Here, the kernel $\kappa(x,X_{i^\prime},D_{\mathsf{s}}, \xi_{\mathsf{s}})$ is non-zero if and only if $X_{i^\prime}$ is one of the $k$ closest points to $x$ among the points in the subsample $D_{\mathsf{s}}$.
\end{ex}

\begin{ex}\label{ex: random forest}
Random forest regression, introduced by \cite{breiman2001random}, is another example of a kernel with the structure \eqref{eq: subsampled kernel def}. In this case, each pair $(D_{\mathsf{s}}, \xi_{\mathsf{s}})$ generates some partition of $\mathcal{X}$. The kernel $\kappa(x,x^\prime,D_{\mathsf{s}}, \xi_{\mathsf{s}})$ is non-zero if and only if $x$ and $x^\prime$ are in the same element of the partition generated by $(D_{\mathsf{s}}, \xi_{\mathsf{s}})$. Often, such partitions are constructed with recursive algorithms, e.g., the ``CART'' algorithm of \cite{breiman2017classification}.
\end{ex}

We impose the following restrictions on the kernels under consideration. 

\begin{assumption}[Honesty and Positive Symmetry]\label{assu: kernel restriction}$\text{ }$\\
\noindent(i) The kernel $\kappa(\cdot,\cdot,D_{\mathsf{s}},\xi)$ is Honest in the sense that
\begin{equation}
\kappa(x,X_i,D_{\mathsf{s}},\xi_{\mathsf{s}}) \ci m(D_i; \theta, g) \mid X_i, D_{\mathsf{s}_{-i}}~,
\end{equation}
where $\ci$ denotes conditional independence and $\mathsf{s}_{-i}$ denotes the set $\mathsf{s}\setminus\{i\}$.

\noindent (ii) The kernel $\kappa(\cdot,\cdot,D_{\mathsf{s}},\xi)$ is positive and satisfies the restriction $\sum_{i\in\mathsf{s}}\kappa(\cdot, X_i, D_{\mathsf{s}},\xi_{\mathsf{s}}) = 1$ almost surely. Moreover, the conditional expectation  $\mathbb{E}\left[\kappa(\cdot, X_i, D_{\mathsf{s}},\xi_{\mathsf{s}})\mid D_{\mathsf{s}}\right]$ is invariant to permutations of the data $D_{\mathsf{s}}$.  
\end{assumption}

\noindent The ``Honesty'' condition stipulated in Part (i) of \cref{assu: kernel restriction} imposes the restriction that any part of the data $D_i$ that can affect the value of the moment $m(D_i; \theta, g)$ cannot affect the value of the kernel $\kappa(x,X_i,D_{\mathsf{s}},\xi_{\mathsf{s}})$. This condition was introduced in \cite{athey2016recursive}. Honesty is often achieved through kernel construction schemes based on sample-splitting; see  \cite{wager2018estimation} and \cite{athey2019generalized} for further discussion. Part (ii) of \cref{assu: kernel restriction} imposes several weak regularity conditions. 

The following two quantities restrict the ``size'' and ``variability'' of the chosen kernel.
\begin{defn}[Shrinkage and Incrementality]\label{def: shrinkage and incrementality}$ $\\Let $\mathsf{s}$ be an arbitrary element of $\mathcal{S}_{n,b}$ and let $l$ be any element of $\mathsf{s}$.\\
(i) We say that the kernel $\kappa(\cdot,\cdot,D_{\mathsf{s}},\xi_{\mathsf{s}})$ has a uniform shrinkage rate $\varepsilon_{b}$ if
\begin{equation}
\sup_{P \in \mathbf{P}} \sup_{j \in [d]} \mathbb{E}\left[ \max\left\{ \|X_i - x^{(j)} \|_\infty : \kappa(x^{(j)},X_i,D_{\mathsf{s}},\xi_{\mathsf{s}}) > 0 \right\} \right] \leq \varepsilon_{b}~.
\end{equation}

\noindent (ii) We say that the kernel $\kappa(\cdot,\cdot,D_{\mathsf{s}},\xi_{\mathsf{s}})$ is uniformly incremental if
\begin{equation}\label{eq: incrementality def}
\inf_{P \in \mathbf{P}} \inf_{j \in [d]} \Var\left( \mathbb{E}\left[ \sum_{i\in\mathsf{s}} 
 \kappa(x^{(j)}, X_i, D_{\mathsf{s}}, \xi_{\mathsf{s}}) m(D_i; \theta, g) \mid  D_l = D \right] \right) \gtrsim b^{-1}
\end{equation}
where $D$ is an independent random variable with distribution $P$.
\end{defn}
\noindent The shrinkage rate of a kernel $\kappa(\cdot,\cdot,D_{\mathsf{s}},\xi_{\mathsf{s}})$ is analogous to the bandwidth of a classic deterministic kernel. The incrementality restriction ensures that the chosen kernel is not overly dependent on a single data point. Both notions were introduced by \cite{wager2018estimation} and have been characterized explicitly for various widely applied subsampled kernel estimators.\footnote{The terminology ``shrinkage'' was introduced in \cite{oprescu2019orthogonal}, and is not intended to connote (explicit) regularization.}

\begin{ex}[continues=ex: knn]
In the case of honest subsampled $k$-NN regression, \cite{khosravi2019non} show that $\varepsilon_{b} \lesssim b^{-1/p}$, where $p$ is the intrinsic dimension of the measure of the covariates $X_i$. See e.g., \cite{kpotufe2011k} for further discussion. In turn, \cite{khosravi2019non} and \cite{peng2022rates} show that the kernels associated with both honest and non-honest variants of $k$-NN regression are incremental, up to logarithmic factors that  depend on the dimension of the covariates.\hfill$\blacksquare$
\end{ex}

\begin{ex}[continues=ex: random forest]
Analogously, for honest random forest regression, under suitable regularity conditions, \cite{wager2018estimation} establish that $\varepsilon_{b} \lesssim b^{-c/p}$, where $p$ is the dimension of the domain of $X_i$. See e.g., \cite{wager2018estimation} and \cite{oprescu2019orthogonal} for further discussion. Bounds adaptive to the intrinsic dimension of the measure of $X_i$ are given in \cite{huo2023adaptation}. \cite{wager2018estimation} and \cite{peng2022rates} give simple conditions under which the kernel associated with subsampled, honest, random forest regression is uniformly incremental, again up to dimension-dependent logarithmic factors.\hfill$\blacksquare$
\end{ex}

\subsection{Moment Restrictions\label{sec: nuisance and moment}}

We restrict attention to conditional moments that satisfy a Neyman orthogonality condition \citep{chernozhukov2018double}. We assume that the nuisance parameter $g_0$ is a collection of $h$ real-valued functions $g_0 = (g_0^{(k)})_{k=1}^h$, each having domain $\mathcal{Z}$ given by the support of the covariates $Z_i$.  Define the norm
\begin{equation}\label{eq: nuisance norm}
\| g - g_0 \|_{2,\infty} = \sup_{k\in [h]} \sup_{j\in [d]} \left(\mathbb{E} \left[ (g^{(k)}(Z_i) - g_0^{(k)}(Z_i))^2 \mid X_i = x^{(j)} \right] \right)^{1/2}
\end{equation}
for any $g = (g^{(k)})_{k=1}^h$. The nuisance parameter $g_0$ takes values in the space $\mathcal{G}$. Throughout, for a functional $F$ on $\mathcal{F}$, we use the notation
\[
\partial_f F(f)[h] = \frac{\text{d}}{\text{d}t}F(f+th)\big\vert_{t=0}
\quad\text{and}\quad
\partial_{f,f} F(f)[h]=\frac{\text{d}^2}{\text{d}t^2}F(f+th)\big\vert_{t=0}
\]
to denote first- and second- order directional derivatives, respectively.

\begin{defn}[Local Neyman Orthogonality]\label{def: orthogonal}
We say that a conditional moment $M(\cdot;\theta_0, g_0)$ is uniformly locally Neyman orthogonal if
\begin{equation}
\partial_g M(x^{(j)}; \theta_0, g_0)[g-g_0] = 0
\end{equation}
for all $P$ in $\mathbf{P}$ and $x^{(j)}$ in $\bm{x}^{(d)}$.
\end{defn}
\noindent The use of Neyman orthogonal moments ensures that the bias induced by the estimation of the nuisance parameter $g_0$ with $\hat{g}_n$ is small \citep{chernozhukov2018double}.

In addition to Neyman orthogonality, we require several smoothness restrictions on the function $m(\cdot;\theta, g)$. In the main text, to ease exposition, we impose the following linearity and boundedness restriction.
\begin{assumption}[Moment Linearity and Boundedness]\label{assu: moment linearity}
The moment function $m(\cdot;\theta, g)$ satisfies the linear representation
\begin{equation}
m(D_i;\theta, g) = m^{(1)}(D_i; g)\cdot \theta + m^{(2)}(D_i; g)
\end{equation}
for some known functions $m^{(1)}(\cdot;g)$ and $m^{(2)}(\cdot; g)$. Moreover, the absolute value of the function $m(\cdot;\theta, g)$ is bounded by the constant $(|\theta|+1)\phi$ for some $\phi \geq 1$ almost surely. 
\end{assumption}
\noindent The linearity restriction entailed in \cref{assu: moment linearity} is inessential and is imposed for the sake of simplicity.\footnote{In \cref{sec: generic}, we show that moment linearity can be replaced by the high-level assumption that $\hat{\theta}_n(\bm{x}^{(d)})$ is consistent for $\theta_0(\bm{x}^{(d)})$. This state of affairs is standard in $M$-estimation problems \citep[see e.g., ][]{newey1994large}. As our running examples use linear moments, and sufficient conditions for the consistency of $\hat{\theta}_n(\bm{x}^{(d)})$ have been established \citep[see e.g., Assumption 4.1 and Theorem 4.3 of ][]{oprescu2019orthogonal}), we omit a detailed consideration of nonlinear moments.} Like in \cref{cor: general u stat clt}, the boundedness restriction can be weakened to a slightly more involved assumption stated in terms of sub-exponential norms. Again, we impose boundedness to ease exposition.

We maintain the following additional mild smoothness restrictions.
\begin{assumption}[Moment Smoothness]\label{assu: moment smoothness}$\text{ }$\\
\noindent \textbf{(i)} The moment $M(\cdot; \theta, g_0)$ is uniformly second order smooth, in the sense that
\begin{equation}\label{eq: orth smooth}
\sup_{P\in\mathbf{P}}  \sup_{j\in[d]} \big\vert \partial_{g,g} M(x^{(j)}; \theta_0, g_0)[g-g_0]\big\vert \lesssim \|g - g_0\|^2_{2,\infty}
\end{equation}
for each $g$ in $\mathcal{G}$.

\noindent \textbf{(ii)} The variogram 
\begin{equation*}
V(x ; g) = \mathbb{E}\left[ (m(D_i ;\theta_0(x), g) - m(D_i ;\theta_0(x), g_0))^2 \mid X = x\right]
\end{equation*}
is uniformly Lipschitz in its first component, in the sense that
\begin{equation}
\sup_{P\in\mathbf{P}} \sup_{g\in\mathcal{G}} \big\vert V(x ; g) - V(x^\prime ; g) \big\vert \lesssim \|x - x^\prime\|_\infty
\end{equation}
holds for all $x$ and $x^\prime$ in $\mathcal{X}$. Moreover, the variogram satisfies the mean-squared continuity condition
\begin{equation}
\sup_{P\in\mathbf{P}} \sup_{j\in[d]} \vert V(x^{(j)}, g) - V(x^{(j)}, g^\prime)\vert \lesssim \|g - g^\prime\|^2_{2,\infty}
\end{equation}
for each $g$ and $g^\prime$ in $\mathcal{G}$.

\noindent \textbf{(iii)} Define the moments
\begin{align}
M^{(1)}(x;\theta,g) &= \mathbb{E}\left[ m^{(1)}(D_i;g) \mid X_i = x\right] \quad\text{and}\\
M^{(2)}(x;g) &= \mathbb{E}\left[ m^{(2)}(D_i;g) \mid X_i = x\right]~,
\end{align}
associated with the functions  $m^{(1)}(\cdot; g)$ and $m^{(2)}(\cdot; g)$ introduced in \cref{assu: moment linearity}. Both moments are 
uniformly Lipschitz in their first component. That is, it holds that
\begin{equation}
\sup_{P\in\mathbf{P}} \sup_{g\in\mathcal{G}} \big\vert M^{(1)}(x ;g) - M^{(1)}(x^\prime ;g) \big\vert \lesssim \|x - x^\prime\|_\infty
\end{equation}
for all $x$, $x^\prime$, and $\theta$, and analogously for $M^{(2)}(x;g)$. Moreover, the first moment is uniformly Lipschitz in its second component and bounded from below in the sense that
\begin{align}
\sup_{P\in\mathbf{P}} \sup_{j\in[d]} \big\vert M^{(1)}(x^{(j)}; \theta_0,g) - M^{(1)}(x^{(j)};g_0)\big\vert & \lesssim \|g - g_0\|_{2,\infty}\quad\text{and}\label{eq: Lipschitz Jacobian main}\\
\inf_{P\in\mathbf{P}} \inf_{j\in[d]} \big\vert M^{(1)}(x^{(j)};g)\big\vert & \geq c\label{eq: well-posedness main}
\end{align}
for each $g$ and $\theta$ and some positive constant $c$.

\noindent \textbf{(iv)} The parameter space $\Theta$ is a bounded, convex subset of $\mathbb{R}$. Moreover, the target parameter satisfies $\theta_0(x)\in\Theta$ for every $x\in\mathcal{X}$ and every $P\in\mathbf{P}$ and the estimator $\hat{\theta}_n(x)$ is constructed as a solution over $\Theta$.
\end{assumption}

Neyman orthogonal moments satisfying the smoothness restrictions specified in \cref{assu: moment linearity} and \cref{assu: moment smoothness} are available for many widely considered statistical problems. For example, the moment \eqref{eq: CATE moment} is the unique Neyman orthogonal identifying moment for the parameter \eqref{eq: CATE} \citep[see e.g.,][]{hahn1998role, chernozhukov2018double}.\footnote{For estimation of CATEs, Part (i) of \cref{assu: moment smoothness} is implied by the more refined bound
\begin{equation}
\sup_{P\in\mathbf{P}}  \sup_{j\in[d]} \big\vert \partial_{g,g} M(x^{(j)}; \theta_0, g_0)[g-g_0]\big\vert \lesssim \|\mu - \mu_0\|_{2,\infty} \|\beta - \beta_0\|_{2,\infty}~,
\end{equation}
where $\mu$ and $\beta$ are defined in \eqref{eq: CATE nuisance}. This structure yields the celebrated ``Double Robustness'' result for estimation of conditional average treatment effects; see \cite{chernozhukov2018double} and \cite{kennedy2023towards} for further discussion. We impose the more general condition \eqref{eq: orth smooth}, as this bound exhibits many of the same features and will hold for a wider variety of problems.} Additional examples of problems where smooth Neyman orthogonal identifying moments are available include estimation of partially linear regression and partially linear instrumental variable regression \citep{chernozhukov2018double}, local average treatment effects \citep{tan2006regression,frolich2007nonparametric}, dynamic treatment effects \citep{lewis2021double}, and long-term treatment effects identified by surrogate outcomes \citep{athey2020estimating,chen2023semiparametric}.

\subsection{Coverage Accuracy\label{sec: kernel coverage}}

The following theorem gives a bound on the error in the nominal coverage probability of the confidence regions introduced in \cref{def: uniform ci}. We emphasize that the result is applicable to asymptotic regimes where the dimension of the query-vector $\bm{x}^{(d)}$ can be exponentially larger than the sample size $n$.
  
\begin{theorem}\label{eq: asymptotic validity subsampled kernel}
Suppose that the kernel $\kappa(\cdot,\cdot,D_{\mathsf{s}},\xi_{\mathsf{s}})$ satisfies \cref{assu: kernel restriction}, has uniform shrinkage rate $\varepsilon_{b}$, and is uniformly incremental. Additionally, suppose that the Neyman orthogonal moment function $M(\cdot;\theta_0, g_0)$ satisfies \cref{assu: moment linearity} and \cref{assu: moment smoothness} and that  $r$ has been chosen to satisfy $n \leq b \sqrt{r}$. If the estimator  $\hat{g}_n$ is statistically independent of the data $\mathbf{D}_n$ and satisfies the moment bound
\begin{align}
\sup_{P\in\mathbf P}
\mathbb E\left[
\|\hat g_n-g_0\|_{2,\infty}^q
\right]^{1/q}
\leq q \left(\frac{b}{n}\right)^{1/4} \delta_{n,g}
\label{eq: g rate reduce}
\end{align}
for some sequence $\delta_{n,g}$ and each $q\geq 2\vee2\log(dn)$, then the confidence region given in \cref{def: uniform ci} satisfies
\begin{align} 
& \sup_{P\in{\mathbf{P}}}\big\vert P\left\{ \theta_0(\bm{x}^{(d)}) \in \hat{\mathcal{C}}(\bm{x}^{(d)})  \right\} - (1 - \alpha) \big\vert \nonumber \\
& \quad\quad\quad\quad
\lesssim 
\left(\frac{b\phi^4\log^8(dn)}{n}\right)^{1/4}
+
\left(
\delta_{n,g}^2
+
\sqrt{\frac{n}{b}}\varepsilon_b
\right)\log^3(dn)~.\label{eq: kernel coverage}
\end{align}
\end{theorem}

\begin{remark}
To expedite exposition, in stating \cref{eq: asymptotic validity subsampled kernel}, we have assumed that the nuisance parameter estimator $\hat{g}_n$ is computed on a separate sample, independent of the data $\mathbf{D}_n$. This can be achieved by randomly splitting the available data into two subsamples. The first can be used to construct the nuisance parameter estimate $\hat{g}_n$; the second to construct the confidence region $\hat{\mathcal{C}}(\bm{x}^{(d)})$. There are practical issues with this strategy. First, the region $\hat{\mathcal{C}}(\bm{x}^{(d)})$  might be sensitive to the choice of the split of the data \citep{ritzwoller2023reproducible}. Second, by splitting the data, researchers incur a potentially meaningful loss in statistical precision. In \cref{app: stability} we give additional conditions under which the nuisance parameter estimator $\hat{g}_n$ can be computed using the same data used to construct the region $\hat{\mathcal{C}}(\bm{x}^{(d)})$.\footnote{Our analysis builds on \cite{chen2022debiased}, who give a pointwise analysis of unconditional moment estimators.} If these conditions are not plausible for a given application, in \cref{app: cross}, we detail a modified version of the confidence region given in \cref{def: uniform ci} that is constructed with cross-splitting.\hfill$\blacksquare$
\end{remark}

\begin{remark}
The bound \eqref{eq: kernel coverage} can be interpreted as a bias-variance decomposition. The first term in \eqref{eq: kernel coverage} results from a Berry-Esseen-type bound, analogous to \cref{cor: general u stat clt}, on the accuracy of a normal approximation to \eqref{eq: U expo}. The term involving the kernel shrinkage $\varepsilon_b$ is a remnant of a bound on the supremum of the bias of the estimator $\hat{\theta}_n(\bm{x}^{(d)})$.\hfill$\blacksquare$
\end{remark}

\begin{remark}
\cref{eq: asymptotic validity subsampled kernel} follows from an application of an abstract result stated in \cref{sec: generic}. The proof is given in \cref{app: kernel verify pf}. The abstract result applies to conditional moment estimators that are not necessarily constructed with subsampled kernels. Rather, the result holds under the high-level assumption that the estimator $\hat{\theta}_n(\bm{x}^{(d)})$ is approximately linear, with a sufficiently small remainder term, and has sufficiently small bias and stochastic equicontinuity. 
\hfill$\blacksquare$
\end{remark}

The proof of \cref{eq: asymptotic validity subsampled kernel} has three main steps. First, through a standard series of expansions \citep[see e.g.,][]{chernozhukov2018double, oprescu2019orthogonal}, we show that the root $R_n(\bm{x}^{(d)})$ can be approximated by
\begin{align}
W_{n,b}(\bm{x}^{(d)}) & = {n \choose b}^{-1} \sum_{\mathsf{s}\in\mathcal{S}_{n,b}} \mathbb{E}\left[u(\bm{x}^{(d)}; D_{\mathsf{s}}, \xi_{\mathsf{s}}, \theta_0, g_0)\mid D_{\mathsf{s}}\right]~,\quad\text{where}\label{eq: U expo}\\
u(x; D_\mathsf{s}, \xi_{\mathsf{s}}, \theta, g) 
&= - M^{(1)}(x;g_0)^{-1}\sum_{i\in\mathsf{s}} 
\big( \kappa(x, X_i, D_{\mathsf{s}}, \xi_{\mathsf{s}}) m(D_i; \theta, g) \nonumber \\
& \quad\quad\quad\quad\quad\quad\quad\quad\quad\quad
- \mathbb{E}\big[ \kappa(x, X_i, D_{\mathsf{s}}, \xi_{\mathsf{s}}) m(D_i; \theta, g)\big]\big)~,\nonumber
\end{align}
with a remainder term given by the second term in \eqref{eq: kernel coverage}. The quantity \eqref{eq: U expo} can be recognized as a $U$-statistic of order $b$. Second, we apply \cref{lem: U stat linearization} to show that this quantity is approximately linear. Written differently, we establish that the root $R_n(\bm{x}^{(d)})$ satisfies a linear representation of the form \eqref{eq: exact linear}, up to a small remainder term. It immediately follows that the bootstrap root $R^*_n(\bm{x}^{(d)})$ satisfies the linear representation \eqref{eq: half as weight}, up to a small remainder term. Finally, we conclude by applying suitable central limit theorems \citep{chernozhuokov2022improved} to the linear terms \eqref{eq: exact linear} and \eqref{eq: half as weight}.

The broadened scope of \cref{lem: U stat linearization} is an essential input into making \cref{eq: asymptotic validity subsampled kernel} operational. To see this, recall that for many honest subsampled kernel estimators, the shrinkage rate $\varepsilon_b$ satisfies the bound $\varepsilon_b \lesssim b^{-c/p}$ for some small constant $c$ and some integer $p$ measuring the dimension of the covariates $X_i$. Suppose that
\begin{equation}
b = n^{\gamma_b}\quad\text{and}\quad \|\hat{g}_n - g_0\|_{2,\infty} \lesssim n^{-\gamma_g}~,
\end{equation}
with probability greater than $1-1/n$, for some constants $\gamma_b$ and $\gamma_g$ between $0$ and $1$. In this case, ignoring logarithmic factors and other constants, the bound \eqref{eq: kernel coverage} can be re-expressed as
\begin{equation}
n^{\frac{1}{4}(\gamma_b - 1)} + n^{\frac{1}{2}(1-\gamma_b-4\gamma_g)} + n^{\frac{1}{2}(1-\gamma_b(1+c/p)) }~.
\end{equation}
In other words, the confidence region defined in \cref{def: uniform ci} is consistent if
\begin{equation}\label{eq: consistence conditions}
1 \leq \gamma_b+ 4\gamma_g  \quad\text{and}\quad 1 \leq \gamma_b(1+c/p)~,
\end{equation}
and so the bound \eqref{eq: kernel coverage} only converges if 
\begin{align}
    \gamma_b \geq \frac{p}{p + c} ~.
\end{align}
Written differently, it is essential to accommodate subsample exponents $\gamma_b$ very close to one, i.e., the regime $b = o(n)$. This is exactly the range enabled by \cref{lem: U stat linearization}.

\subsection{Simulation\label{sec: simulation}}

We now measure the performance  of the confidence region formulated in \cref{def: uniform ci}. We apply a method for simulation design proposed by \cite{athey2021using}. In particular, we calibrate a simulation to the \cite{banerjee2015multifaceted} data using a Generative Adversarial Network (GAN) \citep{goodfellow2014generative}. In effect, we construct a data generating process that approximates the \cite{banerjee2015multifaceted} data, where we know the true value of the CATE $\theta_0(x)$ queried at each value $x$ used to construct the grids displayed in \cref{fig: cate,fig: half}. Further details are given in \cref{app: simulation}. 

Measurements of performance are taken as two parameters vary. First, we consider several values of the sample size $n$. In particular, we consider settings with $n = h\cdot n_0$, for $h$ in $\{1, 2.5, 5, 7.5\}$, where $n_0$ is the sample size of the \cite{banerjee2015multifaceted} data. Second, we vary the proportion $b/n$. We consider three regimes: $b/n = 0.05$, $b/n = (2/(h+1))0.05$, and $b/n=(1/h)0.05$. Observe that $b$ increases in proportion to $n$ in the first regime and that $b$ is constant as $n$ varies in the third regime. The second regime resides between these two extremes. 

\begin{figure}[t]
\begin{centering}
\caption{Performance}
\label{fig: performance}
\medskip{}
\begin{tabular}{c}
\includegraphics[scale=0.38]{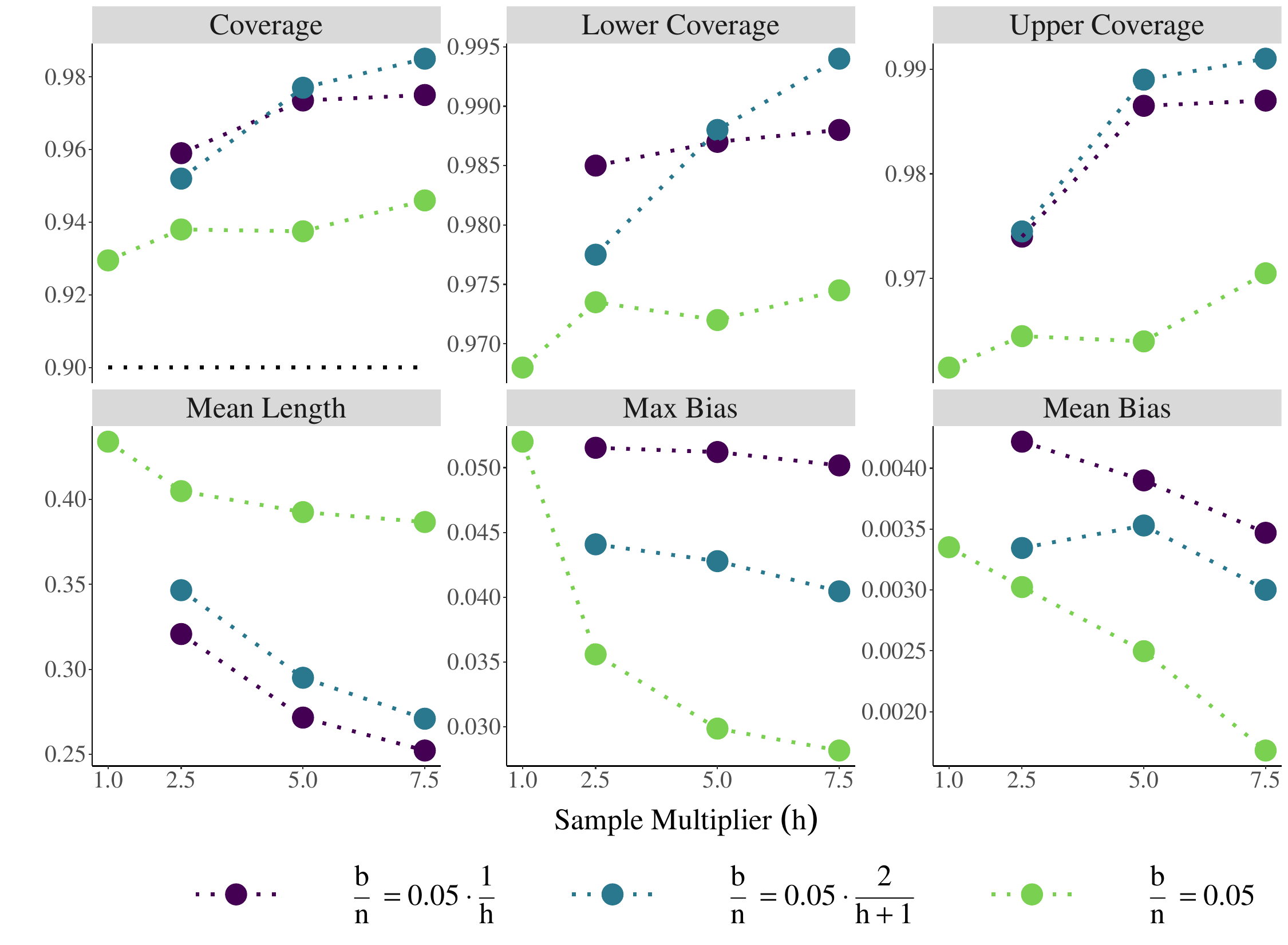}
\end{tabular}
\par\end{centering}
\medskip{}
\justifying
{\footnotesize{}Notes: \cref{fig: performance} displays several measurements of the performance of the confidence intervals formulated in \cref{def: uniform ci} in a simulation calibrated to the \cite{banerjee2015multifaceted} data. The nominal level is $\alpha = 0.1$; a horizontal dotted line is displayed at the nominal coverage $1-\alpha$ in the first panel. The confidence bounds considered are constructed analogously to the confidence bounds displayed in \cref{fig: half}. The $x$-axis of each panel is the sample multiplier $h$. The color of each measurement varies with $b/n$.}{\footnotesize\par}
\end{figure}

\cref{fig: performance} displays measurements of the coverage and width of the confidence region formulated in \cref{def: uniform ci}, in addition to measurements of the bias of the estimator $\hat{\theta}_n(\bm{x}^{(d)})$. Here, we consider the setting where the nuisance parameter estimator $\hat{g}_n$ is computed using the same data used to construct the region $\hat{\mathcal{C}}(\bm{x}^{(d)})$. The first row displays measurements of the coverage of the confidence region, the coverage of the lower bound (i.e., Panel B of \cref{fig: half}), and the coverage of the upper bound (i.e., Panel A of \cref{fig: half}). The nominal level is $\alpha = 0.1$. Throughout, the confidence region is somewhat conservative.

The second row of  \cref{fig: performance} illustrates a bias-variance trade-off with the subsample size $b$. The first panel displays measurements of the average width of the confidence region. Here, the average is taken over both simulation draws and the query-vector $\mathbf{x}^{(d)}$. The width of the confidence region is increasing in the proportion $b/n$ and is essentially constant if $b/n$ is constant as $n$ increases. By contrast, the second and third panels display the maximum and average bias of the estimator $\hat{\theta}_n(\bm{x}^{(d)})$, again taken over the query-vector $\mathbf{x}^{(d)}$. The bias is decreasing in the proportion $b/n$ and is essentially constant if $b$ is constant as $n$ varies.

\end{spacing}
\begin{spacing}{1.2}
\bibliographystyle{apalike}
\bibliography{references.bib}
\end{spacing}
\newpage

\begin{appendix}
\renewcommand\thefigure{\thesection.\arabic{figure}}    

\begin{center}
\large{\it Supplemental Appendix to:}
\vskip0.2cm
\begin{spacing}{1}
\Large{\textbf{Order-Explicit Linearization of High-Dimensional $U$-Statistics\protect\daggerfootnote{\textit{Date}: \today}}}\\\vspace{1em}
\begin{tabular}[t]{c@{\extracolsep{4em}}c} 
\large{David M. Ritzwoller} &  \large{Vasilis Syrgkanis}\vspace{-0.7em}\\ \vspace{-1em}
\small{Stanford University} & \small{Stanford University} \\ \vspace{-0.7em}
\end{tabular}%
\end{spacing}
\end{center}
\begin{spacing}{1.13}
\DoToC
\end{spacing}
\thispagestyle{empty}
\setcounter{page}{0}
\setcounter{figure}{0}   
\newpage

\begin{spacing}{1.4}
\normalsize 

\section{Proofs for Results Supporting \cref{lem: U stat linearization}}\label{app: proofs for u-stat}

\subsection{Proof of \cref{lem: symm and khint}}

\subsubsection{Part (i)} 

The result follows from an iterative application of the following symmetrization inequality. 

\begin{lemma}[\cite{vershynin2018high}, Exercise 6.4.5]\label{lem: symmetrization}
    Let $Z_1,\ldots,Z_n$ be independent, mean-zero, real-valued random variables. If $V_1,\ldots,V_n$ denote an independent sequence of i.i.d.\ Rademacher random variables, then the inequality
    \begin{align}
        \mathbb{E}\left[
        \bigg\vert \sum_{i=1}^n Z_i \bigg\vert^q
        \right]^{1/q}
        \leq 
        2 \mathbb{E}\left[
        \bigg\vert \sum_{i=1}^n V_i Z_i \bigg\vert^q
        \right]^{1/q}
    \end{align}
    holds for each $q \geq 1$. 
\end{lemma}

\noindent In particular, observe that we can write
\begin{align}
    \sum_{i_1=1}^{\lfloor n/m\rfloor} \cdots \sum_{i_m=1}^{\lfloor n/m\rfloor}
    g(D_{i_1}^{(1)}, \ldots, D_{i_m}^{(m)})
    = \sum_{i_1=1}^{\lfloor n/m\rfloor} Z^{(1)}_{i_1}
\end{align}
where the summands
\begin{align}
    Z^{(1)}_{i} = \sum_{i_2=1}^{\lfloor n/m\rfloor} \cdots \sum_{i_m=1}^{\lfloor n/m\rfloor}
    g(D_{i}^{(1)}, D_{i_2}^{(2)}, \ldots, D_{i_m}^{(m)})~.
\end{align}
are independently distributed conditional on the replications $\mathbf{D}_n^{(2)},\ldots,\mathbf{D}_n^{(m)}$. Moreover, the complete degeneracy condition \eqref{eq: completely degenerate} implies that
\begin{align}
    & \mathbb{E}[ Z^{(1)}_{i} \mid \mathbf{D}_n^{(2)},\ldots,\mathbf{D}_n^{(m)}] \nonumber \\\
    & = 
    \sum_{i_2=1}^{\lfloor n/m\rfloor} \cdots \sum_{i_m=1}^{\lfloor n/m\rfloor}
    \mathbb{E}[
    g(D_{i}^{(1)}, D_{i_2}^{(2)}, \ldots, D_{i_m}^{(m)})
    \mid D_{i_2}^{(2)}, \ldots, D_{i_m}^{(m)}] = 0~.
\end{align}
Thus, by applying \cref{lem: symmetrization} conditional on $\mathbf{D}_n^{(2)},\ldots,\mathbf{D}_n^{(m)}$ and taking expectations, we obtain
\begin{align}
    & \mathbb{E}\left[ 
        \bigg\vert \frac{1}{\lfloor n/m\rfloor^m}
        \sum_{i_1=1}^{\lfloor n/m\rfloor} \cdots \sum_{i_m=1}^{\lfloor n/m\rfloor}
    g(D_{i_1}^{(1)}, \ldots, D_{i_m}^{(m)}) \bigg \vert^q 
        \right]^{1/q} \nonumber \\
        & \leq \frac{2}{\lfloor n/m\rfloor^m}
        \mathbb{E}\left[
        \bigg\vert 
        \sum_{i_1=1}^{\lfloor n/m\rfloor} \cdots \sum_{i_m=1}^{\lfloor n/m\rfloor} V_{i_1}^{(1)}
    g(D_{i_1}^{(1)}, \ldots, D_{i_m}^{(m)}) \bigg \vert^q
        \right]^{1/q}~.
\end{align}
Hence, by iterating the same argument $m-1$ more times, first conditionally on all blocks except $\mathbf{D}_n^{(2)}$, then on all blocks except $\mathbf{D}_n^{(3)}$, and so on, we can conclude that
\begin{align}
    & \mathbb{E}\left[ 
        \bigg\vert 
        \frac{1}{\lfloor n/m\rfloor^m}
        \sum_{i_1=1}^{\lfloor n/m\rfloor} \cdots \sum_{i_m=1}^{\lfloor n/m\rfloor}
    g(D_{i_1}^{(1)}, \ldots, D_{i_m}^{(m)}) \bigg \vert^q 
        \right]^{1/q} \nonumber \\
        & \leq \frac{2^m}{\lfloor n/m\rfloor^m}
        \mathbb{E}\left[
        \bigg\vert 
        \sum_{i_1=1}^{\lfloor n/m\rfloor} \cdots \sum_{i_m=1}^{\lfloor n/m\rfloor} V_{i_1}^{(1)}\cdots V_{i_m}^{(m)}
    g(D_{i_1}^{(1)}, \ldots, D_{i_m}^{(m)}) \bigg \vert^q
        \right]^{1/q}~,
\end{align}
as required.\hfill\qed

\subsubsection{Part (ii)}

The result can be obtained as a specialization of the following standard moment inequality for homogeneous Rademacher chaos, often referred to as the Bonami inequality.  

\begin{lemma}[Theorem 3.2.2, \cite{de2012decoupling}]\label{lem: chaos moment bound} Fix a collection of real-valued quantities $\{ z_\mathsf{s} : \mathsf{s} \in \mathcal{S}_{n,b} \}$ and let $V_1,\ldots,V_n$ denote an independent collection of Rademacher random variables. Consider the homogeneous Rademacher chaos of order $b$, given by
\begin{equation}
Z_{b} = \sum_{\mathsf{s} \in \mathcal{S}_{n,b}} V_\mathsf{s} z_{\mathsf{s}}~,
\end{equation}
where $V_{\mathsf{s}} = \Pi_{i\in\mathsf{s}} V_i$ for each subset $\mathsf{s}$ in $[n]$. The moment inequality
\begin{equation}
\mathbb{E} \left[ \vert Z_b \vert^q \right] \leq q^{bq/2} (\Delta_b)^{q/2}~,
\quad\text{where}\quad
\Delta_b = \sum_{s\in\mathcal{S}_{n,b}} (z_{\mathsf{s}})^2~,
\end{equation}
holds for every $q>2$.
\end{lemma}

\noindent To apply \cref{lem: chaos moment bound}, we require some further notation. In particular, let $\bar{n} = \lfloor n /m\rfloor \cdot m$, define the injection
\begin{align}
    h : [m] \times [\lfloor n /m\rfloor] \to [\bar{n}]
    \quad\text{by}\quad h(l,i) = (l-1) \lfloor n /m\rfloor + i~,
\end{align}
and enumerate the sets
\begin{align}
    \mathsf{s}(i_1, \ldots, i_m) = \{h(1,i_1),\ldots,h(m,i_m)\} \in{\mathcal{S}}_{\bar{n},m}
\end{align}
for each $(i_1, \ldots, i_m)$ in $[\lfloor n /m\rfloor]^m$. Observe that the mapping $(i_1, \ldots, i_m) \mapsto \mathsf{s}(i_1, \ldots, i_m)$ is a bijection. Thus, we can define the array
\begin{align}
    z_{\mathsf{s}} = 
    \begin{cases}
        g_{i_1,\ldots,i_m}~,&
        \text{if }\mathsf{s}=\mathsf{s}(i_1, \ldots, i_m)\text{ for some }(i_1,\ldots,i_m)\in[\lfloor n /m\rfloor]^m~,\\
        0~,&\text{otherwise}~,
    \end{cases}
\end{align}
and the sequence of i.i.d.\ Rademacher variables
\begin{align}
    V_{h(l,i)} = V_i^{(l)}~,
\end{align}
for each $l\in[m]$ and $i\in[\lfloor n /m\rfloor]$~, such that
\begin{align}
\sum_{i_1=1}^{\lfloor n/m\rfloor} \cdots \sum_{i_m=1}^{\lfloor n/m\rfloor}
        V_{i_1}^{(1)} \cdots V_{i_m}^{(m)} g_{(i_1, ...,i_m)}
        = 
        \sum_{\mathsf{s} \in \mathcal{S}_{\bar{n},m}}
        V_{\mathsf{s}} z_{\mathsf{s}}  
\end{align}
almost surely. Therefore, \cref{lem: chaos moment bound} implies that
\begin{align}
    & \mathbb{E}\left[
        \bigg\vert
        \sum_{i_1=1}^{\lfloor n/m\rfloor} \cdots \sum_{i_m=1}^{\lfloor n/m\rfloor}
        V_{i_1}^{(1)} \cdots V_{i_m}^{(m)} g_{(i_1, ...,i_m)}
        \bigg\vert^q
        \right]^{1/q} \nonumber \\
        & = 
        \mathbb{E}\left[
        \bigg\vert
        \sum_{\mathsf{s} \in \mathcal{S}_{\bar{n},m}}
        V_{\mathsf{s}} z_{\mathsf{s}}
        \bigg\vert^q
        \right]^{1/q} \nonumber \\
        & \leq q^{m/2} \left( \sum_{\mathsf{s} \in \mathcal{S}_{\bar{n},m}}
        z_{\mathsf{s}}^2 \right)^{1/2}
        \leq q^{m/2} \left(\sum_{i_1=1}^{\lfloor n/m\rfloor} \cdots \sum_{i_m=1}^{\lfloor n/m\rfloor}
        g_{(i_1, ...,i_m)}^2 \right)^{1/2}
\end{align}
for each $q > 2$, as required.\hfill\qed

\subsection{Proof of \cref{lem: bound parts}}

\subsubsection{Part (i)}

It will suffice to establish the following more general statement. Fix integers $m \geq 1$ and $k \geq 1$ and let $g : \mathcal{D}^m \to \mathbb{R}_{+}$ be a symmetric, non-negative kernel. Define the quantity
\begin{align}
    A_{m,k}(g) = \frac{1}{k^m}
        \sum_{i_1=1}^{k} \cdots \sum_{i_m=1}^{k}
        g(D_{i_1}^{(1)},\ldots,D_{i_m}^{(m)})~.
\end{align}
The inequality
\begin{align}
    \mathbb{E}\left[
        \vert
        A_{m,k}(g)
        \vert^{q}
        \right]^{1/q} \nonumber
        \leq C^m \sum_{t=0}^m \left(\frac{q}{k}\right)^t
        \Gamma_{t,q}(g)\label{eq: general k bound to show}
\end{align}
holds for each $q \geq 1 \vee \log(k)$. 

We proceed by induction on $m$. To verify the base case, associated with $m=1$, it suffices to show that
\begin{align}
\mathbb{E}\left[
        \vert
        A_{1,k}(g)
        \vert^{q}
        \right]^{1/q}
        = 
    \mathbb{E}\left[
        \bigg\vert
        \frac{1}{k}
        \sum_{i_1=1}^{k}
        g(D_{i_1}^{(1)})
        \bigg\vert^{q}
        \right]^{1/q} \nonumber
        &\lesssim \Gamma_{0,q}(g)
        + \frac{q}{k}\Gamma_{1,q}(g) \nonumber \\
        & = \mathbb{E}[g(D_{i_1}^{(1)})] + \frac{q}{k} 
        \mathbb{E}\left[ 
        \vert g(D_{i_1}^{(1)}) \vert^q
        \right]^{1/q}
\end{align}
holds for each $q \geq \log(k)$. This bound follows immediately from the following specialization of Rosenthal's inequality for non-negative random variables.
\begin{lemma}\label{lem: large q rosenthal}
    Let $Z_1,\ldots,Z_k$ be i.i.d.\ non-negative random variables. Then, the inequality
    \begin{align}
        \mathbb{E}\left[ 
        \bigg\vert \frac{1}{k} \sum_{i=1}^k Z_i \bigg\vert^q
        \right]^{1/q}
        \lesssim 
        \mathbb{E}[Z_1] + \frac{q}{k} 
        \mathbb{E}\left[ 
        \vert Z_1 \vert^q
        \right]^{1/q}
    \end{align}
    holds for each $q \geq 1 \vee \log(k)$.
\end{lemma}

\begin{proof}
    Rosenthal's inequality for non-negative random variables, stated as Theorem 15.10 in \cite{boucheron2013concentration}, implies that
    \begin{align}
        \mathbb{E}\left[ 
        \bigg\vert \sum_{i=1}^k Z_i \bigg\vert^q
        \right]^{1/q}
        \lesssim 
        k \mathbb{E}[Z_1] + q 
        \mathbb{E}\left[ \max_{i \in [k]}
        \vert Z_i \vert^q
        \right]^{1/q}~.\label{eq: rosenthal state}
    \end{align}
    As $q \geq \log(k)$, we have that $k^{1/q} \leq e$, and therefore
    \begin{align}
        \mathbb{E}\left[ \max_{i \in [k]}
        \vert Z_i \vert^q
        \right]^{1/q}
        \leq k^{1/q}
        \mathbb{E}\left[ 
        \vert Z_i \vert^q \right]^{1/q}
        \leq e \mathbb{E}\left[ 
        \vert Z_i \vert^q \right]^{1/q}~.\label{eq: max bound rosenthal}
    \end{align}
    The result follows by plugging \eqref{eq: max bound rosenthal} into \eqref{eq: rosenthal state} and dividing by $k$.\hfill
\end{proof}

Thus, it suffices to verify the inductive step. That is, assume that the bound \eqref{eq: general k bound to show} has been established for kernels of order $m-1$. Observe that we can write
\begin{align}
    A_{m,k}(g) = 
    \frac{1}{k^m}
        \sum_{i_1=1}^{k} \cdots \sum_{i_m=1}^{k}
        g(D_{i_1}^{(1)},\ldots,D_{i_m}^{(m)})
        &= 
        \frac{1}{k} \sum_{{i_1}=1}^k
        Z^{(1)}_{i_{1}}
\end{align}
where
\begin{align}
    Z^{(1)}_{i_1} = \frac{1}{k^{m-1}}
    \sum_{i_2=1}^k \cdots \sum_{i_m=1}^k
    g(D^{(1)}_{i_1}, D^{(2)}_{i_2},\ldots,D^{(m)}_{i_m})~.
\end{align}
Conditional on the replicates $\mathbf{D}_n^{(2)},\ldots,\mathbf{D}_n^{(m)}$, the variables $Z^{(1)}_{1},\ldots,Z^{(1)}_k$ are i.i.d.\ and non-negative. Therefore, \cref{lem: large q rosenthal} implies that
\begin{align}
    &\mathbb{E}\left[
    \vert 
        A_{m,k}(g) \vert^q
    \mid \mathbf{D}_n^{(2)},\ldots,\mathbf{D}_n^{(m)}\right]^{1/q} \nonumber \\
    & 
    \lesssim 
    \mathbb{E}[Z^{(1)}_{1} \mid \mathbf{D}_n^{(2)},\ldots,\mathbf{D}_n^{(m)}] 
    + \frac{q}{k} 
        \mathbb{E}\left[ 
        \vert Z^{(1)}_{1} \vert^q
        \mid \mathbf{D}_n^{(2)},\ldots,\mathbf{D}_n^{(m)}\right]^{1/q}~.
\end{align}
Hence, Minkowski's inequality implies that
\begin{align}
    &\mathbb{E}\left[
    \vert A_{m,k}(g) \vert^q
    \right]^{1/q}
    \lesssim 
    \mathbb{E}\left[ 
    \mathbb{E}[Z^{(1)}_{1} \mid \mathbf{D}_n^{(2)},\ldots,\mathbf{D}_n^{(m)}]^q \right]^{1/q}
    + \frac{q}{k} 
        \mathbb{E}\left[ 
        \vert Z^{(1)}_{1} \vert^q \right]^{1/q}\label{eq: apply inductive rosenthal}
\end{align}
as $q \geq 1$. Now, by defining the $m-1$ order kernels
\begin{align}
    g_0(d_2,\ldots,d_m) = \mathbb{E}[g(D_1,d_2,\ldots,d_m)]
    \quad\text{and}\quad
    g_d(d_2,\ldots,d_m) = g(d,d_2,\ldots,d_m)
\end{align}
we can identify
\begin{align}
    \mathbb{E}[Z^{(1)}_{1} \mid \mathbf{D}_n^{(2)},\ldots,\mathbf{D}_n^{(m)}]
    = A_{m-1,k}(g_0)
    \quad\text{and}\quad
    Z^{(1)}_{1} = A_{m-1,k}(g_{D^{(1)}_{i_1}})~,
\end{align}
respectively. Hence, the inductive hypothesis implies that
\begin{align}
    \mathbb{E}\left[ 
    \mathbb{E}[Z^{(1)}_{1} \mid \mathbf{D}_n^{(2)},\ldots,\mathbf{D}_n^{(m)}]^q \right]^{1/q}
    &\leq
    C^{m-1} \sum_{t=0}^{m-1} \left( \frac{q}{k} \right)^t \Gamma_{t,q}(g_0)\quad\text{and}\nonumber \\
    \mathbb{E}\left[ 
    \vert Z^{(1)}_{1} \vert^q \right]^{1/q}
    & \leq
    C^{m-1} \sum_{t=0}^{m-1} \left( \frac{q}{k} \right)^t \mathbb{E}\left[ \Gamma_{t,q}(g_{D^{(1)}_{i_1}}) ^q\right]^{1/q}~.\label{eq: apply inductive general}
\end{align}
By recognizing
\begin{align}
    \Gamma_{t,q}(g_0)
    & =
    \mathbb{E}\left[
    \mathbb{E}[g(D_1,\ldots,D_m)\mid D_2,\ldots,D_{t+1}]^q
    \right]^{1/q}
    = \Gamma_{t,q}(g)\nonumber\quad\text{and}
    \\
    \mathbb{E}\left[
    \Gamma_{t,q}(g_{D_1^{(1)}})^q
    \right]^{1/q}
    & =
    \mathbb{E}\left[
    \mathbb{E}[g(D_1,\ldots,D_m)\mid D_1,\ldots,D_{t+1}]^q
    \right]^{1/q}
    = \Gamma_{t+1,q}(g)
\end{align}
for each $0 \leq t \leq m-1$ and plugging the bounds \eqref{eq: apply inductive general} into \eqref{eq: apply inductive rosenthal}, we find that
\begin{align}
    \mathbb{E}\left[
    |A_{m,k}(g)|^q
    \right]^{1/q} 
    & \leq
    C^m \sum_{t=0}^{m-1}
    \left( \frac{q}{k}\right)^t
    \Gamma_{t,q}(g)
    + C^m \frac{q}{k}
    \sum_{t=0}^{m-1}
    \left( \frac{q}{k}\right)^t
    \Gamma_{t+1,q}(g) \nonumber \\
    & \leq C^m \sum_{t=0}^m \left( \frac{q}{k}\right)^t \Gamma_{t,q}(g)~,
\end{align}
as required.\hfill\qed

\subsubsection{Part (ii)}

We begin by stating two facts that will be applied at several points throughout the ensuing argument. First, we record the following property of sub-exponential random variables.

\begin{lemma}[\citet{vershynin2018high}, Proposition 2.7.1]$ $\\
\label{lem: psi1 moment growth}Let $Z$ be a real-valued random variable. Then the moment inequality
\begin{align}
    \mathbb{E}[|Z|^q]^{1/q}
    \leq
    C q \|Z\|_{\psi_1}
\end{align}
holds for each $q\geq 1$~.
\end{lemma}

\noindent Second, we note that if $Z^\prime$ is an independent copy of $Z$, then the triangle inequality implies that
\begin{align}
    \| Z - Z' \|_{\psi_1} \leq \|Z\|_{\psi_1} + \|Z'\|_{\psi_1} = 2\|Z\|_{\psi_1}~.
\end{align}
Thus, by Jensen's inequality, we obtain the bound
\begin{align}
    \| f^{(t)}(D_{[t]}) \|_{\psi_1}
    \leq \| \nabla_1 \cdots \nabla_t f(D_{[b]}) \| \leq 2^t \phi\label{eq: Hoeff Orlicz bound}
\end{align}
for each integer $0 \leq t \leq b$.

Now, our interest is in the quantity
\begin{align}
    \Gamma_{t,q}((f^{(m)})^2)
    & = \mathbb{E}
    \left[
    \mathbb{E}[ f^{(m)}(D_{[m]})^2 \mid D_1, \ldots,D_t]^q
    \right]^{1/q}
\end{align}
for each integer $0 \leq t \leq m$. We first consider the case $0 \leq t \leq m - 1$. To this end, for each deterministic collection $(d_1,\ldots,d_t)$ in $\mathcal{D}^t$, define the symmetric, order $b-t$ kernel
\begin{align}
    & g_{d_1,\ldots,d_t}(d_{t+1},\ldots,d_b) \nonumber \\
    & =
    (-1)^t
    \mathbb{E}\left[
    \nabla_1\cdots\nabla_t f(D_{[b]})
    \mid D_1=d_1,\ldots,D_t=d_t,D_{t+1}=d_{t+1},\ldots,D_b=d_b
    \right]
\end{align}
and let $g_{d_1,\ldots,d_t}^{(1)},\ldots,g_{d_1,\ldots,d_t}^{(b-t)}$ denote its Hoeffding projections, as defined in Part (i) of \cref{lem: Diff H decomposition}. By Part (i) of \cref{lem: Diff H decomposition}, these projections are uncorrelated, and so the decomposition
\begin{align}
    \mathbb{E}\left[
    g_{d_1,\ldots,d_t}(D_{t+1},\ldots,D_b)^2
    \right]  &=
    \mathbb{E}[ g_{d_1,\ldots,d_t}(D_{t+1},\ldots,D_{t+l})]^2 \nonumber \\
    & + 
    \sum_{l=1}^{b-t} {b-t \choose l}
    \mathbb{E}\left[
    \bigg(g_{d_1,\ldots,d_t}^{(l)}(D_{t+1},\ldots,D_{t+l})\bigg)^2
    \right]\label{eq: hoeffding applied to projection}
\end{align}
holds. Now, observe that
\begin{align}
    & g_{d_1,\ldots,d_t}^{(m-t)}(d_{t+1},\ldots,d_m) \\
    & =
    (-1)^{m-t}
    \mathbb{E}\left[
    \nabla_{t+1}\cdots\nabla_m g_{d_1,\ldots,d_t}(D_{t+1},\ldots,D_b)
    \mid D_{t+1}=d_{t+1},\ldots,D_m=d_m
    \right] \nonumber \\
    & = (-1)^m
    \mathbb{E}\left[
    \nabla_1\cdots\nabla_m f(D_{[b]})
    \mid D_1=d_1,\ldots,D_m=d_m
    \right] \nonumber \\
    & = f^{(m)}(d_1,\ldots,d_m)~,
\end{align}
where the first and last equalities follow by definition and the second equality follows from the fact that the stochastic difference operators act on distinct coordinates, and thereby commute. As a consequence, the decomposition \eqref{eq: hoeffding applied to projection} implies that
\begin{align}
    & {b-t \choose m-t}
    \mathbb{E}\left[
    f^{(m)}(D_1,\ldots,D_t,D_{t+1},\ldots,D_m)^2
    \mid D_1, \ldots, D_t\right] \nonumber \\
    & \leq
    \mathbb{E}\left[
    g_{D_1,\ldots,D_t}(D_{t+1},\ldots,D_b)^2
    \mid D_1, \ldots, D_t\right]
\end{align}
holds almost surely. Therefore, we obtain the bound
\begin{align}
    \Gamma_{t,q}((f^{(m)})^2)
    & \leq {b-t \choose m-t}^{-1}
    \mathbb{E}
    \left[
    \mathbb{E}\left[
    g_{D_1,\ldots,D_t}(D_{t+1},\ldots,D_b)^2
    \mid D_1, \ldots, D_t\right]^{q}
    \right]^{1/q}~.
\end{align}
Now, Jensen's inequality for conditional expectation and \cref{lem: psi1 moment growth} imply that
\begin{align}
    & \mathbb{E}
    \left[
    \mathbb{E}\left[
    g_{D_1,\ldots,D_t}(D_{t+1},\ldots,D_b)^2
    \mid D_1, \ldots, D_t\right]^q
    \right]^{1/q} \nonumber \\
    & \leq
    \mathbb{E}
    \left[
    \left\vert
    g_{D_1,\ldots,D_t}(D_{t+1},\ldots,D_b)
    \right\vert^{2q}
    \right]^{1/q} \nonumber \\
    & \leq (Cq)^2 \| g_{D_1,\ldots,D_t}(D_{t+1},\ldots,D_b)\|_{\psi_1}^2
    \leq (Cq)^2 \phi^2 4^t~,
\end{align}
where the last step follows from \eqref{eq: Hoeff Orlicz bound}. Hence, as
\begin{align}
    {b-t \choose m-t}^{-1}
    \leq {b \choose m}^{-1} \left(\frac{eb}{m}\right)^t~,
\end{align}
we can conclude that
\begin{align}
    \Gamma_{t,q}\bigl((f^{(m)})^2\bigr)
    \leq
    \phi^2 (Cq)^2 C^t
    {b \choose m}^{-1}
    \left( \frac{b}{m} \right)^t
\end{align}
for each $0\leq t\leq m-1$.

It remains to consider the case $t=m$. In this case, we have that
\begin{align}
    \Gamma_{m,q}\bigl((f^{(m)})^2\bigr)
    =
    \mathbb{E}[|f^{(m)}(D_{[m]})|^{2q}]^{1/q}
    \leq
    \phi^2 (Cq)^2 4^m~.
\end{align}
by \cref{lem: psi1 moment growth} and \eqref{eq: Hoeff Orlicz bound}. The bound ${b \choose m}\leq (eb/m)^m$ then implies that
\begin{align}
    4^m
    \leq
    C^m {b \choose m}^{-1}
    \left( \frac{b}{m} \right)^m~,
\end{align}
completing the proof.\hfill\qed

\subsection{Proof of \cref{cor: general u stat bernstein}\label{sec: app cor p1}}

The result follows by considering each of the terms in the decomposition 
\begin{align}
& U_{n,b}(u) - \mathbb{E}[u(D_{[b]})] \nonumber \\
& =  \frac{b}{n} \sum_{i=1}^n u^{(1)}(D_i) + \left(U_{n,b}(u) - \mathbb{E}[u(D_{[b]})] - \frac{b}{n} \sum_{i=1}^n u^{(1)}(D_i)\right)~.\label{eq: resid}
\end{align}
The first quantity can be handled through the application of the following standard Bernstein-type bound, stated, e.g., in \cite{song2019approximating}.

\begin{lemma}[Lemma A.2, \cite{song2019approximating}]$ $\\
\label{lem: orlicz large deviation}Let $Z_{1},\ldots,Z_{n}$ be independent,
centered, random vectors in $\mathbb{R}^{d}$. Define the quantity 
\begin{equation}
\sigma^2 = \max_{j\in[d]} \sum_{i=1}^n \mathbb{E}\left[Z^2_{i,j}\right]
\end{equation}
and assume that $\|Z_{ij}\|_{\psi_{1}}\leq \phi $ for all $i\in[n]$ and $j\in[d]$.
The inequality 
\begin{flalign*}
 & P\left\{ \Big\|\sum_{i=1}^{n}Z_{i}\Big\|_{\infty}
 \geq  C \left(\sigma \log^{1/2}(dg) + \phi \log(nd)\left(\log(nd) + \log(g)\right)\right) \right\}
 \lesssim \frac{1}{g}
\end{flalign*}
holds for any constant $g>0$.
\end{lemma}

\noindent In particular, observe that
\begin{align}
    \max_{j \in [d]} \sum_{i=1}^n \frac{b^2}{n^2}\mathbb{E}[ u^{(1)}(D_i)^2]
    & = \frac{b^2}{n} \bar{\sigma}_{b}^2
    \quad\text{and}\quad
    \|\frac{b}{n} u^{(1)}(D_i)\|_{\psi_1} \leq \frac{b}{n}\phi 
\end{align}
Thus, \cref{lem: orlicz large deviation} implies that
\begin{align}
\Big\| \frac{b}{n} \sum_{i=1}^n u^{(1)}(D_i) \Big\|_{\infty} 
\lesssim \frac{b\bar{\sigma}_{b}}{n^{1/2}} \log^{1/2}(nd)
+ \phi \frac{b}{n} \log^2(nd)
\end{align}
with probability greater than $1-C/nd$. In turn, \cref{lem: U stat linearization} implies that
\begin{align}
    \Big\|  U_{n,b}(u) - \mathbb{E}[u(D_{[b]})] - \frac{b}{n} \sum_{i=1}^n u^{(1)}(D_i)\Big\|_{\infty}
    \lesssim 
    \phi \frac{b}{n} \log^2(nd)
\end{align}
with probability greater than $1-C/nd$. Hence, we can conclude that 
\begin{align}
\Big\| U_{n,b}(u) - \mathbb{E}[u(D_{[b]})] \Big\|_{\infty} 
\lesssim \sqrt{\frac{b^2 \bar{\sigma}^2_{b}\log(nd)}{n}}
+ \phi \frac{b}{n} \log^2(nd)
\end{align}
with probability greater than $1-C/nd$, by a union bound.\hfill\qed

\subsection{Proof of \cref{cor: general u stat clt}\label{sec: app cor p2}}

We give the details of the proof of the upper bound encoded in \eqref{eq: general u stat clt}. The lower bound will follow from an analogous argument. Fix a rectangle $\mathsf{R} = [a_l, a_u]$ in $\mathcal{R}$, where $a_l$ and $a_u$ are vectors in $\mathbb{R}^d$ with $a_l\leq a_u$, interpreted component-wise. Define the enlarged rectangle $\mathsf{R}_t=[a_l-t\bm{1}_d,a_u+t\bm{1}_d]$ for each $t>0$ as well as the normalized H\'{a}jek projection
\begin{equation}
\hat{u}^{(1)}(D_i) = \Sigma_b^{-1/2}u^{(1)}(D_i)~.
\end{equation}
Observe that the decomposition
\begin{align}
\sqrt{\frac{n}{b^2}}\Sigma_b^{-1/2} (U_{n,b}(u) - \mathbb{E}[u(D_{[b]})])
& =  \frac{1}{\sqrt{n}}\sum_{i=1}^n \hat{u}^{(1)}(D_i)  \nonumber \\
& + \sqrt{\frac{n}{b^2}}\Sigma_b^{-1/2} \left(U_{n,b}(u) - \mathbb{E}[u(D_{[b]})] - \frac{b}{n} \sum_{i=1}^n u^{(1)}(D_i)\right)\label{eq: resid clt gen}
\end{align}
implies the upper bound
\begin{align}
 & P\left\{
 \sqrt{\frac{n}{b^2}}\Sigma_b^{-1/2} (U_{n,b}(u) - \mathbb{E}[u(D_{[b]})]) 
 \in \mathsf{R}\right\} - P\left\{ \Sigma_b^{-1/2}Z \in \mathsf{R}\right\} \label{eq: general to upper bound}\\
 & \leq \bigg\vert P\left\{\frac{1}{\sqrt{n}} \sum_{i=1}^n \hat{u}^{(1)}( D_i) \in \mathsf{R}_t\right\} - P\left\{ \Sigma_b^{-1/2}Z \in \mathsf{R}_t\right\}\bigg\vert\label{eq: general iid}\\
 & + \bigg\vert P\left\{ \Sigma_b^{-1/2}Z \in \mathsf{R}_t\right\} - P\left\{ \Sigma_b^{-1/2}Z \in \mathsf{R}\right\} \bigg\vert\label{eq: general comparions}\\
& + P\left\{ \big \|
\sqrt{\frac{n}{b^2}}\Sigma_b^{-1/2} \left(U_{n,b}(u) - \mathbb{E}[u(D_{[b]})] - \frac{b}{n} \sum_{i=1}^n u^{(1)}(D_i)\right)
\big \|_\infty> t\right\} ~.\label{eq: general hajek}
\end{align}
We bound the three terms appearing in \eqref{eq: general to upper bound} in succession. 

We begin by bounding the normal approximation term \eqref{eq: general iid} through the application of the following quantitative central limit theorem,
stated as Theorem 2.1 of \citet{chernozhuokov2022improved}.
\begin{lemma}[{\citealp[Theorem 2.1, ][]{chernozhuokov2022improved}}]
\label{lem: baseline clt}Let $X_{1},\ldots,X_{n}$ be a collection
of independent, centered, random vectors in $\mathbb{R}^{d}$ and
let $Z$ be a centered Gaussian random vector with covariance matrix
\begin{equation}
\frac{1}{n}\sum_{i=1}^{n}\mathbb{E}\left[X_{i}X_{i}^{\top}\right].
\end{equation}
If there exist absolute constants $c$, $C_1$, and $\varphi$ such that the bounds
\begin{equation}
\frac{1}{n}\sum_{i=1}^{n}\mathbb{E}\left[X_{i,j}^{2}\right]\geq c~,\quad
\frac{1}{n}\sum_{i=1}^{n}\mathbb{E}\left[X_{i,j}^{4}\right]\leq C_1 \varphi^{2}~,
\quad\text{and}\quad
\|X_{i,j} \|_{\psi_1} \leq \varphi \label{eq: general clt moment bounds}
\end{equation}
hold, then the inequality 
\begin{equation}
\sup_{\mathsf{R}\in\mathcal{R}}
\Big\vert P\left\{ \frac{1}{\sqrt{n}}\sum_{i=1}^{n}X_{i}\in \mathsf{R}\right\} 
-P\left\{ Z\in \mathsf{R}\right\} \Big\vert\leq C_2 \left(\frac{\varphi^{2}\log^{5}(dn)}{n}\right)^{1/4}
\end{equation}
holds for some constant $C_2$ that depends only on $c$ and $C_1$. 
\end{lemma}

\noindent In particular, observe that 
\begin{equation}
\frac{1}{n}\sum_{i=1}^{n}\mathbb{E}\left[\left(\hat{u}^{(1)}(D_{i})\right)^2\right]= \frac{\mathbb{E}\left[\left(u^{(1)}(D_{i})\right)^2\right]}{\sigma^2_{b,j}} = 1
\end{equation}
for each coordinate $j$. Moreover, it holds that
\begin{align}
    \| \hat{u}^{(1)}(x^{(j)}, D_{i}) \|_{\psi_1} \leq (\phi/\underline{\sigma}_b)
\end{align}
for each coordinate $j$ by the boundedness condition \eqref{eq: boundedness intro}. Similarly, it holds that 
\begin{equation}
\frac{1}{n}\sum_{i=1}^{n}\mathbb{E}\left[\left(\hat{u}^{(1)}(x^{(j)},D_{i})\right)^4\right] 
\leq \left(\phi/\underline{\sigma}_b\right)^{2}
\label{eq: standardized}
\end{equation}
for each coordinate $j$, again by boundedness. Consequently, as 
\begin{equation}
\Var(\hat{u}^{(1)}(D_{i}))=\Sigma_b^{-1/2}\Var\left(Z\right)\Sigma_b^{-1/2},
\end{equation}
\cref{lem: baseline clt} implies that the bound
\begin{equation}
\bigg\vert P\left\{\frac{1}{\sqrt{n}} \sum_{i=1}^n \hat{u}^{(1)}(D_i) \in \mathsf{R}_t\right\} - P\left\{ \Sigma_b^{-1/2}Z \in \mathsf{R}_t\right\}\bigg\vert
\lesssim \left(\frac{\phi^{2}\log^{5}(dn)}{\underline{\sigma}_b^{2}n}\right)^{1/4}\label{eq: apply clt intro}
\end{equation}
holds. 

Next, to bound the difference in the Gaussian probabilities \eqref{eq: general comparions}, we apply the following anti-concentration inequality, stated in \citet{chernozhukov2017detailed} and often referred to as Nazarov's inequality. 

\begin{lemma}[{\citealp[Theorem 1, ][]{chernozhukov2017detailed}}]
\label{lem: Nazarov}Let $Z=(Z_{j})_{j=1}^{d}$ be a centered Gaussian
random vector in $\mathbb{R}^{d}$ such that $\mathbb{E}[Z_{j}^{2}]\geq c$
for all $j$ in $[d]$ and some constant $c$. For every $z\in\mathbb{R}^{d}$
and $t>0$, the inequality 
\[
P\left\{ Z\leq z+t\right\} -P\left\{ Z\leq z\right\} \lesssim\frac{t}{c}\sqrt{\log d}
\]
holds. 
\end{lemma}
\noindent In particular, \cref{lem: Nazarov} gives
\begin{equation}
\Big\vert P\left\{ \Sigma_b^{-1/2}Z \in \mathsf{R}_t\right\} - P\left\{ \Sigma_b^{-1/2}Z \in \mathsf{R}\right\} \Big\vert \lesssim t \sqrt{\log(d)}
\end{equation}
for all $t > 0$. By choosing 
\begin{align}
    t = C\frac{1}{\sqrt{\log(dn)}}
    \left(\frac{\phi^{2}\log^{5}(dn)}{\underline{\sigma}_b^{2}n}\right)^{1/4}
\end{align}
we find that the bound
\begin{align}
    \Big\vert P\left\{ \Sigma_b^{-1/2}Z \in \mathsf{R}_t\right\} - P\left\{ \Sigma_b^{-1/2}Z \in \mathsf{R}\right\} \Big\vert \lesssim
    \left(\frac{\phi^{2}\log^{5}(dn)}{\underline{\sigma}_b^{2}n}\right)^{1/4}
    \label{eq: apply nazarov intro}
\end{align}
holds. 

Finally, we bound the term \eqref{eq: general hajek}. Observe that it is without loss of generality to assume that 
\begin{align}
    \frac{\phi^{2}\log^{5}(dn)}{\underline{\sigma}_b^{2}n} < 1
    \label{eq: wlog assumption intro}
\end{align}
as otherwise the bound \eqref{eq: general u stat clt} holds vacuously. Now, observe that \cref{lem: U stat linearization} implies that
\begin{align}
    \bigg\|
    U_{n,b}(u) - \mathbb{E}[u(D_{[b]})]
    - \frac{b}{n} \sum_{i=1}^n u^{(1)}(D_i)
    \bigg\|_\infty \lesssim \phi \frac{b}{n}\log^2(dn)
\end{align}
with probability at least $1 - C/nd$. Consequently, it holds that
\begin{align}
    & \bigg\|
    \sqrt{\frac{n}{b^2}}\Sigma_b^{-1/2} \left(
    U_{n,b}(u) - \mathbb{E}[u(D_{[b]})]
    - \frac{b}{n} \sum_{i=1}^n u^{(1)}(D_i)\right)
    \bigg\|_\infty \nonumber \\
    & \lesssim \phi \frac{1}{\sqrt{n}\underline{\sigma}_b} \log^2(dn)
    =
    \frac{1}{\sqrt{\log(dn)}}
    \left(\frac{\phi^{2}\log^{5}(dn)}{\underline{\sigma}_b^{2}n}\right)^{1/2}
    \label{eq: apply linearization intro}
\end{align}
with probability at least $1 - C/nd$, where the final inequality follows from \eqref{eq: wlog assumption intro}. 

Hence, by plugging the bounds \eqref{eq: apply clt intro}, \eqref{eq: apply nazarov intro}, and \eqref{eq: apply linearization intro} into \eqref{eq: general to upper bound}, we find that 
\begin{align}
 & P\left\{
 \sqrt{\frac{n}{b^2}}\Sigma_b^{-1/2} (U_{n,b}(u) - \mathbb{E}[u(D_{[b]})]) 
 \in \mathsf{R}\right\} - P\left\{ \Sigma_b^{-1/2}Z \in \mathsf{R}\right\} \nonumber\\
 &\lesssim 
 \left(\frac{\phi^{2}\log^{5}(dn)}{\underline{\sigma}_b^{2}n}\right)^{1/4} + \frac{1}{nd} 
 \lesssim 
 \left(\frac{\phi^{2}\log^{5}(dn)}{\underline{\sigma}_b^{2}n}\right)^{1/4}~,
\end{align}
as required.\hfill$\blacksquare$

\subsection{Proof of \cref{rem: lower bound}}

Assume that the components of each observation $D_{i} = (D_{i,j})_{j=1}^d$ are independent and have a standard Gaussian distribution. Consider the kernel
\begin{align}
    u_j(d_1, \ldots,d_b) = \frac{\phi}{16 b} \sum_{1 \leq r < s\leq b} d_{r,j} d_{s,j}
\end{align}
where $d_k = (d_{k,j})_{j=1}^d$ in $\mathbb{R}^d$. We claim that the bound 
\begin{align}\label{eq: lower bound orlicz}
    \| u_j(D_{[b]})\|_{\psi_1} \leq \phi
\end{align}
and the lower bound
\begin{align}
        P\left\{
        \bigg\| U_{n,b}(u) - \mathbb{E}[ u(D_{[b]})]- \frac{b}{n}\sum_{i=1}^n u^{(1)}(D_i) \bigg\|_{\infty}
        \geq c
        \frac{\phi b \log(nd)}{n}
        \right\} \geq \frac{c}{nd}\label{eq: lower bound to show}
    \end{align}
hold for this choice. 

We begin by verifying the bound \eqref{eq: lower bound orlicz}. Consider the quantity
\begin{align}
    Q_{j} = \frac{1}{b} \sum_{1 \leq r < s \leq b} D_{r,j} D_{s,j}~.
\end{align}
Let $A_b \in \mathbb{R}^{b\times b}$ be the symmetric matrix with zeros on the diagonal and off-diagonal entries equal to $1/(2b)$. The eigenvalues of $A_b$ are given by 
\begin{align}
    \lambda_1 = \frac{b-1}{2b}\quad\text{and}\quad
    \lambda_2=\cdots=\lambda_b = -\frac{1}{2b}
\end{align}
where we note that
\begin{align}
    \sum_{l=1}^b \vert \lambda_l \vert = \frac{b-1}{b} \leq 1
    \quad\text{and}\quad
    \sum_{l=1}^b \lambda_l = 0~.\label{eq: mean zero eigenvalues}
\end{align}
Observe that we can write
\begin{align}
    Q_j = D_j^\top A_b D_j~.
\end{align}
Thus, by the orthogonal invariance of the standard Gaussian law, there exist independent standard Gaussian replicates $G_{1,j},\ldots,G_{b,j}$ such that
\begin{align}
    Q_j = \sum_{l=1}^b \lambda_l G_l^2 = \sum_{l=1}^b \lambda_l (G_l^2 - 1)~,
\end{align}
where the second equality follows from \eqref{eq: mean zero eigenvalues}. Consequently, as each replicate $G_j$ satisfies the bound
\begin{align}
    & \mathbb{E}\left[ 
    \exp\left( \frac{\vert G_j^2 -1 \vert}{16}  \right) 
    \right]
    \leq
    \exp(1/16) \mathbb{E}[\exp(G_j^2 / 16)] \nonumber \\
    & \leq
    \exp(1/16) (1 - 1/8)^{-1/2} < 2~,\label{eq: Gaussian orlicz bound}
\end{align}
we can evaluate
\begin{align}
    \mathbb{E}\left[ \exp\left(\frac{\vert Q_j \vert}{16}\right) \right]
    & \leq 
    \mathbb{E}\left[
    \exp\left(\frac{ \sum_{l=1}^b \vert \lambda_l \vert 
    \vert G_l^2 - 1 \vert}{16}\right) \right] \nonumber \\
    &= \prod_{l=1}^b  
    \mathbb{E}\left[
    \exp\left(\frac{
    \vert G_l^2 - 1 \vert}{16}\right) \right]^{\vert \lambda_l \vert }
    \leq 2^{\sum_{l=1}^b \vert \lambda_l \vert } \leq 2~,
\end{align}
where the final equality follows from \eqref{eq: mean zero eigenvalues}. Hence, the bound
\begin{align}
    \| Q_j \|_{\psi_1} \leq 16 
    \quad\text{and thereby}\quad
    \|u_j(D_{[b]}) \|_{\psi_1} \leq \phi
\end{align}
hold by the definition of the $\psi_1$-Orlicz norm.

Next, we verify the bound \eqref{eq: lower bound to show}. First, observe that
\begin{align}
    \mathbb{E}[u_j(D_{[b]})] & = 0
\end{align}
and that 
\begin{align}
    \mathbb{E}[u_j(D_{[b]}) \mid D_1]
    = \frac{\phi}{16b} \left(
    \sum_{l=2}^b D_{1,j} \mathbb{E}[D_{l,j}] + \sum_{2 \leq r < s \leq b}
    \mathbb{E}[D_{r,j} D_{s,j}]
    \right) = 0~.
\end{align}
Thus, the H\'{a}jek projection is given by
\begin{align}
    u^{(1)}(D_1) = \mathbb{E}[u(D_{[b]}) \mid D_1] - \mathbb{E}[u(D_{[b]})] = 0
\end{align}
almost surely. As a consequence, it holds that
\begin{align}
    \bigg\| U_{n,b}(u) - \mathbb{E}[ u(D_{[b]})]- \frac{b}{n}\sum_{i=1}^n u^{(1)}(D_i) \bigg\|_{\infty}
    = \big\| U_{n,b}  \big\|_{\infty}\label{eq: no Hajek}
\end{align}
almost surely.

Now, let $U_{n,b,j}(u)$ denote the $j$th component of $U_{n,b}(u)$. Observe that, as each unordered pair $\{i,k\}$ appears in exactly ${ n -2 \choose b-2}$ subsets $\mathsf{s}$ in $\mathcal{S}_{n,b}$, we can evaluate
\begin{align}
    U_{n,b,j}(u) = {n \choose b}^{-1} \sum_{\mathsf{s} \in \mathcal{S}_{n,b}}
    u_j(D_{\mathsf{s}})
    &= \frac{\phi}{16 b} {n \choose b}^{-1}
    {n-2 \choose b-2} \sum_{1 \leq i < k \leq n} D_{i,j} D_{k,j} \nonumber \\
    & =  \frac{\phi (b-1)}{16n(n-1)}
    \sum_{1 \leq i < k \leq n} D_{i,j} D_{k,j} ~.
\end{align}
By taking an orthonormal change of coordinates in $\mathbb{R}^n$ and identifying
\begin{equation}
    Z_j = \frac{1}{\sqrt{n}} \sum_{i=1}^n D_{i,j} \sim \mathsf{N}(0,1)
\end{equation}
we can see that there exists $W_j \sim \chi^2_{n-1}$ independent of $Z_j$ such that
\begin{align}
    \sum_{i=1}^n D^2_{i,j} = Z_j^2 + W_j~.
\end{align}
Thus, we can write
\begin{align}
    U_{n,b,j}(u) = \frac{\phi (b-1)}{16n(n-1)}
    \sum_{1 \leq i < k \leq n} D_{i,j} D_{k,j}
    & = \frac{\phi (b-1)}{32n(n-1)}\left(
    \left( \sum_{i=1}^n D_{i,j} \right)^2 - \sum_{i=1}^n D_{i,j}^2\right)  \nonumber \\
    & = \frac{\phi (b-1)}{32n(n-1)}\left( n Z_j^2 -(Z_j^2 + W_j)\right) \nonumber \\
    & = \frac{\phi (b-1)}{32n(n-1)}\left( (n-1) Z_j^2 - W_j\right) ~,\label{eq: rep U as Z and W}
\end{align}
almost surely. Thus, the representations \eqref{eq: no Hajek} and \eqref{eq: rep U as Z and W} implies that, on the event
\begin{align}
    \mathcal{A} = \left\{ Z_1^2 \geq \log(e nd) + 2, W_1 \leq 2(n-1) \right\}~,
\end{align}
it holds that 
\begin{align}
\bigg\| U_{n,b}(u) - \mathbb{E}[ u(D_{[b]})]- \frac{b}{n}\sum_{i=1}^n u^{(1)}(D_i) \bigg\|_{\infty}
=
    \big\| U_{n,b}(u)  \big\|_{\infty} \geq \phi \frac{b-1}{32n} \log(end)~.\label{eq: lower bound on event}
\end{align}
Now, observe that
\begin{align}
    P\{ Z_j \geq x \} \geq c x^{-1} \exp(-x^2/2)
    \quad\text{and}\quad
    P\{ W_j \leq 2(n-1) \} \geq 1/2~,
\end{align}
where the first inequality is a standard lower bound for Gaussian random variables and $W_j$ follows from Markov's inequality. Hence, it holds that
\begin{align}
    P\{ \mathcal{A} \}
    &\geq \frac{1}{2} P\{ Z_1^2 \geq \log(e nd) + 2 \} \nonumber \\
    & \gtrsim (\log(e nd)+2)^{-1/2} (e nd)^{-1/2} \gtrsim (nd)^{-1}~.\label{eq: lower bound prob}
\end{align}
The claim is verified by combining \eqref{eq: lower bound on event} and \eqref{eq: lower bound prob}.\hfill$\blacksquare$

\section{An Abstract Bound on Coverage Error\label{sec: generic}}

In this Appendix, we give an abstract bound on the accuracy of the nominal coverage probability for the confidence region introduced in \cref{def: uniform ci}. We make no use of the kernel structure expressed in \eqref{eq: subsampled kernel}. The proof is given in \cref{app: sub pf of generic decomposition}. We require several mild smoothness restrictions on the moment function $M(\cdot;\theta, g)$. In contrast to the set of assumptions specified in \cref{sec: nuisance and moment}, we do not require moment linearity. Instead, we impose the following generalization of Part (iii) of \cref{assu: moment smoothness}. 
\begin{assumption}[Moment Restrictions]\label{assu: app a moment function}$ $\\
The moment function $M(x;\theta_0, g_0)$ is twice continuously differentiable in its second argument. Let
\begin{equation}\label{eq: 1 and 2 moments}
M^{(1)}(x;\theta,g) = \frac{\partial}{\partial \theta^\prime} M(x;\theta^\prime, g) \vert_{\theta^\prime = \theta}
\quad\text{and}\quad
H(x;\theta,g) = \frac{\partial^2}{\partial^2 \theta^\prime} M(x;\theta^\prime, g) \vert_{\theta^\prime = \theta}
\end{equation}
denote the Jacobian and Hessian of $M(\cdot;\theta, g)$ in $\theta$, respectively. The Jacobian $M^{(1)}(x;\theta,g)$ is uniformly Lipschitz in its second argument and bounded from below in the sense that
\begin{align}
\sup_{P\in\mathbf{P}} \sup_{j\in[d]} \big\vert M^{(1)}(x^{(j)};\theta,g) - M^{(1)}(x^{(j)};\theta,g_0)\big\vert & \lesssim \|g - g_0\|_{2,\infty}\quad\text{and}\label{eq: Lipschitz Jacobian}\\
\inf_{P\in\mathbf{P}} \inf_{j\in[d]} \big\vert M^{(1)}(x^{(j)};\theta,g)\big\vert & \geq c\label{eq: well-posedness}
\end{align}
for each $g$ and $\theta$ and some positive constant $c$. The Hessian $H(x;\theta,g)$ is uniformly bounded as $x$, $\theta$, and $g$ vary over their respective domains. 
\end{assumption}

\noindent Moreover, we impose an analogous restriction on the centered empirical moment 
\begin{equation}\label{eq: centered}
\bar{M}_n(x;\theta,g) = M_n(x;\theta,g,\mathbf{D}_n) - \mathbb{E}\left[ M_n(x;\theta,g,\mathbf{D}_n) \right]~,
\end{equation}
where we have made the dependence on $\mathbf{D}_n$ implicit to ease notation.

\begin{assumption}[Empirical Smoothness]\label{assu: empirical smoothness}
The centered empirical moment \eqref{eq: centered} is twice continuously differentiable in its second argument. Let
\begin{align}
\bar{H}_n(x;\theta, g) &= \frac{\partial^2}{\partial^2 \theta^\prime}  \bar{M}_n(x;\theta^\prime, g_0) \vert_{\theta^\prime = \theta}  
\end{align}
denote the Hessian of $\bar{M}_n(\cdot;\theta, g)$ in $\theta$, respectively. The Hessian $\bar{H}_n(x;\theta,g)$ is uniformly bounded almost surely as $x$, $\theta$, and $g$ vary over their respective domains. 
\end{assumption}

Next, we impose a set of high-level restrictions on the structure of the empirical conditional moment \eqref{eq: subsampled kernel}. At times we refer to the normalized statistic
\begin{equation}\label{eq: U def}
W_{n}(x) = -(M^{(1)}(x;g_0))^{-1}\bar{M}_{n}(x;\theta_0,g_0)~.
\end{equation}
First, we impose a condition that ensures that \eqref{eq: U def} is approximately linear.  Fix a generic sequence $\bar{q}_{n,d}$. 

\begin{assumption}[Approximate Linearity]\label{assu: linearity}$ $\\
There exists a function $\bar{u}(\cdot,\cdot)$, a constant $\varphi\geq 1$, and a real-valued sequence $\delta_{n,u}$ such that
\begin{equation}\label{eq: Orlicz bound}
\mathbb{E}\left[\bar{u}(x^{(j)},D_i)\right] = 0~,\quad
\|\bar{u}(x^{(j)},D_i)\|_{\psi_1} \leq \varphi~,
\end{equation}
and
\begin{equation}
\mathbb{E}\left[\bar{u}^4(x^{(j)},D_i)\right] \leq \Var(\bar{u}(x^{(j)},D_i)) \varphi^2
\end{equation}
hold for all $j$ in $[d]$ and $P$ in $\mathbf{P}$. Define the moment $\lambda^2_{j}=\Var(\bar{u}(x^{(j)},D_i))$ and set $\underline{\lambda}^2 = \min_{j\in[d]} \lambda^2_{j}$. The higher-order moment inequality
\begin{align}\label{eq: linearity moment}
    \mathbb{E}\left[ 
    \bigg\vert \sqrt{\frac{n}{\underline{\lambda}^2}}
    \left( W_n(x^{(j)})
    - \frac{1}{n}\sum_{i=1}^n \bar{u}(x^{(j)}, D_i)
    \right)
    \bigg\vert^q 
    \right]^{1/q}
    \leq q^2 \delta_{n,u}~.
\end{align}
holds for all $j$ in $[d]$, $2\vee2\log(nd)\leq q \leq \bar{q}_{n,d}$, and uniformly over $P$ in $\mathbf{P}$, 
\end{assumption}

\begin{remark}
    Compare \cref{assu: linearity} with the bound
    \begin{align}
        \mathbb{E}\left[ 
        \vert U_{n,b}(u_j) 
        - \mathbb{E}[u_j(D_{[b]})]
        - \frac{b}{n} \sum_{i=1}^n u_j^{(1)}(D_i)\vert^{2\log(nd)} \right]^{1/(2\log(dn))} 
        & \lesssim
        \phi \frac{b }{n}\log^2(dn)\label{eq: moment bound example}
    \end{align}
    obtained from \cref{lem: individual degenerate}. The bound \eqref{eq: moment bound example} implies \cref{lem: U stat linearization} by Markov's inequality.\hfill$\blacksquare$
\end{remark}

\noindent Second, we impose several restrictions relating to the estimators $\hat{\theta}_n(\bm{x}^{(d)})$ and $\hat{g}_n$. A closely related collection of conditions is stated in \cite{chernozhukov2018double}.
\begin{assumption}[Bias, Consistency, and Stochastic Equicontinuity]\label{assu: bias}Recall the definition of the object $\underline{\lambda}^2$ introduced in \cref{assu: linearity}. Let 
\begin{align}
\bar{M}^{(1)}_n(x;\theta, g) & = \frac{\partial}{\partial \theta^\prime} \bar{M}_n(x;\theta^\prime, g) \vert_{\theta^\prime = \theta} 
\end{align}
denote the Jacobian of the centered empirical moment $\bar{M}_n(x; \theta, g)$. \\
\textbf{(i)} Define the quantity 
\begin{equation}
\mathsf{Bias}_n(x;\theta, g) = M(x;\theta,g)- \mathbb{E}\left[M_{n}(x;\theta,g,\bm{D}_n)\right]~.
\end{equation}
There exists a sequence $\delta_{n,B}$ such that
\begin{equation}
\sup_{P\in\mathbf{P}} \sup_{g\in\mathcal{G}}
 \sqrt{\frac{n}{\underline{\lambda}^2}} \| \mathsf{Bias}_n(\bm{x}^{(d)};\theta(\bm{x}^{(d)}), g) \|_\infty  \lesssim (1 + \|\theta(\bm{x}^{(d)})\|_\infty) \delta_{n,B}
\end{equation}
uniformly over any vector $\theta(\bm{x}^{(d)}) = \{\theta(x^{(j)})\}_{j=1}^d$. 

\noindent \textbf{(ii)} There exist sequences $\delta_{n,m}$, $\delta_{n,g}$, and $\delta_{n,\theta}$ such that, uniformly over $P$ in $\mathbf{P}$, 
\begin{align}
\left(\frac{n}{\underline{\lambda}^2}\right)^{1/4}
\mathbb{E}
\left[
\big\vert 
\bar{M}^{(1)}_n(x^{(j)};\theta_0(x^{(j)}), g_0)
\big\vert^{q}
\right]^{1/q}
& \leq q \delta_{n,m}
~,\label{eq: concentration}\\
\left(\frac{n}{\underline{\lambda}^2}\right)^{1/4}
\mathbb{E}
\left[
\|\hat{g}_n - g_0 \|_{2,\infty}^q
\right]^{1/q}
& \leq q \delta_{n,g}~,
\quad\text{and}\label{eq: g rate}\\
\left(\frac{n}{\underline{\lambda}^2}\right)^{1/4}
\mathbb{E}
\left[
\big\vert 
\hat{\theta}_n(x^{(j)})  - \theta_0(x^{(j)})
\big\vert^{q}
\right]^{1/q}
& \leq q \delta_{n,\theta}\label{eq: theta rate}
\end{align}
holds for all $j$ in $[d]$ and $2 \vee 2\log(nd)\leq q \leq \bar{q}_{n,d}$.

\noindent \textbf{(iii)}  There exist sequences $\delta_{n,S}$ and $\delta_{n,J}$ such that, uniformly over $P$ in $\mathbf{P}$, 
\begin{align}
\sqrt{\frac{n}{\underline{\lambda}^2}}
\mathbb{E}
\left[
\big\vert 
\bar{M}_n(x^{(j)};\theta_0(x^{(j)}), \hat{g}_n) 
- 
\bar{M}_n(x^{(j)};\theta_0(x^{(j)}), g_0)
\big\vert^{q}
\right]^{1/q}
& \leq q \delta_{n,S}
\end{align}
and
\begin{align}
\sqrt{\frac{n}{\underline{\lambda}^2}}
\mathbb{E}
\left[
\big\vert 
\bar{M}^{(1)}_n(x^{(j)};\theta_0(x^{(j)}), \hat{g}_n) 
- 
\bar{M}^{(1)}_n(x^{(j)};\theta_0(x^{(j)}), g_0)
\big\vert^{q}
\right]^{1/q}
& \leq q \delta_{n,J}~,\label{eq: Jacobian se}
\end{align}
holds for all $j$ in $[d]$ and $2 \vee 2\log(nd)\leq q \leq \bar{q}_{n,d}$.
\end{assumption}

The following theorem gives a non-asymptotic bound on the error in the nominal coverage probability of the confidence regions introduced in \cref{def: uniform ci}.
\begin{theorem}\label{thm: generic decomposition} Collect the error sequences
\begin{align*}
\delta_n  
&= \delta^2_{n,g} + \delta^2_{n,\theta} + \delta^2_{n,m} 
+ \delta_{n,B} + \delta_{n,S}
+  \underline{\lambda}^{1/2}n^{-1/4}  \delta_{n,\theta}\left(\delta_{n,B} + \delta_{n,J}\right) + \delta_{n,u}
\end{align*}
and assume that $\delta_{\varepsilon n} \leq C_\varepsilon \delta_{n}$ for any $0<\varepsilon<1$. Suppose that the Neyman orthogonal moment function $M(x;\theta_0, g_0)$ satisfies \cref{assu: app a moment function} and Part (i) of \cref{assu: moment smoothness} and that the centered empirical moment function $\bar{M}_n(x;\theta_0, g_0)$ satisfies \cref{assu: empirical smoothness}. If \cref{assu: linearity} and \cref{assu: bias} hold, then the confidence region defined in \cref{def: uniform ci} satisfies
\begin{align} \label{eq: asymptotic validity statement}
& \sup_{P\in\mathbf{P}} 
\big\vert P\left\{ \theta_0(\bm{x}^{(d)}) \in \hat{\mathcal{C}}(\bm{x}^{(d)})  \right\} - (1 - \alpha)\big\vert  \nonumber \\
& \quad\quad\quad
\lesssim  \left( \frac{\varphi^2\log^{5}\left(dn\right)}{\underline{\lambda}^{2}n}\right)^{1/4} + \delta_n\log^{3}(dn) + \frac{1}{nd}~.
\end{align} 
\end{theorem}

\subsection{Proof of \cref{thm: generic decomposition}\label{app: sub pf of generic decomposition}}

Throughout, without loss, we will assume that
\begin{equation}
\left(\frac{\varphi^{2}\log^{5}(dn)}{\underline{\lambda}^{2}n}\right)^{1/4} + \delta_n\log^{3}(dn) + \frac{1}{nd} \leq c~,\label{eq: norm without loss}
\end{equation}
as otherwise the desired bound \eqref{eq: asymptotic validity statement} is vacuous. 

It will suffice to show that
\begin{flalign}
& \sup_{z\in\mathbb{R}} \big\vert P\left\{\sqrt{n} \| \hat{\Lambda}_n^{-1/2}R_{n}(\bm{x}^{(d)}) \|_\infty <  z \right\} 
- 
P\left\{\sqrt{n} \| \hat{\Lambda}_n^{-1/2}R^*_{n}(\bm{x}^{(d)}) \|_\infty < z \mid \mathbf{D}_n \right\}  \big\vert \nonumber\\
& \quad\quad\lesssim 
\left(\frac{\varphi^{2}\log^{5}(dn)}{\underline{\lambda}^{2}n}\right)^{1/4} 
+ \delta_n\log^{3}(dn) + \frac{1}{nd} \label{eq: diff in cdfs}
\end{flalign}
with probability greater than $1-C(n^{-1/2}\varphi \underline{\lambda}^{-1} \log^{3/2}(dn) + (nd)^{-1})$. To see this, let $\mathsf{cv}(\gamma)$ denote the $1-\gamma$ quantile of $\sqrt{n} \| \hat{\Lambda}_n^{-1/2}R_{n}(\bm{x}^{(d)}) \|_\infty$ for each $\gamma$ in $(0,1)$ and fix 
\begin{equation}
\beta_{n,d} = \left(\frac{\varphi^{2}\log^{5}(dn)}{\underline{\lambda}^{2}n}\right)^{1/4} + \delta_n\log^{3}(dn) + \frac{1}{nd} ~.
\end{equation}
Observe that \eqref{eq: diff in cdfs} implies that 
\begin{align}
& P\left\{ \sqrt{n} \| \hat{\Lambda}_n^{-1/2}R^*_{n}(\bm{x}^{(d)}) \|_\infty < \mathsf{cv}(\alpha-\beta_{n,d}) \mid \mathbf{D}_n \right\} \nonumber \\
& \geq 
P\left\{ \sqrt{n} \| \hat{\Lambda}_n^{-1/2}R_{n}(\bm{x}^{(d)}) \|_\infty < \mathsf{cv}(\alpha-\beta_{n,d}) \right\} - \beta_{n,d} \geq 1-\alpha\label{eq: cv perturb lower}
\end{align}
and
\begin{align}
& P\left\{ \sqrt{n} \| \hat{\Lambda}_n^{-1/2}R^*_{n}(\bm{x}^{(d)}) \|_\infty < \mathsf{cv}(\alpha+\beta_{n,d}) \mid \mathbf{D}_n \right\} \nonumber \\
& \leq
P\left\{ \sqrt{n} \| \hat{\Lambda}_n^{-1/2}R_{n}(\bm{x}^{(d)}) \|_\infty < \mathsf{cv}(\alpha+\beta_{n,d}) \right\} + \beta_{n,d} \leq 1-\alpha\label{eq: cv perturb upper}
\end{align}
each with probability greater than $1-C(n^{-1/2}\varphi \underline{\lambda}^{-1} \log^{3/2}(dn) + (nd)^{-1})$. Thus, recalling that $\widehat{\mathsf{cv}}(\alpha)$ denotes the $1-\alpha$ quantile of $\sqrt{n} \| \hat{\Lambda}_n^{-1/2}R^*_{n}(\bm{x}^{(d)}) \|_\infty$ conditioned on the data $\mathbf{D}_n$, \eqref{eq: cv perturb lower} and \eqref{eq: cv perturb upper} imply that 
\begin{align}
& P\left\{\mathsf{cv}(\alpha+\beta_{n,d}) < \hat{\mathsf{cv}}(\alpha) < \mathsf{cv}(\alpha-\beta_{n,d})\right\}  \nonumber \\
& \gtrsim 1- \left(\frac{\varphi^2\log^{3}(dn)}{\underline{\lambda}^2n}\right)^{1/2} - \frac{1}{nd} \geq 1- \beta_{n,d}~,\label{eq: cv sandwich}
\end{align}
where the second inequality follows from the fact that the normalization \eqref{eq: norm without loss} implies that
\begin{equation}
\left(\frac{\varphi^2\log^{3}(dn)}{\underline{\lambda}^2n}\right)^{1/2}  
\lesssim 
 \left(\frac{\varphi^2\log^{5}(dn)}{\underline{\lambda}^2n}\right)^{1/4}~.
\end{equation}
Consequently, as we can write 
\begin{align}
P\left\{ \bm{\theta}_0(\bm{x}^{(d)}) \in \hat{\mathcal{C}}(\bm{x}^{(d)}, \mathbf{D}_n)  \right\} 
= 
P\left\{ \sqrt{n} \| \hat{\Lambda}_n^{-1/2}R_{n}(\bm{x}^{(d)}) \|_\infty \leq  \hat{\mathsf{cv}}(\alpha) \right\} 
\end{align}
by definition, the inequality \eqref{eq: cv sandwich} implies that 
\begin{align}
& P\left\{ \bm{\theta}_0(\bm{x}^{(d)}) \in \hat{\mathcal{C}}(\bm{x}^{(d)}, \mathbf{D}_n)  \right\} \nonumber \\
& \leq 
P\left\{ \sqrt{n} \| \hat{\Lambda}_n^{-1/2}R_{n}(\bm{x}^{(d)}) \|_\infty \leq  \mathsf{cv}(\alpha-\beta_{n,d}) \right\}  + \beta_{n,d} 
\lesssim 1- \alpha + \beta_{n,d}
\end{align}
and 
\begin{align}
& P\left\{ \bm{\theta}_0(\bm{x}^{(d)}) \in \hat{\mathcal{C}}(\bm{x}^{(d)}, \mathbf{D}_n)  \right\} \nonumber \\
& \geq
P\left\{ \sqrt{n} \| \hat{\Lambda}_n^{-1/2}R_{n}(\bm{x}^{(d)}) \|_\infty \leq  \mathsf{cv}(\alpha+\beta_{n,d}) \right\}  -\beta_{n,d}
 \gtrsim 1- \alpha - \beta_{n,d}
\end{align}
respectively, as required. 

Hence, the remainder of the proof is devoted to establishing the bound \eqref{eq: diff in cdfs}. To do this, we require some additional notation. Let $\mathcal{R}$ denote the set of hyper-rectangles in $\mathbb{R}^d$. Let $Z$ denote a centered Gaussian random vector with covariance matrix $\Var(\bar{u}(\bm{x}^{(d)}, D_i))$. Let $\Lambda$ be the diagonal matrix with components $\lambda_{j}^2 = \Var(\bar{u}(x^{(j)}, D_i))$. To verify \eqref{eq: diff in cdfs} it will suffice to establish that
\begin{flalign}
&\sup_{\mathsf{R} \in \mathcal{R}} \big\vert P\left\{\sqrt{n} \hat{\Lambda}_n^{-1/2}R_{n}(\bm{x}^{(d)})\in\mathsf{R}\right\} 
                  -P\left\{ \Lambda^{-1/2} Z\in\mathsf{R}\right\} \big\vert \nonumber \\
&\quad\quad \lesssim 
\left(\frac{\varphi^{2}\log^{5}(dn)}{\underline{\lambda}^{2}n}\right)^{1/4} 
+ \delta_n\log^{3}(dn) + \frac{1}{nd} \label{eq: A abs to prove}
\end{flalign}
and that
\begin{flalign}
&\sup_{\mathsf{R} \in \mathcal{R}} \big\vert P\left\{\sqrt{n} \hat{\Lambda}_n^{-1/2}R^*_{n}(\bm{x}^{(d)})\in\mathsf{R} \mid \mathbf{D}_n \right\} 
                  -P\left\{ \Lambda^{-1/2} Z\in\mathsf{R}\right\} \big\vert \nonumber \\
&\quad\quad \lesssim 
\left(\frac{\varphi^{2}\log^{5}(dn)}{\underline{\lambda}^{2}n}\right)^{1/4} 
+ \delta_n\log^{3}(dn) \label{eq: A* abs to prove}
\end{flalign}
with probability greater than $1-Cn^{-1/2}\varphi \underline{\lambda}^{-1} \log^{3/2}(dn) - (nd)^{-1}$. That is, the probability bound \eqref{eq: diff in cdfs} follows from \eqref{eq: A abs to prove} and \eqref{eq: A* abs to prove} by considering hyper-rectangles of the form $\mathsf{R} = [-\infty \bm{1}_{d}, z \bm{1}_{d} ]$ and applying the triangle inequality. 

We provide the details for the proofs of the upper bounds encoded in the absolute inequalities \eqref{eq: A abs to prove} and \eqref{eq: A* abs to prove}, respectively. Analogous lower bounds will follow from the same argument. In particular, fix a rectangle $\mathsf{R}=[a_{l},a_{u}]$ in $\mathcal{R}$,
where $a_{l}$ and $a_{u}$ are vectors in $\mathbb{R}^{d}$ with $a_{l}\leq a_{u}$, interpreted component-wise, and define the enlarged rectangle $\mathsf{R}_{t}=[a_{l}-t\bm{1}_{d},a_{u}+t\bm{1}_{d}]$ for each $t>0$. We obtain upper bounds
\begin{flalign}
  & P\left\{\sqrt{n} \hat{\Lambda}_n^{-1/2}R_{n}(\bm{x}^{(d)})\in\mathsf{R}\right\} 
                  -P\left\{ \Lambda^{-1/2} Z\in\mathsf{R}\right\} \nonumber\\
  &  \quad\quad\leq
       \big\vert P\left\{ \sqrt{n} \Lambda^{-1/2}R_{n}(\bm{x}^{(d)})\in\mathsf{R}_{t}\right\} 
                    -P\left\{ \Lambda^{-1/2} Z\in\mathsf{R}_{t}\right\} \big\vert\label{eq: m estimator clt to bound}\\
 &  \quad\quad\quad
           +\big\vert P\left\{ \Lambda^{-1/2}  Z\in\mathsf{R}_{t}\right\} 
                    -P\left\{ \Lambda^{-1/2} Z\in\mathsf{R}\right\} \big\vert\label{eq: gaussian diff 1}\\
 &  \quad\quad\quad
                   +P\left\{ \sqrt{n} \|(\Lambda^{-1/2}-\hat{\Lambda}_{n}^{-1/2})R_{n}(\bm{x}^{(d)})\|_{\infty}\geq t\right\} ,\label{eq: variance estimation to bound}
\end{flalign}
and similarly
\begin{flalign}
&  P\left\{ \sqrt{n} \hat{\Lambda}_n^{-1/2}R_{n}^{*}(\bm{x}^{(d)})\in\mathsf{R}\mid\mathbf{D}_{n}\right\} 
               - P\left\{  \Lambda^{-1/2} Z\in\mathsf{R}\right\}  \nonumber\\
 & \quad\quad\leq\vert 
                 P\left\{ \sqrt{n} \Lambda^{-1/2}R_{n}^{*}(\bm{x}^{(d)})\in\mathsf{R}_{t}\mid\mathbf{D}_{n}\right\} 
               - P\left\{  \Lambda^{-1/2} Z\in\mathsf{R}_{t}\right\} \vert\label{eq: boot m estimator to bound}\\
 & \quad\quad\quad+\vert 
                  P\left\{  \Lambda^{-1/2} Z\in\mathsf{R}_{t}\right\} 
                - P\left\{ \Lambda^{-1/2}  Z\in\mathsf{R}\right\} \vert\label{eq: gaussian diff 2}\\
 & \quad\quad\quad
               +P\left\{ \sqrt{n} \|(\Lambda^{-1/2}-\hat{\Lambda}_{n}^{-1/2})R_{n}^{*}(\bm{x}^{(d)})\|_{\infty}\geq t\mid\mathbf{D}_{n}\right\}~,\label{eq: variance boot estimation to bound}
\end{flalign}
respectively. It thereby remains to obtain suitable bounds for each term, \eqref{eq: m estimator clt to bound} through \eqref{eq: variance boot estimation to bound}.

To bound the Gaussian approximation errors \eqref{eq: m estimator clt to bound}
and \eqref{eq: boot m estimator to bound}, we apply the following
Theorem, which establishes a generic quantitative central limit for the statistic $R_{n}$ in addition to a generic quantitative conditional central limit theorem for the half-sample bootstrap.
\begin{theorem}
\label{thm: generic m estimation unstudentized}Suppose that the moment function $M(x;\theta_0, g_0)$ satisfies \cref{assu: app a moment function} and Part (i) of \cref{assu: moment smoothness} and that \cref{assu: linearity,assu: bias} hold.

\noindent
\textbf{(i)} The inequality 
\begin{flalign}
&\sup_{R\in\mathcal{R}}\sup_{P\in\mathbf{P}}
\Big\vert P\left\{ \sqrt{n} R_{n}(\bm{x}^{(d)})\in\mathsf{R}\right\} -P\left\{ Z\in\mathsf{R}\right\} \Big\vert \nonumber \\
& \lesssim \frac{\varphi^{1/2}}{\underline{\lambda}^{1/2}} \left(\frac{\log^{5}(dn)}{n}\right)^{1/4}  + \delta_n \log^{5/2}(dn) + \frac{1}{nd}\label{eq: generic estimator clt}
\end{flalign}
holds.

\noindent \textbf{(ii)} If the bootstrap root is constructed with the Half-Sample bootstrap, then the inequality
\begin{flalign}
& \sup_{R\in\mathcal{R}}\sup_{P\in\mathbf{P}}
\Big\vert P\left\{ \sqrt{n} R_{n}^{*}(\bm{x}^{(d)})\in\mathsf{R}\mid\mathbf{D}_{n}\right\} -P\left\{ Z\in\mathsf{R}\right\} \Big\vert \nonumber \\
& \lesssim \frac{\varphi^{1/2}}{\underline{\lambda}^{1/2}}\left(\frac{\log^{5}\left(dn\right)}{n}\right)^{1/4}+ \delta_n \log^{5/2}(dn) \label{eq: generic bootstrap clt}
\end{flalign}
holds with probability greater than $1-C(n^{-1/2}\varphi \underline{\lambda}^{-1} \log^{3/2}(dn) + (nd)^{-1})$. 
\end{theorem}
\noindent In particular, \cref{thm: generic m estimation unstudentized} implies that
\begin{align}
& \big\vert P\left\{ \sqrt{n} \Lambda^{-1/2}R_{n}(\bm{x}^{(d)})\in\mathsf{R}_{t}\right\} 
                    -P\left\{ \Lambda^{-1/2} Z\in\mathsf{R}_{t}\right\} \big\vert \nonumber \\
&=
 \big\vert P\left\{ \sqrt{n} R_{n}(\bm{x}^{(d)})\in \Lambda^{1/2} \mathsf{R}_{t}\right\} 
                    -P\left\{  Z\in  \Lambda^{1/2} \mathsf{R}_{t}\right\} \big\vert \nonumber \\
                    &\quad\quad \lesssim \frac{\varphi^{1/2}}{\underline{\lambda}^{1/2}} \left(\frac{\log^{5}(dn)}{n}\right)^{1/4}  + \delta_n\log^{5/2}(dn) + \frac{1}{nd} \label{eq: generic estimator clt apply}
\end{align}
and analogously 
\begin{align}
& \vert P\left\{ \sqrt{n} \Lambda^{-1/2}R_{n}^{*}(\bm{x}^{(d)})\in\mathsf{R}_{t}\mid\mathbf{D}_{n}\right\} 
      - P\left\{  \Lambda^{-1/2} Z\in\mathsf{R}_{t}\right\} \vert \nonumber \\
& =
 \big\vert P\left\{ \sqrt{n} R^*_{n}(\bm{x}^{(d)})\in \Lambda^{1/2} \mathsf{R}_{t}\mid \mathbf{D}_{n}\right\} 
                    -P\left\{  Z\in  \Lambda^{1/2} \mathsf{R}_{t}\right\} \big\vert \nonumber \\
                    & \quad\quad \lesssim \frac{\varphi^{1/2}}{\underline{\lambda}^{1/2}} \left(\frac{\log^{5}(dn)}{n}\right)^{1/4}  + \delta_n\log^{5/2}(dn) \label{eq: generic bootstrap clt apply}
\end{align}
with probability greater than $1-C(n^{-1/2}\varphi \underline{\lambda}^{-1} \log^{3/2}(dn) + (nd)^{-1})$, as $\Lambda^{1/2} \mathsf{R}_{t}$ is a hyper-rectangle.

To bound the differences in the Gaussian probabilities (\ref{eq: gaussian diff 1})
and (\ref{eq: gaussian diff 2}), we apply \cref{lem: Nazarov}. In particular, we have that 
\begin{equation}
\vert P\left\{ \Lambda^{-1/2} Z\in\mathsf{R}_{t}\right\} -P\left\{ \Lambda^{-1/2} Z\in\mathsf{R}\right\} \vert\lesssim t\sqrt{\log d}\label{eq: nazarov apply app a}
\end{equation}
for all $t>0$.

Finally, we bound the terms (\ref{eq: variance estimation to bound}) and (\ref{eq: variance boot estimation to bound}) resulting from variance estimation. Observe that 
\begin{align}
\|(\Lambda^{-1/2} - \hat{\Lambda}_n^{-1/2}) R_n(\bm{x}^{(d)})\|_{\infty} 
&= \|(I - \hat{\Lambda}_n^{-1/2}\Lambda^{1/2}) \Lambda^{-1/2} R_n(\bm{x}^{(d)})\|_{\infty} \nonumber\\
&\leq \sup_{j\in[d]} 
|\lambda_j/\hat{\lambda}_{n,j} - 1|\cdot \|\Lambda^{-1/2}R_n(\bm{x}^{(d)})\|_{\infty} \label{eq: var perturb in two}
\end{align}
and analogously 
\begin{align}
\|(\Lambda^{-1/2} - \hat{\Lambda}_n^{-1/2}) R^*_n(\bm{x}^{(d)})\|_{\infty} 
\leq \sup_{j\in[d]} |\lambda_j/\hat{\lambda}_{n,j} - 1|\cdot \|\Lambda^{-1/2}R^*_n(\bm{x}^{(d)})\|_{\infty}~.\label{eq: var boot perturb in two}
\end{align}
We give probability bounds for the terms on the right-hand sides of \eqref{eq: var perturb in two} and \eqref{eq: var boot perturb in two}. Observe that the Borell-TIS inequality \citep[e.g., Theorem 2.1.1 of][]{adler2009random} implies that
\[
P\left\{ \|\Lambda^{-1/2}Z\|_{\infty}\geq C \sqrt{\log{dn}}\right\} \le n^{-1}~.
\]
Thus, \cref{thm: generic m estimation unstudentized} implies that
\begin{flalign}
& P\left\{ \sqrt{n}\|\Lambda^{-1/2}R_{n}(\bm{x}^{(d)})\|_{\infty}\geq C \sqrt{\log{dn}}\right\}  \nonumber \\
& \lesssim \left(\frac{\varphi^{2}\log^{5}(dn)}{\underline{\lambda}^{2}n}\right)^{1/4}  + \delta_n\log^{5/2}(dn)  + \frac{1}{nd} \label{eq: BTIS estimator}
\end{flalign}
and that
\begin{flalign}
& P\left\{ \sqrt{n} \|\Lambda^{-1/2}R^*_{n}(\bm{x}^{(d)})\|_{\infty}\geq C \sqrt{\log{dn}}\mid\mathbf{D}_{n}\right\}  \nonumber \\
& \lesssim \left(\frac{\varphi^{2}\log^{5}(dn)}{\underline{\lambda}^{2}n}\right)^{1/4} + \delta_n\log^{5/2}(dn)\label{eq: BTIS boot}
\end{flalign}
with probability greater than $1-C(n^{-1/2}\varphi \underline{\lambda}^{-1} \log^{3/2}(dn) + (nd)^{-1})$. In turn, to bound the discrepancy between the bootstrap variance estimate $\hat{\lambda}_{n,j}$ and $\lambda_j$, we apply the following Lemma.
\begin{lemma}\label{lem: variance accuracy}Suppose that the conditions of \cref{thm: generic decomposition} hold. If the bootstrap root is constructed with the Half-Sample bootstrap, then
\begin{align}
&\sup_{j\in[d]} \bigg\vert \frac{\hat{\lambda}_{n,j}}{\lambda_{j}} - 1\bigg\vert 
\lesssim
\frac{\varphi}{\underline{\lambda}}
\sqrt{\frac{\log(dn)}{n}} 
+
\frac{\varphi^2\log^3(dn)}{\underline{\lambda}^2n}
\nonumber \\
& 
\quad\quad\quad\quad\quad\quad\quad\quad\quad\quad
+\log^2(dn)\delta_n
+
\log^4(dn)\delta_n^2
\end{align}
with probability greater than $1 - C/nd$.
\end{lemma}
\noindent 
To apply \cref{lem: variance accuracy}, observe that 
\begin{align}
    \bigg\vert \frac{\lambda_{j}}{\hat{\lambda}_{n,j}} - 1\bigg\vert
    =
    \frac{\vert \lambda_{j}/\hat{\lambda}_{n,j} - 1\vert}{\hat{\lambda}_{n,j}/ \lambda_{j}}~.
\end{align}
Thus, the event that
\begin{align}
    \sup_{j\in[d]} \bigg\vert \frac{\hat{\lambda}_{n,j}}{\lambda_{j}} - 1\bigg\vert
    \leq \frac{1}{2}
    \quad\text{implies}\quad
    \frac{\hat{\lambda}_{n,j}}{\lambda_{j}} \geq \frac{1}{2}~.\label{eq: good lambda event}
\end{align}
Hence, on this event, we find that
\begin{align}
    \sup_{j\in[d]} \bigg\vert \frac{\lambda_{j}}{\hat{\lambda}_{n,j}} - 1\bigg\vert
    \leq 2
    \sup_{j\in[d]} \bigg\vert \frac{\hat{\lambda}_{n,j}}{\lambda_{j}} - 1\bigg\vert~.\label{eq: switch sides}
\end{align}
Observe now that the normalization \eqref{eq: norm without loss} implies that 
\begin{align}
&\frac{\varphi}{\underline{\lambda}}
\sqrt{\frac{\log(dn)}{n}} 
+
\frac{\varphi^2\log^3(dn)}{\underline{\lambda}^2n}
+\log^2(dn)\delta_n
+
\log^4(dn)\delta_n^2 \nonumber \\
& \lesssim
\left(\frac{\varphi^{2}\log^{5}(dn)}{\underline{\lambda}^{2}n}\right)^{1/4} 
+ \delta_n \log^3(dn) <c
\end{align}
and so the condition \eqref{eq: good lambda event} is satisfied by the event characterized in \cref{lem: variance accuracy}. Thus, by combining the bounds \eqref{eq: var perturb in two}, \eqref{eq: BTIS estimator}, and \eqref{eq: switch sides} with \cref{lem: variance accuracy}, we find that
\begin{align}
&P\bigg\{ \sqrt{n} \|(\Lambda^{-1/2}-\hat{\Lambda}_{n}^{-1/2})R_{n}(\bm{x}^{(d)})\|_{\infty}
\nonumber \\
& \quad\quad\quad \gtrsim 
\bigg(
\frac{\varphi}{\underline{\lambda}}
\sqrt{\frac{\log(dn)}{n}} 
+
\frac{\varphi^2\log^3(dn)}{\underline{\lambda}^2n}
+\log^2(dn)\delta_n
+
\log^4(dn)\delta_n^2
\bigg)\sqrt{\log{dn}}\bigg\}\nonumber\\
&\lesssim
\left(\frac{\varphi^{2}\log^{5}(dn)}{\underline{\lambda}^{2}n}\right)^{1/4}  
+ \delta_n\log^{5/2}(dn) + \frac{1}{nd}~.\label{eq: var bound estimator state}
\end{align}
Similarly, by combining the bounds \eqref{eq: var boot perturb in two}, \eqref{eq: BTIS boot}, and \eqref{eq: switch sides} with Lemma \eqref{lem: variance accuracy}, we find that
\begin{flalign}
&P\bigg\{ \sqrt{n} \|(\Lambda^{-1/2}-\hat{\Lambda}_{n}^{-1/2})R_{n}^{*}(\bm{x}^{(d)})\|_{\infty} \nonumber \\
& \quad\quad\quad \gtrsim
\left(
\frac{\varphi}{\underline{\lambda}}
\sqrt{\frac{\log(dn)}{n}} 
+
\frac{\varphi^2\log^3(dn)}{\underline{\lambda}^2n}
+\log^2(dn)\delta_n
+
\log^4(dn)\delta_n^2
\right)\sqrt{\log{dn}}\mid\mathbf{D}_{n}\bigg\} \nonumber \\
&\lesssim
\left(\frac{\varphi^{2}\log^{5}(dn)}{\underline{\lambda}^{2}n}\right)^{1/4}  
+ \delta_n\log^{5/2}(dn)\label{eq: var bound boot state}
\end{flalign}
with probability greater than $1-C(n^{-1/2}\varphi \underline{\lambda}^{-1} \log^{3/2}(dn) + (nd)^{-1})$.

To put the pieces together, recall the upper bounds \eqref{eq: m estimator clt to bound} and \eqref{eq: boot m estimator to bound}. By setting 
\begin{align}
t =
\left(
\frac{\varphi}{\underline{\lambda}}
\sqrt{\frac{\log(dn)}{n}} 
+
\frac{\varphi^2\log^3(dn)}{\underline{\lambda}^2n}
+\log^2(dn)\delta_n
+
\log^4(dn)\delta_n^2
\right)\sqrt{\log{dn}}~,
\end{align}
the bounds \eqref{eq: generic estimator clt apply}, \eqref{eq: nazarov apply app a}, and \eqref{eq: var bound estimator state} imply that
\begin{flalign}
&
 P\left\{ \sqrt{n} \hat{\Lambda}_{n}^{-1/2}R_{n}(\bm{x}^{(d)})\in\mathsf{R}\right\} 
       -P\left\{\Lambda^{-1/2} Z\in\mathsf{R}\right\} \nonumber\\
& \quad\quad\lesssim 
\left(\frac{\varphi^{2}\log^{5}(dn)}{\underline{\lambda}^{2}n}\right)^{1/4} 
+ \delta_n \log^{5/2}(dn) \nonumber\\
&\quad\quad\quad+
\left(
\frac{\varphi}{\underline{\lambda}}
\sqrt{\frac{\log(dn)}{n}} 
+
\frac{\varphi^2\log^3(dn)}{\underline{\lambda}^2n}
+\log^2(dn)\delta_n
+
\log^4(dn)\delta_n^2
\right)\log(dn)
+ \frac{1}{nd} \nonumber \\
&\quad\quad\lesssim 
\left(\frac{\varphi^{2}\log^{5}(dn)}{\underline{\lambda}^{2}n}\right)^{1/4} 
+ \delta_n \log^3(dn) + \frac{1}{nd}~. \label{eq: final generic clt}
\end{flalign}
Indeed, the second inequality follows from the normalization \eqref{eq: norm without loss}. In particular,
\begin{align}
\frac{\varphi}{\underline{\lambda}}
\sqrt{\frac{\log(dn)}{n}}\log(dn)
+
\frac{\varphi^2\log^3(dn)}{\underline{\lambda}^2n}\log(dn)
&\lesssim
\left(\frac{\varphi^{2}\log^{5}(dn)}{\underline{\lambda}^{2}n}\right)^{1/4}~,
\end{align}
while
\begin{align}
\log^3(dn)\delta_n
+
\log^5(dn)\delta_n^2
&\lesssim
\delta_n\log^3(dn)~.
\end{align}
Similarly, the bounds \eqref{eq: generic bootstrap clt apply}, \eqref{eq: nazarov apply app a}, and \eqref{eq: var bound boot state} imply that
\begin{flalign}
&
 P\left\{ \sqrt{n} \hat{\Lambda}_{n}^{-1/2}R_{n}^{*}(\bm{x}^{(d)})\in\mathsf{R}\mid\mathbf{D}_{n}\right\} 
       -P\left\{ \Lambda^{-1/2} Z\in\mathsf{R}\right\} \nonumber\\
& \quad\quad\lesssim
\left(\frac{\varphi^{2}\log^{5}(dn)}{\underline{\lambda}^{2}n}\right)^{1/4} 
+ \delta_n \log^{5/2}(dn) \nonumber\\
&\quad\quad\quad+
\left(
\frac{\varphi}{\underline{\lambda}}
\sqrt{\frac{\log(dn)}{n}} 
+
\frac{\varphi^2\log^3(dn)}{\underline{\lambda}^2n}
+\log^2(dn)\delta_n
+
\log^4(dn)\delta_n^2
\right)\log(dn) \nonumber \\
& \quad\quad\lesssim 
\left(\frac{\varphi^{2}\log^{5}(dn)}{\underline{\lambda}^{2}n}\right)^{1/4} 
+ \delta_n \log^3(dn) + \frac{1}{nd}~,\label{eq: final generic boot clt}
\end{flalign}
with probability greater than $1-C(n^{-1/2}\varphi \underline{\lambda}^{-1} \log^{3/2}(dn) + (nd)^{-1})$. Hence, the bounds \eqref{eq: final generic clt} and \eqref{eq: final generic boot clt} verify the bounds encoded in \eqref{eq: A abs to prove} and \eqref{eq: A* abs to prove}, as required.

\subsection{Proof of \cref{thm: generic m estimation unstudentized}}

\subsubsection{Part (i)}

Throughout, we take $\theta_0(x)=0$ for all $x$ without loss of generality.  We begin by giving a high-probability bound on the difference
\begin{equation}
\| R_n(\bm{x}^{(d)}) - W_n(\bm{x}^{(d)}) \|_\infty
\end{equation}
which will be needed at a later point in the proof. To this end, let $x$ be any component of the vector $\bm{x}^{(d)}$. By a Taylor expansion about $\theta_0(x)$, we have that 
\begin{align}
M(x; \hat{\theta}_n(x), \hat{g}_n) - M(x ;\theta_0(x), \hat{g}_n)
 & = (\hat{\theta}_n(x) - \theta_0(x)) M^{(1)}(x; \theta_0(x), \hat{g}_n) \nonumber \\
 & + (\hat{\theta}_n(x) - \theta_0(x))^2 H(x; \tilde{\theta}_0(x), \hat{g}_n) \label{eq: taylor expand moment}
\end{align}
for some $\tilde{\theta}_0(x)$ between $\hat{\theta}_n(x)$ and $\theta_0(x)$. Moreover, we can write
\begin{align}
&(\hat{\theta}_n(x) - \theta_0(x)) M^{(1)}(x; \theta_0(x), \hat{g}_n)\nonumber \\
& = (\hat{\theta}_n(x) - \theta_0(x)) M^{(1)}(x; \theta_0(x), g_0)\nonumber \\ 
& + (\hat{\theta}_n(x) - \theta_0(x)) \left(M^{(1)}(x; \theta_0(x), \hat{g}_n)   -  M^{(1)}(x; \theta_0(x), g_0) \right) \label{eq: M 1 in g}
\end{align}
and
\begin{align}
M(x;\hat{\theta}_n(x),\hat{g}_n) - M(x;\theta_0(x),\hat{g}_n) 
& = (M(x;\hat{\theta}_n(x),\hat{g}_n) - M_n(x;\hat{\theta}_n(x),\hat{g}_n,\mathbf{D}_n)) \nonumber\\
& \quad+ (M(x;\theta_0(x),g_0) - M(x;\theta_0(x),\hat{g}_n)) \nonumber\\
& = \left(M(x;\hat{\theta}_n(x),\hat{g}_n) - \mathbb{E}\left[M_n(x;\hat{\theta}_n(x),\hat{g}_n,\mathbf{D}_n)\right]\right) \nonumber\\
& \quad+ \left(\mathbb{E}\left[M_n(x;\hat{\theta}_n(x),\hat{g}_n,\mathbf{D}_n)\right] - M_n(x;\hat{\theta}_n(x),\hat{g}_n,\mathbf{D}_n)\right) \nonumber\\
& \quad+ (M(x;\theta_0(x),g_0) - M(x;\theta_0(x),\hat{g}_n))~.\label{eq: use minimization}
\end{align}
Thus, by the identity
\begin{align}
\bar{M}_n(x;\hat{\theta}_n(x),\hat{g}_n) &= \bar{M}_n(x;\hat{\theta}_n(x),\hat{g}_n) - \bar{M}_n(x;\theta_0(x),\hat{g}_n) \nonumber\\
& + \bar{M}_n(x;\theta_0(x),\hat{g}_n) - \bar{M}_n(x;\theta_0(x),g_0) + \bar{M}_n(x;\theta_0(x),g_0)~,
\end{align}
the equalities \eqref{eq: taylor expand moment}, \eqref{eq: M 1 in g}, and \eqref{eq: use minimization} imply that
\begin{align}
&M^{(1)}(x;\theta_0(x),g_0) (\hat{\theta}_n(x) - \theta_0(x))\nonumber\\
 & = M^{(1)}(x;\theta_0(x),\hat{g}_n)  (\hat{\theta}_n(x) - \theta_0(x)) \nonumber\\
& - \left(M^{(1)}(x;\theta_0(x),\hat{g}_n) - M^{(1)}(x;\theta_0,g_0)\right) (\hat{\theta}_n(x) - \theta_0(x)) \nonumber\\
& = -\bar{M}_n(x;\theta_0(x),g_0)  \label{eq: basic decomposition}\\
& \quad+ \mathsf{Bias}(x;\hat{\theta}_n(x),\hat{g}_n) + \mathsf{Nuis}(x;\theta_0(x),\hat{g}_n) \label{eq: bias and nuis}\\
& \quad+ \mathsf{Stoch}^{(1)}(x; \hat{\theta}_n(x),\hat{g}_n) + \mathsf{Stoch}^{(2)}(x; \hat{g}_n)  \label{eq: stoch terms}\\
& \quad- (\hat{\theta}_n(x) - \theta_0(x))^2 H(x; \tilde{\theta}_0(x), \hat{g}_n) \label{eq: theta squared}\\
& \quad- (\hat{\theta}_n(x) - \theta_0(x)) \left(M^{(1)}(x; \theta_0(x), \hat{g}_n)   -  M^{(1)}(x; \theta_0(x), g_0) \right)\label{eq: theta g term}~
\end{align}
where
\begin{align}
\mathsf{Bias}(x;\hat{\theta}_n(x),\hat{g}_n) &= M(x;\hat{\theta}_n(x),\hat{g}_n) - \mathbb{E}\left[M_n(x;\hat{\theta}_n(x),\hat{g}_n)\right]~,\label{eq: bias def}\\
\mathsf{Nuis}(x;\theta_0(x),\hat{g}_n) &= M(x;\theta_0(x),g_0) - M(x;\theta_0(x),\hat{g}_n)~,\label{eq: nuis def}\\
 \mathsf{Stoch}^{(1)}(x; \hat{\theta}_n(x),\hat{g}_n) &= \bar{M}_n(x;\hat{\theta}_n(x),\hat{g}_n) - \bar{M}_n(x;\theta_0(x),\hat{g}_n)~,\quad\text{and}\label{eq: stoch 1 def}\\
 \mathsf{Stoch}^{(2)}(x; \hat{g}_n)&=  \bar{M}_n(x;\theta_0(x),g_0) - \bar{M}_n(x;\theta_0(x),\hat{g}_n)~,\label{eq: stoch 2 def}
\end{align}
respectively. 

We now give moment bounds for the terms \eqref{eq: bias and nuis}, \eqref{eq: stoch terms}, \eqref{eq: theta squared}, and \eqref{eq: theta g term}. Fix any $q \geq 2 \vee 2\log(nd)$. We repeatedly use the fact that, for any random vector \(Z(\bm{x}^{(d)})=(Z(x^{(j)}))_{j=1}^d\),
\begin{align}
\mathbb{E}\left[\|Z(\bm{x}^{(d)})\|_\infty^q\right]^{1/q}
\leq d^{1/q}\max_{j\in[d]}\mathbb{E}\left[|Z(x^{(j)})|^q\right]^{1/q}
\lesssim \max_{j\in[d]}\mathbb{E}\left[|Z(x^{(j)})|^q\right]^{1/q}~,
\end{align}
where the final inequality follows from the lower bound on \(q\). To handle \eqref{eq: bias and nuis}, observe that Parts (i) and (ii) of \cref{assu: bias} imply that
\begin{align}
&\mathbb{E}\left[
\left|
\sqrt{ \frac{n}{ \underline{\lambda}^2 } }
\left\|
\mathsf{Bias}_n(\bm{x}^{(d)};\hat{\theta}_n(\bm{x}^{(d)}),\hat{g}_n)
\right\|_\infty
\right|^q
\right]^{1/q} \nonumber\\
&\lesssim
\delta_{n,B}
\left(
1+
\mathbb{E}\left[
\|\hat{\theta}_n(\bm{x}^{(d)})\|_\infty^q
\right]^{1/q}
\right) \nonumber\\
&\lesssim
\delta_{n,B}
\left(
1+
q\frac{\underline{\lambda}^{1/2}}{n^{1/4}}\delta_{n,\theta}
\right) 
\lesssim
q^2\left(
\delta_{n,B}
+
\frac{\underline{\lambda}^{1/2}}{n^{1/4}}\delta_{n,\theta}\delta_{n,B}
\right)~. \label{eq: proof bias bound A}
\end{align}
Moreover, a Taylor expansion, Neyman orthogonality, second-order smoothness, i.e., \cref{assu: app a moment function}, and \cref{assu: bias}, Part (ii), give that
\begin{align}
&\mathbb{E}\left[
\left|
\sqrt{ \frac{n}{ \underline{\lambda}^2 } }
\left\|
\mathsf{Nuis}(\bm{x}^{(d)};\theta_0(\bm{x}^{(d)}),\hat{g}_n)
\right\|_\infty
\right|^q
\right]^{1/q} \nonumber\\
&\lesssim
\mathbb{E}\left[
\left|
\sqrt{ \frac{n}{ \underline{\lambda}^2 } }
\|\hat{g}_n-g_0\|_{2,\infty}^2
\right|^q
\right]^{1/q}
=
\mathbb{E}\left[
\left|
\left(\frac{n}{\underline{\lambda}^2}\right)^{1/4}
\|\hat{g}_n-g_0\|_{2,\infty}
\right|^{2q}
\right]^{1/q}
\lesssim q^2\delta_{n,g}^2~. \label{eq: neyman orth bound}
\end{align}
Next, we handle the term \eqref{eq: stoch terms}. By \cref{assu: empirical smoothness}, a Taylor expansion gives
\begin{align}
\mathsf{Stoch}^{(1)}(x; \hat{\theta}_n(x),\hat{g}_n)
&=
(\hat{\theta}_n(x) - \theta_0(x)) \bar{M}_n^{(1)}(x; \theta_0(x), \hat{g}_n) \nonumber\\
&\quad+
(\hat{\theta}_n(x) - \theta_0(x))^2
\bar{H}_n(x; \tilde{\theta}_0(x), \hat{g}_n) \label{eq: taylor expand empirical moment}
\end{align}
for some, potentially different, \(\tilde{\theta}_0(x)\) between \(\hat{\theta}_n(x)\) and \(\theta_0(x)\). To bound this term, observe that
\begin{align}
\vert \bar{M}_n^{(1)}(x; \theta_0(x), \hat{g}_n)  \vert
&\leq
\vert \bar{M}_n^{(1)}(x; \theta_0(x), g_0)\vert \nonumber\\
&\quad+
\vert \bar{M}_n^{(1)}(x; \theta_0(x), \hat{g}_n)
-
\bar{M}_n^{(1)}(x; \theta_0(x), g_0)\vert~.
\end{align}
Hence, \cref{assu: empirical smoothness} and \cref{assu: bias}, Parts (ii) and (iii), together with Holder's inequality, imply that
\begin{align}
&\mathbb{E}\left[
\left|
\sqrt{ \frac{n}{ \underline{\lambda}^2 } }
\left\|
\mathsf{Stoch}^{(1)}(\bm{x}^{(d)}; \hat{\theta}_n(\bm{x}^{(d)}),\hat{g}_n)
\right\|_\infty
\right|^q
\right]^{1/q} \nonumber\\
&\lesssim
q^2\delta_{n,\theta}\delta_{n,m}
+
q^2\frac{\underline{\lambda}^{1/2}}{n^{1/4}}\delta_{n,\theta}\delta_{n,J}
+
q^2\delta_{n,\theta}^2~. \label{eq: Stoch 1 bound}
\end{align}
Moreover, \cref{assu: bias}, Part (iii), gives that
\begin{align}
&\mathbb{E}\left[
\left|
\sqrt{ \frac{n}{ \underline{\lambda}^2 } }
\left\|
\mathsf{Stoch}^{(2)}(\bm{x}^{(d)}; \hat{g}_n)
\right\|_\infty
\right|^q
\right]^{1/q}
\lesssim q\delta_{n,S}
\lesssim q^2\delta_{n,S}~. \label{eq: Stoch 2 bound}
\end{align}
Finally, we handle the terms \eqref{eq: theta squared} and \eqref{eq: theta g term}. \cref{assu: bias}, Part (ii), and \cref{assu: app a moment function} imply that
\begin{align}
&\mathbb{E}\bigg[
\left|
\sqrt{ \frac{n}{ \underline{\lambda}^2 } }
\left\|
(\hat{\theta}_n(\bm{x}^{(d)})-\theta_0(\bm{x}^{(d)}))^2
H(\bm{x}^{(d)};\tilde{\theta}_0(\bm{x}^{(d)}),\hat{g}_n)
\right\|_\infty
\right|^q
\bigg]^{1/q}
\lesssim q^2\delta_{n,\theta}^2~, \label{eq: theta 2 hessian bound}
\end{align}
and, by the Lipschitz property of \(M^{(1)}(\cdot;\theta,g)\) in \(g\),
\begin{align}
&\mathbb{E}\bigg[
\bigg|
\sqrt{ \frac{n}{ \underline{\lambda}^2 } }
\bigg\|
(\hat{\theta}_n(\bm{x}^{(d)})-\theta_0(\bm{x}^{(d)})) \nonumber\\
&
\quad\quad\quad\quad\quad\quad
\left(
M^{(1)}(\bm{x}^{(d)};\theta_0(\bm{x}^{(d)}),\hat{g}_n)
-
M^{(1)}(\bm{x}^{(d)};\theta_0(\bm{x}^{(d)}),g_0)
\right)
\bigg\|_\infty
\bigg|^q
\bigg]^{1/q} \nonumber\\
&
\lesssim q^2\delta_{n,\theta}\delta_{n,g}~. \label{eq: theta jac bound}
\end{align}
Together, the decomposition \eqref{eq: basic decomposition} and the lower-boundedness of the Jacobian \(M^{(1)}(\cdot;\theta,g)\) imply that
\begin{align}\label{eq: reduce to u stat}
&\mathbb{E}\left[
\left|
\sqrt{ \frac{n}{ \underline{\lambda}^2 } }
\left\|
R_n(\bm{x}^{(d)})-W_n(\bm{x}^{(d)})
\right\|_\infty
\right|^q
\right]^{1/q} \\
&\lesssim
q^2\bigg(
\delta_{n,g}^2 + \delta_{n,\theta}^2 + \delta_{n,B} + \delta_{n,S} + \delta_{n,\theta}
\left( \delta_{n,m} + \delta_{n,g} + \underline{\lambda}^{1/2}n^{-1/4}
\left( \delta_{n,B} + \delta_{n,J} \right) \right)
\bigg) \nonumber\\
&\lesssim
q^2\bigg(
\delta_{n,g}^2 + \delta_{n,\theta}^2 + \delta_{n,m}^2 + \delta_{n,B} + \delta_{n,S} +
\underline{\lambda}^{1/2}n^{-1/4}\delta_{n,\theta}
\left( \delta_{n,B} + \delta_{n,J} \right)
\bigg)~,\nonumber
\end{align}
where the final inequality uses \(ab\leq a^2+b^2\), applied to the products
\(\delta_{n,\theta}\delta_{n,m}\) and \(\delta_{n,\theta}\delta_{n,g}\).

With this in place, we turn to the proof of the normal approximation to $R_{n}(\bm{x}^{(d)})$ on hyper-rectangles. Fix a rectangle $\mathsf{R}=[a_{l},a_{u}]$ in $\mathcal{R}$.  For the sake of exposition, we give the details of the proof of the upper bound 
\begin{align}
& P\left\{\sqrt{n} R_{n}(\bm{x}^{(d)})\in\mathsf{R}\right\} -P\left\{\Lambda^{-1/2} Z\in\mathsf{R} \right\} \nonumber \\
& \lesssim 
\left(\frac{\varphi^{2}\log^{5}(dn)}{\underline{\lambda}^{2}n}\right)^{1/4} 
+ \delta_n \log^{5/2}(dn) + \frac{1}{nd}~.
\end{align}
The matching lower bound will follow from a very similar argument. Consider the decomposition
\begin{flalign}
R_n(x) & = \left(W_n(x)-\frac{1}{n}\sum_{i=1}^n \bar{u}(x, D_i)\right) + \frac{1}{n}\sum_{i=1}^n \bar{u}(x, D_i) + \Delta_n(x),\quad\text{where}\label{eq: R decomp}\\
\Delta_n(x) & = R_n(x) - W_n(x)\label{eq: Delta def}~,
\end{flalign}
and we recall that the function $\bar{u}(\cdot,\cdot)$ is defined in \cref{assu: linearity}. Define the normalized functions
\begin{equation*}
\hat{u}(\bm{x}^{(d)}, D_i) =  \Lambda^{-1/2} \bar{u}(\bm{x}^{(d)}, D_i)\quad\text{and}\quad\hat{W}_n(\bm{x}^{(d)}) =  \Lambda^{-1/2} W_n(\bm{x}^{(d)})
\end{equation*}
and the analogously normalized rectangle $\tilde{\mathsf{R}} = [\Lambda^{-1/2}a_l, \Lambda^{-1/2}a_u]$ and the enlarged rectangle  $\tilde{\mathsf{R}}_t = [\Lambda^{-1/2}a_l - \mathbf{1}_dt, \Lambda^{-1/2}a_u + \mathbf{1}_dt]$. Observe that the decomposition \eqref{eq: R decomp} yields the upper bound
\begin{flalign}
 &  P\left\{\sqrt{n} R_{n}(\bm{x}^{(d)})\in\mathsf{R}\right\} -P\left\{Z\in\mathsf{R} \right\} \nonumber \\
 & =  P\left\{\sqrt{n} \Lambda^{-1/2} R_{n}(\bm{x}^{(d)})\in\tilde{\mathsf{R}}\right\} -P\left\{\Lambda^{-1/2}Z\in\tilde{\mathsf{R}} \right\} \nonumber \\
 & \quad \leq\Big\vert P\left\{ \frac{1}{\sqrt{n}}\sum_{i=1}^n \hat{u}(\bm{x}^{(d)}, D_i) \in \tilde{\mathsf{R}}_t  \right\} 
                                 -P\left\{\Lambda^{-1/2}Z\in \tilde{\mathsf{R}}_t  \right\} \Big\vert \label{eq: linear clt}\\
 & \quad\quad +\Big\vert P\left\{\Lambda^{-1/2}Z\in\tilde{\mathsf{R}}_{t}\right\} 
    -P\left\{\Lambda^{-1/2}Z\in\tilde{\mathsf{R}}\right\} \Big\vert\label{eq: normal diff expansion} \\
 & \quad\quad +P\left\{\sqrt{n}\|\hat{W}_n(\bm{x}^{(d)}) - \frac{1}{n} \sum_{i=1}^n \hat{u}(\bm{x}^{(d)}, D_i) \|_\infty \ge \frac{1}{2}t \right\} \label{eq: term linear}\\
 & \quad\quad +P\left\{\sqrt{n} \| \Lambda^{-1/2} \Delta_n(\bm{x}^{(d)})\|_\infty \ge \frac{1}{2}t \right\} \label{eq: term delta}
\end{flalign}
for each $t>0$. We proceed by providing appropriate bounds for the terms \eqref{eq: linear clt} through \eqref{eq: term delta}.

We begin by bounding the normal approximation term through an application of \cref{lem: baseline clt}. In particular, observe that 
\begin{equation}
\frac{1}{n}\sum_{i=1}^{n}\mathbb{E}\left[\hat{u}^{2}(x^{(j)}, D_{i})\right]=1
\end{equation}
by definition and that 
\begin{equation}
\frac{1}{n}\sum_{i=1}^{n}\mathbb{E}\left[\hat{u}^{4}(x^{(j)},D_{i})\right]
\leq \left(\varphi/\underline{\lambda}\right)^{2}\label{eq: fourth bound}
\end{equation}
by \cref{assu: linearity}. Similarly we have that 
\begin{equation}
\| \hat{u}(x^{(j)}, D_{i}) \|_{\psi_1}
\leq (\varphi /\underline{\lambda})
\end{equation}
by \cref{assu: linearity}. Consequently, as 
\begin{equation}
\Var(\hat{u}(\bm{x}^{(d)}, D_{i}))=\Lambda^{-1/2}\Var\left(Z\right)\Lambda^{-1/2},
\end{equation}
by definition, Lemma \ref{lem: baseline clt} implies that the bound
\begin{equation}
\Big\vert P\left\{ \frac{1}{\sqrt{n}}\sum_{i=1}^{n}\hat{u}(\mathbf{x}^{(d)},D_{i})\in \mathsf{\tilde{R}}_t\right\} 
-P\left\{ \Lambda^{-1/2}Z\in \mathsf{\tilde{R}}_t\right\} \Big\vert
\lesssim \left(\frac{\varphi^{2}\log^{5}(dn)}{\underline{\lambda}^{2}n}\right)^{1/4}~.\label{eq: baseline clt applied}
\end{equation}
holds.

In turn, to bound the terms \eqref{eq: normal diff expansion}, we have that
\begin{equation}\label{eq: nazarov basic decompositon}
\Big\vert P\left\{\Lambda^{-1/2}Z\in\tilde{\mathsf{R}}_{t}\right\} 
             -P\left\{\Lambda^{-1/2}Z\in\tilde{\mathsf{R}}\right\} \Big\vert \lesssim t \sqrt{\log d}~.
\end{equation}
by \cref{lem: Nazarov}. To bound the terms \eqref{eq: term linear}, taking $q = 2\log(nd)$, \cref{assu: linearity} implies that
\begin{align}
    \sqrt{n}\|\hat{W}_n(\bm{x}^{(d)}) - \frac{1}{n} \sum_{i=1}^n \hat{u}(\bm{x}^{(d)}, D_i) \|_\infty \lesssim \delta_{n,u} \log^2(dn)
    \label{eq: apply linearity to abstract}
\end{align}
with probability greater than $1-C/nd$, by applying Markov's inequality and a union bound. Likewise, the bound \eqref{eq: reduce to u stat} implies that
\begin{align}
&\sqrt{n} \| \Lambda^{-1/2} \Delta_n(\bm{x}^{(d)})\|_\infty  \nonumber \\
&\lesssim
    (
\delta_{n,g}^2 + \delta_{n,\theta}^2 + \delta_{n,m}^2 + \delta_{n,B} + \delta_{n,S} +
\underline{\lambda}^{1/2}n^{-1/4}\delta_{n,\theta}
( \delta_{n,B} + \delta_{n,J} )
)\log^2(dn)\label{eq: bound U stat approx}
\end{align}
with probability greater than $1-C/nd$, by applying Markov's inequality and a union bound.

Thus, by choosing $t=C\delta_n\log^2(dn)$, the sum of term \eqref{eq: term linear} and term \eqref{eq: term delta} is
upper bounded by $C/nd$. Hence, by plugging this choice of $t$ into \eqref{eq: nazarov basic decompositon}, we can conclude that 
\begin{align}
&\Big\vert P\left\{\sqrt{n} R_{n}(\bm{x}^{(d)})\in\mathsf{R}\right\} -P\left\{ \Lambda^{-1/2} Z\in\mathsf{R}\right\} \Big\vert \nonumber\\
& \lesssim \left(\frac{\varphi^{2}\log^{5}(dn)}{\underline{\lambda}^{2}n}\right)^{1/4} 
+ \delta_n \log^{5/2}(dn) + \frac{1}{nd}~,
\end{align}
as required.\hfill\qed

\subsubsection{Part (ii)}

The result follows from an argument whose structure is similar to the proof of \cref{thm: generic m estimation unstudentized}, Part (i). Again, we take $\theta_0(x)=0$ for all $x$, without loss of generality. We are interested in studying the discrepancy 
\begin{flalign*}
R_{n}^{*}(x) & =\hat{\theta}_\mathsf{h}(x)-\hat{\theta}_n(x) = (\hat{\theta}_\mathsf{h}(x)-\theta_{0}(x))-R_{n}(x).
\end{flalign*}
In terms of the decomposition \eqref{eq: R decomp}, we can write
\begin{flalign} \label{eq: re R decomp}
R_n(x)  = \left(W_n(x) - \frac{1}{n}\sum_{i=1}^n \bar{u}(x, D_i)\right) + \frac{1}{n}\sum_{i=1}^n \bar{u}(x, D_i) + \Delta_n(x)~,
\end{flalign}
where $\Delta_n(x)$ is defined in \eqref{eq: Delta def}. On the other hand, as $\hat{\theta}_\mathsf{h}(x)$ is constructed with a random half-sample $\mathsf{h}$ of the data $\mathbf{D}_{n}$,
we have that 
\begin{flalign} \label{eq: R* decomp boot}
\hat{\theta}_\mathsf{h}(x)-\theta_{0}(x)
  = \left(\frac{2}{n}\sum_{i\in\mathsf{h}} \bar{u}(x, D_i) - W_\mathsf{h}(x)\right) + \frac{2}{n}\sum_{i\in\mathsf{h}} \bar{u}(x, D_i) + \Delta_\mathsf{h}(x)~,
\end{flalign}
where $U_\mathsf{h}(x)$ and $\Delta_\mathsf{h}(x)$ are constructed with the half-sample $\mathsf{h}$. 

The proof of \cref{thm: generic m estimation unstudentized}, Part (i), worked by giving a high probability bound for the first and third term in \eqref{eq: re R decomp} and showing that the second term satisfies a central limit theorem. Here, as we are interested in giving a bound conditioned on the data $\mathbf{D}_n$, we show that the difference between the second terms in \eqref{eq: re R decomp} and \eqref{eq: R* decomp boot} satisfies a central limit theorem on the event that the first and third terms in \eqref{eq: re R decomp} and \eqref{eq: R* decomp boot} satisfy a specified bound, which we show holds with high probability. To this end, let $\hat{W}_\mathsf{h}(\bm{x}^{(d)})$ is defined analogously to $\hat{W}_n(\bm{x}^{(d)})$ and let $\mathcal{F}_n(t)$ and $\mathcal{F}_\mathsf{h}(t)$ denote the events that
\begin{equation}\label{eq: F event}
\sqrt{\frac{n}{\underline{\lambda}^2}} \| \Delta_n(\bm{x}^{(d)})\|_\infty \leq t/4
\quad\text{and}\quad
\sqrt{\frac{n}{\underline{\lambda}^2}} \| \Delta_\mathsf{h}(\bm{x}^{(d)})\|_\infty \leq t/4~,
\end{equation}
respectively. Similarly, let $\mathcal{H}_n(t)$ and $\mathcal{H}_\mathsf{h}(t)$ denote the events that
\begin{flalign}\label{eq: H event}
&\sqrt{n}\| \frac{1}{n}\sum_{i=1}^n \hat{u}(\bm{x}^{(d)}, D_i) - \hat{W}_n(\bm{x}^{(d)}) \|_\infty \leq t/4
\quad\text{and}\\
&\sqrt{n}\| \frac{2}{n}\sum_{i\in\mathsf{h}} \hat{u}(\bm{x}^{(d)}, D_i) - \hat{W}_\mathsf{h}(\bm{x}^{(d)}) \|_\infty \leq t/4\nonumber
\end{flalign}
respectively. Define the event $\mathcal{E}_n(t) = \mathcal{F}_n(t) \cap \mathcal{F}_\mathsf{h}(t) \cap \mathcal{H}_n(t) \cap \mathcal{H}_\mathsf{h}(t)$.  

Fix a hyper-rectangle $\mathsf{R}$ in $\mathcal{R}$. As before, we prove only the requisite upper bound. Recall the definitions of the transformed rectangle $\tilde{\mathsf{R}}$ and the enlarged transformed rectangle $\tilde{\mathsf{R}}_t$. On the event $\mathcal{E}_n(t) $, we have
\begin{flalign}
 & P\left\{\sqrt{n} R_{n}^{*}(\bm{x}^{(d)})\in\mathsf{R}\mid\mathbf{D}_{n}\right\} - P\left\{ Z\in\mathsf{R}\right\} \nonumber \\
 & = P\left\{\sqrt{n} \Lambda^{-1/2} R_{n}^{*}(\bm{x}^{(d)})\in\tilde{\mathsf{R}}\mid\mathbf{D}_{n}\right\} - P\left\{\Lambda^{-1/2} Z\in\tilde{\mathsf{R}}\right\} \nonumber \\
 &\leq P\left\{ \frac{2}{\sqrt{n}}\sum_{i\in\mathsf{h}} \hat{u}(\bm{x}^{(d)}, D_i) - \frac{1}{\sqrt{n}}\sum_{i=1}^n \hat{u}(\bm{x}^{(d)}, D_i)\in\tilde{\mathsf{R}}_{t} \mid\mathbf{D}_{n}\right\} 
                         -P\left\{ \Lambda^{-1/2}Z\in\tilde{\mathsf{R}}_{t}\right\} \nonumber \\
 &+\vert P\left\{  \Lambda^{-1/2}Z\in\tilde{\mathsf{R}}\right\} -P\left\{  \Lambda^{-1/2}Z\in\tilde{\mathsf{R}}_t\right\} \vert
\end{flalign}
for each $t>0$. As the data $D_\mathsf{h}$ are drawn independently and identically with distribution $P$ in $\mathbf{P}$ and we have assumed that $\delta_{n/2}\lesssim \delta_n$, by setting $t = C\delta_n\log^2(dn)$, bounds \eqref{eq: bound U stat approx} and \eqref{eq: apply linearity to abstract} imply that the event $\mathcal{E}_n(t)$ occurs with probability greater than $1-C/nd$. Thus, \cref{lem: Nazarov} implies that 
\begin{flalign}
 & P\left\{\sqrt{n} R_{n}^{*}(\bm{x}^{(d)})\in\mathsf{R}\mid\mathbf{D}_{n}\right\} - P\left\{ Z\in\mathsf{R}\right\} \label{eq: up to u stat}\\
 & \leq
 P\left\{ \frac{2}{\sqrt{n}}\sum_{i\in\mathsf{h}} \hat{u}(\bm{x}^{(d)}, D_i)  - \frac{1}{\sqrt{n}}\sum_{i=1}^n \hat{u}(\bm{x}^{(d)}, D_i)\in\tilde{\mathsf{R}}_{t} \mid\mathbf{D}_{n}\right\} 
 - P\left\{ \Lambda^{-1/2}Z\in\tilde{\mathsf{R}}_{t}\right\}
 \label{eq: to u stat CLT}\\
& +  \delta_n \log^{5/2}(dn) \nonumber
\end{flalign}
with probability greater than $1-C/nd$. Hence, it suffices to bound the term \eqref{eq: to u stat CLT}.

To this end, we apply a coupling argument introduced in \cite{yadlowsky2023evaluating}, which is similar to a Poissonization technique studied in \cite{praestgaard1993exchangeably} (see also Section 3.6.2 of \cite{van1996weak}). In particular, let $V_{i}$ be a random variable taking the value $1$ when $i$
is an element of the subset $\mathsf{h}$ and taking the value $-1$ otherwise. Observe that
\begin{flalign}\label{eq: binomialization}
\frac{2}{n}\sum_{i\in\mathsf{h}} \hat{u}(\bm{x}^{(d)}, D_i) - \frac{1}{n}\sum_{i=1}^n \hat{u}(\bm{x}^{(d)}, D_i) = \frac{1}{n} \sum_{i=1}^n V_i \hat{u}(\bm{x}^{(d)}, D_i)~.
\end{flalign}
Let $\tilde{V}_{1},\ldots,\tilde{V}_{n}$ denote a collection of random
variables valued on $\{-1,1\}$. We define their joint distribution
as follows. Let $Q_{n}$ denote a random variable with distribution
$\mathsf{Bin}\left(n,1/2\right)$. If $Q_{n}\geq n/2$, then choose $Q_{n}-n/2$
indices $i$ in $[n]$ with $V_{i}=-1$ and set $\tilde{V}_{i}=1$.
If $Q_{n}<n/2$, then choose $n/2-Q_{n}$ indices with $V_{i}=1$
and set $\tilde{V}_{i}=-1$. Set $\tilde{V}_{i}=V_{i}$ for all other
units. Observe that the collection $\tilde{V}_{i}$ are independent
and identically distributed Rademacher random variables. With this in place, 
we obtain the decomposition
\begin{equation}\label{eq: binom approx}
\frac{1}{n}\sum_{i=1}^{n}V_{i}\hat{u}(\bm{x}^{(d)}, D_i)
=\frac{1}{n}\sum_{i=1}^{n}\tilde{V}_{i} \hat{u}(\bm{x}^{(d)}, D_i)
+\frac{1}{n}\sum_{i=1}^{n}\left(V_{i}-\tilde{V}_{i}\right)\hat{u}(\bm{x}^{(d)}, D_i)~.
\end{equation}
Let $\mathcal{V}(t^\prime)$ denote the event that
\begin{equation}
\sqrt{n} \Big\| \frac{1}{n}\sum_{i=1}^{n}\left(V_{i}-\tilde{V}_{i}\right)\hat{u}(\bm{x}^{(d)}, D_i) \Big\|_\infty < t^\prime~.
\end{equation}
Fix any rectangle $\mathsf{R}^\prime = [a^\prime_l, a^\prime_u]$ and define the enlarged rectangle $\mathsf{R}^\prime_{t^\prime} = [a^\prime_l - t^\prime\mathbf{1}_d,a^\prime_u + t^\prime\mathbf{1}_d]$.
On the event $\mathcal{V}(t^\prime)$, the decomposition \eqref{eq: binom approx} implies that 
\begin{flalign}
&  P\left\{ \frac{1}{\sqrt{n}}\sum_{i=1}^{n}V_{i}\hat{u}(\bm{x}^{(d)}, D_i) \in \mathsf{R}^\prime \mid \mathbf{D}_n \right\} 
- P\left\{  \Lambda^{-1/2} Z\in\mathsf{R}^\prime \right\}  \nonumber\\
& \leq  P\left\{\frac{1}{\sqrt{n}}\sum_{i=1}^{n}\tilde{V}_{i} \hat{u}(\bm{x}^{(d)}, D_i) \in \mathsf{R}^\prime_{t^\prime} \mid \mathbf{D}_n \right\} 
- P\left\{  \Lambda^{-1/2} Z\in \mathsf{R}^\prime_{t^\prime} \right\} \label{eq: to mult clt}\\
& \quad\quad + \bigg \vert P\left\{  \Lambda^{-1/2} Z\in\mathsf{R}^\prime_{t^\prime}\right\} -P\left\{  \Lambda^{-1/2} Z\in\mathsf{R}^\prime\right\} \bigg \vert\label{eq: binom nazarov}
\end{flalign}
for all $t>0$. \cref{lem: Nazarov} implies that \eqref{eq: binom nazarov} is less than $t^\prime\sqrt{\log(d)}$. To handle the term \eqref{eq: to mult clt}, we apply the following quantitative central limit theorem, stated as Lemma 4.6 in \citet{chernozhuokov2022improved}.
\begin{lemma}[{\citealp[Lemma 4.6,  ][]{chernozhuokov2022improved}}]
\label{lem: multiplier bootstrap clt}Consider the setting and assumptions
of \cref{lem: baseline clt}. Let $\bm{X}_{n}=(X_{1},\ldots,X_{n})$
collect the observed data and let $\tilde{V}_{1},\ldots,\tilde{V}_{n}$ be a collection
of independent Rademacher random variables. We have that
\begin{align}
& \sup_{\mathsf{R}\in\mathcal{R}}\Big\vert P\left\{ n^{-1/2}\sum_{i=1}^{n}X_{i}\in \mathsf{R}\right\} 
-P\left\{ n^{-1/2}\sum_{i=1}^{n}\tilde{V}_{i}X_{i}\in \mathsf{R}\mid\bm{X}_{n}\right\} \Big\vert \\
& \quad\quad\quad\quad\quad\quad\quad\quad\quad\quad\quad
\leq C_3 \left(\frac{\varphi^{2}\log^{5}(dn)}{n}\right)^{1/4}.
\end{align}
with probability greater than
\begin{equation}
1-C\frac{\varphi \log^{3/2}(dn)}{n^{1/2}}
\end{equation}
for some constant $C_3$ that depends only on the constants $C_2$ and $c$ defined in the statement of \cref{lem: baseline clt}. 
\end{lemma}

\noindent Thus, on the event $\mathcal{V}(t')$, \cref{lem: baseline clt} and \cref{lem: multiplier bootstrap clt} imply that
\begin{flalign*}
 P\left\{ \frac{1}{\sqrt{n}}\sum_{i=1}^{n}V_{i}\hat{u}(\bm{x}^{(d)}; D_i) \in \mathsf{R}^\prime \mid \mathbf{D}_n \right\} 
- P\left\{ \Lambda^{-1/2} Z\in\mathsf{R}^\prime\right\} 
\lesssim \left(\frac{\varphi^{2}\log^{5}(dn)}{\underline{\lambda}^2 n}\right)^{1/4} + t'\sqrt{\log(d)}~,
\end{flalign*}
with probability greater than $1-Cn^{-1/2}\underline{\lambda}^{-1}\varphi \log^{3/2}(dn)$. 

Hence, it suffices to give a high probability bound on $\mathcal{V}(t')$. To this end, observe that
\begin{flalign}
G_{n} & =\Big\|\frac{1}{\sqrt{n}}\sum_{i=1}^{n}\left(V_{i}-\tilde{V}_{i}\right)\hat{u}(\bm{x}^{(d)}; D_i)\Big\|_{\infty}
\end{flalign}
is equidistributed with
\begin{align}
    \Big\|\frac{2}{\sqrt{n}}\sum_{i=1}^{\vert Q_{n} - n/2\vert}\hat{u}(\bm{x}^{(d)};D_{i})\Big\|_{\infty}~.
\end{align}
Consider the decomposition
\begin{flalign}
 P\left\{ G_{n}\geq t' \right\} 
 & \leq P\left\{ G_{n}\geq t',\vert Q_{n} - n/2 \vert \leq\delta\frac{n}{2}\right\} +P\left\{ \vert Q_{n} - n/2 \vert \geq\delta\frac{n}{2}\right\} \nonumber \\
 & \leq P\left\{ \max_{1\leq k\leq\delta\frac{n}{2}}\Big\|\frac{2}{\sqrt{n}}\sum_{i=1}^{k}\hat{u}(\bm{x}^{(d)};D_{i})\Big\|_{\infty}\geq t'\right\} 
 +P\left\{ \vert Q_{n} - n/2\vert \geq\delta\frac{n}{2}\right\} ,\label{eq: set up for maximal}
\end{flalign}
for some $\delta>0$ to be chosen. Observe that
\begin{equation}
P\left\{ \vert Q_{n} - n/2\vert \geq\frac{\delta n}{2}\right\} \le2\exp\left(-\frac{\delta^{2}n}{6}\right)\label{eq: multiplicative chernoff}
\end{equation}
by the multiplicative Chernoff bound. To bound the first term in (\ref{eq: set up for maximal}), we combine \cref{lem: orlicz large deviation} with the following L\'{e}vy type inequality for independent random vectors, due to \cite{montgomery1993comparison}. 

\begin{lemma}[Theorem 1.1.5, \cite{de2012decoupling}] \label{lem: independent levy}
Let $Z_{1},\ldots,Z_{n}$ be independent random vectors in $\mathbb{R}^{d}$. There exist universal constants $C_1$ and $C_2$ such that 
\begin{equation}
P\left\{ \max_{1\leq k \leq n} \| \sum_{i=1}^k Z_i \|_\infty > t \right\} \leq C_1 P\left\{ \| \sum_{i=1}^n Z_i \|_\infty > \frac{t}{C_2} \right\} 
\end{equation}
for all $t>0$. 
\end{lemma}
\noindent
In particular, as 
\begin{equation*}
\|\frac{2}{\sqrt{n}}\hat{u}(x^{(j)}; D_{i}) \|_{\psi_{1}}
\leq\frac{2}{\sqrt{n}}\frac{\varphi}{\underline{\lambda}}~,
\end{equation*}
and 
\begin{equation}
\max_{j \in [d]} \sum_{k=1}^{\delta n / 2} \mathbb{E}\left[ \left(\frac{2}{\sqrt{n}}\hat{u}(x^{(j)}; D_{i}) \right)^2\right] = 2\delta~,
\end{equation}
\cref{lem: orlicz large deviation} and \cref{lem: independent levy} imply that
\begin{align}\label{eq: apply levy orlicz}
P\left\{ \max_{1\leq k\leq\delta\frac{n}{2}}\Big\|\frac{2}{\sqrt{n}}\sum_{i=1}^{k}\hat{u}\left(\bm{x}^{(d)};D_{i}\right)\Big\|_{\infty}
\geq C \left(\delta \log^{1/2}(dn) + \frac{2}{\sqrt{n}}\frac{\varphi}{\underline{\lambda}} \log^2(dn)\right) \right\} \lesssim \frac{1}{n}~.
\end{align}
Now, the choice 
\begin{equation}
\delta=C \sqrt{\frac{\log n}{n}}\label{eq: delta choice}
\end{equation}
gives 
\begin{equation}
P\left\{ \vert Q_{n} - n/2 \vert \geq\frac{\delta n}{2}\right\} \lesssim \frac{1}{n}\label{eq: binomial choice}
\end{equation}
by (\ref{eq: multiplicative chernoff}). Plugging this choice into \eqref{eq: apply levy orlicz} yields
\begin{flalign}\label{eq: HJ choice}
P\left\{ \max_{1\leq k\leq\delta\frac{n}{2}}\Big\|\frac{2}{\sqrt{n}}\sum_{i=1}^{k}\hat{u}\left(\bm{x}^{(d)};D_{i}\right)\Big\|_{\infty}
\geq C \frac{1}{\sqrt{n}}\frac{\varphi}{\underline{\lambda}} \log^2(dn)  \right\} \lesssim \frac{1}{n}~.
\end{flalign}
Hence, we find that the inequality
\begin{align}
& P\bigg\{ \Big\|\frac{1}{\sqrt{n}}\sum_{i=1}^{n}\left(V_{i}-\tilde{V}_{i}\right)\hat{u}(\bm{x}^{(d)}, D_{i})\Big\|_{\infty} \nonumber \\
& 
\quad\quad\quad\quad\geq C 
\left(\frac{\varphi^2\log^{4}\left(dn\right)}{\underline{\lambda}^2 n}\right)^{1/2} \bigg\}  \lesssim \frac{1}{n}
\end{align}
holds for all $n$ sufficiently large. 

Thus, by setting
\begin{equation*}
t' = C \left( \frac{\varphi^2\log^{4}\left(dn\right)}{\underline{\lambda}^2 n}\right)^{1/2}~,
\end{equation*}
we find that
\begin{align}
& P\left\{ \frac{1}{\sqrt{n}}\sum_{i=1}^{n}V_{i}\hat{u}(\bm{x}^{(d)}, D_i) \in \mathsf{R}^{\prime} \mid \mathbf{D}_n \right\} 
- P\left\{ \Lambda^{-1/2} Z\in \mathsf{R}^\prime \right\}  \nonumber\\
& \quad\quad\quad\quad
\lesssim
 \left(\frac{\varphi^{2}\log^{5}(dn)}{\underline{\lambda}^2 n}\right)^{1/4}
 + \sqrt{\log (d)}\left( \frac{\varphi^2 \log^{4}\left(dn\right)}{\underline{\lambda}^2 n}\right)^{1/2}
\lesssim
\left( \frac{\varphi^2 \log^{5}\left(dn\right)}{\underline{\lambda}^2 n}\right)^{1/4}~,\label{eq: binom end}
\end{align}
with probability greater than $1-C n^{-1/2}\underline{\lambda}^{-1}\varphi \log^{3/2}(dn)$. Putting the pieces together, the inequalities \eqref{eq: up to u stat} and \eqref{eq: binom end} imply that 
\begin{flalign*}
& P\left\{\sqrt{n} R_{n}^{*}(\bm{x}^{(d)})\in\mathsf{R}\mid\mathbf{D}_{n}\right\} - P\left\{  Z\in\mathsf{R}\right\}
\lesssim 
\left( \frac{\varphi^2 \log^{5}\left(dn\right)}{\underline{\lambda}^2 n}\right)^{1/4}+ \delta_n\log^{5/2}(dn)
\end{flalign*}
with probability greater than $1-C(n^{-1/2}\varphi \underline{\lambda}^{-1} \log^{3/2}(dn) + (nd)^{-1})$, as required.\hfill\qed

\subsection{Proof of \cref{lem: variance accuracy}}

Again, we take $\theta_0(x)=0$ for all $x$, without loss of generality. Recall from the proof of \cref{thm: generic m estimation unstudentized}, Part (ii), that
 $V_{i}$ is a random variable taking the value $1$ when $i$
is an element of the subset $\mathsf{h}$, and taking the value $-1$ otherwise, and that 
\begin{flalign}\label{eq: re binomialization}
\frac{2}{n}\sum_{i\in\mathsf{h}} \bar{u}(x, D_i) - \frac{1}{n}\sum_{i=1}^n \bar{u}(x, D_i) = \frac{1}{n} \sum_{i=1}^n V_i \bar{u}(x, D_i)~.
\end{flalign}
Define the objects
\begin{align}\label{eq: Q term def}
T_n(x) &= \frac{1}{n}\sum_{i=1}^n \bar{u}(x, D_i) - W_n(x)\quad\text{and}\quad
T_\mathsf{h}(x)= \frac{2}{n}\sum_{i\in\mathsf{h}} \bar{u}(x, D_i) - W_\mathsf{h}(x)~.
\end{align}
and recall the objects $ \Delta_\mathsf{h}(x^{(j)})$ and $\Delta_n(x^{(j)})$ introduced in \eqref{eq: Delta def} and \eqref{eq: R* decomp boot}, respectively. 

Observe that we can write
\begin{align}
\hat{\lambda}^2_{n,j} 
& = n\mathbb{E}_{V}\left[\left(R^*_n(x^{(j)})\right)^2 \right]\\
& = n\mathbb{E}_{V}\left[\left(\frac{1}{n} \sum_{i=1}^n V_i \bar{u}(x^{(j)}, D_i)  - T_\mathsf{h}(x^{(j)}) + T_n(x^{(j)}) + \Delta_\mathsf{h}(x^{(j)}) -  \Delta_n(x^{(j)})\right)^2\right]~,\nonumber 
\end{align}
where the notation $\mathbb{E}_{V}\left[\cdot\right]$ denotes that the expectation is evaluated only over the random variables $V_1,\ldots,V_n$. Thus, by defining the quantities
\begin{align}
    \bar{\lambda}^2_{n,j}
    & = n \mathbb{E}_{V}\left[
    \left(
    \frac{1}{n} \sum_{i=1}^n V_i \bar{u}(x^{(j)}, D_i)
    \right)^2
    \right]\quad\text{and}\quad\\
    \Gamma_{n,j} & =
    n\mathbb{E}_{V}\left[\left( T_n(x^{(j)}) - T_\mathsf{h}(x^{(j)}) + \Delta_\mathsf{h}(x^{(j)}) -  \Delta_n(x^{(j)})\right)^2\right]\nonumber
\end{align}
we obtain the bound
\begin{align}
    \vert \hat{\lambda}^2_{n,j} - \lambda^2_j \vert 
    & \leq 
    \vert \bar{\lambda}_{n,j}^2 - \lambda^2_j \vert  + 2 \bar{\lambda}_{n,j} \Gamma^{1/2}_{n,j} + \Gamma_{n,j}\label{eq: lambda cauchy-schwarz}
\end{align}
by expanding the square and applying the Cauchy-Schwarz inequality. 

Now, we obtain a large deviation bound for $\Gamma_{n,j}$. Observe that \cref{assu: linearity} and \eqref{eq: reduce to u stat} imply that
\begin{align}
\mathbb{E}\left[
\left(
\frac{n}{\underline{\lambda}^2}
T_n(x^{(j)})^2
\right)^{q/2}
\right]
&\lesssim q^{2q}\delta_n^q
\quad\text{and}\quad
\mathbb{E}\left[
\left(
\frac{n}{\underline{\lambda}^2}
\Delta_n(x^{(j)})^2
\right)^{q/2}
\right]
\lesssim q^{2q}\delta_n^q~. \label{eq: Tn Deltan moments}
\end{align}
respectively. Likewise, Jensen's inequality implies that
\begin{align}
&\mathbb{E}\left[
\left(
\mathbb{E}_{V}\left[
\frac{n}{\underline{\lambda}^2}
T_{\mathsf{h}}(x^{(j)})^2
\right]
\right)^{q/2}
\right] \leq
\mathbb{E}\left[
\mathbb{E}_{V}\left[
\left|
\sqrt{\frac{n}{\underline{\lambda}^2}}
T_{\mathsf{h}}(x^{(j)})
\right|^q
\right]
\right]
\lesssim q^{2q}\delta_n^q\label{eq: Th conditional moment}
\end{align}
and analogously
\begin{align}
&\mathbb{E}\left[
\left(
\mathbb{E}_{V}\left[
\frac{n}{\underline{\lambda}^2}
\Delta_{\mathsf{h}}(x^{(j)})^2
\right]
\right)^{q/2}
\right]
\lesssim q^{2q}\delta_n^q~. \label{eq: Deltah conditional moment}
\end{align}
Thus, by taking \(q=2\vee 2\log(nd)\) and applying Markov and a union bound, we find that
\begin{align}
&\sup_{j\in[d]} \frac{n}{\underline{\lambda}^2}
\bigg(
\mathbb{E}_{V}\left[
T_{\mathsf{h}}(x^{(j)})^2
\right]
+
\mathbb{E}_{V}\left[
\Delta_{\mathsf{h}}(x^{(j)})^2
\right]  \nonumber \\
&\quad\quad\quad\quad\quad\quad\quad\quad
+
T_n(x^{(j)})^2
+
\Delta_n(x^{(j)})^2 \bigg)
\lesssim
\log^4(dn)\delta_n^2 \label{eq: Gamma pieces}
\end{align}
with probability at least \(1-C/nd\). Consequently, we find that
\begin{align}
    \sup_{j \in [d]} \Gamma_{n,j}
    \lesssim
    \underline{\lambda}^2 \log^4(dn)\delta_n^2 \label{eq: Gamma final bound}
\end{align}
with probability at least \(1 - C/nd\).

Next, we derive a large deviation bound for the difference $\vert \bar{\lambda}_{n,j}^2 - \lambda^2_j \vert$. Observe that we can evaluate
\begin{flalign}
\mathbb{E}\left[V_i V_{i^\prime}\right] 
&= \frac{1}{2} \mathbb{E}\left[V_i \mid V_{i^\prime} = 1\right] - \frac{1}{2} \mathbb{E}\left[V_{i^\prime} \mid V_i = -1\right]\\
&= \frac{1}{2}\left(\frac{n/2-1}{n-1} - \frac{n/2}{n-1}\right) - \frac{1}{2} \left(\frac{n/2}{n-1} - \frac{n/2-1}{n-1}\right) =  -\frac{1}{n-1} \nonumber 
\end{flalign}
and thereby
\begin{flalign}
\bar{\lambda}^2_{n,j} 
&= \frac{1}{n}\sum_{i=1}^n \bar{u}^2(x^{(j)}, D_i) + \frac{1}{n}\sum_{i=1}^n \sum_{i^\prime \neq i} \mathbb{E}\left[V_i V_{i^\prime}\right] \bar{u}(x^{(j)}, D_i)  \bar{u}(x^{(j)}, D_{i^\prime}) \nonumber \\
& = \frac{1}{n}\sum_{i=1}^n \bar{u}^2(x^{(j)}, D_i) - \frac{1}{n}\frac{1}{n-1} \sum_{i=1}^n \sum_{i^\prime \neq i} \bar{u}(x^{(j)}, D_i)  \bar{u}(x^{(j)}, D_{i^\prime})~.\label{eq: bar lambda de randomize}
\end{flalign}
We now bound the two terms appearing in \eqref{eq: bar lambda de randomize}. First, we apply the following
Rosenthal type inequality.
\begin{lemma}[{\citealp[Theorem 9, ][]{boucheron2005moment}}]
\label{lem: Rosenthal centered}Let $X_{1},\ldots,X_{b}$ be independent
centered random variables. For any integer $q\geq2$, we have that
\[
\|\frac{1}{b}\sum_{i=1}^b X_{i}\|_{q}\lesssim\sqrt{\frac{\Var(X_{i})q}{b}}+\frac{q}{b}\|\max_{i\in[b]}| X_{i}| \|_{q}.
\]
\end{lemma}

\noindent In particular, the variables
\begin{align}
\bar{u}^2(x^{(j)},D_i)-\lambda_j^2
\end{align}
are independent and mean-zero. Moreover, \cref{assu: linearity} implies that
\begin{align}
\mathbb{E}\left[
\left(
\bar{u}^2(x^{(j)},D_i)-\lambda_j^2
\right)^2
\right]
\leq
\mathbb{E}\left[\bar{u}^4(x^{(j)},D_i)\right]
\leq
\lambda_j^2\varphi^2~.
\end{align}
In turn, the bound $\|\bar{u}(x^{(j)},D_i)\|_{\psi_1}\leq\varphi$ gives
\begin{align}
\left\|
\max_{i\in[n]}
\left|
\bar{u}^2(x^{(j)},D_i)-\lambda_j^2
\right|
\right\|_q
\lesssim
q^2\varphi^2
\end{align}
for every \(q\geq 2\vee2\log(nd)\). Therefore, \cref{lem: Rosenthal centered} implies that
\begin{align}
\mathbb{E}\left[
\left|
\frac{1}{n}\sum_{i=1}^n
\left(
\bar{u}^2(x^{(j)},D_i)-\lambda_j^2
\right)
\right|^q
\right]^{1/q}
&\lesssim
\lambda_j\varphi\sqrt{\frac{q}{n}}
+
\frac{\varphi^2q^3}{n}~. \label{eq: first term variance moment}
\end{align}
By taking $q=2\vee2\log(nd)$, applying Markov and a union bound gives
\begin{align}
\sup_{j\in[d]}
\frac{1}{\lambda_j^2}
\left|
\frac{1}{n}\sum_{i=1}^n \bar{u}^2(x^{(j)},D_i)-\lambda_j^2
\right|
&\lesssim
\frac{\varphi}{\underline{\lambda}}
\sqrt{\frac{\log(dn)}{n}}
+
\frac{\varphi^2\log^3(dn)}{\underline{\lambda}^2n} \label{eq: first term variance bound}
\end{align}
with probability greater than \(1-C/nd\).

To handle the second term, we apply the following sub-exponential formulation of the \cite{hanson1971bound} concentration inequality for quadratic forms, due to \cite{gotze2021concentration}.
\begin{lemma}[Proposition 1.1, \cite{gotze2021concentration}]\label{lem: Hanson Wright}
Let $X_1,\ldots,X_n$ be independent, centered, random variables satisfying $\mathbb{E}\left[X_i^2\right] = \sigma_{i}^2$ and $\|X_i\|_{\psi_1}\leq\varphi$. If $A=(a_{i,i^\prime})$ is any symmetric $n\times n$ matrix, then the inequality 
\begin{align}
& P\left\{ \big\vert \sum_{i=1}^n \sum_{i^\prime=1}^n a_{i,i^\prime} X_i X_{i^\prime} - \sum_{i=1}^n \sigma_i^2 a_{i,i} \big\vert \geq t \right\} \nonumber\\
& \leq
2 \exp\left( -\frac{1}{C} \min \left(\frac{t^2}{\varphi^4 \|A\|^2_{F}}, \frac{t^{1/2}}{\varphi \|A\|_{\mathsf{op}}^{1/2} }\right) \right)
\end{align}
holds for any $t\geq0$, where $\| \cdot \|_F$ and $\| \cdot \|_{\mathsf{op}}$ denote the Frobenius and $\ell_2$ operator norms, respectively.
\end{lemma}

\noindent In particular, let $A$ denote the $n \times n$ matrix with zeroes on the diagonal and $(n(n-1))^{-1}$ in every other entry. In this case
\begin{align}
\|A\|^2_{F} & = \frac{1}{n}\frac{1}{n-1}\quad\text{and}\quad\|A\|^{1/2}_{\mathsf{op}} = \frac{1}{\sqrt{n}}
\end{align}
and so \cref{lem: Hanson Wright} implies that
\begin{equation}
\sup_{j\in[d]} \frac{1}{\lambda_j^2}
\Big\vert \frac{1}{n}\frac{1}{n-1} 
\sum_{i=1}^n \sum_{i^\prime\neq i} \bar{u}(x^{(j)}, D_i)  \bar{u}(x^{(j)}, D_{i^\prime})
\Big\vert
\lesssim \frac{\varphi^2}{\underline{\lambda}^2n} \log^2(dn)
\label{eq: second term variance bound}
\end{equation}
with probability greater than $1-C/nd$. 

Hence, by combining \eqref{eq: bar lambda de randomize}, \eqref{eq: first term variance bound}, and \eqref{eq: second term variance bound}, we find that
\begin{align}
\sup_{j\in[d]}
\left|
\frac{\bar{\lambda}_{n,j}^2}{\lambda_j^2}-1
\right|
&\lesssim
\frac{\varphi}{\underline{\lambda}}
\sqrt{\frac{\log(dn)}{n}}
+
\frac{\varphi^2\log^3(dn)}{\underline{\lambda}^2n} \label{eq: bar lambda relative bound}
\end{align}
with probability greater than \(1-C/nd\). Thus, by plugging \eqref{eq: Gamma final bound} and \eqref{eq: bar lambda relative bound} into the bound \eqref{eq: lambda cauchy-schwarz}, we obtain
\begin{align}
\sup_{j\in[d]}
\left|
\frac{\hat{\lambda}_{n,j}^2}{\lambda_j^2}-1
\right|
&\leq
\sup_{j\in[d]}
\left|
\frac{\bar{\lambda}_{n,j}^2}{\lambda_j^2}-1
\right|
+
2\sup_{j\in[d]}\frac{\bar{\lambda}_{n,j}}{\lambda_j}
\sup_{j\in[d]}\frac{\Gamma_{n,j}^{1/2}}{\lambda_j}
+
\sup_{j\in[d]}\frac{\Gamma_{n,j}}{\lambda_j^2} \nonumber\\
&\lesssim
\frac{\varphi}{\underline{\lambda}}
\sqrt{\frac{\log(dn)}{n}}
+
\frac{\varphi^2\log^3(dn)}{\underline{\lambda}^2n}
+
\log^2(dn)\delta_n
+
\log^4(dn)\delta_n^2 \label{eq: lambda square final bound}
\end{align}
with probability greater than $1-C/nd$, as required.\hfill\qed

\section{Proof of \cref{eq: asymptotic validity subsampled kernel}\label{app: kernel verify pf}}

\subsection{Proof of \cref{eq: asymptotic validity subsampled kernel}}

The result follows from an application of \cref{thm: generic decomposition}. We begin by verifying the requisite assumptions. \cref{thm: generic decomposition} requires that the moment function $M(\cdot;\theta_0,g_0)$ is Neyman Orthogonal and satisfies \cref{assu: app a moment function}. Neyman Orthogonality is also imposed in the conditions of \cref{eq: asymptotic validity subsampled kernel}, and so is satisfied. \cref{assu: app a moment function} requires that the moment function $M(x;\theta_0,g_0)$ is twice continuously differentiable and has a Jacobian function 
\begin{equation}
M^{(1)}(x; g) 
= \frac{\partial}{\partial \theta^\prime} M(x;\theta,g) \mid_{\theta^\prime = \theta} 
\end{equation}
that satisfies the bounds \eqref{eq: Lipschitz Jacobian} and \eqref{eq: well-posedness}. Twice continuous differentiability 
follows immediately from  \cref{assu: moment linearity}, which states that the moment function $m(\cdot;\theta,g)$ satisfies the linear representation
\begin{equation}\label{eq: linear in online app}
m(d;\theta, g) = m^{(1)}(d;g)\cdot \theta + m^{(2)}(d;g)~,
\end{equation}
with bounded functions $m^{(1)}(\cdot;g)$ and $m^{(2)}(\cdot;g)$. In particular, the representation \eqref{eq: linear in online app} implies that the Jacobian can be written
\begin{equation}
M^{(1)}(x; \theta, g) 
= \mathbb{E}[ m^{(1)}(D_i;g) \mid X_i = x]
\end{equation}
and that the Hessian can be written
\begin{equation}
H(x; \theta, g) 
= \frac{\partial^2}{\partial^2 \theta^\prime} M(x;\theta,g_0) \mid_{\theta^\prime = \theta} = 0~.
\end{equation}
Here, our ability to exchange integration and differentiation follows from boundedness of $m(\cdot;\theta, g)$. Consequently, the bounds \eqref{eq: Lipschitz Jacobian} and \eqref{eq: well-posedness} follow from the bounds \eqref{eq: Lipschitz Jacobian main} and \eqref{eq: well-posedness main} stated as Part (iii) of \cref{assu: moment smoothness}. 

\cref{thm: generic decomposition} also requires that the centered empirical moment $\bar{M}_n(x; \theta, g)$ satisfy  \cref{assu: empirical smoothness}. \cref{assu: empirical smoothness} imposes the restriction that $\bar{M}_n(x; \theta, g)$ is twice continuously differentiable and that its Hessian
\begin{align}
\bar{H}_n(x;\theta, g)&= \frac{\partial^2}{\partial^2 \theta^\prime}  \bar{M}_n(x;\theta^\prime, g_0) \vert_{\theta^\prime = \theta}  
\end{align}
is uniformly bounded almost surely as $x$, $\theta$, and $g$ vary over their respective domains. Recall that, in our case, we have that
\begin{equation}
\bar{M}_n(x; \theta,g) = \sum_{i=1}^n \left( K(x,X_i)m(D_i; \theta, g) - \mathbb{E}\left[K(x,X_i) m(D_i; \theta, g)\right]\right)~.
\end{equation}
Twice continuous differentiability again follows from \cref{assu: moment linearity}, i.e., the linear representation \eqref{eq: linear in online app}. In turn, we have that $\bar{H}_n(x;\theta, g) = 0$ almost surely, verifying the uniform boundedness condition. Moreover, for future reference, the empirical Jacobian is given by 
\begin{equation}
\bar{M}_n^{(1)}(x; g) = \sum_{i=1}^n \left( K(x,X_i)m^{(1)}(D_i; g) - \mathbb{E}\left[K(x,X_i) m^{(1)}(D_i;  g)\right]\right)
\end{equation}
again by \cref{assu: moment linearity}.

We now quantify the generic sequence $\delta_n^{u}$ defined in \cref{assu: linearity}. Recall the normalized statistic $W_n(x)$ defined in \eqref{eq: U def}. By the definition \eqref{eq: subsampled kernel def}, this quantity can be written 
\begin{align}
W_{n}(x) & = \frac{1}{r} \sum_{q=1}^r u(x; D_{\mathsf{s}_q}, \xi_{\mathsf{s}_q}, \theta_0, g_0)~,\quad\text{where}\nonumber\\
u(x; D_\mathsf{s}, \xi_{\mathsf{s}}, \theta, g) 
&= - M^{(1)}(x;g_0)^{-1}\sum_{i\in\mathsf{s}} 
\big( \kappa(x, X_i, D_{\mathsf{s}}, \xi_{\mathsf{s}}) m(D_i; \theta, g) \nonumber \\
& \quad\quad\quad\quad\quad\quad\quad\quad\quad\quad
- \mathbb{E}\big[ \kappa(x, X_i, D_{\mathsf{s}}, \xi_{\mathsf{s}}) m(D_i; \theta, g)\big]\big)~.\nonumber
\end{align}
Define the de-randomized kernel function and H\'{a}jek projection 
\begin{align}
\tilde{u}(x; D)          &= \mathbb{E}\left[u(x; D_{\mathsf{s}}, \xi_{\mathsf{s}}, \theta_0(x), g_0) \mid D_\mathsf{s} = D \right]\quad\text{and}\label{eq: derandom kernel}\\
\tilde{u}^{(1)}(x; D)  &= \mathbb{E}\left[u(x; D_{\mathsf{s}}, \xi_{\mathsf{s}}, \theta_0(x), g_0) \mid i\in\mathsf{s}, D_i = D \right]~,\label{eq: hajek}
\end{align}
respectively, in addition to the quantities
\begin{equation}
\sigma^2_{b,j} = \Var(\tilde{u}^{(1)}(x^{(j)}, D_i))\quad\text{and}\quad \underline{\sigma}^2_b = \min_{j \in [d]} \sigma^2_{b,j}~.
\end{equation}
The result below follows from an application of \cref{lem: U stat linearization}. 
\begin{lemma}[Asymptotic Linearity]\label{lem: u stat linearity} Suppose that the de-randomized kernel function \eqref{eq: derandom kernel} is bounded by $\phi \geq 1$ almost surely and is invariant to permutations of its second argument. Suppose that $b$ and $r$ are chosen to satisfy $n\leq \sqrt{r} b$. If the inequality
\begin{equation}\label{eq: asymp lin conditions}
\frac{(1+\|\theta_0(\bm{x}^{(d)})\|_\infty) \phi b\log(dn)}{n} \leq C
\end{equation}
holds, then
\begin{align}
&\mathbb{E}\left[
\bigg\vert
\sqrt{\frac{n}{b^2 \underline{\sigma}^2_b}}
\left(
W_{n}(x^{(j)})-\frac{b}{n}\sum_{i=1}^n\tilde{u}^{(1)}(x^{(j)};D_i)
\right)
\bigg\vert^q
\right]^{1/q} \nonumber \\
& \lesssim
q^2
\left(\frac{(1+\|\theta_0(\bm{x}^{(d)})\varphi^2\|_\infty)^2}{\underline{\sigma}^2_b n}\right)^{1/2}~,\label{eq: asymp lin statement}
\end{align}
for each $j\in[d]$ and each $2\vee2\log(nd)\leq q\leq cn/b$.
\end{lemma}

\noindent In particular, observe that \cref{assu: moment linearity} implies that the de-randomized kernel function \eqref{eq: derandom kernel} is bounded by $\phi$, up to a constant that depends only on $\mathbf{P}$. Permutation invariance again follows by assumption. We may assume that the first condition in \eqref{eq: asymp lin conditions} holds, as otherwise the desired bound is vacuously true. Consequently, \cref{assu: linearity} holds with the choice
\begin{equation}
\delta_{n,u} = C \left( \frac{ b \phi^2}{n} \right)^{1/2}
\end{equation}
by incrementality and the fact that $\|\theta_0(\bm{x}^{(d)})\|_\infty$ is bounded as $P$ varies over $\mathbf{P}$.

Next, we quantify the sequences introduced in \cref{assu: bias}. The lemma below follows from arguments similar to those used in \cite{wager2018estimation} and \cite{oprescu2019orthogonal}. 

\begin{lemma}[Bias, Consistency, and Stochastic Equicontinuity] \label{lem: kernel bcs}$ $\\
Suppose that the conditions of \cref{eq: asymptotic validity subsampled kernel} are satisfied.

\noindent \textbf{(i)} The bound
\begin{align}
 & \sqrt{\frac{n}{b^2\underline{\sigma}_b^2}}
\| \mathsf{Bias}_n(\bm{x}^{(d)};\theta(\bm{x}^{(d)}), \hat{g}_n) \|_\infty
\lesssim
(1+\|\theta(\bm{x}^{(d)})\|_\infty)\delta_{n,B}~,
\quad\text{where} \label{eq: bias eq}\\
 & \delta_{n,B}
=
\left(\frac{n}{b^2\underline{\sigma}_b^2}\right)^{1/2}\varepsilon_b
\nonumber
\end{align}
holds.

\noindent \textbf{(ii)} The bound
\begin{align}
\left(\frac{n}{b^2\underline{\sigma}_b^2}\right)^{1/4}
\mathbb{E}\left[
\left|
\bar{M}^{(1)}_n(x^{(j)}; g_0)
\right|^q
\right]^{1/q}
&\lesssim
q\delta_{n,m}~,
\quad\text{where}\quad
\delta_{n,m}
=
\left(\frac{\phi^4}{\underline{\sigma}_b^2 n}\right)^{1/4}
\label{eq: concentration eq}
\end{align}
holds for each $q\geq 2\vee2\log(dn)$ satisfying
\begin{equation}
\frac{\phi^2bq^2}{n}\leq c
\label{eq: part ii normalization}
\end{equation}
and each $j\in[d]$.

\noindent \textbf{(iii)} The bounds
\begin{align}
&\sqrt{\frac{n}{b^2\underline{\sigma}_b^2}}
\mathbb{E}\left[
\big\vert
\bar{M}_n(x^{(j)};\theta_0(x^{(j)}), \hat{g}_n)
-
\bar{M}_n(x^{(j)};\theta_0(x^{(j)}), g_0)
\big\vert^q
\right]^{1/q}
\lesssim
q\delta_{n,S}~,\quad\text{where}\nonumber\\
\delta_{n,S}
&=
\left(\frac{n}{b^2\underline{\sigma}_b^2}\right)^{1/2}
\bigg(
\left(\frac{b}{n}\right)^{1/4}\delta_{n,g}
+
\varepsilon_b
+
(1+\|\theta_0(\bm{x}^{(d)})\|_\infty)\phi\sqrt{\frac{b}{n}}
\bigg)~,
\label{eq: stochastic equicontinuity moment}
\end{align}
and
\begin{align}
&\sqrt{\frac{n}{b^2\underline{\sigma}_b^2}}
\mathbb{E}\left[
\big\vert
\bar{M}^{(1)}_n(x^{(j)};\hat{g}_n)
-
\bar{M}^{(1)}_n(x^{(j)};g_0)
\big\vert^q
\right]^{1/q}
\lesssim
q\delta_{n,J}~,\quad\text{where}\nonumber\\
\delta_{n,J}
&=
\left(\frac{n}{b^2\underline{\sigma}_b^2}\right)^{1/2}
\bigg(
\left(\frac{b}{n}\right)^{1/4}\delta_{n,g}
+
\varepsilon_b
+
(1+\|\theta_0(\bm{x}^{(d)})\|_\infty)\phi\sqrt{\frac{b}{n}}
\bigg)
\label{eq: jacobian stochastic equicontinuity moment}
\end{align}
hold for each $q\geq 2\vee2\log(dn)$ satisfying
\begin{equation}
\frac{\phi^2(1+\|\theta_0(\bm{x}^{(d)})\|_\infty)^2bq^2}{n}
+
\varepsilon_b
\leq c
\label{eq: part iii normalization}
\end{equation}
and each $j\in[d]$.

\noindent \textbf{(iv)} The bound
\begin{align}
&\left(\frac{n}{b^2\underline{\sigma}_b^2}\right)^{1/4}
\mathbb{E}\left[
\big\vert
\hat{\theta}_n(x^{(j)})-\theta_0(x^{(j)})
\big\vert^q
\right]^{1/q}
\lesssim
q\delta_{n,\theta}~,\quad\text{where}\nonumber\\
\delta_{n,\theta}
&=
\left(\frac{n}{b^2\underline{\sigma}_b^2}\right)^{1/4}
\bigg(
\left(\frac{b}{n}\right)^{1/4}\delta_{n,g}
+
(1+\|\theta_0(\bm{x}^{(d)})\|_\infty)\varepsilon_b
+
(1+\|\theta_0(\bm{x}^{(d)})\|_\infty)\phi\sqrt{\frac{b}{n}}
\bigg)
\label{eq: theta moment rate}
\end{align}
holds for each $q\geq 2\vee2\log(dn)$ satisfying 
\begin{equation}
\varepsilon_b\leq 1~,\quad
q\left(\frac{b}{n}\right)^{1/4}\delta_{n,g}\leq 1~,
\quad\text{and}\quad
\frac{\phi^2(1+\|\theta_0(\bm{x}^{(d)})\|_\infty)^2bq^2}{n}+\varepsilon_b\leq 1
\label{eq: part iv normalizations}
\end{equation}
and each $j\in[d]$.

\end{lemma} 

\noindent \noindent Observe that the choice
\[
\bar{u}(x, D_i) = b\cdot \tilde{u}^{(1)}(x, D_i)~,
\]
suggested by \cref{lem: u stat linearity} implies that
\begin{equation}\label{eq: b bounds lambda sq}
b \phi^2 \gtrsim \underline{\lambda}^2 = b^2 \underline{\sigma}^2_b \gtrsim b~,
\end{equation}
by \cref{assu: moment linearity} and incrementality. We set $\bar{q}_{n,d}=C\log(dn)$. We may assume that there exists some \(c\leq 1\) such that
\begin{equation}\label{eq: final normalizations}
\left(\delta_{n,g}^2+\sqrt{\frac{n}{b}}\varepsilon_b\right)\log^3(dn)\leq c
\quad\text{and}\quad
\frac{b\phi^4\log^8(dn)}{n}\leq c~,
\end{equation}
as otherwise the bound is vacuously true. Consequently, all of the conditions of \cref{lem: kernel bcs} are satisfied for every \(q\leq \bar{q}_{n,d}\). Thus, \cref{lem: kernel bcs} and \cref{lem: u stat linearity} imply that we can set
\begin{align}
\delta_{n,u}
&=
C\left(\frac{\phi^2}{\underline{\sigma}_b^2n}\right)^{1/2}
\lesssim
\left(\frac{b\phi^2}{n}\right)^{1/2}~,\\
\delta_{n,B}
&  = C \sqrt{\frac{n}{b^2 \underline{\sigma}^2_b}} \varepsilon_{b}
\lesssim \sqrt{\frac{n}{b}} \varepsilon_{b}~,\\
\delta_{n,m}
& =
C\left(\frac{\phi^4}{\underline{\sigma}_b^2n}\right)^{1/4}
\lesssim
\left(\frac{b\phi^4}{n}\right)^{1/4}~,\\
\delta_{n,\theta}
& =
C \left(\frac{n}{b^2 \underline{\sigma}^2_b}\right)^{1/4}
\left(
\left(\frac{b}{n}\right)^{1/4}\delta_{n,g}
+
\varepsilon_b
+
\phi\sqrt{\frac{b}{n}}
\right) \nonumber\\
&\lesssim
\delta_{n,g}
+
\left(\frac{n}{b}\right)^{1/4}\varepsilon_b
+
\left(\frac{b}{n}\right)^{1/4}\phi~,\quad\text{and}\\
\delta_{n,S}=\delta_{n,J}
& =
C\log^{1/2}(dn)\left(\frac{b}{n}\right)^{1/4}\delta_{n,g}
+
C\sqrt{\frac{n}{b}}\varepsilon_b
+
C\phi\sqrt{\frac{b}{n}}\log(dn)~,
\end{align}
respectively, where we have repeatedly used the fact that \(\|\theta_0(\bm{x}^{(d)})\|_\infty\) is uniformly bounded.

With these results in place, we apply \cref{thm: generic decomposition}. We begin by giving a suitable upper bound for the sequence
\begin{equation}
\delta_n
=
\delta^2_{n,g}
+
\delta^2_{n,\theta}
+
\delta^2_{n,m}
+
\delta_{n,B}
+
\delta_{n,S}
+
\delta_{n,u}
+
\underline{\lambda}^{1/2}n^{-1/4}
\delta_{n,\theta}
\left(
\delta_{n,B}
+
\delta_{n,J}
\right)
\end{equation}
introduced in the statement of \cref{thm: generic decomposition}. Observe that
\begin{align}
\frac{\underline{\lambda}^{1/2}}{n^{1/4}}(\delta_{n,B}+\delta_{n,J})\delta_{n,\theta}
&\lesssim
\frac{b^{1/4}}{n^{1/4}}(\delta_{n,B}+\delta_{n,J})\delta_{n,\theta}
\lesssim
\delta_{n,B}+\delta_{n,J}~,
\end{align}
as otherwise the desired bound would be vacuous, where the first inequality follows from \eqref{eq: b bounds lambda sq}. Consequently, we find that, if \(\delta_{n,J}=\delta_{n,S}\), then
\begin{equation}
\delta_n
\lesssim
\delta^2_{n,g}
+
\delta^2_{n,\theta}
+
\delta^2_{n,m}
+
\delta_{n,B}
+
\delta_{n,S}
+
\delta_{n,u}~.
\end{equation}
Plugging in the choices specified above gives
\begin{align}
\delta_n
&\lesssim
\delta_{n,g}^2
+
\sqrt{\frac{n}{b}}\varepsilon_b
+
\sqrt{\frac{b}{n}}\phi^2
+
\left(\frac{b}{n}\right)^{1/4}\delta_{n,g}\log^{1/2}(dn)
+
\phi\sqrt{\frac{b}{n}}\log(dn) \label{eq: first delta express}\\
&\lesssim
\delta_{n,g}^2
+
\sqrt{\frac{n}{b}}\varepsilon_b
+
\phi^2\sqrt{\frac{b}{n}}\log(dn)~. \label{eq: delta final}
\end{align}
where, to get the second inequality, we apply the elementary bound
\[
\left(\frac{b}{n}\right)^{1/4}\delta_{n,g}\log^{1/2}(dn)
\lesssim
\delta_{n,g}^2
+
\sqrt{\frac{b}{n}}\log(dn)
\]
and use \(\phi\geq1\). Hence, the inequality \eqref{eq: delta final}, the Orlicz-norm bound
\begin{equation*}
\| \bar{u}(x^{(j)},D) \|_{\psi_1}
\leq b\cdot (1+|\theta_0(x^{(j)})|)\phi
\lesssim b\phi~,
\end{equation*}
and \cref{thm: generic decomposition} imply that
\begin{align}
& \sup_{P\in\mathbf{P}}
\big\vert
P\left\{ \theta_0(\bm{x}^{(d)})\in\hat{\mathcal{C}}(\bm{x}^{(d)}) \right\}
-
(1-\alpha)
\big\vert \nonumber\\
&\lesssim
\left(\frac{b\phi^2\log^5(dn)}{n}\right)^{1/4}
+
\left(
\delta_{n,g}^2
+
\sqrt{\frac{n}{b}}\varepsilon_b
+
\phi^2\sqrt{\frac{b}{n}}\log(dn)
\right)\log^3(dn)
+
\frac{1}{nd} \nonumber\\
&\lesssim
\left(\frac{b\phi^4\log^8(dn)}{n}\right)^{1/4}
+
\left(
\delta_{n,g}^2
+
\sqrt{\frac{n}{b}}\varepsilon_b
\right)\log^3(dn)~,
\label{eq: asymptotic validity statement apply}
\end{align}
as required. \hfill\qed

\subsection{Proof of \cref{lem: u stat linearity}} 

To ease notation, we define the parameter
\[
\phi(\theta_0) = (1+\|\theta_0(x^{(j)})\|)\phi
\]
and drop dependence on $\theta(x)$ and $g$ when writing $u(x; D_{\mathsf{s}},\xi_{\mathsf{s}}, \theta(x), g)$, as these values will be taken to be $\theta_0(x)$ and $g_0$ throughout. We are interested in studying the discrepancy
\begin{equation}
 W_n(x^{(j)}) - \frac{b}{n}\sum_{i=1}^n \tilde{u}^{(1)}(x^{(j)}; D_i)~.
\end{equation}
Define the quantities
\begin{flalign}
\hat{U}_{n}(x^{(j)})
& =\frac{1}{N_{b}}\sum_{\mathsf{s}\in\mathcal{S}_{n,b}}
u(x^{(j)}; D_{\mathsf{s}},\xi_{\mathsf{s}})
\quad\text{and}\quad
\bar{U}_{n}(x^{(j)}) = \frac{1}{N_{b}}\sum_{\mathsf{s}\in\mathcal{S}_{n,b}}
\tilde{u}(x^{(j)}; D_\mathsf{s})\label{eq: complete derandom}
\end{flalign}
for some collection of independent random variables $\bm{\xi} = (\xi_{\mathsf{s}})_{\mathsf{s}\in\mathcal{S}_{n,b}}$
having the same distribution as $\xi$. The statistics $\hat{U}_{n}(x^{(j)})$
and $\bar{U}_{n}(x^{(j)})$ are the complete, randomized and de-randomized,
$U$-statistics associated with the randomized $d$-dimensional kernel $u(x^{(j)};\cdot,\cdot)$,
respectively. Consider the decomposition
\begin{align}
& \sqrt{\frac{n}{b^2 \underline{\sigma}^2_b}}  (W_n(x^{(j)}) - \frac{b}{n}\sum_{i=1}^n \tilde{u}^{(1)}(x^{(j)}; D_i)) \nonumber\\
& \quad\quad= \sqrt{\frac{n}{b^2 \underline{\sigma}^2_b}} (W_{n}(x^{(j)})-\hat{U}_{n}(x^{(j)}))
+\sqrt{\frac{n}{b^2 \underline{\sigma}^2_b}} (\hat{U}_{n}(x^{(j)})-\bar{U}_{n}(x^{(j)})) \nonumber \\
& \quad\quad\quad+\sqrt{\frac{n}{b^2 \underline{\sigma}^2_b}} (\bar{U}_{n}(x^{(j)})-\frac{b}{n}\sum_{i=1}^{n}\tilde{u}^{(1)}(x^{(j)}; D_{i}))~.\label{eq: W tilde decomposition}
\end{align}
We bound each of the three terms in \eqref{eq: W tilde decomposition} in succession.

A higher-order moment bound for the third term in \eqref{eq: W tilde decomposition} is obtained by applying the same argument used to verify \cref{lem: u stat linearity}. In particular, Part (ii) of \cref{lem: Diff H decomposition} and \cref{lem: u stat linearity} imply that
\begin{align}
        \mathbb{E}\left[ 
        \bigg\vert \bar{U}_n(x^{(j)}) 
        - \frac{b}{n} \sum_{i=1}^n \tilde{u}^{(1)}(x^{(j)}; D_i) \bigg\vert^{q} \right]^{1/q} 
        & \lesssim
        \phi q^2 \frac{b}{n}\label{eq: moment bound apply in c}
\end{align}
for each $2 \vee 2\log(nd) \leq q\leq cn/b$. Analogous bounds for the first two terms in \eqref{eq: W tilde decomposition} are obtained through the application of the following Lemma. 

\begin{lemma}\label{eq: apply bernstein to randomness}
Let $\mathbf{D}_n = (D_i)_{i=1}^n$ and  $\bm{\xi} = (\xi_{\mathsf{s}})_{\mathsf{s}\in\mathcal{S}_{n,b}}$ denote two independent collections of independent and identically distributed random variables. Consider the real-valued function $u(D_{\mathsf{s}}, \xi_{\mathsf{s}})$. Assume that the absolute value of each component of $u(D_{\mathsf{s}}, \xi_{\mathsf{s}})$ is bounded by the constant $\phi\geq1$ almost surely. 

\noindent \textbf{(i)} If the bound
\begin{equation}
q^{1/2}\frac{\phi b }{n} \leq 1
\end{equation}
holds for some $q\geq 2\vee 2\log(dn)$, then the bound
\begin{align}
&\mathbb{E}\left[
\bigg\vert\
{n \choose b}^{-1}\sum_{\mathsf{s}\in\mathcal{S}_{n,b}}
 \left(u(D_{\mathsf{s}},\xi_{\mathsf{s}})
 -
 \mathbb{E}\left[u(D_{\mathsf{s}},\xi_{\mathsf{s}}) \mid \mathbf{D}_{\mathsf{s}} \right]\right)
\bigg\vert^q
\right]^{1/q} \lesssim q^{1/2} \frac{\phi b}{n}\label{eq: just data lemma state}
\end{align}
holds for all $b>2$. 

\noindent \textbf{(ii)} If, in addition, the sets $(\mathsf{s}_q)_{q=1}^r$ are drawn independently and identically from  $\mathcal{S}_{n,b}$ and $n\leq \sqrt{r} b$, then
\begin{align}
&\mathbb{E}\left[
\bigg\vert
 \frac{1}{r}\sum_{q=1}^{r} 
 \left(u(D_{\mathsf{s}_{q}},\xi_{\mathsf{s}_{q}})
 -
 \mathbb{E}\left[u(D_{\mathsf{s}_{q}},\xi_{\mathsf{s}_{q}}) \mid \mathbf{D}_{n}, \bm{\xi} \right]\right)
\bigg\vert^q
\right]^{1/q} \lesssim q^{1/2} \frac{ b \phi }{n}\label{eq: random lemma state}
\end{align}
holds.
\end{lemma}

\noindent In particular, observe that 
\[
\hat{U}_{n}(x^{(j)})
=\frac{1}{N_{b}}\sum_{\mathsf{s}\in\mathcal{S}_{n,b}}u(x^{(j)};D_{\mathsf{s}},\xi_{\mathsf{s}})
=\mathbb{E}[u(x^{(j)};D_{\mathsf{s}_{g}},\xi_{\mathsf{s}_{g}})\mid\mathbf{D}_{n},\bm{\xi}]
\]
and that therefore we can write 
\[
W_{n}(x^{(j)})-\hat{U}_{n}(x^{(j)})
=\frac{1}{r}\sum_{q=1}^{r}Z_{q},
\quad\text{with}\quad
Z_{q}=u(x^{(j)};D_{\mathsf{s}_{q}},\xi_{\mathsf{s}_{q}})
-\mathbb{E}[u(x^{(j)};D_{\mathsf{s}_{q}},\xi_{\mathsf{s}_{q}})\mid\mathbf{D}_{n},\bm{\xi}]~.
\]
Consequently, by the normalization \(q^{1/2}\phi(\theta_0)b/n\leq1\), Part (ii) of \cref{eq: apply bernstein to randomness} implies that
\begin{align} 
&\mathbb{E}\left[
\bigg\vert
\sqrt{\frac{n}{b^2 \underline{\sigma}^2_b}}
\left(W_{n}(x^{(j)})-\hat{U}_{n}(x^{(j)})\right)
\bigg\vert^q
\right]^{1/q} 
\lesssim
\sqrt{\frac{n}{b^2 \underline{\sigma}^2_b}} 
q^{1/2}\frac{b\phi(\theta_0)}{n}
=
\left(\frac{\phi(\theta_0)^2q}{\underline{\sigma}^2_b n}\right)^{1/2}~. \label{eq: data and rand term linear}
\end{align}
In turn, we can similarly write 
\[
\hat{U}_{n}(x^{(j)})-\bar{U}_{n}(x^{(j)})
=\frac{1}{N_{b}}\sum_{\mathsf{s}\in\mathcal{S}_{n,b}}Z_{\mathsf{s}}
\quad\text{with}\quad 
Z_{\mathsf{s}}
=u(x^{(j)};D_{\mathsf{s}},\xi_{\mathsf{s}})
-\mathbb{E}\left[u(x^{(j)};D_{\mathsf{s}},\xi_{\mathsf{s}})\mid D_{\mathsf{s}}
\right]~.
\]
Thus, again by the normalization \(q^{1/2}\phi(\theta_0)b/n\leq1\), Part (i) of \cref{eq: apply bernstein to randomness} implies that
\begin{align}
&\mathbb{E}\left[
\bigg\vert
\sqrt{\frac{n}{b^2 \underline{\sigma}^2_b}}
\left(\hat{U}_{n}(x^{(j)})-\bar{U}_{n}(x^{(j)})\right)
\bigg\vert^q
\right]^{1/q} 
\lesssim
\left(\frac{\phi(\theta_0)^2q}{\underline{\sigma}^2_b n}\right)^{1/2}~. \label{eq: just data term linear}
\end{align}
Putting the pieces together, the bounds \eqref{eq: moment bound apply in c}, \eqref{eq: data and rand term linear}, and \eqref{eq: just data term linear} imply that
\begin{align}
&\mathbb{E}\left[
\bigg\vert
\sqrt{\frac{n}{b^2 \underline{\sigma}^2_b}}
\left(
W_{n}(x^{(j)})-\frac{b}{n}\sum_{i=1}^n\tilde{u}^{(1)}(x^{(j)};D_i)
\right)
\bigg\vert^q
\right]^{1/q} 
\lesssim
q^2
\left(\frac{\phi(\theta_0)^2}{\underline{\sigma}^2_b n}\right)^{1/2}~,\label{eq: apply three bounds in asymp lin}
\end{align}
for each $j\in[d]$ and each $2\vee2\log(nd)\leq q\leq cn/b$, as required.\hfill\qed

\subsection{Proof of \cref{lem: kernel bcs}} 

Throughout, we use the shorthand $N_b = {n \choose b}$.

\subsubsection{Part (i)}
Observe that
\begin{align*}
\mathbb{E}\left[M_n(x; \theta, g) \right] 
&= \mathbb{E}\left[\sum_{i=1}^n K(x, X_i) m(D_i; \theta, g) \right]\\
&= \frac{1}{N_b} \sum_{\mathsf{s}\in\mathcal{S}_{n,b}} \sum_{i\in \mathsf{s}} \mathbb{E}\left[\kappa(x, X_i, \mathsf{s}, \xi) m(D_i; \theta, g) \right]\\
&= \frac{1}{N_b} \sum_{\mathsf{s}\in\mathcal{S}_{n,b}} \sum_{i\in \mathsf{s}} 
\mathbb{E}\left[ \mathbb{E}\left[\kappa(x, X_i, \mathsf{s}, \xi) \mid X_i, D_{\mathsf{s}_{-i}}\right] \mathbb{E}\left[ m(D_i; \theta, g)\mid X_i, D_{\mathsf{s}_{-i}}\right]   \right]\tag{Honesty}\\
&= \frac{1}{N_b} \sum_{\mathsf{s}\in\mathcal{S}_{n,b}} \sum_{i\in \mathsf{s}} 
\mathbb{E}\left[ \kappa(x, X_i, \mathsf{s}, \xi) \mathbb{E}\left[ m(D_i; \theta, g)\mid X_i\right] \right]\\
&= \frac{1}{N_b} \sum_{\mathsf{s}\in\mathcal{S}_{n,b}} \sum_{i\in \mathsf{s}} 
\mathbb{E}\left[ \kappa(x, X_i, \mathsf{s}, \xi) M(X_i; \theta, g) \right]~.
\end{align*}
Therefore, the normalization 
\begin{equation}\label{eq: proof normalization}
\sum_{i \in \mathsf{s}}  \kappa(x, X_i, \mathsf{s}, \xi) = 1
\end{equation}
implies that
\begin{equation*}
\mathsf{Bias}_n(x; \theta, g) 
= \frac{1}{N_b} \sum_{\mathsf{s}\in\mathcal{S}_{n,b}} \sum_{i\in \mathsf{s}} 
\mathbb{E}\left[ \kappa(x, X_i, \mathsf{s}, \xi) \left(M(X_i; \theta, g) - M(x; \theta, g)\right)\right]~.
\end{equation*}
By the boundedness part of \cref{assu: moment linearity} and Part (iii) of \cref{assu: moment smoothness}, we find that
\begin{equation}\label{eq: proof bias bound}
\mathsf{Bias}_n(x; \theta, g) \lesssim (1 + \vert\theta\vert) \mathbb{E} \left[ \kappa(x, X_i, \mathsf{s}, \xi) \|X_i - x\|_\infty \right] \leq (1 + \vert\theta\vert) \varepsilon_{b}~,
\end{equation}
where the final inequality follows from the definition of the shrinkage $\varepsilon_b$ and the normalization \eqref{eq: proof normalization}.\hfill\qed

\subsubsection{Part (ii)} Define the function
\begin{equation*}
J(x; D_\mathsf{s},\xi_\mathsf{s})
=
\sum_{i\in \mathsf{s}}
\left(
\kappa(x,X_i,\mathsf{s},\xi_\mathsf{s})m^{(1)}(D_i;g_0)
-
\mathbb{E}\left[\kappa(x,X_i,\mathsf{s},\xi_\mathsf{s})m^{(1)}(D_i;g_0)\right]
\right)
\end{equation*}
and observe that
\begin{equation}
\bar{M}_n^{(1)}(x;g_0)
=
\frac{1}{r}\sum_{q=1}^r J(x;D_{\mathsf{s}_q},\xi_{\mathsf{s}_q})~.
\end{equation}
Consider the decomposition
\begin{equation}
\bar{M}_n^{(1)}(x;g_0)=\tilde{A}(x)+\hat{A}(x)+\bar{A}(x)~,
\end{equation}
where
\begin{align}
\tilde{A}(x)
&=
\frac{1}{r}\sum_{q=1}^r
\left(
J(x;D_{\mathsf{s}_q},\xi_{\mathsf{s}_q})
-
\mathbb{E}\left[J(x;D_{\mathsf{s}_q},\xi_{\mathsf{s}_q})\mid\mathbf{D}_n,\bm{\xi}\right]
\right)~,\label{eq: tilde A}\\
\hat{A}(x)
&=
\frac{1}{N_b}\sum_{\mathsf{s}\in\mathcal{S}_{n,b}}
\left(
J(x;D_{\mathsf{s}},\xi_\mathsf{s})
-
\mathbb{E}\left[J(x;D_{\mathsf{s}},\xi_{\mathsf{s}})\mid D_\mathsf{s}\right]
\right)~,\quad\text{and}\label{eq: hat A}\\
\bar{A}(x)
&=
\frac{1}{N_b}\sum_{\mathsf{s}\in\mathcal{S}_{n,b}}
\mathbb{E}\left[J(x;D_{\mathsf{s}},\xi_{\mathsf{s}})\mid D_\mathsf{s}\right]~,\label{eq: bar A}
\end{align}
respectively. We again apply \cref{eq: apply bernstein to randomness} to bound \eqref{eq: tilde A} and \eqref{eq: hat A}. In particular, by \cref{assu: moment linearity}, the normalization \eqref{eq: part ii normalization}, and the restriction $n\leq b\sqrt{r}$, \cref{eq: apply bernstein to randomness} implies that, for each \(j\in[d]\),
\begin{align}
\mathbb{E}\left[
\left|
\tilde{A}(x^{(j)})
\right|^q
\right]^{1/q}
&\lesssim
q^{1/2}\frac{b\phi}{n}
\quad\text{and}\label{eq: Q tilde bound bias}\\
\mathbb{E}\left[
\left|
\hat{A}(x^{(j)})
\right|^q
\right]^{1/q}
&\lesssim
q^{1/2}\frac{b\phi}{n}\label{eq: Q hat bound bias}
\end{align}
respectively. In turn, to bound the term \eqref{eq: bar A}, write
\[
\bar{J}_{j}(D_{\mathsf{s}})
=
\mathbb{E}\left[J(x^{(j)};D_{\mathsf{s}},\xi_{\mathsf{s}})\mid D_{\mathsf{s}}\right]~.
\]
Observe that \(\bar{J}_{j}(\cdot)\) is symmetric, mean-zero, and bounded by a constant multiple of \(\phi\). Moreover,
\[
\bar{A}(x^{(j)})=U_{n,b}(\bar{J}_{j})~.
\]
Let \(\bar{J}_{j}^{(\ell)}\) denote the \(\ell\)-th Hoeffding projection of \(\bar{J}_{j}\). By Part (ii) of \cref{lem: Diff H decomposition}, we have
\begin{align}
\bar{A}(x^{(j)})
&=
\frac{b}{n}\sum_{i=1}^n \bar{J}_{j}^{(1)}(D_i)
+
\sum_{\ell=2}^{b}{b\choose \ell}U_{n,\ell}(\bar{J}_{j}^{(\ell)})~. \label{eq: A bar Hoeffding}
\end{align}
We bound the two terms in \eqref{eq: A bar Hoeffding} in succession. By Hoeffding orthogonality,
\[
b\,\mathbb{E}\left[\left(\bar{J}_{j}^{(1)}(D_i)\right)^2\right]
\leq
\mathbb{E}\left[\bar{J}_{j}(D_{[b]})^2\right]
\lesssim
\phi^2~.
\]
Moreover, \(|\bar{J}_{j}^{(1)}(D_i)|\lesssim\phi\) almost surely. Therefore, \cref{lem: Rosenthal centered} gives
\begin{align}
\mathbb{E}\left[
\left|
\frac{b}{n}\sum_{i=1}^n \bar{J}_{j}^{(1)}(D_i)
\right|^q
\right]^{1/q}
&\lesssim
b\left(
\sqrt{\frac{q}{n}\mathbb{E}\left[\left(\bar{J}_{j}^{(1)}(D_i)\right)^2\right]}
+
\frac{q}{n}
\left\|\max_{i\in[n]}\left|\bar{J}_{j}^{(1)}(D_i)\right|\right\|_q
\right) \nonumber\\
&\lesssim
\phi\sqrt{\frac{bq}{n}}
+
\phi\frac{bq}{n}
\lesssim
q\phi\sqrt{\frac{b}{n}}~. \label{eq: A bar first projection}
\end{align}
Next, \cref{lem: u stat linearity} and Minkowski's inequality imply that
\begin{align}
\mathbb{E}\left[
\left|
\sum_{\ell=2}^{b}{b\choose \ell}U_{n,\ell}(\bar{J}_{j}^{(\ell)})
\right|^q
\right]^{1/q}
&\leq
\sum_{\ell=2}^{b}
\mathbb{E}\left[
\left|
{b\choose \ell}U_{n,\ell}(\bar{J}_{j}^{(\ell)})
\right|^q
\right]^{1/q} \nonumber\\
&\lesssim
\phi q\sum_{\ell=2}^{b}
\left(
\frac{Cqb}{n}
\right)^{\ell/2}
\lesssim
\phi q^2\frac{b}{n}~. \label{eq: A bar higher order}
\end{align}
where the final inequality follows from the geometric series formula and the fact that \(bq/n\leq c\), which is implied by \eqref{eq: part ii normalization}. Since \(bq^2/n\leq c\) is also implied by \eqref{eq: part ii normalization}, the right-hand side of \eqref{eq: A bar higher order} is bounded by \(q\phi\sqrt{b/n}\). Combining \eqref{eq: A bar Hoeffding}, \eqref{eq: A bar first projection}, and \eqref{eq: A bar higher order}, we obtain
\begin{align}
\mathbb{E}\left[
\left|
\bar{A}(x^{(j)})
\right|^q
\right]^{1/q}
&\lesssim
q\phi\sqrt{\frac{b}{n}}~.\label{eq: Q bar bound bias}
\end{align}
Thus, the bounds \eqref{eq: Q tilde bound bias}, \eqref{eq: Q hat bound bias}, and \eqref{eq: Q bar bound bias} imply that
\begin{align}
\mathbb{E}\left[
\left|
\bar{M}_n^{(1)}(x^{(j)};g_0)
\right|^q
\right]^{1/q}
&\lesssim
q\phi\sqrt{\frac{b}{n}}~. \label{eq: M 1 concentrate}
\end{align}
Multiplying both sides of \eqref{eq: M 1 concentrate} by
\[
\left(\frac{n}{b^2\underline{\sigma}_b^2}\right)^{1/4}
\]
gives
\begin{align}
\left(\frac{n}{b^2\underline{\sigma}_b^2}\right)^{1/4}
\mathbb{E}\left[
\left|
\bar{M}_n^{(1)}(x^{(j)};g_0)
\right|^q
\right]^{1/q}
&\lesssim
q\phi\frac{1}{\underline{\sigma}_b^{1/2}n^{1/4}}
=
q\left(\frac{\phi^4}{\underline{\sigma}_b^2n}\right)^{1/4}~,
\end{align}
as required.\hfill\qed

\subsubsection{Part (iii)}

We give the details of the proof of the stated moment bound on the discrepancy
\begin{equation}
\bar{M}_n(x^{(j)};\theta_0(x^{(j)}), \hat{g}_n)
-
\bar{M}_n(x^{(j)};\theta_0(x^{(j)}), g_0)
\end{equation}
only. The argument giving the analogous bound associated with the term
\(\bar{M}^{(1)}_n(x^{(j)};\cdot)\) is analogous. Define the functions
\begin{align}
F(x; D_{\mathsf{s}}, \xi_{\mathsf{s}}, g)
&=
\sum_{i\in\mathsf{s}}
f(x, X_i; D_{\mathsf{s}}, \xi_{\mathsf{s}}, g)
\quad\text{and}\quad
\bar{F}(x; D_{\mathsf{s}}, g)
=
\mathbb{E}_{\xi_{\mathsf{s}}}\left[
F(x; D_{\mathsf{s}}, \xi_{\mathsf{s}}, g)
\right]~,
\end{align}
where
\begin{align}
f(x, X_i; D_{\mathsf{s}}, \xi_{\mathsf{s}}, g)
&=
\kappa(x, X_i, D_{\mathsf{s}}, \xi_{\mathsf{s}})
\left(
m(D_i;\theta_0,g)-m(D_i;\theta_0,g_0)
\right) \nonumber\\
&\quad
-
\mathbb{E}\left[
\kappa(x, X_i, D_{\mathsf{s}}, \xi_{\mathsf{s}})
\left(
m(D_i;\theta_0,g)-m(D_i;\theta_0,g_0)
\right)
\right]
\end{align}
and the notation \(\mathbb{E}_{A}[\cdot]\) indicates that we are evaluating the expectation over the randomness in \(A\). Define the quantities
\begin{align}
\tilde{F}_n(x;g)
&=
\frac{1}{r}\sum_{q=1}^{r}
\left(
F(x;D_{\mathsf{s}_q},\xi_{\mathsf{s}_q},g)
-
\mathbb{E}_{\mathsf{s}}\left[
F(x;D_{\mathsf{s}},\xi_{\mathsf{s}},g)
\mid \mathbf{D}_n,\bm{\xi}
\right]
\right)~,\label{eq: tilde G}\\
\hat{F}_n(x;g)
&=
\frac{1}{N_b}\sum_{\mathsf{s}\in\mathcal{S}_{n,b}}
\left(
F(x;D_{\mathsf{s}},\xi_{\mathsf{s}},g)
-
\bar{F}(x;D_{\mathsf{s}},g)
\right)~,\quad\text{and}\label{eq: hat G}\\
\bar{F}_n(x;g)
&=
\frac{1}{N_b}\sum_{\mathsf{s}\in\mathcal{S}_{n,b}}
\bar{F}(x;D_{\mathsf{s}},g)~,\label{eq: bar G}
\end{align}
and consider the decomposition
\begin{align}
\bar{M}_n(x;\theta_0(x),g)
-
\bar{M}_n(x;\theta_0(x),g_0)
=
\tilde{F}_n(x;g)+\hat{F}_n(x;g)+\bar{F}_n(x;g)~.\label{eq: G decomp}
\end{align}
We again apply \cref{eq: apply bernstein to randomness} to bound \eqref{eq: tilde G} and \eqref{eq: hat G}. In particular, by \cref{assu: moment linearity}, the normalization \eqref{eq: part iii normalization}, and the restriction \(n\leq b\sqrt{r}\), \cref{eq: apply bernstein to randomness} implies that
\begin{align}
\mathbb{E}\left[
\left|
\tilde{F}_n(x^{(j)};\hat{g}_n)
\right|^q
\right]^{1/q}
&\lesssim
q^{1/2}\frac{b(1+\|\theta_0(\bm{x}^{(d)})\|_\infty)\phi}{n}
\quad\text{and}\label{eq: W tilde bound bias}\\
\mathbb{E}\left[
\left|
\hat{F}_n(x^{(j)};\hat{g}_n)
\right|^q
\right]^{1/q}
&\lesssim
q^{1/2}\frac{b(1+\|\theta_0(\bm{x}^{(d)})\|_\infty)\phi}{n}~.\label{eq: W hat bound bias}
\end{align}
Here, we have used the statistical independence of \(\hat{g}_n\) and \(\mathbf{D}_n\), together with the fact that the preceding bounds hold uniformly over \(g\in\mathcal{G}\).

To bound the term \eqref{eq: bar G}, first observe that
\begin{equation}
\mathbb{E}\left[\bar{F}(x;D_{\mathsf{s}},g)\right]=0
\quad\text{and}\quad
\left|\bar{F}(x;D_{\mathsf{s}},g)\right|
\lesssim
(1+\|\theta_0(\bm{x}^{(d)})\|_\infty)\phi
\end{equation}
by the boundedness part of \cref{assu: moment linearity}. In turn, observe that
\begin{align}
\mathbb{E}\left[
\left(
\bar{F}(x;D_{\mathsf{s}},g)
\right)^2
\right]
&\leq
\mathbb{E}\left[
\sum_{i\in\mathsf{s}}
\kappa(x,X_i,D_{\mathsf{s}},\xi_{\mathsf{s}})
\mathbb{E}\left[
\left(
m(D_i;\theta_0,g)-m(D_i;\theta_0,g_0)
\right)^2
\mid X_i
\right]
\right]\nonumber\\
&\lesssim
\mathbb{E}\left[
\sum_{i\in\mathsf{s}}
\kappa(x,X_i,D_{\mathsf{s}},\xi_{\mathsf{s}})
V(x,g)
\right]
+
\varepsilon_b\nonumber\\
&\lesssim
\|g-g_0\|_{2,\infty}^2+\varepsilon_b~,
\label{eq: F variance bound}
\end{align}
where the first inequality follows from Honesty and Jensen's inequality, the second inequality follows from \cref{assu: moment smoothness}, Part (ii), and the definition of the shrinkage rate \(\varepsilon_b\), and the third inequality follows from \cref{assu: moment smoothness}, Part (ii), and the normalization that \(\sum_{i\in\mathsf{s}}\kappa(x,X_i,D_{\mathsf{s}},\xi)=1\) almost surely.

Let \(\bar{F}^{(\ell)}(x;\cdot,g)\) denote the \(\ell\)-th Hoeffding projection of the kernel \(\bar{F}(x;\cdot,g)\). By Part (ii) of \cref{lem: Diff H decomposition}, we have
\begin{align}
\bar{F}_n(x;g)
&=
\frac{b}{n}\sum_{i=1}^{n}\bar{F}^{(1)}(x;D_i,g)
+
\sum_{\ell=2}^{b}{b\choose \ell}U_{n,\ell}\left(\bar{F}^{(\ell)}(x;\cdot,g)\right)~. \label{eq: F Hoeffding}
\end{align}
By Hoeffding orthogonality and \eqref{eq: F variance bound},
\begin{align}
b\mathbb{E}\left[
\left(
\bar{F}^{(1)}(x;D_i,g)
\right)^2
\right]
&\leq
\mathbb{E}\left[
\left(
\bar{F}(x;D_{\mathsf{s}},g)
\right)^2
\right]
\lesssim
\|g-g_0\|_{2,\infty}^2+\varepsilon_b~.
\end{align}
Moreover,
\[
\left|\bar{F}^{(1)}(x;D_i,g)\right|
\lesssim
(1+\|\theta_0(\bm{x}^{(d)})\|_\infty)\phi
\]
almost surely. Hence, \cref{lem: Rosenthal centered} gives
\begin{align}
&\mathbb{E}\left[
\left|
\frac{b}{n}\sum_{i=1}^{n}\bar{F}^{(1)}(x^{(j)};D_i,g)
\right|^q
\right]^{1/q} \nonumber\\
&\quad\quad\lesssim
\sqrt{\frac{bq}{n}}
\left(
\|g-g_0\|_{2,\infty}
+
\varepsilon_b^{1/2}
\right)
+
\frac{b}{n}(1+\|\theta_0(\bm{x}^{(d)})\|_\infty)\phi q~. \label{eq: F first order bound}
\end{align}
Next, \cref{lem: u stat linearity} and Minkowski's inequality imply that
\begin{align}
&\mathbb{E}\left[
\left|
\sum_{\ell=2}^{b}{b\choose \ell}U_{n,\ell}\left(\bar{F}^{(\ell)}(x^{(j)};\cdot,g)\right)
\right|^q
\right]^{1/q} \nonumber\\
&\quad\quad\leq
\sum_{\ell=2}^{b}
\mathbb{E}\left[
\left|
{b\choose \ell}U_{n,\ell}\left(\bar{F}^{(\ell)}(x^{(j)};\cdot,g)\right)
\right|^q
\right]^{1/q} \nonumber\\
&\quad\quad\lesssim
(1+\|\theta_0(\bm{x}^{(d)})\|_\infty)\phi q
\sum_{\ell=2}^{b}
\left(
\frac{Cqb}{n}
\right)^{\ell/2}
\lesssim
(1+\|\theta_0(\bm{x}^{(d)})\|_\infty)\phi q^2\frac{b}{n}~,
\label{eq: F higher order bound}
\end{align}
where the final inequality follows from the geometric series formula and \eqref{eq: part iii normalization}. Combining \eqref{eq: F Hoeffding}, \eqref{eq: F first order bound}, and \eqref{eq: F higher order bound}, we find that, for each fixed \(g\in\mathcal{G}\),
\begin{align}
\mathbb{E}\left[
\left|
\bar{F}_n(x^{(j)};g)
\right|^q
\right]^{1/q}
&\lesssim
\sqrt{\frac{bq}{n}}
\left(
\|g-g_0\|_{2,\infty}
+
\varepsilon_b^{1/2}
\right)
+
\frac{b}{n}(1+\|\theta_0(\bm{x}^{(d)})\|_\infty)\phi q^2~. \label{eq: G bar bound}
\end{align}
Consequently, by the elementary bound
\[
\sqrt{\frac{bq}{n}}\varepsilon_b^{1/2}
\leq
\frac{1}{2}\frac{bq}{n}
+
\frac{1}{2}\varepsilon_b
\]
and the fact that \(\hat{g}_n\) is statistically independent of \(\mathbf{D}_n\), the rate condition on \(\hat{g}_n\) gives
\begin{align}
\mathbb{E}\left[
\left|
\bar{F}_n(x^{(j)};\hat{g}_n)
\right|^q
\right]^{1/q}
&\lesssim
q^{3/2}\sqrt{\frac{b}{n}}\left(\frac{b}{n}\right)^{1/4}\delta_{n,g}
+
\sqrt{\frac{bq}{n}}\varepsilon_b^{1/2} \nonumber\\
&\quad\quad+
\frac{b}{n}(1+\|\theta_0(\bm{x}^{(d)})\|_\infty)\phi q^2 \nonumber\\
&\lesssim
q^{3/2}\sqrt{\frac{b}{n}}\left(\frac{b}{n}\right)^{1/4}\delta_{n,g}
+
\varepsilon_b
+
\frac{b}{n}(1+\|\theta_0(\bm{x}^{(d)})\|_\infty)\phi q^2~.
\label{eq: se general g bar bound}
\end{align}
Putting the pieces together, the decomposition \eqref{eq: G decomp} and the bounds \eqref{eq: W tilde bound bias}, \eqref{eq: W hat bound bias}, and \eqref{eq: se general g bar bound} imply that
\begin{align}
&\mathbb{E}\left[
\left|
\bar{M}_n(x^{(j)};\theta_0(x^{(j)}),\hat{g}_n)
-
\bar{M}_n(x^{(j)};\theta_0(x^{(j)}),g_0)
\right|^q
\right]^{1/q} \nonumber\\
&\quad\quad\lesssim
q^{3/2}\sqrt{\frac{b}{n}}\left(\frac{b}{n}\right)^{1/4}\delta_{n,g}
+
\varepsilon_b
+
\frac{b}{n}(1+\|\theta_0(\bm{x}^{(d)})\|_\infty)\phi q^2~,
\end{align}
where we have absorbed the bounds \eqref{eq: W tilde bound bias} and \eqref{eq: W hat bound bias} into the final term. By the normalization \eqref{eq: part iii normalization}, we have
\[
q^{1/2}\sqrt{\frac{b}{n}}\lesssim 1
\quad\text{and}\quad
q(1+\|\theta_0(\bm{x}^{(d)})\|_\infty)\phi\sqrt{\frac{b}{n}}\lesssim 1~.
\]
Therefore,
\begin{align}
&\mathbb{E}\left[
\left|
\bar{M}_n(x^{(j)};\theta_0(x^{(j)}),\hat{g}_n)
-
\bar{M}_n(x^{(j)};\theta_0(x^{(j)}),g_0)
\right|^q
\right]^{1/q} \nonumber\\
&\quad\quad\lesssim
q\bigg(
\left(\frac{b}{n}\right)^{1/4}\delta_{n,g}
+
\varepsilon_b
+
(1+\|\theta_0(\bm{x}^{(d)})\|_\infty)\phi\sqrt{\frac{b}{n}}
\bigg)~.
\end{align}
Multiplying both sides by
\[
\sqrt{\frac{n}{b^2\underline{\sigma}_b^2}}
\]
gives \eqref{eq: stochastic equicontinuity moment}.\hfill$\qed$

\subsubsection{Part (iv)}

Recall the decomposition
\begin{align}
&M^{(1)}(x;\theta_0(x),g_0) (\hat{\theta}_n(x) - \theta_0(x))\label{eq: basic decomp consist}\\
& = -\bar{M}_n(x;\theta_0(x),g_0)  \label{eq: base u stat}\\
& \quad+ \mathsf{Bias}_n(x;\hat{\theta}_n(x),\hat{g}_n) + \mathsf{Nuis}(x;\theta_0(x),\hat{g}_n) \label{eq: bias and nuis consist}\\
& \quad+ \mathsf{Stoch}^{(1)}(x; \hat{\theta}_n(x),\hat{g}_n) + \mathsf{Stoch}^{(2)}(x; \hat{g}_n)  \label{eq: stoch terms consist}\\
& \quad- (\hat{\theta}_n(x) - \theta_0(x)) \left(M^{(1)}(x; \theta_0(x), \hat{g}_n)   -  M^{(1)}(x; \theta_0(x), g_0) \right)\label{eq: theta g term consistency}~
\end{align}
stated as \eqref{eq: theta g term} in the Proof of \cref{thm: generic m estimation unstudentized}, Part (i), where we note that the various terms appearing in \eqref{eq: basic decomp consist} are defined in \eqref{eq: bias def} through \eqref{eq: stoch 2 def}. Here, we have used the fact that the Hessian $H(x; \theta, g) = 0$ almost surely, by \cref{assu: moment linearity}. We also use Part (iv) of \cref{assu: moment smoothness}, which implies that $\hat{\theta}_n(x)$ and $\theta_0(x)$ are uniformly bounded.

First, observe that, by an argument identical to the argument used to establish Part (ii) of this Lemma, i.e., \eqref{eq: M 1 concentrate}, we have that
\begin{align}
\mathbb{E}\left[
\big\vert
\bar{M}_n(x^{(j)};\theta_0(x^{(j)}),g_0)
\big\vert^q
\right]^{1/q}
&\lesssim
q(1+\|\theta_0(\bm{x}^{(d)})\|_\infty)\phi
\sqrt{\frac{b}{n}}~. \label{eq: M concentrate consist}
\end{align}
Moreover, as before, Parts (iii) and (iv) of \cref{assu: moment smoothness} imply that
\begin{align}
&\mathbb{E}\left[
\bigg\vert
(\hat{\theta}_n(x^{(j)})-\theta_0(x^{(j)}))
\left(M^{(1)}(x^{(j)};\theta_0(x^{(j)}),\hat{g}_n)
-
M^{(1)}(x^{(j)};\theta_0(x^{(j)}),g_0)
\right)
\bigg\vert^q
\right]^{1/q} \nonumber\\
&\quad\quad\lesssim
\mathbb{E}\left[
\|\hat{g}_n-g_0\|_{2,\infty}^q
\right]^{1/q}
\lesssim
q\left(\frac{b}{n}\right)^{1/4}\delta_{n,g}~. \label{eq: product term consist}
\end{align}
In turn, observe that Part (i) of this Lemma and Part (iv) of \cref{assu: moment smoothness} give that
\begin{align}
\left|
\mathsf{Bias}_n(x^{(j)};\hat{\theta}_n(x^{(j)}),\hat{g}_n)
\right|
&\lesssim
(1+\|\theta_0(\bm{x}^{(d)})\|_\infty)\varepsilon_b~. \label{eq: bias consist}
\end{align}
and that Neyman orthogonality gives
\begin{align}
\mathbb{E}\left[
\left|
\mathsf{Nuis}(x^{(j)};\theta_0(x^{(j)}),\hat{g}_n)
\right|^q
\right]^{1/q}
&\lesssim
\mathbb{E}\left[
\|\hat{g}_n-g_0\|_{2,\infty}^{2q}
\right]^{1/q} \nonumber\\
&\lesssim
q^2\sqrt{\frac{b}{n}}\delta_{n,g}^2
\lesssim
q\left(\frac{b}{n}\right)^{1/4}\delta_{n,g}~,
\label{eq: nuisance consist}
\end{align}
where the final inequality follows from the normalization \(q(b/n)^{1/4}\delta_{n,g}\leq1\).

Moreover, as before, \cref{assu: empirical smoothness} and a Taylor expansion gives
\begin{align}
\vert \mathsf{Stoch}^{(1)}(x; \hat{\theta}_n(x),\hat{g}_n) \vert
& =
\vert  (\hat{\theta}_n(x) - \theta_0(x)) \bar{M}_n^{(1)}(x; \theta_0(x), \hat{g}_n) \vert \nonumber\\
& \leq
\vert \hat{\theta}_n(x)-\theta_0(x)\vert
\vert \bar{M}_n^{(1)}(x;\theta_0(x),g_0)\vert \nonumber\\
&\quad+
\vert \hat{\theta}_n(x)-\theta_0(x)\vert
\vert
\bar{M}_n^{(1)}(x;\theta_0(x),\hat{g}_n)
-
\bar{M}_n^{(1)}(x;\theta_0(x),g_0)
\vert~,
\end{align}
where we have used the fact that the empirical Hessian $\bar{H}_n(x; \tilde{\theta}_0(x), \hat{g}_n) = 0$ almost surely, by \cref{assu: moment linearity}. Consequently, Part (iv) of \cref{assu: moment smoothness}, Part (ii) of this Lemma, and Part (iii) of this Lemma imply that
\begin{align}
&\mathbb{E}\left[
\left|
\mathsf{Stoch}^{(1)}(x^{(j)};\hat{\theta}_n(x^{(j)}),\hat{g}_n)
\right|^q
\right]^{1/q} \nonumber\\
&\quad\quad\lesssim
q(1+\|\theta_0(\bm{x}^{(d)})\|_\infty)\phi\sqrt{\frac{b}{n}} \nonumber\\
&\quad\quad\quad+
q^{3/2}\sqrt{\frac{b}{n}}\left(\frac{b}{n}\right)^{1/4}\delta_{n,g}
+
\varepsilon_b
+
\frac{b}{n}(1+\|\theta_0(\bm{x}^{(d)})\|_\infty)\phi q^2 \nonumber\\
&\quad\quad\lesssim
q(1+\|\theta_0(\bm{x}^{(d)})\|_\infty)\phi\sqrt{\frac{b}{n}}
+
q\left(\frac{b}{n}\right)^{1/4}\delta_{n,g}
+
q(1+\|\theta_0(\bm{x}^{(d)})\|_\infty)\varepsilon_b~. \label{eq: stoch 1 bound consist}
\end{align}
Here, the final inequality follows from the normalizations \eqref{eq: part iv normalizations}. Similarly, Part (iii) of this Lemma gives
\begin{align}
&\mathbb{E}\left[
\left|
\mathsf{Stoch}^{(2)}(x^{(j)};\hat{g}_n)
\right|^q
\right]^{1/q} \nonumber\\
&\quad\quad\lesssim
q^{3/2}\sqrt{\frac{b}{n}}\left(\frac{b}{n}\right)^{1/4}\delta_{n,g}
+
\varepsilon_b
+
\frac{b}{n}(1+\|\theta_0(\bm{x}^{(d)})\|_\infty)\phi q^2 \nonumber\\
&\quad\quad\lesssim
q\left(\frac{b}{n}\right)^{1/4}\delta_{n,g}
+
q(1+\|\theta_0(\bm{x}^{(d)})\|_\infty)\varepsilon_b
+
q(1+\|\theta_0(\bm{x}^{(d)})\|_\infty)\phi\sqrt{\frac{b}{n}}~. \label{eq: se consist}
\end{align}
Again, the final inequality follows from the normalization \eqref{eq: part iv normalizations}.

Consequently, plugging the bounds \eqref{eq: M concentrate consist}, \eqref{eq: product term consist}, \eqref{eq: bias consist}, \eqref{eq: nuisance consist}, \eqref{eq: stoch 1 bound consist}, and \eqref{eq: se consist} into the decomposition \eqref{eq: basic decomp consist}, applying Minkowski's inequality, and using the fact that $M^{(1)}(x;\theta_0(x),g_0)$ is bounded from below by Part (iii) of \cref{assu: moment smoothness}, we find that
\begin{align}
&\mathbb{E}\left[
\big\vert
\hat{\theta}_n(x^{(j)})-\theta_0(x^{(j)})
\big\vert^q
\right]^{1/q} \nonumber\\
&\quad\quad\lesssim
q\left(\frac{b}{n}\right)^{1/4}\delta_{n,g}
+
q(1+\|\theta_0(\bm{x}^{(d)})\|_\infty)\varepsilon_b
+
q(1+\|\theta_0(\bm{x}^{(d)})\|_\infty)\phi\sqrt{\frac{b}{n}}~. \label{eq: theta moment consistency unnormalized}
\end{align}
Multiplying both sides of \eqref{eq: theta moment consistency unnormalized} by
\[
\left(\frac{n}{b^2\underline{\sigma}_b^2}\right)^{1/4}
\]
gives
\begin{align}
\left(\frac{n}{b^2\underline{\sigma}_b^2}\right)^{1/4}
\mathbb{E}\left[
\big\vert
\hat{\theta}_n(x^{(j)})-\theta_0(x^{(j)})
\big\vert^q
\right]^{1/q}
&\lesssim
q\delta_{n,\theta}~,
\end{align}
as required.
\hfill\qed

\subsection{Proof of \cref{eq: apply bernstein to randomness}}

Throughout, we use the shorthand $N_b={n\choose b}$. We begin by considering the quantity
\begin{equation}
\frac{1}{N_b}\sum_{\mathsf{s}\in\mathcal{S}_{n,b}} Z_{\mathsf{s}},
\quad\text{where}\quad
Z_{\mathsf{s}}
=
u(D_{\mathsf{s}},\xi_{\mathsf{s}})
-
\mathbb{E}\left[u(D_{\mathsf{s}},\xi_{\mathsf{s}})\mid \mathbf{D}_{\mathsf{s}}\right]~.
\end{equation}
Conditioned on the data $\mathbf{D}_n$, the observations $Z_{\mathsf{s}}$, $\mathsf{s}\in\mathcal{S}_{n,b}$, are centered and mutually independent. Moreover, each component of $Z_{\mathsf{s}}$ is bounded by $2\phi$ almost surely. Hence, by  \cref{lem: Rosenthal centered}, we obtain
\begin{align}
\mathbb{E}\left[
\bigg\vert
\frac{1}{N_b}\sum_{\mathsf{s}\in\mathcal{S}_{n,b}} Z_{\mathsf{s},j}
\bigg\vert^q
\mid \mathbf{D}_n
\right]^{1/q}
&\lesssim
\phi\left(
\sqrt{\frac{q}{N_b}}
+
\frac{q}{N_b}
\right)~. \label{eq: complete randomized moment scalar}
\end{align}
As $N_b^{-1/2}\lesssim b/n$ for $b>2$, we have that
\begin{align}
\phi\left(
\sqrt{\frac{q}{N_b}}
+
\frac{q}{N_b}
\right)
&\lesssim
\frac{b\phi q^{1/2}}{n}
+
\phi q\frac{b^2}{n^2}
\lesssim
\frac{b\phi q^{1/2}}{n}~,
\end{align}
where the last inequality follows from the normalization $\phi b q^{1/2}/n\leq1$ and the fact that $\phi\geq1$. Taking expectations over $\mathbf{D}_n$ verifies \eqref{eq: just data lemma state}.

Next, consider the quantity
\begin{equation}
\frac{1}{r}\sum_{q=1}^r Z_q,
\quad\text{where}\quad
Z_q
=
u(D_{\mathsf{s}_q},\xi_{\mathsf{s}_q})
-
\mathbb{E}\left[
u(D_{\mathsf{s}_q},\xi_{\mathsf{s}_q})
\mid \mathbf{D}_n,\bm{\xi}
\right]~.
\end{equation}
Conditioned on $\mathbf{D}_n$ and $\bm{\xi}$, the observations $Z_q$, $q\in[r]$, are centered and mutually independent. Moreover, each component of $Z_q$ is bounded by $2\phi$ almost surely. Thus, the same argument used above gives
\begin{align}
&\mathbb{E}\left[
\bigg\vert
\frac{1}{r}\sum_{q=1}^r Z_q
\bigg\vert^q
\mid \mathbf{D}_n,\bm{\xi}
\right]^{1/q} \lesssim
\phi\left(
\sqrt{\frac{q}{r}}
+
\frac{q}{r}
\right)~. \label{eq: incomplete randomized moment vector}
\end{align}
Now, the restriction $n\leq\sqrt{r}b$ implies that
\begin{align}
\phi\left(
\sqrt{\frac{q}{r}}
+
\frac{q}{r}
\right)
&\lesssim
\frac{b\phi q^{1/2}}{n}
+
\phi q\frac{b^2}{n^2}
\lesssim
\frac{b\phi q^{1/2}}{n}~,
\end{align}
where the final inequality again follows from the normalization $\phi b q^{1/2}/n\leq1$ and the fact that $\phi\geq1$. Taking expectations over $\mathbf{D}_n$ and $\bm{\xi}$ verifies \eqref{eq: random lemma state}.\hfill\qed

\section{Additional Results and Discussion\label{app: additional}} 

\subsection{Stochastic Equicontinuity without Sample Splitting\label{app: stability}}

In the main text, we assume that the nuisance parameter estimator $\hat{g}_n$ is computed on a separate sample, independent of the data $\mathbf{D}_n$. In this section, we give additional conditions under which the nuisance parameter estimator $\hat{g}_n$ can be computed on the same data used to construct the estimator $\hat{\theta}_n(\bm{x}^{(d)})$. Our analysis builds on a similar result given in \cite{chen2022debiased}, who give a pointwise analysis of unconditional moment estimators. By contrast, we give a high-dimensional analysis of conditional moment estimators constructed with subsampled kernels. 

In the proof of \cref{eq: asymptotic validity subsampled kernel}, the statistical independence of the estimator $\hat{g}_n$ and the data $\mathbf{D}_n$ is only needed in the proof of \cref{lem: kernel bcs}, Part (iii). Thus, our objective is to prove an analogous result, without imposing this restriction. In particular, consider the functions
\begin{flalign*}
a(x;D_{i},D_{\mathsf{s}},g) & =\kappa(x,X_{i};D_{\mathsf{s}})m(D_{i},g)\quad\text{and}\quad A(x;g)=\mathbb{E}\left[\kappa(x,X_{i};D_{\mathsf{s}})m(D_{i},g)\right],
\end{flalign*}
where $\kappa(\cdot,\cdot;D_{\mathsf{s}})$ is a kernel function and
$m(D_{i},g)$ is a moment function. We are interested in the complete,
deterministic, $U$-statistic 
\begin{flalign}
 & {n \choose b}^{-1} \sum_{\mathsf{s}\in\mathcal{S}_{n,b}}F(\bm{x}^{(d)};D_{\mathsf{s}},\hat{g}_{n})\label{eq: u stat}
\end{flalign}
where the kernel function $F(x;D_{\mathsf{s}},g)$ is given by
\begin{flalign*}
F(x;D_{\mathsf{s}},g) & =\sum_{i\in\mathsf{s}}f(x;D_{i},D_{\mathsf{s}},g),\\
f(x;D_{i},D_{\mathsf{s}},g) & =a(x;D_{i},D_{\mathsf{s}},g)-a(x;D_{i},D_{\mathsf{s}},g_{0})-\left(A(x;g)-\mathbb{E}\left[A(x;g_{0})\right]\right),
\end{flalign*}
and $\hat{g}_{n}$ is some estimator of the nuisance parameter $g_{0}$
computed with the data $\mathbf{D}_{n}$. That is, for the sake of
simplicity, we have restricted our attention to complete, deterministic
subsampled kernel estimators. Incomplete, random, subsampled
kernel estimators can be accommodated through the same methods applied
repeatedly throughout the arguments supporting \cref{eq: asymptotic validity subsampled kernel}.

\subsubsection{Regularity and Stability}

To state our additional conditions, we require some additional notation. Let $\mathbf{D}_{n}^{\prime}=(D_{i}^{\prime})_{i=1}^{n}$ denote an
independent copy of $\mathbf{D}_{n}$. Let $\hat{g}_{n}^{(-\mathsf{s})}$
denote a version of the estimator $\hat{g}_{n}$ formed with all observations
in $\mathbf{D}_{n}$, except that the observations $D_{\mathsf{s}}$
are replaced by the observations $D_{\mathsf{s}}^{\prime}$. Define the norm
\[
\|g-\tilde{g}\|_{2q,\infty}=\max_{k\in[h]}\max_{j\in[d]}\left(\mathbb{E}\left[\left(g^{(k)}(Z_{i})-\tilde{g}^{(k)}(Z_{i})\right)^{2q}\mid X_{i}=x^{(j)}\right]\right)^{1/2q}
\]
for any nuisance parameters $g=(g^{(k)})_{k=1}^{h}$ and $\tilde{g}=(\tilde{g}^{(k)})_{k=1}^{h}$
in the space $\mathcal{G}$ and any positive integer $q>0$.

First, we require that moment $m(D_{i},g)$ satisfies a pair of smoothness conditions.
\begin{assumption}
\label{assu: mean-square-continuity}Define the higher-order
variogram 
\[
V^{(q)}(x;g,g^{\prime})=\mathbb{E}\left[\left(m(D_{i},g)-m(D_{i},g^{\prime})\right)^{2q}\mid X_{i}=x\right]
\]
for each $g$ and $g^{\prime}$ in $\mathcal{G}$. For each integer
$q>0$, the Lipschitz condition 
\begin{equation}
\vert V^{(q)}(x;g,g^{\prime})-V^{(q)}(x^{\prime};g,g^{\prime})\vert\lesssim\|x-x^{\prime}\|_{\infty}^{q}\label{eq: Lipschitz variogram}
\end{equation}
holds for each $x$ and $x^{\prime}$ in $\mathcal{X}$ and the mean-square
continuity condition 
\begin{equation}
\sup_{j\in[d]}V^{(q)}(x^{(j)};g,g^{\prime})\lesssim\|g-g^{\prime}\|_{2q,\infty}^{2q}\label{eq: mean square continuity}
\end{equation}
holds for each $g$ and $g^{\prime}$ in $\mathcal{G}$. 
\end{assumption}
\noindent 
\cref{assu: mean-square-continuity} is a higher-order analogue to Part (ii) of \cref{assu: moment smoothness}.

Second, we impose the following higher-order shrinkage condition.
\begin{assumption}
\label{assu: higher-order shrinkage} Fix a sequence $\varepsilon_b$. The kernel $\kappa(\cdot,\cdot,D_{\mathsf{s}},\xi_{\mathsf{s}})$ satisfies the higher-order shrinkage bound
\begin{equation}
\sup_{P \in \mathbf{P}} \sup_{j \in [d]} \mathbb{E}\left[ \max\left\{ \|X_i - x^{(j)} \|_\infty : \kappa(x^{(j)},X_i,D_{\mathsf{s}},\xi_{\mathsf{s}}) > 0 \right\}^q \right]^{1/q} \leq \varepsilon_{b}~.
\end{equation}
each integer \(q\geq 2\vee2\log(nd)\).
\end{assumption}

Third, we impose the following higher-order moment stability conditions.
\begin{assumption}
\label{assu: stability}There exists some increasing sequence
$\gamma(q)$, which may depend on the parameters $n$, $b$, and $d$,
such that the $L_{q}$-norm stability bounds
\begin{flalign}
\|m(D_{i},\hat{g}_{n})-m(D_{i},\hat{g}_{n}^{(-[b])})\|_{q} & \lesssim q\gamma(q)\frac{b}{n}\quad\text{and}\label{eq: stability bound D_1}\\
\|m(D_{1}^{\prime\prime},\hat{g}_{n})-m(D_{1}^{\prime\prime},\hat{g}_{n}^{(-[b])})\|_{q} & \lesssim q\gamma(q)\frac{b}{n}\label{eq: stability bound D}
\end{flalign}
hold, where $D_{1}^{\prime\prime}$ is an independent copy of $D_{1}$. 
\end{assumption}
\noindent 
\cref{assu: stability} quantifies the sensitivity of the moment $m(D_{1},\hat{g}_{n})$ to re-sampling $b$ elements of the data $\mathbf{D}_n$. Analogous conditions are studied in \cite{abou2019exponential} and \cite{chen2022debiased}. We discuss moments and estimators that satisfy \cref{assu: stability} in \cref{sec: verify}.

\subsubsection{Stochastic Equicontinuity}

The following Theorem gives a suitable bound for the $U$-statistic \eqref{eq: u stat}.

\begin{theorem}\label{thm: no split stochastic equicontinuity}
Suppose that the kernel \(\kappa(x,X_i;D_{\mathsf{s}})\) satisfies \cref{assu: kernel restriction} and \cref{assu: higher-order shrinkage}, that the moment function \(m(D_i,g)\) satisfies \cref{assu: mean-square-continuity}, and that the nuisance parameter estimator satisfies the higher-order moment bound
\begin{equation}
\sup_{P\in\mathbf{P}}
\mathbb{E}\left[
\|\hat{g}_n-g_0\|_{2q,\infty}^{2q}
\right]^{1/(2q)}
\lesssim
q\left(\frac{b}{n}\right)^{1/4}\delta_{n,g}
\label{eq: D nuisance moment}
\end{equation}
for some sequence \(\delta_{n,g}\). If \cref{assu: stability} holds, then the bound
\begin{align}
&\mathbb{E}\left[
\left|
\frac{1}{N_b}
\sum_{\mathsf{s}\in\mathcal{S}_{n,b}}
F(x^{(j)};D_{\mathsf{s}},\hat{g}_n)
\right|^q
\right]^{1/q} \nonumber\\
&\quad\quad\lesssim
q^{3/2}\sqrt{\frac{b}{n}}
\left(
\left(\frac{b}{n}\right)^{1/4}\delta_{n,g}
+
\varepsilon_b^{1/2}
+
b^{1/q}\gamma(2q)
\right)
\label{eq: no split se moment}
\end{align}
holds uniformly over \(P\in\mathbf{P}\), for each \(j\in[d]\) and each even integer \(q\geq 2\vee2\log(nd)\).
\end{theorem}

\begin{remark}
Compare the bound \eqref{eq: no split se moment} to the rate given in \cref{lem: kernel bcs}, Part (iii). Taking \(q=\bar{q}_{n,d}\asymp\log(dn)\) and writing
\[
\gamma_{n,b}^{\star}
=
b^{1/\bar{q}_{n,d}}\gamma(2\bar{q}_{n,d})~,
\]
the bound \eqref{eq: no split se moment} gives the same-sample analogue of the stochastic equicontinuity rate
\[
\sqrt{\frac{b}{n}}\log^{3/2}(dn)
\left(
\left(\frac{b}{n}\right)^{1/4}\delta_{n,g}
+
\varepsilon_b^{1/2}
+
\gamma_{n,b}^{\star}
\right)~.
\]
The first term is comparable to the corresponding term in \cref{lem: kernel bcs}, Part (iii), up to the additional \(\log^{1/2}(dn)\) factor generated by the generalized Efron--Stein inequality. Moreover, the elementary inequality
\[
\sqrt{\frac{b\log(dn)}{n}}\varepsilon_b^{1/2}
\leq
\frac{b\log(dn)}{n}
+
\varepsilon_b
\]
allows the shrinkage term to be compared directly with the \(\varepsilon_b\) term appearing in \cref{lem: kernel bcs}, Part (iii), under the same small-order normalizations used in the proof of \cref{eq: asymptotic validity subsampled kernel}. Thus, relative to the sample-split argument, the principal additional contribution is the stability term
\[
\sqrt{\frac{b}{n}}\log^{3/2}(dn)\gamma_{n,b}^{\star}~.
\]
Consequently, under the conditions of \cref{thm: no split stochastic equicontinuity}, if this stability contribution is of smaller order than the stochastic equicontinuity rate in \cref{lem: kernel bcs}, Part (iii), a result analogous to \cref{eq: asymptotic validity subsampled kernel} will hold without sample splitting. We discuss conditions under which \(\gamma_{n,b}^{\star}\) decreases in \cref{sec: verify}.\hfill{}$\blacksquare$
\end{remark}

\begin{remark}
\cref{thm: no split stochastic equicontinuity} follows from an argument similar to the argument developed in \cite{chen2022debiased}. There is one important difference. \cite{chen2022debiased} apply a ``double-centering trick,'' due to \cite{kumar2013near}, to get a variance bound. To obtain a high-dimensional bound, with a logarithmic dependence on the dimension $d$, we replace this step with an application of a generalized Efron-Stein inequality, due to \cite{boucheron2005moment}.\hfill{}$\blacksquare$
\end{remark}

\subsubsection{\label{sec: verify}Verifying \cref{thm: no split stochastic equicontinuity}}

In this section, we study conditions under which \cref{assu: stability} is satisfied. First, we impose an additional smoothness condition on the
moment function $m(D_{i},g)$. Throughout, we write the nuisance parameter estimator $\hat{g}_{n}$ as $\hat{g}_{n}=(\hat{g}_{k,n})_{k=1}^{h}$. 
\begin{assumption}
\label{assu: finite moment bound}The inequalities
\begin{equation}
\|m(D_i,\hat g_n)-m(D_i,\hat g_n^{(-[b])})\|_{2q}
\lesssim
\max_{k\in[h]}
\mathbb{E}\left[
\sup_{z\in\mathcal Z}
\left|
\hat g_{k,n}(z)-\hat g_{k,n}^{(-[b])}(z)
\right|^{2q}
\right]^{1/(2q)}
\label{eq: same finte approx}
\end{equation}
and 
\begin{equation}
\|m(D_i^{\prime\prime},\hat g_n)-m(D_i^{\prime\prime},\hat g_n^{(-[b])})\|_{2q}
\lesssim
\max_{k\in[h]}
\mathbb{E}\left[
\left|
\hat g_{k,n}(Z_i^{\prime\prime})
-
\hat g_{k,n}^{(-[b])}(Z_i^{\prime\prime})
\right|^{2q}
\right]^{1/(2q)}
\label{eq: independent approx}
\end{equation}
hold for any positive integer $q$. 
\end{assumption}
\noindent The bound (\ref{eq: same finte approx})
stipulates that the higher-order moments of $m(D_{i},\hat{g}_{n})-m(D_{i},\hat{g}_{n}^{(-[b])})$
are smaller than the supremum of the higher-order moments of $\hat{g}_{k,n}(z)-\hat{g}_{k,n}^{(-[b])}(z)$
over the space $\mathcal{Z}$. The bound (\ref{eq: independent approx})
is weaker, as we only need to control the distance $\hat{g}_{k,n}(z)-\hat{g}_{k,n}^{(-[b])}(z)$
at $z=Z_{i}^{\prime}$. 

Bounds of the form specified by \cref{assu: finite moment bound} are satisfied by many standard choices of moment function. For example, consider the moment function 
\begin{equation}
m(D_{i},g)=\mu(1,Z_{i})-\mu(0, Z_{i})+\beta(W_{i},Z_{i})(Y_{i}-\mu(W_{i},Z_{i}))\label{eq: aipw se}
\end{equation}
where the nuisance function $g$ collects the moment parameters $g=(\mu,\beta)$. Observe that \eqref{eq: aipw se} is analogous to the moment function \eqref{eq: CATE moment} used as a running example in the main text.
\begin{lemma}
\label{lem: continuity}If the quantities 
\begin{equation}
\vert\hat{\beta}_{n}(W_{i},Z_{i})\vert\quad\text{and}\quad\vert Y_{i}-\hat{\mu}_{n}(W_{i},Z_{i})\vert\label{eq: bounded quantities}
\end{equation}
are bounded almost surely, then it holds that 
\begin{flalign}
 & \|m(D_{i},\hat{g}_{n})-m(D_{i},\hat{g}_{n}^{(-[b])})\|_{2p}\nonumber \\
 & \lesssim\max\bigg\{\max_{w\in\left\{ 0,1\right\} }\left\{ \|\hat{\mu}_{n}(w,Z_{i})-\hat{\mu}_{n}^{(-[b])}(w,Z_{i})\|_{2p}\right\} ,\label{eq: orlicz continuity}\\
 & \quad\quad\quad\quad\quad\quad\|\hat{\mu}_{n}(W_{i},Z_{i})-\hat{\mu}_{n}^{(-[b])}(W_i,Z_{i})\|_{2p},\|\hat{\beta}_{n}(W_{i},Z_{i})-\hat{\beta}_{n}^{(-[b])}(W_{i},Z_{i})\|_{2p}\bigg\},\quad\text{and}\nonumber \\
 & \|m(D_{i}^{\prime\prime},\hat{g}_{n})-m(D_{i}^{\prime\prime},\hat{g}_{n}^{(-[b])})\|_{2p}\nonumber \\
 & \lesssim\max\bigg\{\max_{w\in\left\{ 0,1\right\} }\left\{ \|\hat{\mu}_{n}(w,Z_{i}^{\prime\prime})-\hat{\mu}_{n}^{(-[b])}(w,Z_{i}^{\prime\prime})\|_{2p}\right\} ,\label{eq: orlicz continuity Z}\\
 & \quad\quad\quad\quad\quad\quad\|\hat{\mu}_{n}(W_{i},Z_{i}^{\prime\prime})-\hat{\mu}_{n}^{(-[b])}(w,Z_{i}^{\prime\prime})\|_{2p},\|\hat{\beta}_{n}(W_{i},Z_{i}^{\prime\prime})-\hat{\beta}_{n}^{(-[b])}(W_{i},Z_{i}^{\prime\prime})\|_{2p}\bigg\},\nonumber 
\end{flalign}
respectively, where $D_{i}^{\prime\prime}$ is an independent copy
of $D_{i}$ and $Z_{i}^{\prime\prime}$ is an independent copy of
$Z_{i}$. 
\end{lemma}

Now, we consider the higher-order stability of the nuisance parameter estimator $\hat{g}_{n}(z)$. For the sake of simplicity, we assume that $\hat{g}_{n}(z)$ is scalar valued and takes the form of a complete, deterministic, $U$-statistic
\begin{equation}\label{eq: g hat as u stat}
\hat{g}_{n}(z)=
{n \choose b'}^{-1} \sum_{\mathsf{s}\in\mathcal{S}_{n,b^{\prime}}}u_{z}(D_{\mathsf{s}}),
\end{equation}
where $u_{z}(D_{\mathsf{s}})$ is some deterministic kernel function
of order $b^{\prime}$. Again, let $\hat{g}^{(-[b])}_{n}(z)$ be constructed analogously to $\hat{g}_{n}(z)$ using the $\mathbf{D}_{n}$, but replacing $D_{[b]}$ with an independent copy
$D_{[b]}^{\prime}$. 
\begin{lemma}
\label{lem: kernel stability}Suppose that the kernel
function $u_{z}(D_{\mathsf{s}})$ satisfies the bound $\vert u_{z}(D_{\mathsf{s}})\vert\leq\phi$
almost surely for each $z$ in $\mathcal{Z}$. If the estimator $\hat{g}_{n}(z)$ is given by \eqref{eq: g hat as u stat}, then it  holds that 
\begin{flalign}
\|\hat{g}_{n}(z)-\hat{g}^{(-[b])}_{n}(z)\|_{q} & \lesssim\frac{\sqrt{b^{\prime}b}}{n}q\phi+ \frac{b'}{n}  q^2 \phi .\label{eq: kernel stability state}
\end{flalign}
for each $z$ in $\mathcal{Z}$ and $2\log(dn)\leq q \leq c n/b'$
\end{lemma}
\begin{remark}
Under \cref{assu: finite moment bound}, \cref{lem: kernel stability} implies that
\begin{equation}
\|m(D_{1}^{\prime\prime},\hat{g}_{n})-m(D_{1}^{\prime\prime},\hat{g}_{n}^{(-[b])})\|_{q}
\lesssim
\frac{\sqrt{b'b}}{n}q\phi
+
\frac{b'}{n}q^2\phi .
\label{eq: moment bound apply kernel stability}
\end{equation}
Thus, relative to \cref{assu: stability}, we can take
\[
\gamma(q)
\lesssim
\phi\left(
\sqrt{\frac{b'}{b}}
+
q\frac{b'}{b}
\right).
\]
Consequently, at the choice \(q\asymp\log(dn)\), invoked in \cref{thm: no split stochastic equicontinuity}, this sequence is small provided
\[
\frac{b'}{b}\to0
\quad\text{and}\quad
\log(dn)\frac{b'}{b}\to0~.
\]
In particular, \cref{lem: kernel stability} suggests that the stability contribution in \eqref{eq: no split se moment} is negligible if \(b'\) is sufficiently small relative to \(b\).\hfill{}$\blacksquare$
\end{remark}
\begin{remark}
\cref{assu: stability} entails two stability bounds, \eqref{eq: stability bound D_1} and \eqref{eq: stability bound D}. Under \cref{assu: finite moment bound}, \cref{lem: kernel stability} quantifies the sequence $\gamma(q)$ introduced on the right-hand side of \eqref{eq: stability bound D_1}. By the discussion above, if $b^{\prime}=o(b)$, this sequence is sufficiently small for a result analogous to \cref{lem: kernel bcs}, Part (iii) to hold. Handling the bound \eqref{eq: stability bound D} requires developing a more refined argument. In particular, under \cref{assu: finite moment bound}, to bound the right-hand-side of \eqref{eq: stability bound D_1}, we must account for the supremum over the space $\mathcal{Z}$ within the expectation. Such a generalization should be reasonably straightforward, potentially through a chaining argument. Roughly speaking, we should expect the supremum to contribute a term like $\log(p)$ to the stability, where $p$ is the dimension of $\mathcal{Z}$. In this case, in settings where $p$ is not high-dimensional, this would not generate any issues. This should be contrasted with analogous suprema taken over the parameter space of a decision tree (which are generated if stochastic equicontinuity is controlled with a union bound). Here the dimension of the parameter space is roughly $2^{\text{Depth of Tree}}$, which may be large enough to make a material difference. However, in order to operationalize this intuition, we would need to place further restrictions sufficient for the smoothness of the H\'{a}jek projection of $\hat{g}_n(z)$. We leave this for future work. \hfill{}$\blacksquare$
\end{remark}

\subsection{Cross-Fitting\label{app: cross}}

\cref{eq: asymptotic validity subsampled kernel} holds under the assumption that the nuisance parameter estimator $\hat{g}_n$ is computed on a data set that is statistically independent of the data used to construct the confidence region $\hat{\mathcal{C}}(\bm{x}^{(d)})$. This condition can be achieved by splitting the data into two parts, at the cost of reducing statistical precision and  introducing  randomness independent of the observed data. In \cref{app: stability}, we introduce further restrictions that allow the nuisance parameter estimator $\hat{g}_n$ and the confidence region $\hat{\mathcal{C}}(\bm{x}^{(d)})$ to be computed using the same data. However, in practice, these conditions are more stringent, and it may be unclear whether they are satisfied for a given nuisance parameter estimator $\hat{g}_n$.

In this section, we detail an alternative procedure for constructing a confidence region similar to the confidence region specified in \cref{def: uniform ci}, whose validity will hold under the same conditions imposed in \cref{eq: asymptotic validity subsampled kernel}. 
The procedure is based on a cross-fit estimator.  To introduce this estimator, we require some additional notation. Let $\mathcal{R}_{n,k}$ denote the set of partitions of $[n]$ into $k$ equally sized and mutually exclusive subsets. That is, for each $\mathsf{r} = (\mathsf{s}_1,\ldots,\mathsf{s}_k)$ in $\mathcal{R}_{n,k}$, the sets $\mathsf{s}_1,\ldots,\mathsf{s}_k$ are mutually exclusive, have union equal to $[n]$, and are each of size $n/k$. Throughout, we set $q=n/k$ and let $\tilde{\mathsf{s}}$ denote the complement of the set $\mathsf{s}$ in $[n]$. Finally, for each subset $\mathsf{s}$ of $[n]$, let $\hat{g}_{\mathsf{s}}$ denote a version of the nuisance parameter estimator $\hat{g}_n$ computed with the data $D_{\mathsf{s}}$. 

The cross-fit estimator is computed as follows. For each subset $\mathsf{s}$ in $\mathcal{S}_{n,q}$, let the estimator $\hat{\theta}_\mathsf{s}(\bm{x}^{(d)})$ be constructed by solving the empirical conditional moment equality \eqref{eq: subsampled kernel} using the data $D_{\mathsf{s}}$  and the nuisance parameter estimator $\hat{g}_{\tilde{\mathsf{s}}}$. That is, the nuisance parameter estimator is computed on the sample of data whose indices are not in $\mathsf{s}$. The $k$-fold cross-fit estimator is given by
\begin{equation}\label{eq: k fold cross fit estimator}
\hat{\theta}_{\mathsf{r}}(\bm{x}^{(d)}) = \frac{1}{k} \sum_{\mathsf{s} \in \mathsf{r}} \hat{\theta}_\mathsf{s}(\bm{x}^{(d)})~,
\end{equation}
where $\mathsf{r}$ denotes a random element of $\mathcal{R}_{n,k}$.

It is somewhat unclear how to implement the half-sample bootstrap with the estimator $\hat{\theta}_\mathsf{r}(\bm{x}^{(d)})$. We propose the following computationally efficient variant. Let $\mathcal{H}(\mathsf{s})$ denote the set of half-samples of the set $\mathsf{s}$, i.e., the set of subsets of $\mathsf{s}$ that contain exactly half of its elements. We say that the collection $\mathsf{H} = (\mathsf{h}_l)_{l=1}^k$ is a half-sample of the partition $\mathsf{r} = (\mathsf{s}_l)_{l=1}^k$ if each $\mathsf{h}_l$ is an element of $\mathcal{H}(\mathsf{s}_l)$, i.e., if each $\mathsf{h}_l$ is a half-sample of $\mathsf{s}_l$. For each $l$ in $[k]$, let the estimator $\hat{\theta}_{\mathsf{h}_l}(\bm{x}^{(d)})$  be constructed by solving the empirical conditional moment equality \eqref{eq: subsampled kernel} using the data $D_{\mathsf{h}_l}$ and the nuisance parameter estimator $\hat{g}_{\tilde{\mathsf{s}}_l}$, i.e., using the same nuisance parameter estimates used to construct \eqref{eq: k fold cross fit estimator}. The half-sample $k$-fold cross-split estimator is given by
\begin{equation}\label{eq: k fold half-split estimator}
\hat{\theta}_{\mathsf{H}}(\bm{x}^{(d)}) = \frac{1}{k} \sum_{\mathsf{h} \in \mathsf{H}} \hat{\theta}_\mathsf{h}(\bm{x}^{(d)})~.
\end{equation}
In this way, once the $k$-fold cross-fit estimator \eqref{eq: k fold cross fit estimator} has been computed, the nuisance parameter estimates $\hat{g}_{\tilde{\mathsf{s}}_1}, \ldots, \hat{g}_{\tilde{\mathsf{s}}_k}$ do not need to be recomputed to construct the half-sample $k$-fold cross-split estimator \eqref{eq: k fold half-split estimator}. Simultaneous confidence intervals can then be constructed analogously to the intervals introduced in \cref{def: uniform ci} by using the half-sample $k$-fold cross-split bootstrap root
\begin{equation} \label{eq: half-sample k-fold root}
R^*_n(\bm{x}^{(d)}) = \hat{\theta}_{\mathsf{H}}(\bm{x}^{(d)}) - \hat{\theta}_{\mathsf{r}}(\bm{x}^{(d)})
\end{equation}
in place of the half-sample bootstrap root \eqref{eq: bootstrap root}. In other words, when implementing the half-sample bootstrap based on an estimator constructed with $k$-fold cross-fitting, one can avoid recomputing nuisance parameters if the half-samples are ``stratified'' across the $k$-folds. The error bound given in \cref{eq: asymptotic validity subsampled kernel} will generalize to the confidence region based on the bootstrap root \eqref{eq: half-sample k-fold root} through a 
straightforward argument, so long as $k$ is bounded. 
 
\subsection{Data\label{sec: data}}

In this appendix, we document our treatment of the \cite{banerjee2015multifaceted} data.\footnote{The data from \cite{banerjee2015multifaceted} were acquired from \url{https://dataverse.harvard.edu/dataset.xhtml?persistentId=doi:10.7910/DVN/NHIXNT} on 
September 10, 2021.} The data from the graduation program implemented in Pakistan are considered in \cite{chen2023semiparametric}. Here, we consider the data from the graduation program implemented in Ghana, as it has a larger sample size. 

The data record measurements of many pre-treatment and post-treatment outcomes for 2,606 individuals in the northern region of Ghana. Baseline survey measurements were made prior to the allocation of treatment. A multifaceted treatment was randomly allocated to 632 of the individuals. \cite{banerjee2015multifaceted} consider data from two endline surveys, made two years and three years after the initial asset transfer, respectively. For the purpose of this paper, we consider only data from the baseline survey, records of the treatment allocation, and measurements from the first endline survey. We omit data from 164 attrited individuals, none of whom were assigned to the treatment. 

The covariate vector $Z_i$ is composed of measurements of 16 pre-treatment outcomes. Four of these outcomes are associated with consumption: total monthly consumption and total monthly consumption on food, non-food, and durable commodities. Each consumption variable is measured in 2014 US dollars. We transform total monthly consumption to logs, base 10. Three of the variables are associated with assets, each of which is an index constructed from survey data measuring total assets, total productive assets, and total household assets. We transform the total assets measurement to logs, base 10. Five of the outcomes are associated with food security. These consist of four binary variables indicating different aspects of food security, e.g. did a child skip a meal, in addition to an index aggregating these measurements.\footnote{There are $2$ individuals with missing values for the food security index. We impute these values with the median values of the food security index.} The final four variables are associated with finance and income: the total amount of outstanding loans, the total amount of savings, income from agriculture, and total income from business. The covariate vector $X_i$ collects the total monthly consumption and assets for each individual. The outcome $Y_i$ measures the total assets two years after the initial asset transfer. Again, we transform these measurements to logs, base 10.

In \cref{fig: cate} and \cref{fig: half}, we set the subsample proportion $b/n$ equal to 0.05. In constructing the nuisance parameter estimate, we set $b/n$ = 0.025. We use $r=200$ bootstrap replicates throughout. We use 20,000 trees to construct \cref{fig: cate} and 2,000 trees in each bootstrap replicate to construct \cref{fig: half} and throughout the simulation. 

\subsection{Simulation Details\label{app: simulation}}

\subsubsection{Calibration\label{app: calibration}}
We calibrate a simulation to the \cite{banerjee2015multifaceted} data using a collection of GANs. This approach to simulation design was proposed by \cite{athey2021using}. Roughly speaking, a GAN is a pair of neural networks. The objective of the first network, the generator, is to generate data that looks like the \cite{banerjee2015multifaceted} data. The objective of the second network, the discriminator, is to discriminate between the real \cite{banerjee2015multifaceted} data and the data generated by the generator. These networks compete iteratively until convergence. The idea is that, after convergence, the generator is a good proxy for the true process that generated the  \cite{banerjee2015multifaceted} data. We use the ``WGAN'' package associated with \cite{athey2021using}.

To calibrate our simulation, we estimate three GANs. The first GAN estimates the distribution of the covariates $X_i$, i.e., baseline consumption and baseline total assets. The second GAN estimates the distribution of $Z_i$ conditional on $X_i$. Recall that $Z_i$ collects all baseline covariates, other than the covariates in $X_i$. The third GAN estimates the distribution of $Y_i$ conditioned on $Z_i$, $X_i$, and $W_i$. To generate an observation $D_i$, we generate $X_i$, generate $Z_i$ conditioned on $X_i$, and generate the potential outcomes $Y_i(1)$ and $Y_i(0)$ from the estimated distributions of $Y_i$ conditioned on $Z_i$, $X_i$, and $W_i = 1$ and $W_i = 0$, respectively. The treatment indicator $W_i$ is drawn i.i.d., Bernoulli with the observed probability in the \cite{banerjee2015multifaceted} data and we set $Y_i = Y_i(W_i)$. In this way, we know the true treatment effect $Y_i(1) - Y_i(0)$ for each unit in our simulation. We draw 10 million observations $D_i$ with this process. In the simulation, datasets of various sizes are sampled from this collection. 

\begin{figure}
\begin{centering}
\caption{Calibrated CATEs}
\label{fig: simulation cate}
\medskip{}
\begin{tabular}{c}
\includegraphics[scale=0.4]{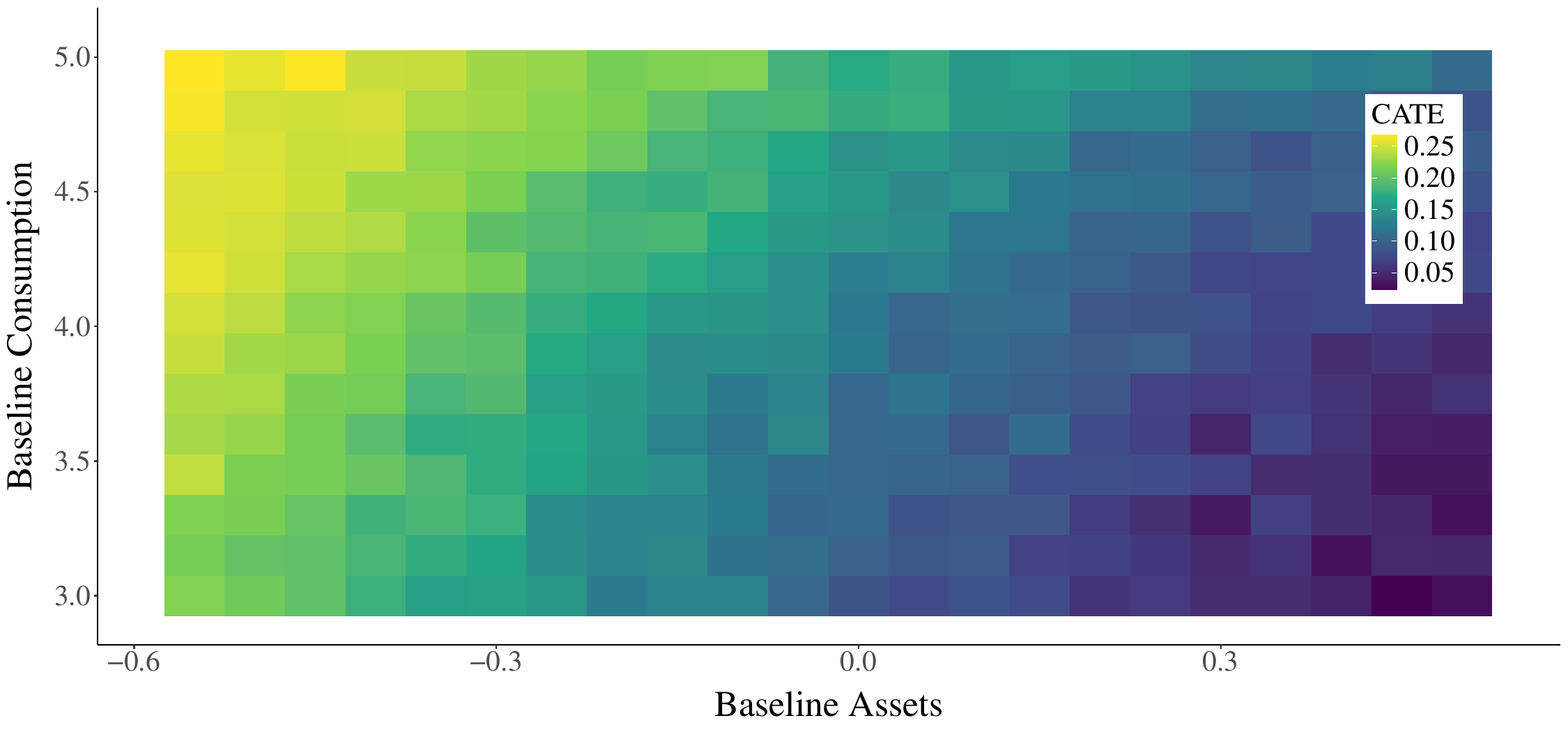}\tabularnewline
\end{tabular}
\par\end{centering}
\medskip{}
\justifying
{\footnotesize{}Notes: \cref{fig: simulation cate} displays the ``true'' value of CATE used in our calibrated simulation.}{\footnotesize\par}
\end{figure}

We use a related procedure to determine the true CATE $\theta_0(x)$ at each value $x$ in the query-vector $\mathbf{x}^{(d)}$ (i.e., the centers of each of the rectangles displayed in \cref{fig: cate}). Specifically, for each $x$ in $\mathbf{x}^{(d)}$, we draw 100,000 observations from the distribution of $Z_i$ conditioned on $X_i = x$. We then draw observations $Y_i(1)$ and $Y_i(0)$ for each of these replicates, and compute the average of the true treatment effects $Y_i(1) - Y_i(0)$. \cref{fig: simulation cate} displays these pseudo-true values of the CATE $\theta_0(x)$. Our simulation design captures many of the same features of the data recovered by GRF, but gives a somewhat smoother picture of the CATE. 

\begin{figure}
\caption{Validation}
\label{fig:validation}
\medskip{}
\begin{centering}
\begin{tabular}{c}
\includegraphics[scale=0.25]{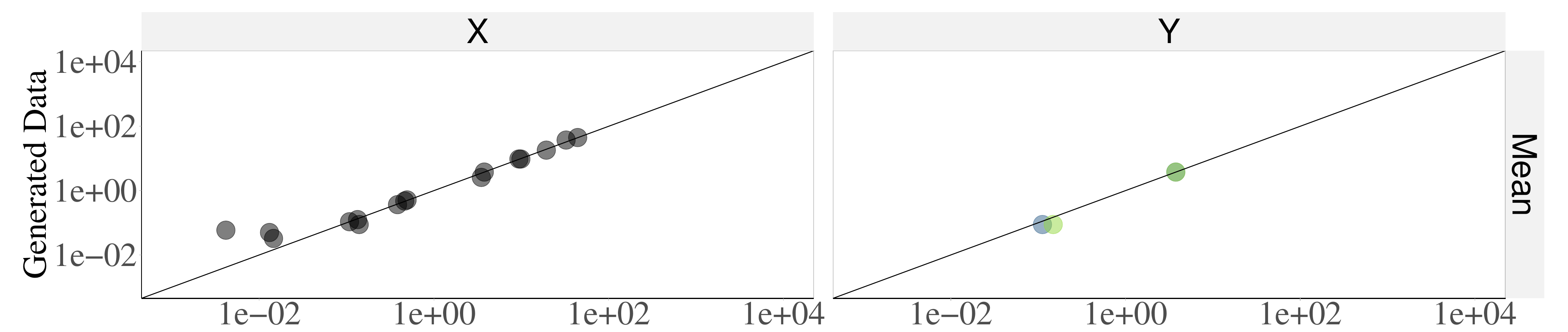}\tabularnewline
\includegraphics[scale=0.25]{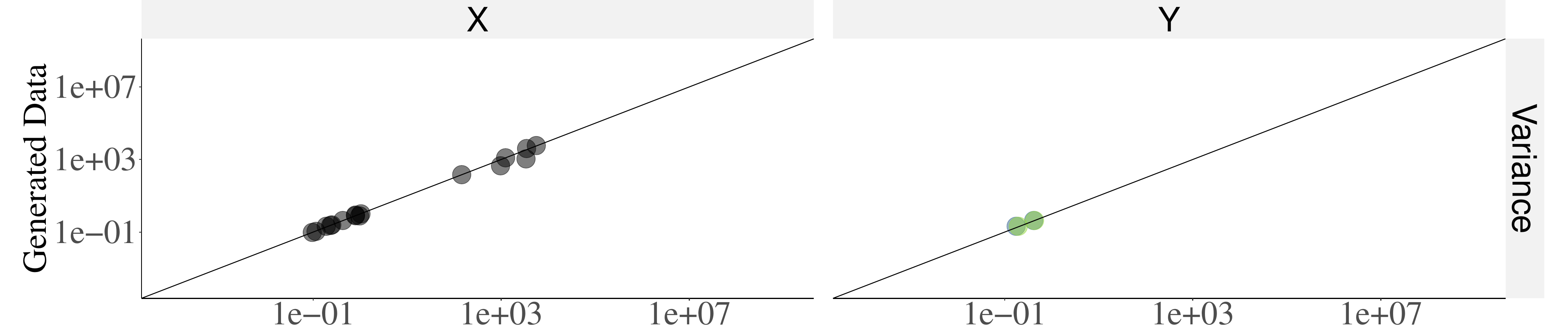}\tabularnewline
\includegraphics[scale=0.25]{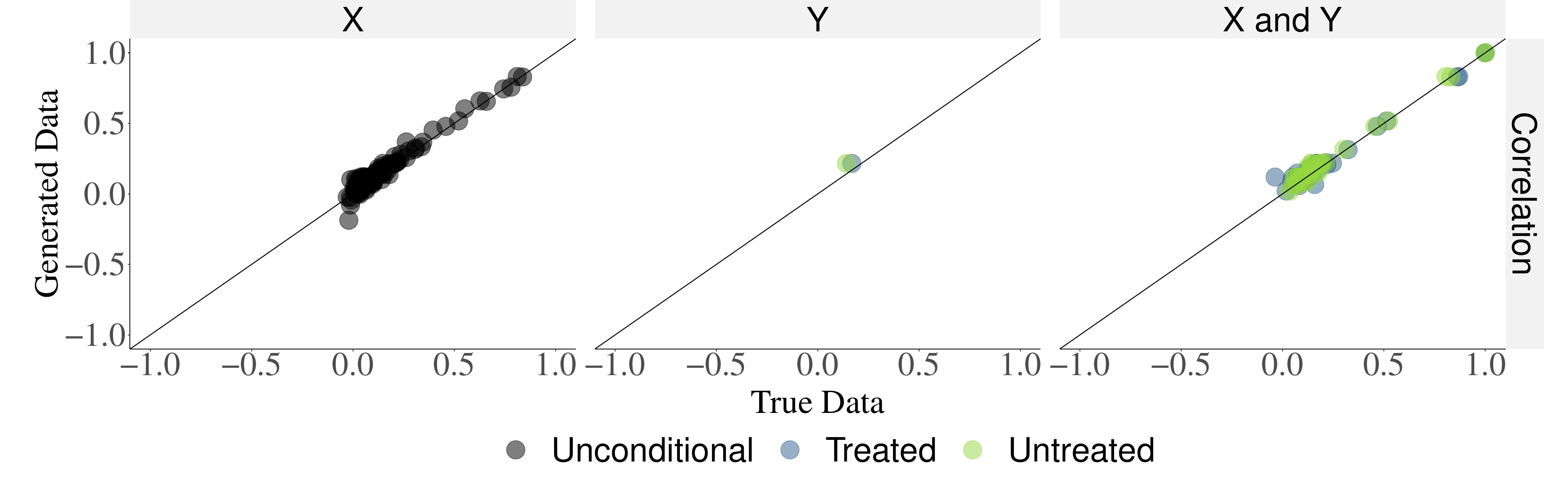}\tabularnewline
\end{tabular}
\par\end{centering}
\medskip{}
\justifying
{\footnotesize{}Notes: \cref{fig:validation} displays scatterplots comparing the moments of the data from \cite{banerjee2015multifaceted} to the GAN-generated simulation data. Columns differentiate between different types of variables. Rows differentiate between different types of moments. The x-axis of each sub-panel measures the moments of the true data. The y-axis of each sub-panel measures the moments of the generated data. The x and y axes in the first two rows are displayed in log-scale. A forty-five degree line is displayed in all sub-panels. Blue and green dots denote moments conditioned on treatment being set to one and zero, respectively. Black dots denote unconditioned moments.}{\footnotesize\par}
\end{figure}

\begin{figure}[h]
\begin{centering}
\caption{Covariate Density}
\label{fig: density}
\medskip{}
\begin{tabular}{c}
\textit{Panel A: Observed Covariates}\tabularnewline
\includegraphics[scale=0.4]{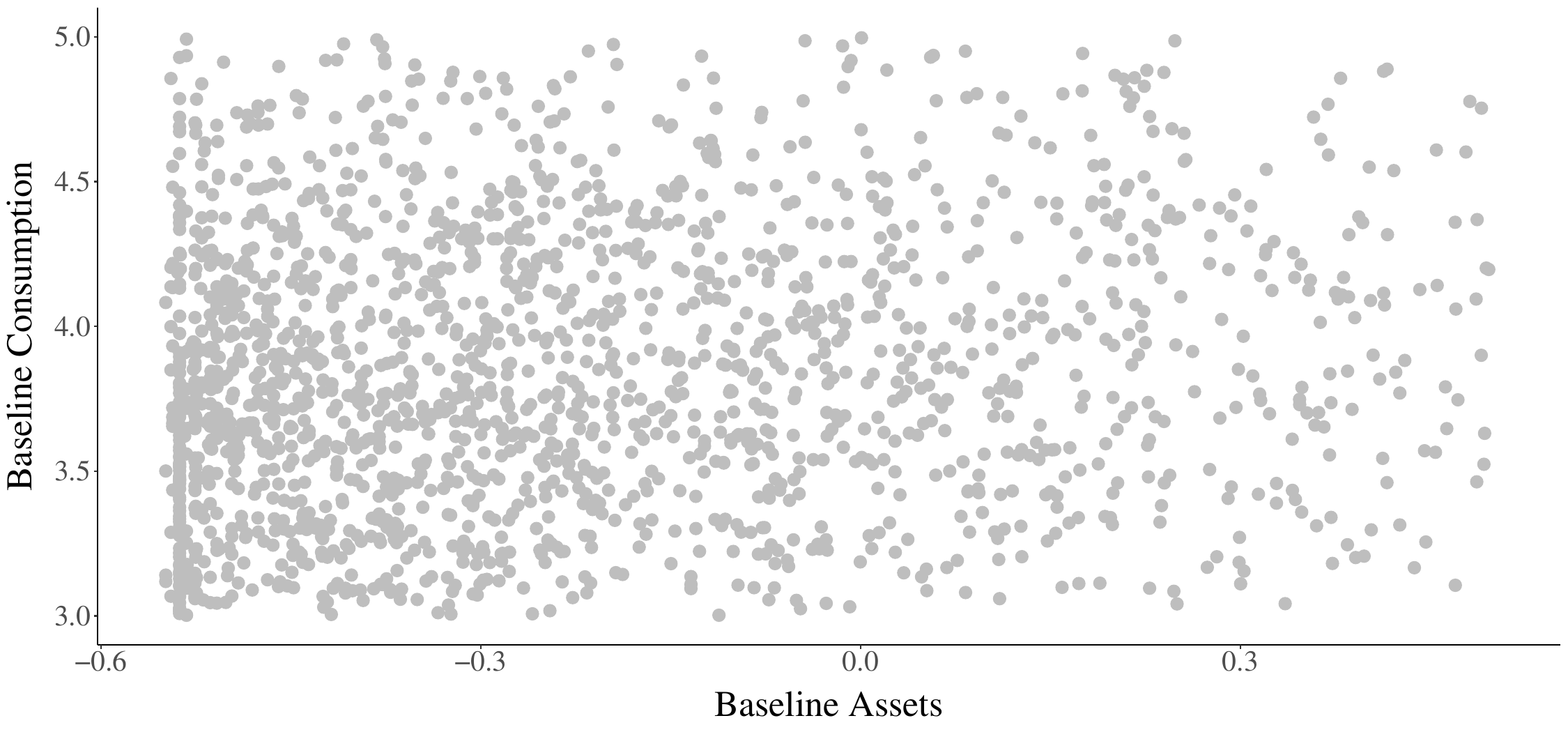}\tabularnewline
\textit{Panel B: Simulation Covariate Density}\tabularnewline
\includegraphics[scale=0.4]{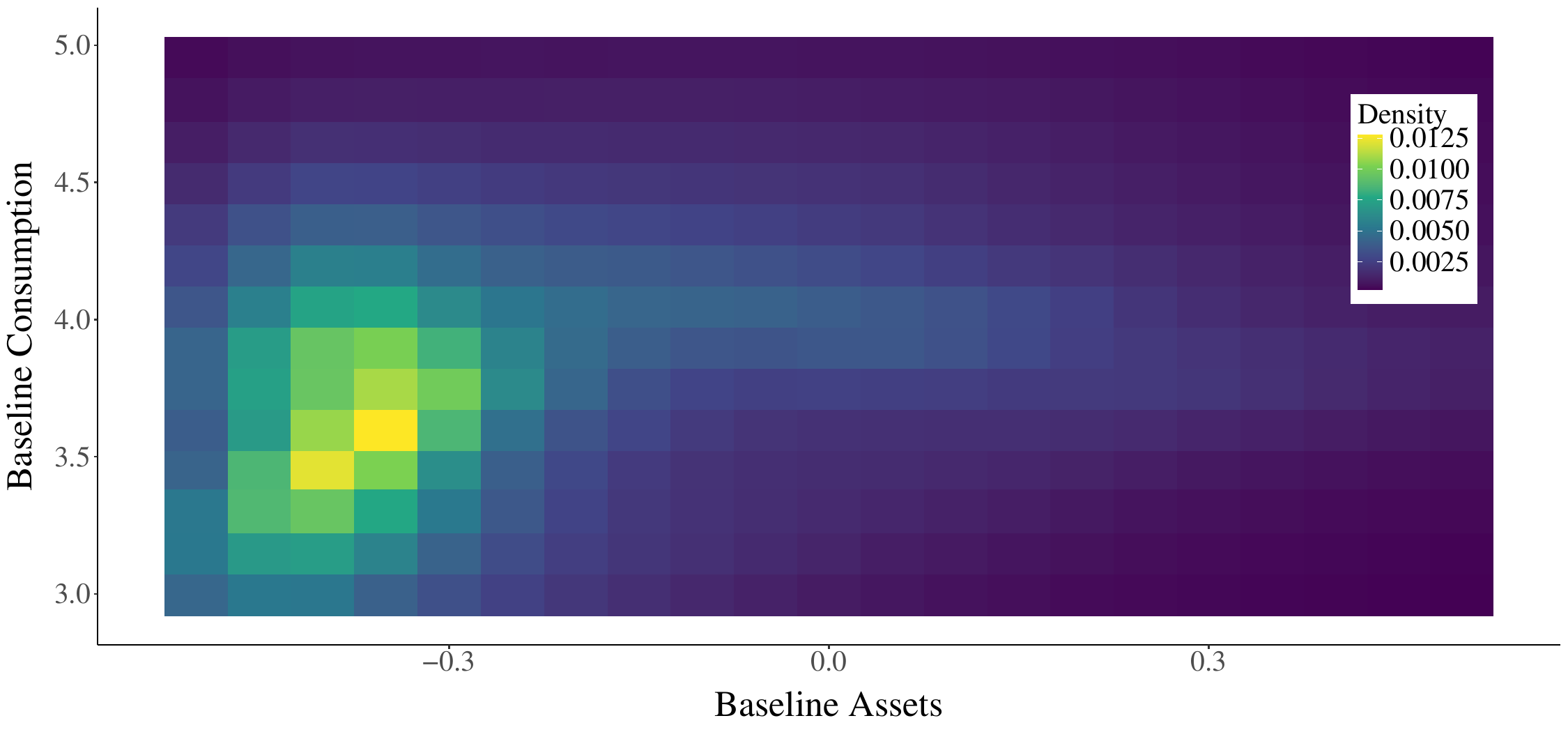}\tabularnewline
\end{tabular}
\par\end{centering}
\medskip{}
\justifying
{\footnotesize{}Notes: Panel A of \cref{fig: density} displays a scatter plot of the observed values of baseline consumption and baseline assets in the \cite{banerjee2015multifaceted} data. The horizontal and vertical axes display the baseline monthly consumption, normalized to 2014 dollars on a logarithmic scale base 10, and an index for baseline assets, respectively. Panel B displays a heat-map giving the density of the joint distribution of baseline consumption and baseline assets associated with our calibrated simulation.}{\footnotesize\par}
\end{figure}

\cref{fig:validation} displays a scatterplot comparing the moments from the \cite{banerjee2015multifaceted} data to those generated by our calibrated simulation. The distributions match quite closely. \cref{fig: density} compares a scatter plot of the observed values of baseline consumption and baseline assets in the \cite{banerjee2015multifaceted} data with a heat-map of the distribution of these covariates in our simulation. The limits of the horizontal and vertical axes in this Figure match \cref{fig: cate,fig: half} displayed in the main text. Some observations fall outside of the limits of this figure. The quartiles of baseline log consumption are 3.33, 3.76, and 4.20. The quartiles of baseline assets are -0.45, -0.71, and 0.03.

\subsubsection{Additional Results\label{app: sim results}}

\begin{figure}[t]
\begin{centering}
\caption{CATE Estimates, $2$-Fold Cross-Fitting}
\label{fig: cross cate}
\medskip{}
\begin{tabular}{c}
\includegraphics[scale=0.4]{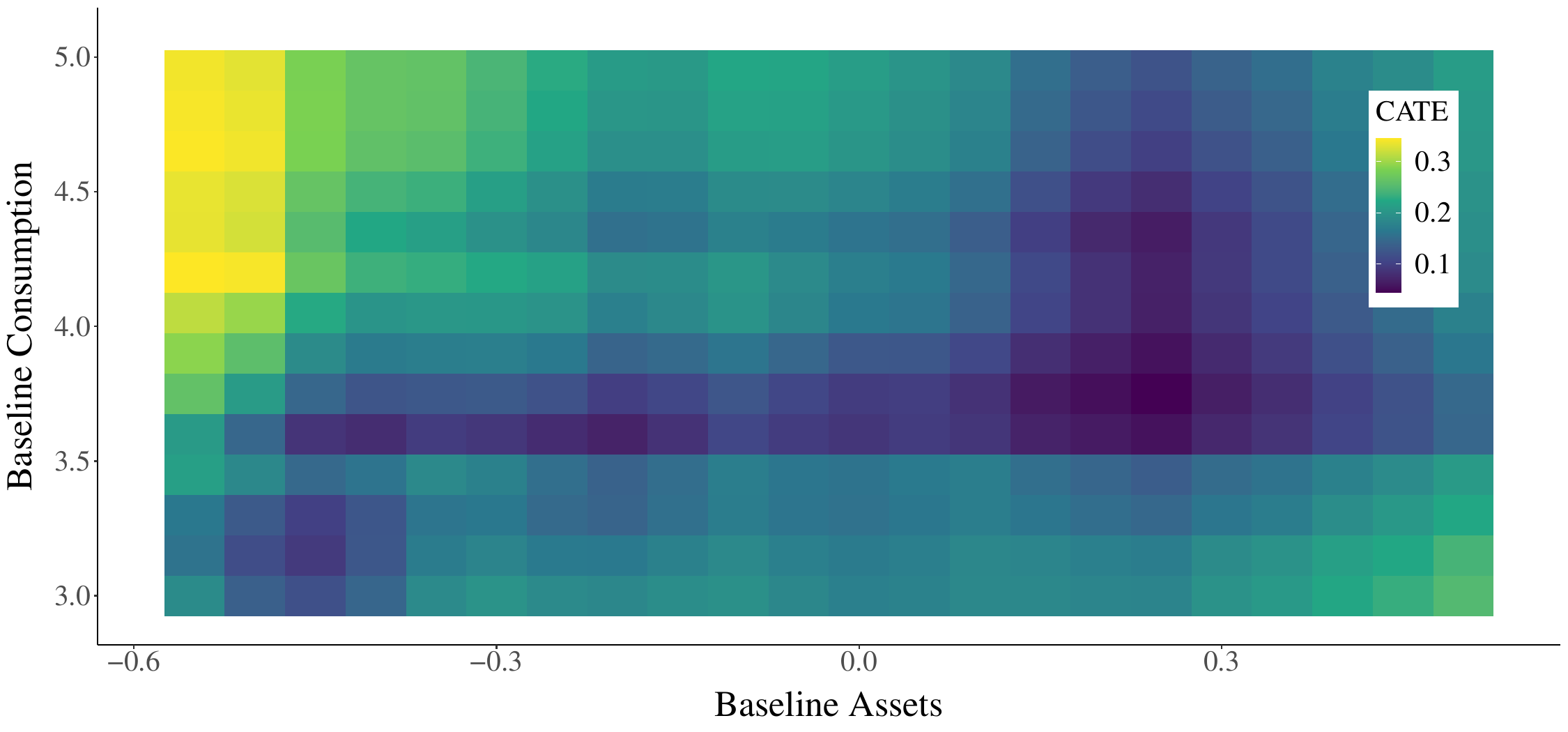}\tabularnewline
\end{tabular}
\par\end{centering}
\medskip{}
\justifying
{\footnotesize{}Notes: \cref{fig: cross cate} displays a heat map giving CATE estimates for the intervention studied in \cite{banerjee2015multifaceted} on post-treatment assets, using the $2$-fold cross-split estimator \eqref{eq: k fold cross fit estimator}. Other features of the display are analogous to \cref{fig: cate}.}{\footnotesize\par}
\end{figure}

\begin{figure}[t]
\begin{centering}
\caption{Half-Sample Confidence Region Lower Bound, $2$-Fold Cross-Fitting}
\label{fig: half cross}
\medskip{}
\begin{tabular}{c}
\includegraphics[scale=0.40]{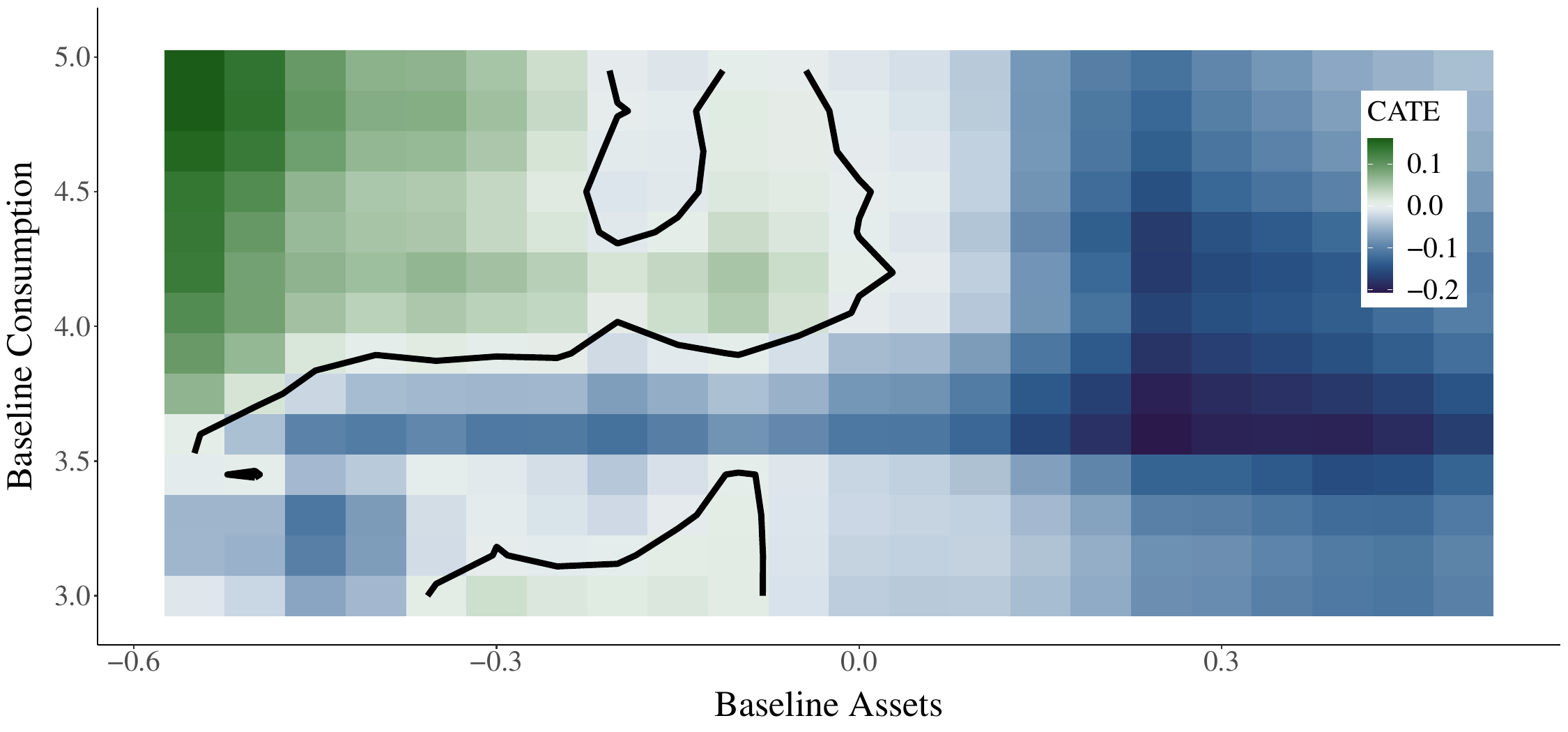}\tabularnewline
\end{tabular}
\par\end{centering}
\medskip{}
\justifying
{\footnotesize{}Notes: \cref{fig: half cross} displays a heat map giving half-sample lower confidence bound for the CATE of the intervention studied in \cite{banerjee2015multifaceted} on post-treatment total assets.  Other features of the display are analogous to \cref{fig: half}.}{\footnotesize\par}
\end{figure}

\cref{fig: cross cate} and \cref{fig: half cross} give versions of \cref{fig: cate} and \cref{fig: half}, constructed using the $2$-fold cross-split estimator \eqref{eq: k fold cross fit estimator} and the $2$-fold cross-split bootstrap root \eqref{eq: half-sample k-fold root}. The estimates and confidence bounds are quantitatively and qualitatively very similar. 

\begin{figure}[t]
\begin{centering}
\caption{Performance, $2$-Fold Cross-Fitting}
\label{fig: cross performance}
\medskip{}
\begin{tabular}{c}
\includegraphics[scale=0.38]{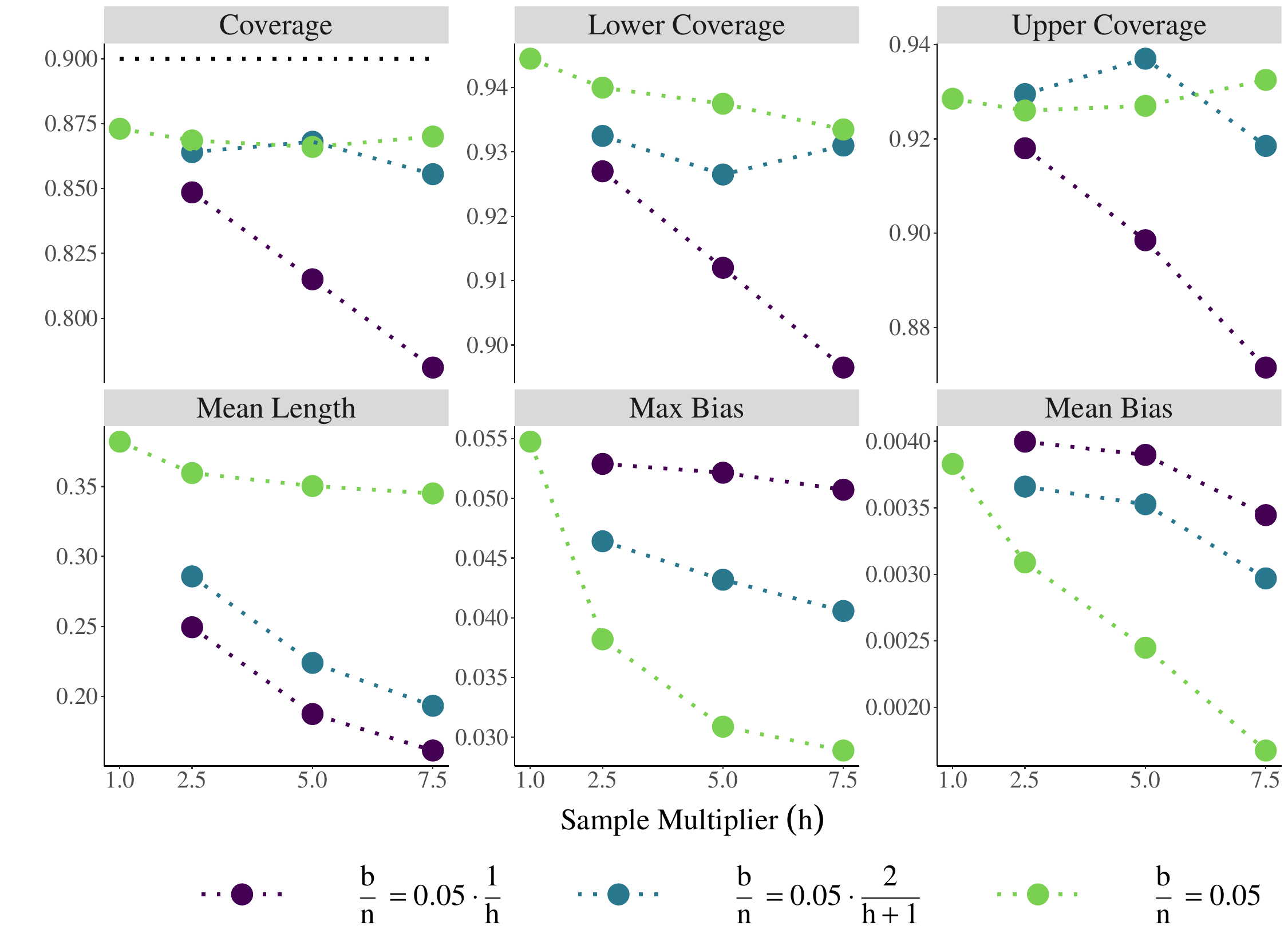}
\end{tabular}
\par\end{centering}
\medskip{}
\justifying
{\footnotesize{}Notes: \cref{fig: cross performance} displays several measurements of the performance of the confidence intervals formulated in \cref{def: uniform ci}, constructed using the $2$-fold cross-split estimator \eqref{eq: k fold cross fit estimator} and the $2$-fold cross-split bootstrap root \eqref{eq: half-sample k-fold root}, in a simulation calibrated to the \cite{banerjee2015multifaceted} data. Other features of the display are analogous to \cref{fig: performance}.}{\footnotesize\par}
\end{figure}

\cref{fig: cross performance} displays results for the simulation presented in \cref{sec: simulation}, analogous to \cref{fig: performance}, for the $2$-fold cross-split estimator \eqref{eq: k fold cross fit estimator} and the $2$-fold cross-split bootstrap root \eqref{eq: half-sample k-fold root}. The confidence region is very slightly anti-conservative, so long as $b/n$ is decreasing as $n$ increases. The measurements of bias and variance exhibit patterns very similar to the patterns displayed in \cref{fig: performance}.

\section{Proofs Supporting \cref{app: additional}\label{app: bin boot proofs}}

\subsection{Proof of \cref{thm: no split stochastic equicontinuity}}

Throughout, we let $x$ denote an arbitrary element of $\bm{x}^{(d)}$ and use the shorthand $N_b = {n \choose b}$. We are interested in giving a higher-order moment bound for the quantity
\begin{flalign}
\frac{1}{N_b}\sum_{\mathsf{s}\in\mathcal{S}_{n,b}}F(x;D_{\mathsf{s}},\hat{g}_{n})
&=
\frac{1}{N_b}\sum_{\mathsf{s}\in\mathcal{S}_{n,b}}
\sum_{i\in\mathsf{s}}
\bigg(
a(x;D_i,D_{\mathsf{s}},\hat g_n)
-
a(x;D_i,D_{\mathsf{s}},g_0) \nonumber\\
&\quad\quad\quad\quad\quad\quad
\quad\quad-
\left(
A(x;\hat g_n)-A(x;g_0)
\right)
\bigg)~.
\label{eq: object of interest}
\end{flalign}
We begin by decomposing \eqref{eq: object of interest} into three terms that will be easier to handle in isolation. In particular, by the representation \eqref{eq: Hoeff perm rep} the quantity \eqref{eq: object of interest} can be re-expressed as
\begin{flalign}
\frac{1}{N_{b}}\sum_{\mathsf{s}\in\mathcal{S}_{n,b}}F(x;D_{\mathsf{s}},\hat{g}_{n}) & =\frac{1}{n!}\sum_{\pi\in\mathcal{P}_{n}}\bigg\lfloor\frac{n}{b}\bigg\rfloor^{-1}\sum_{l=1}^{\lfloor n/b\rfloor}F_{1}(x;\mathsf{s}_{\pi,l})\label{eq: F_1}\\
 & -\frac{1}{n!}\sum_{\pi\in\mathcal{P}_{n}}\bigg\lfloor\frac{n}{b}\bigg\rfloor^{-1}\sum_{l=1}^{\lfloor n/b\rfloor}F_{2}(x;\mathsf{s}_{\pi,l})\label{eq: F_2}\\
 & +\frac{1}{n!}\sum_{\pi\in\mathcal{P}_{n}}\bigg\lfloor\frac{n}{b}\bigg\rfloor^{-1}\sum_{l=1}^{\lfloor n/b\rfloor}F_{3}(x;\mathsf{s}_{\pi,l})\label{eq: F_3}
\end{flalign}
where
\begin{flalign*}
F_{1}(x;\mathsf{s}) & =\sum_{i\in\mathsf{s}}a(x;D_{i},D_{\mathsf{s}},\hat{g}_{n})-a(x;D_{i},D_{\mathsf{s}},\hat{g}_{n}^{(-\mathsf{s})})\\
F_{2}(x;\mathsf{s}) & =\sum_{i\in\mathsf{s}}A(x;\hat{g}_{n})-A(x;\hat{g}_{n}^{(-\mathsf{s})}),\quad\text{and}\\
F_{3}(x;\mathsf{s}) & =\sum_{i\in\mathsf{s}}a(x;D_{i},D_{\mathsf{s}},\hat{g}_{n}^{(-\mathsf{s})})-a(x;D_{i},D_{\mathsf{s}},g_{0})\\
 & \quad\quad\quad\quad\quad\quad\quad\quad\quad\quad\quad\quad-(A(x;\hat{g}_{n}^{(-\mathsf{s})})-A(x;g_{0})),
\end{flalign*}
respectively. We give suitable probability bounds for the terms in
\eqref{eq: F_1} through \eqref{eq: F_3}. 

We begin by considering the term (\ref{eq: F_1}). Fix any integer
$q\geq1$. We show that
\begin{equation}
\Big\|\frac{1}{n!}\sum_{\pi\in\mathcal{P}_{n}}\bigg\lfloor\frac{n}{b}\bigg\rfloor^{-1}\sum_{l=1}^{\lfloor n/b\rfloor}F_{1}(x;\mathsf{s}_{\pi,l})\Big\|_{2q}\lesssim2qn^{-1}b\gamma(2q)b^{1/2q}~.\label{eq: posited F 1 norm bound}
\end{equation}
To this end, observe that 
\begin{flalign}
\|\frac{1}{n!}\sum_{\pi\in\mathcal{P}_{n}}\bigg\lfloor\frac{n}{b}\bigg\rfloor^{-1}\sum_{l=1}^{\lfloor n/b\rfloor}F_{1}(x;\mathsf{s}_{\pi,l})\|_{2q} & \leq\|\sum_{i=1}^{b}\kappa(x,X_{i};D_{[b]})(m(D_{i},\hat{g}_{n})-m(D_{i},\hat{g}_{n}^{(-[b])}))\|_{2q}\nonumber \\
 & \leq\|\max_{i\in[b]}\vert m(D_{i},\hat{g}_{n})-m(D_{i},\hat{g}_{n}^{(-[b])})\vert^{2}\|_{q}^{1/2},\label{eq: jensen and kernel}
\end{flalign}
where the first inequality follows from Jensen's inequality and the
second inequality follows from \cref{assu: kernel restriction}, Part
(ii). Lemma 2.2.2 of \citet{van1996weak}, implies that, if $A_{1},\ldots,A_{n}$
are any collection of real-valued random variables, then
\begin{equation}
\|\max_{i\in[n]}\vert A_{i} \vert \|_{q}\lesssim n^{1/q}\max_{i\in[n]}\|\vert A_{i} \vert \|_{q}.\label{eq: L_q maximal}
\end{equation}
Thus, we have that 
\begin{alignat*}{1}
\|\max_{i\in[b]}\vert m(D_{i},\hat{g}_{n})-m(D_{i},\hat{g}_{n}^{(-[b])})\vert ^{2}\|_{q}^{1/2} & \leq b^{1/(2q)}\|m(D_{i},\hat{g}_{n})-m(D_{i},\hat{g}_{n}^{(-[b])})\|_{2q}.\\
 & \lesssim2qn^{-1}b\gamma(2q)b^{1/2q}
\end{alignat*}
where the second inequality follows from Assumption \ref{assu: stability}. Hence, the condition (\ref{eq: posited F 1 norm bound}) holds.

We now turn to the term \eqref{eq: F_2}. In this case, we show that
\begin{equation}
\Big\|\frac{1}{n!}\sum_{\pi\in\mathcal{P}_{n}}\bigg\lfloor\frac{n}{b}\bigg\rfloor^{-1}
\sum_{l=1}^{\lfloor n/b\rfloor}F_{2}(x;\mathsf{s}_{\pi,l})\Big\|_{2q}
\lesssim
2q n^{-1}b\gamma(2q)b^{1/2q}~.
\label{eq: posited F 2 norm bound}
\end{equation}
To this end, let \(\mathbf{D}_{n}^{\prime\prime}=(D_i^{\prime\prime})_{i=1}^n\) be another independent copy of \(\mathbf{D}_n\). Observe that
\begin{flalign}
&\Big\|
\frac{1}{n!}\sum_{\pi\in\mathcal{P}_{n}}\bigg\lfloor\frac{n}{b}\bigg\rfloor^{-1}
\sum_{l=1}^{\lfloor n/b\rfloor}F_{2}(x;\mathsf{s}_{\pi,l})
\Big\|_{2q} \nonumber\\
&\quad\leq
\left\|
\sum_{i=1}^{b}
\left(
A(x;\hat g_n)-A(x;\hat g_n^{(-[b])})
\right)
\right\|_{2q} \nonumber\\
&\quad\leq
\left\|
\sum_{i=1}^{b}
\kappa(x,X_i^{\prime\prime};D_{[b]}^{\prime\prime})
\left(
m(D_i^{\prime\prime},\hat g_n)
-
m(D_i^{\prime\prime},\hat g_n^{(-[b])})
\right)
\right\|_{2q} \nonumber\\
&\quad\leq
\left\|
\max_{i\in[b]}
\left|
m(D_i^{\prime\prime},\hat g_n)
-
m(D_i^{\prime\prime},\hat g_n^{(-[b])})
\right|^2
\right\|_{q}^{1/2} \nonumber\\
&\quad\leq
b^{1/(2q)}
\left\|
m(D_i^{\prime\prime},\hat g_n)
-
m(D_i^{\prime\prime},\hat g_n^{(-[b])})
\right\|_{2q} \nonumber\\
&\quad\lesssim
2qn^{-1}b\gamma(2q)b^{1/2q}~,
\end{flalign}
where the first two inequalities follow from Jensen's inequality, the third inequality from \cref{assu: kernel restriction}, Part (ii), the fourth inequality follows from \eqref{eq: L_q maximal}, and the final inequality follows from the independent-copy stability bound in \cref{assu: stability}. Hence, the condition (\ref{eq: posited F 2 norm bound}) holds.

Finally, we consider the term \eqref{eq: F_3}. The argument here is
somewhat more involved. To simplify exposition, let $k=\lfloor n/b\rfloor$
and define the sets $\mathsf{s}_{l}=\left\{ (l-1)b+1,\ldots,bl\right\} $
for each $l$ in $1,\ldots,k$. It will suffice to show that 
\begin{equation}
\left\|
\frac{1}{k}\sum_{l=1}^{k}F_{3}(x;\mathsf{s}_{l})
\right\|_{q}
\lesssim
\sqrt{\frac{1}{k}}
\left(
\left(\frac{1}{k}\right)^{1/4}\delta_{n,g}
+
\varepsilon_{b}^{1/2}
+
b^{1/q}\gamma(q)
\right)q^{3/2}
\label{eq: F_3 L_q to show}
\end{equation}
as Jensen's inequality implies that 
\begin{equation}
\left\|
\frac{1}{n!}\sum_{\pi\in\mathcal{P}_{n}}\bigg\lfloor\frac{n}{b}\bigg\rfloor^{-1}
\sum_{l=1}^{\lfloor n/b\rfloor}F_{3}(x;\mathsf{s}_{\pi,l})
\right\|_{q}
\leq
\left\|
\frac{1}{k}\sum_{l=1}^{k}F_{3}(x;\mathsf{s}_{l})
\right\|_{q}~.
\label{eq: F 3 L_q Jensen}
\end{equation}
To establish the bound \eqref{eq: F_3 L_q to show}, we apply the
following generalized Efron-Stein inequality, due to \citet{boucheron2005moment}.
See also Chapter 15.2 of \citet{boucheron2013concentration}.
\begin{lemma}[{\citealp[Theorem 2, ][]{boucheron2005moment}}]
\label{thm: generalized efron stein}
Let $X=(X_{i})_{i=1}^{n}$ be
a sequence of independent random variables and let $X^{\prime}=(X_{i}^{\prime})_{i=1}^{n}$
denote an independent copy of $X$. Let $X^{(-i)}$ be constructed
by taking $X$ and replacing $X_{i}$ with $X_{i}^{\prime}$. Consider
the random variable $f(X)$, where $f(\cdot)$ is any real-valued
function. It holds that
\begin{flalign*}
\|f(X)-\mathbb{E}\left[f(X)\right]\|_{q}
& \lesssim
\sqrt{q}
\left\|
\sum_{i=1}^{n}
\left(f(X)-f(X^{(-i)})\right)^{2}
\right\|_{q/2}^{1/2}
\end{flalign*}
for any $q\geq2$. 
\end{lemma}

\noindent Before applying \cref{thm: generalized efron stein}, observe that
\begin{equation}
\mathbb{E}\left[
\frac{1}{k}\sum_{l=1}^{k}F_{3}(x;\mathsf{s}_{l})
\right]=0~.
\label{eq: F3 mean zero}
\end{equation}
Indeed, conditional on \(\hat{g}_{n}^{(-\mathsf{s}_{l})}\), the block
\(D_{\mathsf{s}_{l}}\) is independent of \(\hat{g}_{n}^{(-\mathsf{s}_{l})}\).
Thus, by Honesty we find,
\begin{align}
&\mathbb{E}\bigg[
\sum_{i\in\mathsf{s}_{l}}
(
a(x;D_i,D_{\mathsf{s}_{l}},\hat{g}_{n}^{(-\mathsf{s}_{l})})
-
a(x;D_i,D_{\mathsf{s}_{l}},g_0)
- \nonumber \\
& \quad\quad\quad\quad
(
A(x;\hat{g}_{n}^{(-\mathsf{s}_{l})})
-
A(x;g_0)
)
)
\mid|
\hat{g}_{n}^{(-\mathsf{s}_{l})}
\bigg]=0~.
\end{align}
Taking expectations verifies \eqref{eq: F3 mean zero}.

To apply \cref{thm: generalized efron stein}, let
$\hat{g}_{n}^{(-\mathsf{s}_{l},-\mathsf{s}_{r})}$ denote a version
of the estimator $\hat{g}_{n}$ formed with all observations in $\mathbf{D}_{n}$,
except that the observations $D_{\mathsf{s}_{l}}$ and $D_{\mathsf{s}_{r}}$
are replaced by the observations $D_{\mathsf{s}_{l}}^{\prime}$ and
$D_{\mathsf{s}_{r}}^{\prime\prime}$, respectively. Additionally,
let $\mathbf{D}_{n}^{\prime\prime\prime}$ be an additional, independent,
copy of $\mathbf{D}_{n}$ and let $\tilde{g}_{n}^{(-\mathsf{s}_{l})}$
denote a version of the estimator $\hat{g}_{n}$ formed with all observations
in $\mathbf{D}_{n}$, except that the observations $D_{\mathsf{s}_{l}}$
are replaced by the observations $D_{\mathsf{s}_{l}}^{\prime\prime\prime}$.
Let the quantity $F_{3}^{(-r)}(x;\mathsf{s}_{l})$ be given by 
\begin{flalign*}
F_{3}^{(-r)}(x;\mathsf{s}_{l})
&=
\sum_{i\in\mathsf{s}_{l}}
\bigg(
a(x;D_{i},D_{\mathsf{s}_{l}},\hat{g}_{n}^{(-\mathsf{s}_{l},-\mathsf{s}_{r})})
-
a(x;D_{i},D_{\mathsf{s}_{l}},g_{0})\\
&\quad\quad\quad\quad\quad\quad
-
\left(
A(x;\hat{g}_{n}^{(-\mathsf{s}_{l},-\mathsf{s}_{r})})
-
A(x;g_{0})
\right)
\bigg)
\end{flalign*}
if \(r\neq l\), and by 
\begin{flalign*}
F_{3}^{(-l)}(x;\mathsf{s}_{l})
&=
\sum_{i\in\mathsf{s}_{l}}
\bigg(
a(x;D_{i}^{\prime\prime},D_{\mathsf{s}_{l}}^{\prime\prime},\tilde{g}_{n}^{(-\mathsf{s}_{l})})
-
a(x;D_{i}^{\prime\prime},D_{\mathsf{s}_{l}}^{\prime\prime},g_{0})\\
&\quad\quad\quad\quad\quad\quad
-
\left(
A(x;\tilde{g}_{n}^{(-\mathsf{s}_{l})})
-
A(x;g_{0})
\right)
\bigg)
\end{flalign*}
otherwise. Combining \eqref{eq: F3 mean zero} with \cref{thm: generalized efron stein} implies that
\begin{flalign}
 & \left\|
 \frac{1}{k}\sum_{l=1}^{k}F_{3}(x;\mathsf{s}_{l})
 \right\|_{q}\nonumber \\
 & \leq
 \sqrt{q}
 \bigg\|
 \sum_{l=1}^{k}
 \left(
 \frac{1}{k}
 \left(
 F_{3}(x;\mathsf{s}_{l})
 -
 F_{3}^{(-l)}(x;\mathsf{s}_{l})
 \right)
 +
 \frac{1}{k}
 \sum_{r\neq l}
 \left(
 F_{3}(x;\mathsf{s}_{r})
 -
 F_{3}^{(-l)}(x;\mathsf{s}_{r})
 \right)
 \right)^{2}
 \bigg\|_{q/2}^{1/2}~.
 \label{eq: apply generalized efron-stein}
\end{flalign}
\noindent We obtain a suitable bound for \eqref{eq: apply generalized efron-stein}
through the application of the following Lemma.
\begin{lemma}
\label{lem: V momen bound}
Suppose that the conditions of \cref{thm: no split stochastic equicontinuity} hold. Then
\begin{flalign}
 & \bigg\|
 \sum_{l=1}^{k}
 \left(
 \frac{1}{k}
 \left(
 F_{3}(x^{(j)};\mathsf{s}_{l})
 -
 F_{3}^{(-l)}(x^{(j)};\mathsf{s}_{l})
 \right)
 +
 \frac{1}{k}
 \sum_{r\neq l}
 \left(
 F_{3}(x^{(j)};\mathsf{s}_{r})
 -
 F_{3}^{(-l)}(x^{(j)};\mathsf{s}_{r})
 \right)
 \right)^{2}
 \bigg\|_{q/2}^{1/2}\nonumber \\
 & \lesssim
 \sqrt{\frac{1}{k}}
 \left(
 \left(\frac{1}{k}\right)^{1/4}\delta_{n,g}
 +
 \varepsilon_{b}^{1/2}
 +
 b^{1/q}\gamma(q)
 \right)q
 \label{eq: V moment bound state}
\end{flalign}
for each \(j\in[d]\) and each even integer \(q\).
\end{lemma}
\noindent Consequently, the bound \eqref{eq: apply generalized efron-stein}
and \cref{lem: V momen bound} imply \eqref{eq: F_3 L_q to show}.

Putting the pieces together, by the decomposition \eqref{eq: F_1}--\eqref{eq: F_3}, Minkowski's inequality, and the bounds \eqref{eq: posited F 1 norm bound}, \eqref{eq: posited F 2 norm bound}, and \eqref{eq: F_3 L_q to show}, we find that
\begin{align}
&\left\|
\frac{1}{N_b}
\sum_{\mathsf{s}\in\mathcal{S}_{n,b}}
F(x;D_{\mathsf{s}},\hat{g}_n)
\right\|_q \nonumber\\
&\quad\quad\lesssim
q\frac{b}{n}\gamma(2q)b^{1/(2q)}
+
q^{3/2}\sqrt{\frac{1}{k}}
\left(
\left(\frac{1}{k}\right)^{1/4}\delta_{n,g}
+
\varepsilon_b^{1/2}
+
b^{1/q}\gamma(q)
\right) \nonumber\\
&\quad\quad\lesssim
q^{3/2}\sqrt{\frac{b}{n}}
\left(
\left(\frac{b}{n}\right)^{1/4}\delta_{n,g}
+
\varepsilon_b^{1/2}
+
b^{1/q}\gamma(2q)
\right)~,
\label{eq: final D1 moment bound}
\end{align}
where the final inequality follows from \(k=\lfloor n/b\rfloor\), the monotonicity of \(\gamma(\cdot)\), and the fact that \(b\leq n\).\hfill$\blacksquare$

\subsection{Proof of Lemma \ref{lem: V momen bound}}

Throughout, we let $x$ denote an arbitrary element of $\bm{x}^{(d)}$ and use the shorthand $N_b = {n \choose b}$.
Observe that
\begin{flalign}
 & \bigg\|\sum_{l=1}^{k}\left(\frac{1}{k}\left(F_{3}(x;\mathsf{s}_{l})-F_{3}^{(-l)}(x^{(j)};\mathsf{s}_{l})\right)+\frac{1}{k}\sum_{r\neq l}\left(F_{3}(x;\mathsf{s}_{r})-F_{3}^{(-l)}(x;\mathsf{s}_{r})\right)\right)^{2}\bigg\|_{q/2}\label{eq: V moment decomp}\\
 & \leq\sum_{l=1}^{k}\bigg\|\left(\frac{1}{k}\left(F_{3}(x;\mathsf{s}_{l})-F_{3}^{(-l)}(x^{(j)};\mathsf{s}_{l})\right)+\frac{1}{k}\sum_{r\neq l}\left(F_{3}(x;\mathsf{s}_{r})-F_{3}^{(-l)}(x;\mathsf{s}_{r})\right)\right)^{2}\bigg\|_{q/2}\nonumber \\
 & \lesssim\bigg\|\sum_{l=1}^{k}\frac{1}{k^{2}}\left(F_{3}(x;\mathsf{s}_{l})-F_{3}^{(-l)}(x^{(j)};\mathsf{s}_{l})\right)^{2}+\frac{1}{k}\sum_{l=1}^{k}\sum_{r\neq l}\left(F_{3}(x;\mathsf{s}_{r})-F_{3}^{(-l)}(x;\mathsf{s}_{r})\right)^{2}\|_{q/2}\nonumber \\
 & \lesssim\frac{1}{k}\|F_{3}(x;\mathsf{s}_{l})\|_{q}^{2}\label{eq: p moment bound}\\
 & +k\|F_{3}(x;\mathsf{s}_{r})-F_{3}^{(-l)}(x;\mathsf{s}_{r})\|_{q}^{2},\label{eq: p moment stability bound}
\end{flalign}
where the first inequality follows from the triangle inequality, the
second inequality follows from Cauchy-Schwarz and Jensen's inequality,
and the final inequality follows from the triangle inequality. 

Consider the term \eqref{eq: p moment bound}. Observe that
\begin{flalign}
 & \mathbb{E}\left[\vert F_{3}(x;D_{\mathsf{s}_{1}})\vert^{q}\right]\nonumber \\
 & =
\mathbb{E}\left[
\bigg\vert
\sum_{i\in\mathsf{s}_{1}}
\bigg(
a(x;D_i,D_{\mathsf{s}_{1}},\hat{g}_{n}^{(-\mathsf{s}_{1})})
-
a(x;D_i,D_{\mathsf{s}_{1}},g_0)
-
\left(A(x;\hat{g}_{n}^{(-\mathsf{s}_{1})})-A(x;g_0)\right)
\bigg)
\bigg\vert^q
\right]\nonumber\\
&\lesssim
\mathbb{E}\left[
\bigg\vert
\sum_{i\in\mathsf{s}_{1}}
\left(
a(x;D_i,D_{\mathsf{s}_{1}},\hat{g}_{n}^{(-\mathsf{s}_{1})})
-
a(x;D_i,D_{\mathsf{s}_{1}},g_0)
\right)
\bigg\vert^q
\right]\label{eq: unconditional term a}\\
&\quad+
\mathbb{E}\left[
\bigg\vert
\sum_{i\in\mathsf{s}_{1}}
\left(
A(x;\hat{g}_{n}^{(-\mathsf{s}_{1})})-A(x;g_0)
\right)
\bigg\vert^q
\right]\label{eq: conditional term A}
\end{flalign}
by the triangle inequality. We begin by bounding the term \eqref{eq: unconditional term a}. Observe that
\begin{flalign}
& \mathbb{E}\left[
\bigg\vert
\sum_{i\in\mathsf{s}_{1}}
\kappa(x,X_i;D_{\mathsf{s}_{1}})
\left(
m(D_i,\hat{g}_{n}^{(-\mathsf{s}_{1})})-m(D_i,g_0)
\right)
\bigg\vert^q
\right]\nonumber\\
&\leq
\mathbb{E}\left[
\sum_{i\in\mathsf{s}_{1}}
\kappa(x,X_i;D_{\mathsf{s}_{1}})
\left|
m(D_i,\hat{g}_{n}^{(-\mathsf{s}_{1})})-m(D_i,g_0)
\right|^q
\right]\nonumber\\
&\leq
\mathbb{E}\left[
\mathbb{E}\left[
\sum_{i\in\mathsf{s}_{1}}
\kappa(x,X_i;D_{\mathsf{s}_{1}})
\mathbb{E}\left[
\left|
m(D_i,\hat{g}_{n}^{(-\mathsf{s}_{1})})-m(D_i,g_0)
\right|^q
\mid X_i,\hat{g}_{n}^{(-\mathsf{s}_{1})}
\right]
\mid \hat{g}_{n}^{(-\mathsf{s}_{1})}
\right]
\right]\nonumber\\
&\lesssim
\mathbb{E}\left[
\|\hat{g}_{n}^{(-\mathsf{s}_{1})}-g_0\|_{q,\infty}^q
\right]
+
\varepsilon_b^{q/2},\label{eq: unconditional A label}
\end{flalign}
where the first inequality follows from Jensen's inequality, the second inequality follows from Honesty, i.e., \cref{assu: kernel restriction}, Part (i), and the final inequality follows from \cref{assu: mean-square-continuity}, \cref{assu: higher-order shrinkage}, and \cref{assu: kernel restriction}, Part (ii).

To handle the second term \eqref{eq: conditional term A}, let \(\mathbf{D}_{n}^{\prime\prime}=(D_i^{\prime\prime})_{i=1}^n\) be another independent copy of \(\mathbf{D}_n\). By Jensen's inequality,
\begin{flalign}
&\mathbb{E}\left[
\bigg\vert
\sum_{i\in\mathsf{s}_{1}}
\left(
A(x;\hat{g}_{n}^{(-\mathsf{s}_{1})})-A(x;g_0)
\right)
\bigg\vert^q
\right]\nonumber\\
&\leq
\mathbb{E}\left[
\bigg\vert
\sum_{i\in\mathsf{s}_{1}}
\kappa(x,X_i^{\prime\prime};D_{\mathsf{s}_{1}}^{\prime\prime})
\left(
m(D_i^{\prime\prime},\hat{g}_{n}^{(-\mathsf{s}_{1})})-m(D_i^{\prime\prime},g_0)
\right)
\bigg\vert^q
\right]\nonumber\\
&\lesssim
\mathbb{E}\left[
\|\hat{g}_{n}^{(-\mathsf{s}_{1})}-g_0\|_{q,\infty}^q
\right]
+
\varepsilon_b^{q/2},\label{eq: condition A label}
\end{flalign}
where the final inequality follows from the same argument used to obtain \eqref{eq: unconditional A label}. Hence, the bounds \eqref{eq: unconditional A label} and \eqref{eq: condition A label} imply that
\begin{flalign}
\|F_{3}(x;\mathsf{s}_{1})\|_{q}^{2}
&\lesssim
\left(
q^q\left(\frac{b}{n}\right)^{q/4}\delta_{n,g}^q
+
\varepsilon_b^{q/2}
\right)^{2/q}\nonumber\\
&\lesssim
q^2
\left(
\left(\frac{b}{n}\right)^{1/2}\delta_{n,g}^2
+
\varepsilon_b
\right),\label{eq: 2p bound no stable}
\end{flalign}
where we have used the rate condition \eqref{eq: D nuisance moment} and the fact that \(\hat{g}_{n}^{(-\mathsf{s}_{1})}\) and \(\hat{g}_n\) are identically distributed.

Next, consider the term \eqref{eq: p moment stability bound}. For \(r\neq l\), observe that
\begin{flalign}
&\mathbb{E}\left[
\left|
F_{3}(x;\mathsf{s}_{r})-F_{3}^{(-l)}(x;\mathsf{s}_{r})
\right|^q
\right]\nonumber\\
&=
\mathbb{E}\left[
\bigg\vert
\sum_{i\in\mathsf{s}_{r}}
\left(
a(x;D_i,D_{\mathsf{s}_{r}},\hat{g}_{n}^{(-\mathsf{s}_{r})})
-
a(x;D_i,D_{\mathsf{s}_{r}},\hat{g}_{n}^{(-\mathsf{s}_{r},-\mathsf{s}_{l})})
\right)
\right.\nonumber\\
&\quad\quad\quad\quad\left.
-
\sum_{i\in\mathsf{s}_{r}}
\left(
A(x;\hat{g}_{n}^{(-\mathsf{s}_{r})})
-
A(x;\hat{g}_{n}^{(-\mathsf{s}_{r},-\mathsf{s}_{l})})
\right)
\bigg\vert^q
\right]\nonumber\\
&\lesssim
\mathbb{E}\left[
\bigg\vert
\sum_{i\in\mathsf{s}_{r}}
\left(
a(x;D_i,D_{\mathsf{s}_{r}},\hat{g}_{n}^{(-\mathsf{s}_{r})})
-
a(x;D_i,D_{\mathsf{s}_{r}},\hat{g}_{n}^{(-\mathsf{s}_{r},-\mathsf{s}_{l})})
\right)
\bigg\vert^q
\right]\label{eq: stability term unconditional}\\
&\quad+
\mathbb{E}\left[
\bigg\vert
\sum_{i\in\mathsf{s}_{r}}
\left(
A(x;\hat{g}_{n}^{(-\mathsf{s}_{r})})
-
A(x;\hat{g}_{n}^{(-\mathsf{s}_{r},-\mathsf{s}_{l})})
\right)
\bigg\vert^q
\right]\label{eq: stability term conditional}
\end{flalign}
by the triangle inequality. To bound the quantity \eqref{eq: stability term unconditional}, observe that
\begin{flalign}
&\mathbb{E}\left[
\bigg\vert
\sum_{i\in\mathsf{s}_{r}}
\kappa(x,X_i;D_{\mathsf{s}_{r}})
\left(
m(D_i,\hat{g}_{n}^{(-\mathsf{s}_{r})})
-
m(D_i,\hat{g}_{n}^{(-\mathsf{s}_{r},-\mathsf{s}_{l})})
\right)
\bigg\vert^q
\right]\nonumber\\
&\leq
\mathbb{E}\left[
\max_{i\in[b]}
\left|
m(D_i^{\prime\prime},\hat{g}_{n})
-
m(D_i^{\prime\prime},\hat{g}_{n}^{(-[b])})
\right|^q
\right]\nonumber\\
&\leq
b\mathbb{E}\left[
\left|
m(D_i^{\prime\prime},\hat{g}_{n})
-
m(D_i^{\prime\prime},\hat{g}_{n}^{(-[b])})
\right|^q
\right]\nonumber\\
&\lesssim
b\left(qn^{-1}b\gamma(q)\right)^q,\label{eq: var bound stability term unconditional}
\end{flalign}
where the first inequality follows from Jensen's inequality, \cref{assu: kernel restriction}, Part (ii), and exchangeability, and the final inequality follows from the independent-copy stability bound in \cref{assu: stability}.

Similarly, to bound the quantity \eqref{eq: stability term conditional}, let \(\mathbf{D}_n^{\prime\prime}\) be an independent copy of \(\mathbf{D}_n\). By Jensen's inequality,
\begin{flalign}
&\mathbb{E}\left[
\bigg\vert
\sum_{i\in\mathsf{s}_{r}}
\left(
A(x;\hat{g}_{n}^{(-\mathsf{s}_{r})})
-
A(x;\hat{g}_{n}^{(-\mathsf{s}_{r},-\mathsf{s}_{l})})
\right)
\bigg\vert^q
\right]\nonumber\\
&\leq
\mathbb{E}\left[
\bigg\vert
\sum_{i\in\mathsf{s}_{r}}
\kappa(x,X_i^{\prime\prime};D_{\mathsf{s}_{r}}^{\prime\prime})
\left(
m(D_i^{\prime\prime},\hat{g}_{n}^{(-\mathsf{s}_{r})})
-
m(D_i^{\prime\prime},\hat{g}_{n}^{(-\mathsf{s}_{r},-\mathsf{s}_{l})})
\right)
\bigg\vert^q
\right]\nonumber\\
&\leq
\mathbb{E}\left[
\max_{i\in[b]}
\left|
m(D_i^{\prime\prime},\hat{g}_{n})
-
m(D_i^{\prime\prime},\hat{g}_{n}^{(-[b])})
\right|^q
\right]\nonumber\\
&\lesssim
b\left(qn^{-1}b\gamma(q)\right)^q,\label{eq: var bound stability term conditional}
\end{flalign}
where the final inequality again follows from the independent-copy stability bound in \cref{assu: stability}. Consequently, \eqref{eq: var bound stability term unconditional} and \eqref{eq: var bound stability term conditional} imply that
\begin{flalign}
\|F_{3}(x;\mathsf{s}_{r})-F_{3}^{(-l)}(x;\mathsf{s}_{r})\|_{q}^{2}
&\lesssim
b^{2/q}\left(qn^{-1}b\gamma(q)\right)^2~.\label{eq: 2p bound stable}
\end{flalign}
Putting the pieces together, \eqref{eq: V moment decomp}, \eqref{eq: 2p bound no stable}, and \eqref{eq: 2p bound stable} imply that
\begin{flalign*}
&\bigg\|\sum_{l=1}^{k}\left(\frac{1}{k}\left(F_{3}(x;\mathsf{s}_{l})-F_{3}^{(-l)}(x;\mathsf{s}_{l})\right)
+
\frac{1}{k}\sum_{r\neq l}\left(F_{3}(x;\mathsf{s}_{r})-F_{3}^{(-l)}(x;\mathsf{s}_{r})\right)\right)^2
\bigg\|_{q/2}^{1/2}\\
&\quad\lesssim
\left[
\frac{q^2}{k}
\left(
\left(\frac{b}{n}\right)^{1/2}\delta_{n,g}^2
+
\varepsilon_b
\right)
+
k b^{2/q}
\left(qn^{-1}b\gamma(q)\right)^2
\right]^{1/2}\\
&\quad\lesssim
\sqrt{\frac{1}{k}}
\left(
\left(\frac{1}{k}\right)^{1/4}\delta_{n,g}
+
\varepsilon_b^{1/2}
+
b^{1/q}\gamma(q)
\right)q~,
\end{flalign*}
where the final inequality uses \(k=\lfloor n/b\rfloor\).\hfill{}$\blacksquare$

\subsection{Proof of \cref{lem: continuity}}

Consider the decomposition
\begin{flalign*}
 & m(D_i,\hat{g}_{n})-m(D_i,\hat{g}_{n}^{(-[b])})\\
 & =
 \left(\hat{\mu}_{n}(Z_i,1)-\hat{\mu}_{n}^{(-[b])}(Z_i,1)\right)
 -
 \left(\hat{\mu}_{n}(Z_i,0)-\hat{\mu}_{n}^{(-[b])}(Z_i,0)\right)\\
 &\quad
 -\hat{\beta}_{n}^{(-[b])}(W_i,Z_i)
 \left(\hat{\mu}_{n}(Z_i,W_i)-\hat{\mu}_{n}^{(-[b])}(Z_i,W_i)\right)\\
 &\quad
 +
 \left(Y_i-\hat{\mu}_{n}(Z_i,W_i)\right)
 \left(\hat{\beta}_{n}(W_i,Z_i)-\hat{\beta}_{n}^{(-[b])}(W_i,Z_i)\right).
\end{flalign*}
We have that
\begin{flalign*}
 & \|m(D_i,\hat{g}_{n})-m(D_i,\hat{g}_{n}^{(-[b])})\|_{2p}\\
 & \leq
 \|\hat{\mu}_{n}(Z_i,1)-\hat{\mu}_{n}^{(-[b])}(Z_i,1)\|_{2p}
 +
 \|\hat{\mu}_{n}(Z_i,0)-\hat{\mu}_{n}^{(-[b])}(Z_i,0)\|_{2p}\\
 &\quad
 +
 \|\hat{\beta}_{n}^{(-[b])}(W_i,Z_i)
 (\hat{\mu}_{n}(Z_i,W_i)-\hat{\mu}_{n}^{(-[b])}(Z_i,W_i))\|_{2p}\\
 &\quad
 +
 \|(Y_i-\hat{\mu}_{n}(Z_i,W_i))
 (\hat{\beta}_{n}(W_i,Z_i)-\hat{\beta}_{n}^{(-[b])}(W_i,Z_i))\|_{2p}\\
 &\lesssim
 \|\hat{\mu}_{n}(Z_i,1)-\hat{\mu}_{n}^{(-[b])}(Z_i,1)\|_{2p}
 +
 \|\hat{\mu}_{n}(Z_i,0)-\hat{\mu}_{n}^{(-[b])}(Z_i,0)\|_{2p}\\
 &\quad
 +
 \|\hat{\mu}_{n}(Z_i,W_i)-\hat{\mu}_{n}^{(-[b])}(Z_i,W_i)\|_{2p}
 +
 \|\hat{\beta}_{n}(W_i,Z_i)-\hat{\beta}_{n}^{(-[b])}(W_i,Z_i)\|_{2p},
\end{flalign*}
where the first inequality follows from the triangle inequality and
the second inequality follows from the boundedness condition \eqref{eq: bounded quantities}, applied to both the original and resampled nuisance estimators. This verifies the condition \eqref{eq: orlicz continuity}. An identical argument verifies the condition \eqref{eq: orlicz continuity Z}.
\hfill{}$\blacksquare$

\subsection{Proof of \cref{lem: kernel stability}}

To ease notation, we omit dependence on the evaluation point $z$.
Recall the definition of the Hájek projection
\[
u^{(1)}(D)
=
\mathbb{E}\left[u(D_{[b']})\mid D_1=D\right]
-
\mathbb{E}\left[u(D_{[b']})\right],
\]
where $D$ is an independent copy of $D_1$. Let
\begin{flalign*}
\nu^2
&=
\Var(u(D_{[b']}))
\quad\text{and}\quad
\sigma_{b'}^2
=
\Var(u^{(1)}(D))
\end{flalign*}
denote the kernel variance and Hájek projection variance, respectively. Consider the decomposition
\begin{flalign}
\hat{g}_{n}(\mathbf{D}_{n})-\hat{g}_{n}(\mathbf{D}_{n}^{(-[b])})
&=
\frac{b'}{n}\sum_{i=1}^{b}\left(u^{(1)}(D_i)-u^{(1)}(D_i')\right)\label{eq: diff of Hajek}\\
&\quad+
\left(\hat{g}_{n}(\mathbf{D}_{n})-\mathbb{E}[u(D_{[b']})]
-\frac{b'}{n}\sum_{i=1}^{n}u^{(1)}(D_i)\right)\label{eq: H proj resid one}\\
&\quad-
\left(\hat{g}_{n}(\mathbf{D}_{n}^{(-[b])})-\mathbb{E}[u(D_{[b']})]
-\frac{b'}{n}
\left(
\sum_{i=1}^{b}u^{(1)}(D_i')
+
\sum_{i=b+1}^{n}u^{(1)}(D_i)
\right)\right)~.\label{eq: H proj resid 2}
\end{flalign}
The result is obtained by giving higher-order moment bounds for each
of the terms \eqref{eq: diff of Hajek}, \eqref{eq: H proj resid one},
and \eqref{eq: H proj resid 2}.

We begin by considering the terms \eqref{eq: H proj resid one} and
\eqref{eq: H proj resid 2}. Observe that the decomposition \eqref{eq: break up Hajek}, \cref{lem: individual degenerate}, and the triangle inequality imply that
\begin{flalign}
 \left\|
 \hat{g}_{n}(\mathbf{D}_{n})-\mathbb{E}[u(D_{[b']})]
 -
 \frac{b'}{n}\sum_{i=1}^{n}u^{(1)}(D_i)
 \right\|_{q}
& \leq
\phi q\sum_{m=2}^{b'}
\left(\frac{Cqb'}{n}\right)^{m/2}
\lesssim
\phi q^2\frac{b'}{n}
\label{eq: proj bound 1}
\end{flalign}
for \(q\leq cn/b'\). Likewise,
\begin{flalign}
&\left\|
\hat{g}_{n}(\mathbf{D}_{n}^{(-[b])})-\mathbb{E}[u(D_{[b']})]
-
\frac{b'}{n}
\left(
\sum_{i=1}^{b}u^{(1)}(D_i')
+
\sum_{i=b+1}^{n}u^{(1)}(D_i)
\right)
\right\|_{q} \nonumber\\
&\quad\quad\leq
\phi q\sum_{m=2}^{b'}
\left(\frac{Cqb'}{n}\right)^{m/2}
\lesssim
\phi q^2\frac{b'}{n}~.
\label{eq: proj bound 2}
\end{flalign}
To handle the term \eqref{eq: diff of Hajek}, we apply \cref{lem: Rosenthal centered}. In particular, Hoeffding orthogonality gives
\[
b'\sigma_{b'}^2
\leq
\Var(u(D_{[b']}))
\lesssim
\phi^2,
\]
and so \(\sigma_{b'}^2\lesssim \phi^2/b'\). Therefore,
\begin{flalign}
&\left\|
\frac{b'}{n}\sum_{i=1}^{b}\left(u^{(1)}(D_i)-u^{(1)}(D_i')\right)
\right\|_{q} \nonumber\\
&=
\frac{b'b}{n}
\left\|
\frac{1}{b}\sum_{i=1}^{b}
\left(u^{(1)}(D_i)-u^{(1)}(D_i')\right)
\right\|_{q}\nonumber\\
&\lesssim
\frac{b'b}{n}
\left(
\sqrt{\frac{\sigma_{b'}^2 q}{b}}
+
q\frac{\phi}{b}
\right)
\lesssim
\frac{\sqrt{b'b}}{n}q\phi
+
\frac{b'}{n}q\phi~.
\label{eq: apply rosenthal stability}
\end{flalign}
Putting the pieces together, we find that
\[
\|\hat{g}_{n}(\mathbf{D}_{n})-\hat{g}_{n}(\mathbf{D}_{n}^{(-[b])})\|_{q}
\lesssim
\frac{\sqrt{b'b}}{n}q\phi
+
\frac{b'}{n}q^2\phi
\]
by combining the bounds \eqref{eq: proj bound 1}, \eqref{eq: proj bound 2},
and \eqref{eq: apply rosenthal stability}, as required.
\hfill{}$\blacksquare$

\section{Supplemental Figures\label{app: supplemental figures}}

\clearpage 

\end{spacing}
\end{appendix}
\end{document}